\documentclass[a4paper,12pt]{article}
\pdfoutput=1
\usepackage{graphicx,subfigure,amsmath,amssymb,multirow}
\usepackage{cite}
\usepackage{longtable}
\usepackage{color}

\renewcommand{\baselinestretch}{1.5}
\newlength{\dinwidth}
\newlength{\dinmargin}
\def\be{\begin{equation}}
\def\ee{\end{equation}}
\def\ba{\begin{eqnarray}}
\def\ea{\end{eqnarray}}
 \def\la{ \langle}
  \def\ra{ \rangle}
     \def\e{ \epsilon}
      \def\r{ \gamma}
       \def\lbd{\lambda}
        \def \d {{\rm d}}
           \def\w{\omega}
            \def\u{\mu}
              \def\a{\alpha}
  \def\b{\beta}
\def\v{\nu}
     \def\ve{ \varepsilon}
          \def\qb{{ \bf q}_\bot}
           \def\kb{{ \bf k}_\bot}

\allowdisplaybreaks

\begin{document}
\title{\bf  Form factors of $V'\to V''$ transition within the light-front quark models}
\author{Qin Chang$^{a,b}$\footnote{q\_chang@foxmail.com},  Li-Ting Wang$^{a}$ and Xiao-Nan Li$^{a,c}$\\
{ $^a$\small Institute of Particle and Nuclear Physics, Henan Normal University, Henan 453007, China}\\
{ $^b$\small Institute of Particle Physics and Key Laboratory of Quark and Lepton Physics~(MOE) }\\[-0.2cm]
{ \small Central China Normal University, Wuhan, Hubei 430079, China}\\
{ $^c$\small School of physics and electronic engineering, Anyang Normal University, Henan 455000, China}
%{ $^b$\small Institute of Particle Physics and Key Laboratory of Quark and Lepton Physics~(MOE) }\\[-0.2cm]
%{ \small Central China Normal University, Wuhan, Hubei 430079, China}
}
\date{}

\maketitle

\begin{abstract}
In this paper, we  calculate the matrix element and form factors of vector-to-vector~($V'\to V''$) transition within the standard light-front~(SLF) and covariant light-front~(CLF) quark models~(QMs), and investigate the self-consistency and Lorentz covariance of the CLF QM within two types of correspondences between  the  manifest covariant Bethe-Salpeter  approach and the LF approach.  The zero-mode and valence contributions to the form factors of $V'\to V''$ transition in the CLF QM and their relation to the SLF results are analyzed,   and the main conclusions obtained via the decay constants of vector and axial-vector mesons and the form factors of $P\to V$  transition in the previous works are confirmed again. Furthermore, we present our numerical predictions for the form factors of $c\to (q,s)$~($q=u,d$) induced $D^*\to (K^*\,,\rho)$, $D^*_s\to (\phi\,,K^*)$, $J/\Psi\to (D^*_s\,,D^*)$,  $B^*_c\to (B^*_s\,,B^*)$ transitions and $b\to (q,s,c)$ induced $B^*\to (D^*\,,K^*\,,\rho)$, $B^*_s\to (D^*_s\,,\phi\,,K^*)$, $B^*_c\to (J/\Psi\,,D^*_s\,,D^*)$, $\Upsilon(1S)\to (B_c^*\,,B_s^*\,,B^*)$ transitions, which  can be applied further to the relevant phenomenological studies of meson decays.

\end{abstract}

\newpage
\section{Introduction }
The mesonic transition form factors are important ingredients  in the study of weak and electromagnetic decays of mesons. There are many  approaches for evaluating  these nonperturbative quantities,  for instance, Wirbel-Stech-Bauer model~\cite{Wirbel:1985ji}, lattice calculations~\cite{Daniel:1990ah}, vector meson dominance model~\cite{Ametller:1993we,Gao:1999qn}, perturbative QCD  with some nonperturbative inputs~\cite{Lepage:1980fj,Li:1992nu}, QCD sum rules~\cite{Shifman:1978bx,Shifman:1978by} and light-front quark models~(LF QMs)~\cite{Terentev:1976jk,Berestetsky:1977zk,Cheng:1997au,Carbonell:1998rj,Jaus:1999zv}. In this paper, we will calculate the form factors of $V'\to V''$ transition~($V'$ and $V''$ denote vector mesons) within the standard and the covariant light-front approaches.

The standard light-front quark model~(SLF QM) proposed by Terentev and Berestetsky~\cite{Terentev:1976jk,Berestetsky:1977zk} is a relativistic quark model based on the LF formalism~\cite{Dirac:1949cp} and LF quantization of QCD~\cite{Brodsky:1997de}. It provides a conceptually simple and phenomenologically feasible framework for the determination of nonperturbative quantities. However,   the matrix element evaluated in this approach lacks manifest Lorentz covariance and the zero-mode contributions can not be determined explicitly.  In order to fill these gaps, many efforts have been made in the past years~\cite{Cheng:1997au,Carbonell:1998rj,Jaus:1999zv,Karmanov:1991fv,Karmanov:1994ck,Karmanov:1996un,Choi:1998nf}. In Ref.~\cite{Carbonell:1998rj}, a method based on the covariant  LF framework is developed to identify and separate the $\w$-dependent spurious contributions, where $\w$ is the light-like four-vector used to define  light-front by $\w\cdot x=0$ and the $\w$-dependent contributions may violate the covariance.  With such an approach, the $\w$-independent physical contributions can be well determined, while the effects of zero-mode are not fully considered. In Ref.~\cite{Jaus:1999zv}, a basically different technique is developed by Jaus to deal with the covariance and zero-mode problems with the help of a manifestly covariant Bethe-Saltpeter~(BS) approach as a guide to the calculation. In such a covariant light-front quark model~(CLF QM), the zero-mode contributions can be well determined, and the result of the matrix element is expected to be covariant because  the spurious contributions can be eliminated  by the inclusion of zero-mode contributions~\cite{Jaus:1999zv}. 

The SLF and CLF QMs have been widely used for the determination of nonperturbative quantities, such as  form factor, decay constant and distribution amplitude, as well as the other features, of hadrons, which are further applied to phenomenological researches~\cite{Jaus:1989au,Jaus:1989av,Jaus:1991cy,Jaus:1996np,Cheng:1996if,ODonnell:1996sya,Cheung:1996qt,Choi:1996mq,Choi:1997iq,Choi:1998jd,Ji:2000fy,DeWitt:2003rs,Choi:2007yu,Choi:2007se,Barik:1997qq,Hwang:2000ez,Hwang:2010hw,Hwang:2001zd,Hwang:2009cu,Geng:2001de,Geng:2016pyr,Chang:2016ouf,Chang:2017sdl,Chang:2018aut,Chang:2018mva,Bakker:2000pk,Bakker:2002mt,Bakker:2003up,Choi:2004ww,Choi:2009ai,Choi:2009ym,Choi:2010ha,Choi:2010zb,Choi:2011xm,Choi:2010be,Choi:2014ifm,Choi:2017uos,Choi:2017zxn,Ryu:2018egt,Hwang:2001hj,Hwang:2001wu,Hwang:2010iq,Cheung:2014cka,Wang:2008ci,Wang:2008xt,Shen:2008zzb,Wang:2009mi,Cheng:2017pcq,Kang:2018jzg,Verma:2011yw,Shi:2016gqt,Wang:2018duy,Jaus:2002sv,Cheng:2003sm,Cheng:2004cc,Cheng:2004ew}.  For the weak decays, the form factors of $P\to (P\,,V)$ transitions have been calculated within the SLF and the CLF QMs in Refs.~\cite{Jaus:1989au,Jaus:1989av,Geng:2001de} and Refs.~\cite{Jaus:1999zv,Jaus:2002sv,Cheng:2003sm, Verma:2011yw}, respectively; besides, the form factors of $P\to (S\,,A\,,T)$ transitions are studied within the CLF QM~\cite{Cheng:2003sm, Verma:2011yw}. In addition, the SLF approach is also used to evaluate the form factors of $\text{baryon}\to\text{baryon}$ processes with help of diquark picture~\cite{Wei:2009np,Ke:2017eqo,Wang:2017mqp,Zhao:2018zcb,Zhao:2018mrg,Chua:2018lfa}. However, the form factors of $V'\to V''$ transition have not been fully investigated. With the rapid development of particle physics experiment, some weak decays of vector mesons are  hopeful to be observed by LHC and SuperKEKB/Belle-II experiments {\it et al.} in the future due to the high luminosity~\cite{Abe:2010gxa,Bediaga:2012py,Aaij:2014jba}. The theoretical evaluation of the form factors of $V'\to V''$ transition can provide some useful references and  essential inputs for the relevant phenomenological studies. In Ref.~\cite{Jaus:2002sv}, the angular condition for $\la V'' | \r^\u |V'\ra$ are studied, but only the electromagnetic transition form factors ($V'= V''=\rho$) are obtained. In this paper, the form factors  related to the current matrix elements, $\la V'' |  \r^\u  |V'\ra$ and $\la V'' |  \r^\u\r_5 |V'\ra$, will be calculated within the SLF and CLF QMs, and moreover, the self-consistency and covariance of  CLF QM and the effects of zero-mode will be analyzed in detail.

The manifest covariance is a remarkable feature of the CLF QM relative to the SLF QM~\cite{Jaus:1999zv}. However, it should be noted that although the main $\w$ dependences are associated with the $C$ functions\footnote{In the calculation of a matrix element within the CLF QM, after $p_1'^-$ integration is carried out, the terms related to $\hat{p}'_1$ in the integrand have to be decomposed in terms of $P=p'+p''$, $q=p'-p''$ and $\w$, where $p'$ and  $p''$ are the momenta of initial and final states, respectively~\cite{Jaus:1999zv}. Correspondingly, three types of coefficients are introduced. The A functions are the coefficients of the terms irrelevant to $\w$, while the coefficients combined with $\w$ include B and C functions, in which only the C functions are associated with  the zero-mode contributions. One may refer to Ref.~\cite{Jaus:1999zv} for detailed explanation.   \label{fn:abc}   } and can be eliminated by the zero-mode contributions, there are still some residual $\w$ dependences due to the nonvanishing spurious contributions associated with  $B$ functions\textsuperscript{\ref {fn:abc}}, which  are unfortunately nonzero within the traditional correspondence scheme between the  covariant BS  model and the LF QM~(named as type-I scheme~\cite{Choi:2013mda}),  and therefore violate the strict covariance of CLF results~\cite{Jaus:1999zv,Jaus:2002sv,Chang:2018zjq,Chang:2019mmh}. Besides, the self-consistency is another challenge to the  CLF QM.
For instance, the authors of Ref.~\cite{Cheng:2003sm} find that the CLF results for $f_V$ obtained respectively via  longitudinal ($\lbd=0$) and  transverse ($\lbd=\pm$) polarization states are  inconsistent with each other,  $[f_V]_{\rm CLF}^{\lbd=0}\neq [f_V]_{\rm CLF}^{\lbd=\pm}$,  because the former receives an additional contribution characterized by the $B_1^{(2)}$ function.

In order to recover the self-consistency of CLF QM, the authors of Ref.~\cite{Choi:2013mda} present a modified correspondence between the  covariant BS  approach and the LF approach~(named as type-II scheme~\cite{Choi:2013mda}), which requires an additional $M\to M_0$ replacement relative to the traditional  type-I correspondence scheme. Within this  modified correspondence scheme,  $[f_V]_{\rm CLF}^{\lbd=0}\dot{=} [f_V]_{\rm CLF}^{\lbd=\pm}$~\cite{Choi:2013mda}  is obtained. In our previous works~\cite{Chang:2018zjq,Chang:2019mmh}, the problems of self-consistency and covariance are studied via $f_{P,V,A}$ and form factors of $P\to (P,V)$ transitions. It is found that such two problems have the same origin, and can be resolved simultaneously by employing  type-II correspondence scheme because the contributions associated with  $B_1^{(2)}$ and $B_3^{(3)}$ functions vanish numerically~\cite{Chang:2018zjq,Chang:2019mmh}.  Moreover, it is also found that ~\cite{Choi:2013mda,Chang:2018zjq,Chang:2019mmh}
\begin{align}
[{\cal Q}]_{\rm SLF}=[{\cal Q}]_{\rm val.}\doteq[{\cal Q}]_{\rm CLF},\label{eq:sc}
\end{align}
 where ${\cal Q}$ denotes $f_{P,V,A}$ and form factors of $P\to (P,V)$ transitions, the subscript ``val.'' denotes the valence contribution in the CLF QM, and the symbol ``$\doteq$''  denotes that the two quantities are equal to each other numerically.
The form factors of $V'\to V''$ transition involves much more $B$ functions and thus may present much stricter test on the CLF QM, as well as above-mentioned  findings. In this paper, these issues will be studied in detail.

Our paper is organized as follows. In sections 2 and 3, we would like to review briefly the SLF and the CLF QMs, respectively, for convenience of discussion, and then present our theoretical results for the form factors of  $V'\to V''$ transition. In section 4, the self-consistency and covariance of CLF results are discussed in detail, and  the zero-mode and the valence contribution in the CLF QM and their relations to the SLF results are analyzed. After that, we present our numerical results for some $c\to (q,s)$ and $b\to (q,s,c)$~($q=u,d$)  induced transitions. Finally, our summary is made in section 5.

\section{ Theoretical results in the SLF QM}
\label{sec:2}
\subsection{General formalism }
In this section, in order to clarify the convention and notation used in this paper,  we would like to review briefly the framework of SLF QM.  One  may refer to, for instance,  Refs.~\cite{Jaus:1989au,Jaus:1989av} for details. The form factors of $V'\to V''$ transition  can be  defined as~\cite{Wang:2007ys}
\begin{align}
\label{eq:defFV}
\la   V''(\e'', p'')| \bar{q}''_1 \r_{\u} q'_1  |V'(\e', p')\ra
=&-(\e' \cdot \e''^{*})\Big[P_{\u}\,V_1(q^2)-q_{\u}\,V_2(q^2)\Big]\nonumber\\
&+\frac{(\e' \cdot q)(\e''^{*}\cdot q)}{M'^2-M''^2}\bigg[\Big(P^{\v}-\frac{M'^2-M''^2}{q^2}q^\v\Big) \,V_3(q^2)\nonumber\\
&+\frac{M'^2-M''^2}{q^2}q^\v\,V_4(q^2)\bigg]\nonumber\\
&-(\e' \cdot q)\,\e''^*_{\u} \,V_5(q^2)+(\e''^* \cdot q)\,\e'_{\u}\,V_6(q^2)\,,
\end{align}
\begin{align}
\label{eq:defFA}
\la   V''(\e'', p'')|\bar{q}''_1\r_{\u} \r_{5}  q'_1 |V'(\e', p')\ra
=&-i\ve_{\u\v\a\b}\e'^{\a}\e''^{*\b}\bigg[\Big(P^{\v}-\frac{M'^2-M''^2}{q^2}q^\v\Big)\,A_1(q^2)\nonumber\\
&+\frac{M'^2-M''^2}{q^2}q^{\v}\,A_2(q^2)\bigg]\nonumber\\
&-\frac{i\,\ve_{\u\v\a\b} P^{\a}q^{\b}}{M'^2-M''^2}\,\Big[\e''^{*}\cdot q \,\e'^{\v} A_3(q^2) -\e' \cdot q \,\e''^{*\v} A_4(q^2)\Big]\nonumber\\
\end{align}
where, $P=p'+p''$, $q=p'-p''$ and $\ve_{0123}=-1$.
The main work of LF approach is to evaluate the current matrix element,
\begin{align}\label{eq:amp1}
{\cal B} \equiv \la  V''(\e'',p'') | \bar{q}''_1 (k_1'')\Gamma q'_1(k_1') |V'(\e',p') \ra \,,\qquad \Gamma=\r_\u\,,\r_\u\r_5\,,
\end{align}
which will be further used to extract the form factors.

In the framework of SLF QM, a meson bound-state  consisting a quark $q_1$ and antiquark $\bar{q}_2$ with a total momentum $p$ can be written as
\begin{align}
|M(p)\ra =  \sum_{h_1,h_2} \int \frac{\d^3 \tilde{k}_1}{(2\pi)^32\sqrt{k_1^+}} \frac{\d^3 \tilde{k}_2}{(2\pi)^32\sqrt{k_2^+}} (2\pi)^3 \delta^3 ({\tilde{p}-\tilde{k}_1-\tilde{k}_2}) \Psi_{h_1,h_2}(\tilde{k}_1,\tilde{k}_2)|q_1({k}_1,h_1)\ra|\bar{q}_2({k}_2,h_2)\ra\,,
\label{eq:Fockexp}
\end{align}
where, $\tilde{p}=(p^+,\mathbf{p_\bot})$ and  $\tilde{k}_{1,2}=(k_{1,2}^+,\mathbf{k}_{1,2\bot})$ are the on-mass-shell LF momenta, $\Psi_{h_1,h_2}(\tilde{k}_1,\tilde{k}_2)$ is the momentum-space wavefunction~(WF), and the one particle state is defined as $|q_1({k}_1,h_1)\ra=\sqrt{2k_1^+}\,b_{h1}^{\dagger}(k_1)|0\ra$ and $ |\bar{q}_2({k}_2,h_2)\ra=\sqrt{2k_2^+}\,d_{h2}^{\dagger}(k_2)|0\ra$.
The momenta of $q_1$ and $\bar{q}_{2}$ can be written in terms of the internal  LF relative momentum variables $(x,{\mathbf{ k}_{\bot}})$ as
\begin{align}\label{eq:momk1}
k_1^+=xp^+\,,\quad\, \mathbf{k}_{1\bot}=x\mathbf{p}_{\bot}+\mathbf{k}_{\bot} \,,\qquad  k_2^+=\bar{x}p^+ \,,\quad\, \mathbf{k}_{2\bot}=\bar{x}\mathbf{p}_{\bot}-\mathbf{k}_{\bot}\,,
\end{align}
where, $\bar{x}=1-x$, $\mathbf{k}_{\bot}=(k^x\,,k^y)$ and $\mathbf{p}_{\bot}=(p^x\,,p^y)$.

 The momentum-space WF in Eq.~\eqref{eq:Fockexp} satisfies the normalization condition and can be expressed as
\begin{align}
\label{eq:LFWFP2}
\Psi_{h_1,h_2}(x,\mathbf{k}_{\bot})=S_{h_1,h_2}(x,\mathbf{k}_{\bot}) \psi(x,\mathbf{k}_{\bot}) \,,
\end{align}
where, $\psi(x,\mathbf{k}_{\bot})$ is the radial WF and responsible for describing the momentum distribution of the constituent quarks in the bound-state; $S_{h_1,h_2}(x,\mathbf{k}_{\bot})$  is the spin-orbital WF and responsible for constructing a state of definite spin $(S,S_z)$ out of the LF helicity $(h_1,h_2)$ eigenstates.  For the former, we adopt the Gaussian type WF
\begin{align}
\label{eq:RWFs}
\psi_s(x,\mathbf{k}_{\bot}) =4\frac{\pi^{\frac{3}{4}}}{\beta^{\frac{3}{2}}} \sqrt{ \frac{\partial k_z}{\partial x}}\exp\left[ -\frac{k_z^2+\mathbf{k}_\bot^2}{2\beta^2}\right]\,,
\end{align}
where, $k_z$ is the relative momentum in $z$-direction and can be written as
\begin{align}
 k_z=(x-\frac{1}{2})M_0+\frac{m_2^2-m_1^2}{2 M_0}\,,
\end{align}
with the invariant mass $M_0^2=\frac{m_1^2+\mathbf{k}_{\bot}^2}{x}+\frac{m_2^2+\mathbf{k}_{\bot}^2}{\bar{x}}$. Besides the Gaussian type WF,  there are some other choices, which will be discussed later. The spin-orbital WF, $S_{h_1,h_2}(x,\mathbf{k}_{\bot}) $, can be obtained by the interaction-independent Melosh transformation.  It is convenient to use its covariant form, which can be further reduced by using the equation of motion on spinors and finally written as~\cite{Jaus:1989av,Cheng:2003sm}
\begin{align}\label{eq:defS2}
S_{h_1,h_2}=\frac{\bar{u}(k_1,h_1)\Gamma' v(k_2,h_2)}{\sqrt{2} \hat{M}_0}\,,
\end{align}
where, $\hat{M}_0^2=M_0^2-(m_1-m_2)^2$. For the vector state, one shall take
\begin{align}
& \Gamma'_V=-\not\!\hat{\epsilon}+\frac{\hat{\epsilon}\cdot (k_1-k_2)}{D_{ V,{\rm LF}}}\,,\quad D_{V,{\rm LF}}=M_0+m_1+m_2\,,
\label{eq:vSLF}
\end{align}
where,
\begin{align}
\hat{\epsilon}^{\mu}_{\lbd=0}=&\frac{1}{M_0}\left(p^+,\frac{-M_0^2+\mathbf{p}_{\bot}^2}{p^+},\mathbf{p}_{\bot}\right)\,,\\
\hat{\epsilon}^{\mu}_{\lbd=\pm}=&\left(0,\frac{2}{p^+}\boldsymbol{\epsilon}_{\bot}\cdot \mathbf{p}_{\bot}, \boldsymbol{\epsilon}_{\bot}\right)\,,
\quad \boldsymbol{\epsilon}_{\bot}\equiv \mp \frac{(1,\pm i)}{\sqrt{2}}\,.
\end{align}

In practice, for the $V'(p')\to V''(p'')$ transition, we shall take the convenient Drell-Yan-West frame, $q^+=0$,  where $q\equiv p'-p''=k_1'-k_1''$ is the momentum  transfer.
 %It implies that the form factors are known only for space-like momentum transfer, $q^2=-\mathbf{q}_\bot^2\leqslant 0$, and the ones in the time-like region need  an additional $q^2$ extrapolation.
 In addition,  we also take  a Lorentz frame where $\mathbf{p}_{\bot}'=0$ for convenience of calculation. In this frame, the momenta of constituent quarks in initial and final states are written as
\begin{align}\label{eq:momk2}
\tilde{k}_1'=(xp'^+,\mathbf{k}_{\bot}')  \,,\quad  \tilde{k}_1''=(xp'^+, x\mathbf{p}_{\bot}'' +\mathbf{k}_{\bot}'') \,,\quad   \tilde{k}_2=(\bar{x}p'^+ ,-\mathbf{k}_{\bot}' ) =(\bar{x}p'^+ ,\bar{x}\mathbf{p}_{\bot}'' -\mathbf{k}_{\bot}'' )\,,
\end{align}
where, $\mathbf{p}_{\bot}''=-\mathbf{q}_{\bot}$  and $\mathbf{k}_{\bot}''=\mathbf{k}_{\bot}'-\bar{x}\mathbf{q}_{\bot}$.

Finally, equipping Eq.~\eqref{eq:amp1} with the formulas given above and making some simplifications, we obtain
\begin{eqnarray}
{\cal B}_{\rm  SLF}=\sum_{h'_1,h''_1,h_2} \int  \frac{\d x \,\d^2{ \mathbf{k}_\bot'}}{(2\pi)^3\,2x}  {\psi''}^{*}(x,\mathbf{k}_{\bot}''){\psi'}(x,\mathbf{k}_{\bot}')
S''^{\dagger}_{h''_1,h_2}(x,\mathbf{k}_{\bot}'')\,C_{h''_1,h'_1}(x,\mathbf{k}_{\bot}',\mathbf{k}_{\bot}'')\,S'_{h'_1,h_2}(x,\mathbf{k}_{\bot}')\,,
\label{eq:B}
\end{eqnarray}
where $C_{h''_1,h'_1}(x,\mathbf{k}_{\bot}',\mathbf{k}_{\bot}'') \equiv  \bar{u}_{h''_1}(x,\mathbf{k}_{\bot}'')  \Gamma   u_{h'_1}(x,\mathbf{k}_{\bot}')$.
% It should be noted that, in the derivation of   Eq.~\eqref{eq:B}, we  the form factor extracted via Eq.~\eqref{eq:B} is based on the $q^+=0$ frame, which implies a space-like momentum transfer,  $q^2=-q_\bot^2\leqslant 0$; while, the one in the time-like space require an additional $q^2$ extrapolation.

\subsection{Theoretical results }
Using the formulas given in the last subsection, one can obtain the expression of ${\cal B}_{\rm  SLF}^\u$ for the $V'\to V''$ transition. In the SLF QM, in order to extract the form factors, one has to take explicit $\mu$, $\lambda'$ and $\lambda''$. In this work, for convenience of  calculation, we take the strategy as follows: (i) We take $\u=+$ firstly and then use  ${\cal B}_{\rm  SLF}^+$ with $(\lbd',\lbd'')=(-,+)$, $(+,+)$, $(+,0)$ and $(0,+)$ to extract $V_{3}$, $V_{1}$, $V_{5}$ and $V_{6}$, respectively; (ii)  We multiply  both sides of Eq.~\eqref{eq:defFV} by $\e'^{\u*}$, and then use ${\cal B}_{\rm  SLF}\cdot \e'^{*}$ with $(\lbd',\lbd'')=(-,+)$ and $(+,+)$ to extract  $V_{4}$ and $V_{2}$, respectively. (iii)  For $A_1$, $A_2$, $A_3$ and $A_4$, we take $\lbd'=\lbd''=\pm$, and multiply  both sides of Eq.~\eqref{eq:defFA} by $q^\u$, $P^\u$,  $\e''^{\u*}$ and $\e'^{\u}$, respectively. After some derivations and simplifications,   we  finally obtain the SLF results for the form factors of $V'\to V''$ transition written as
\begin{equation}\label{eq:FSLF}
[{\cal F}(q^2)]_{\rm  SLF}=\int\frac{\d x\,\d^2{\bf k_\bot'}}{(2\pi)^3\,2x}\frac{{\psi''}^*(x,{\bf k''_\bot})\,{\psi'}(x,{\bf k_\bot'})}{2\hat {M}'_0\hat {M}''_0}\,{\cal \widetilde{F}}^{\rm SLF}(x,{\bf k}_\bot',q^2)\,,
\end{equation}
where, ${\cal F}$ denotes $V_{1-6}$ and $A_{1-4}$, and the integrands are
\begin{align}
\widetilde{V}_1^{\rm SLF}=&2\bigg\{4x\frac{(\kb'\cdot \qb)^2}{\qb^2}+\frac{(\bar x-x)^2}{\bar{x}}\kb'^2-\kb'\cdot\qb+\frac{1}{\bar x}(\bar x m'_1+xm_2)(\bar{x}m''_1+xm_2)\nonumber\\
&-\frac{4}{D'_{ V,{\rm LF}}}\Big[\kb'^2-\frac{(\kb'\cdot \qb)^2}{\qb^2}\Big]\Big[(\bar x-x)m'_1-m''_1-2xm_2\Big]\nonumber\\
&+\frac{4}{D''_{ V,{\rm LF}}}\Big[\kb'^2-\frac{(\kb'\cdot \qb)^2}{\qb^2}\Big]\Big[m'_1-(\bar x-x)m''_1+2xm_2\Big]\nonumber\\
&+\frac{8}{\bar{x}D'_{ V,{\rm LF}}D''_{ V,{\rm LF}}}\Big[\kb'^2-\frac{(\kb'\cdot \qb)^2}{\qb^2}\Big]\Big[\kb'\cdot\kb''+(\bar x m'_1-xm_2)(\bar x m_1''-xm_2)\Big]
\bigg\}\,,\\
%------------------------------
\widetilde{V}_2^{\rm{ SLF}}=&\frac{4}{x\bar x}\bigg\{-\frac{\kb'\cdot \qb}{\qb^2}(\bar xm'_1+xm_2)^2-(1-2x\bar x)\frac{\kb''\cdot \qb}{\qb^2}{\bf k}_\bot'^2+x^2\kb'\cdot \kb''\nonumber\\
&+(\bar xm'_1+xm_2)(\bar x^2m'_1+x\bar xm''_1+xm_2)
\nonumber\\
&-\frac{x\bar x}{D'_{ V,{\rm LF}}}\Big[2\frac{\kb'\cdot \qb}{\qb^2}{\bf k}_\bot'^2(m'_1+m_2)+{\kb'\cdot \qb}(\bar xm'_1-xm_2)-{\bf k}_\bot'^2 \big((1+2\bar x)m'_1+m''_1+2\bar xm_2\big)\Big]\nonumber\\
&+\frac{1}{D''_{ V,{\rm LF}}}\frac{\kb''\cdot \qb}{\qb^2}\Big[2\bar x(\bar xm'_1+xm_2){\bf k'_\bot\cdot q_\bot}-{\bf k}_\bot'^2\big(m'_1-(1-2x\bar x)m''_1+2x\bar xm_2\big)\nonumber\\
&-(\bar xm'_1+xm_2)\big((m'_1-m''_1)(xm_2-\bar xm''_1)-\bar xq^2\big)\Big]
\nonumber\\
&-\frac{1}{D'_{ V,{\rm LF}}D''_{ V,{\rm LF}}}\Big[\frac{\kb''\cdot \qb}{\qb^2}{\bf k}_\bot'^2\Big(2{\bf k'_\bot\cdot\bar k''_\bot}-\bar{x}q^2+(\bar{x}m'_1-xm_2)^2+(\bar{x}m''_1-xm_2)^2\nonumber\\
&+x\bar x(m'_1-m''_1)^2\Big)-{\bf k}'_\bot\cdot{\bf k}''_\bot\Big({\bf k}_\bot'^2+(xm_2-\bar xm'_1)^2\Big)
\Big]
\bigg\}
\nonumber\\
&-\widetilde{V}_1^{\rm{ SLF}}(x,{\bf k'_\bot},q^2)-\widetilde{V}_6^{\rm{ SLF}}(x,{\bf k'_\bot},q^2)+\frac{q^2}{2(M'^2-M''^2)}\Big[\widetilde{V}_3^{\rm{ SLF}}(x,{\bf k'_\bot},q^2)+\widetilde{V}_4^{\rm{ SLF}}(x,{\bf k'_\bot},q^2)\Big]\,,\\
%------------------------------
\widetilde{V}_3^{\textup{\scriptsize SLF}}=&\frac{4(M'^2-M''^2)}{q^2}\bigg\{-2x\Big[\kb'^2 +\bar x\kb'\cdot \qb-2\frac{(\kb'\cdot \qb)^2}{ \qb^2}\Big]
 \nonumber\\
&-\frac{1}{D'_{ V,{\rm LF}}}\Big[\kb' \cdot\qb(\bar x- x)(\bar xm'_1-xm_2)+\Big(\kb'^2-2\frac{(\kb'\cdot \qb)^2}{ \qb^2}\Big)\Big((\bar x- x)m'_1-m''_1-2xm_2\Big)\Big]\nonumber\\
&+\frac{1}{D''_{ V,{\rm LF}}}\Big[\bar x(\bar xm'_1+xm_2)q^2+{\bf k'_\bot\cdot q_\bot}\Big(2\bar xm'_1-\bar x(\bar x-x)m''_1+x(2\bar x+1)m_2\Big)\nonumber\\
&+\Big(\kb'^2 -2\frac{(\kb'\cdot \qb)^2}{ \qb^2}\Big)\Big(m'_1-(\bar x-x)m''_1+2xm_2\Big)\Big]\nonumber\\
&+\frac{2}{\bar{x}D'_{ V,{\rm LF}}D''_{ V,{\rm LF}}}\Big[\kb'^2 +\bar x\kb'\cdot \qb-2\frac{(\kb'\cdot \qb)^2}{ \qb^2}\Big]
\Big[\kb'\cdot \kb''+(\bar x m'_1-xm_2)(\bar{x} m''_1-xm_2)\Big]
\bigg\}\,,\\
%------------------------------
\widetilde{V}_4^{\rm SLF}=&-8\bigg\{(1-2x){\bf k}'^2_\bot+\frac{2({\bf k'_\bot\cdot q_\bot})^2}{{\bf q}_\bot^2} -\frac{\bf k'_\bot\cdot q_\bot}{{\bf q}_\bot^2}\Big[2{\bf k}'^2_\bot-\bar x q^2+(m'_1-m''_1)^2\Big]\nonumber\\
&+(m'_1-m''_1)(\bar xm'_1+xm_2)\nonumber\\
&+\frac{1}{x\bar xD'_{ V,{\rm LF}}}\Big[2\bar x(\bar xm'_1-xm_2)\frac{({\bf k'_\bot\cdot q_\bot})^2}{{\bf q}_\bot^2}-\bar x{\bf k}'^2_\bot\big(x(\bar{x}-x)m'_1+xm''_1-2x^2m_2\big)\nonumber\\
&+\frac{\bf k'_\bot\cdot q_\bot}{{\bf q}_\bot^2}(\bar xm'_1-xm_2)\big((m'_1-m''_1)(\bar xm'_1+xm_2)+\bar x^2q^2\big)\nonumber\\
&+\frac{\bf k'_\bot\cdot q_\bot}{{\bf q}_\bot^2} {\bf k}'^2_\bot\big((2x\bar x-1)m'_1+m''_1+2x\bar xm_2\big)
\Big]\nonumber\\
&-\frac{1}{x\bar xD''_{ V,{\rm LF}}}\frac{\bf k''_\bot\cdot q_\bot}{{\bf q}_\bot^2}\Big[2\bar x(\bar xm'_1+xm_2){\bf k'_\bot\cdot q_\bot}-{\bf k}_\bot'^2(m'_1-m''_1+2x\bar xm''_1+2x\bar xm_2)\nonumber\\
&-(\bar xm'_1+xm_2)\big((m'_1-m''_1)(xm_2-\bar xm''_1)-\bar xq^2\big)\Big]
\nonumber\\
&+\frac{1}{x\bar xD'_{ V,{\rm LF}}D''_{ V,{\rm LF}}}\Big[\frac{\bf k''_\bot\cdot q_\bot}{{\bf q}_\bot^2}{\bf k}_\bot'^2\Big(2{\bf k'_\bot\cdot k''_\bot}-\bar {x}(1-2x)q^2+(\bar{x}m'_1-xm_2)^2+(\bar{x}m''_1-xm_2)^2\nonumber\\
&+x\bar x(m'_1-m''_1)^2\Big)-\Big(\frac{2({\bf k'_\bot\cdot q_\bot})^2}{{\bf q}_\bot^2}-{\bf k}_\bot'^2-\bar x{\bf k'_\bot\cdot q_\bot}\Big)\big({\bf k}_\bot'^2+(xm_2-\bar xm'_1)^2\big)
\Big]\bigg\}
\nonumber\\
&+\left (1+\frac{q^2}{M'^2-M''^2}\right)\widetilde{V}_3^{\rm{ SLF}}(x,{\bf k'_\bot},q^2)+2\left[\widetilde{V}_5^{\rm{ SLF}}(x,{\bf k'_\bot},q^2)-\widetilde{V}_6^{\rm{ SLF}}(x,{\bf k'_\bot},q^2)\right]\,,\\
%------------------------------
\widetilde{V}_5^{\rm{ SLF}}=&\frac{4M''}{ M_0''}\bigg\{2(x-\bar{x}) \frac{(\kb'\cdot \qb)^2}{\qb^2}-(m''_1-m_2)(\bar xm'_1+xm_2)\nonumber\\
&- \frac{\kb'\cdot \qb}{\qb^2}\Big[(x-\bar x) (x{M''_0}^2-\bar xq^2)+\frac{x-\bar x}{\bar{x}} \kb'^2+\frac{x}{\bar{x}}m_2^2-m'_1m''_1+(m'_1-m''_1)m_2\Big]
\nonumber\\
&+\frac{2}{\bar xD'_{ V,{\rm LF}} }\frac{\kb'\cdot \qb}{\qb^2}\Big[(xm_2-\bar xm'_1)(m''_1m_2+\bar xxM_0''^2-\bar x^2q^2)\nonumber\\
&+{\bf k}'^2_\bot\big(m''_1-\bar{x}m'_1-\bar{x}m_2\big)+\bar x{\bf k'_\bot\cdot q_\bot}\big(2\bar xm'_1-m''_1+(\bar x-x)m_2\big)\Big]
\nonumber\\
&-\frac{1}{x\bar xD''_{ V,{\rm LF}} }\Big[\bar xm'_1+xm_2-\frac{\kb'\cdot \qb}{\qb^2}\big( m'_1+(x-\bar{x})m''_1+2xm_2\big)\Big] \Big[(x-\bar x){\bf k}_\bot''^2-\bar x^2m_1''^2+x^2m_2^2\Big]
\nonumber\\
&+\frac{2}{x\bar x^2D'_{ V,{\rm LF}}D''_{ V,{\rm LF}} }\frac{\kb'\cdot \qb}{\qb^2} \Big[{\bf  k}_\bot'\cdot {\bf  k}_\bot''+(xm_2-\bar xm'_1)(xm_2-\bar xm''_1)\Big]\Big[(x-\bar{x}){\bf  k}_\bot''^2-\bar x^2m_1''^2+x^2m_2^2\Big]
\bigg\}
\nonumber\\
&+2\widetilde{V}_1^{\rm{ SLF}}(x,{\bf k'_\bot},q^2)+\left(1-\frac{q^2}{M'^2-M''^2}\right)\widetilde{V}_3^{\rm{ SLF}}(x,{\bf k'_\bot},q^2)\,,\\
%------------------------------
\widetilde{V}_6^{\rm{ SLF}}=&\frac{4M'}{ M_0'}\bigg\{(\bar x-x)({\bf k}'^2_\bot+x\bar xM_0'^2+m'_1m_2)\nonumber\\
&+\frac{\kb'\cdot \qb}{\qb^2}\Big[x(x-\bar x)M_0'^2+\frac{x-\bar x}{\bar{x}}(m_2^2+{\bf k}'^2_\bot)-(m_2-m_1')(m_2+m_1'')\Big]
\nonumber\\
&-\frac{1}{\bar xD'_{ V,{\rm LF}}} \big(m_2^2+{\bf k}'^2_\bot-\bar x^2M_0'^2\big)\Big[(x-\bar x)(xm_2-\bar xm'_1)+\frac{\kb'\cdot \qb}{\qb^2}\big(m''_1+(x-\bar{x})m'_1+2xm_2)\big)\Big]\nonumber\\
&+\frac{2}{\bar xD''_{ V,{\rm LF}}}\frac{\kb''\cdot \qb}{\qb^2}\Big[(\bar xm''_1-xm_2)(m'_1m_2+x\bar{x}M_0'^2+\kb'^2)-\kb'\cdot \kb''(m'_1-m_2)\Big]\nonumber\\
&+\frac{2}{\bar x^2D'_{ V,{\rm LF}}D''_{ V,{\rm LF}}}\frac{\kb''\cdot \qb}{\qb^2}(m_2^2+{\bf k}'^2_\bot-\bar x^2M_0'^2)\Big[{\bf k}'^2_\bot+\bar x{\bf k'_\bot\cdot q}_\bot+(xm_2-\bar xm'_1)(xm_2-\bar xm''_1)\Big]
\bigg\}\nonumber\\
&+2\widetilde{V}_1^{\rm{ SLF}}(x,{\bf k'_\bot},q^2)-\left(1-\frac{q^2}{M'^2-M''^2}\right)\widetilde{V}_3^{\rm{ SLF}}(x,{\bf k'_\bot},q^2)\,,
%------------------------------
\end{align}
%==============
\begin{align}
\widetilde{A}_{1}^{\rm{ SLF}}=&\frac{2}{x\bar x({M'}^{2}-{M''}^{2})}\bigg\{
(x-\bar x){\bf k}_\bot'\cdot {\bf k}_\bot''\left[\bar x{\bf q}_\bot^2-2{\bf k_\bot'\cdot q_\bot}+x({M'}^{2}-{M''}^{2})\right]\nonumber\\
&+{\bf k'_\bot\cdot q_\bot}\left(\bar x^2m_1'^{2}+\bar x^2m_1''^{2}-2x^2m_2^2\right)+\bar x{\bf q}_\bot^2(-\bar x^2 {m'_1}^{2}+x^2m_2^2)\nonumber\\
&+x(\bar x m'_1+xm_2)(\bar x m''_1+xm_2)({M'}^{2}-{M''}^{2})\nonumber\\
&-\frac{1}{D_{ V,{\rm LF}}'}\Big[2\bar x({\bf k_\bot'\cdot q_\bot})^2(xm''_1+xm_2+xm'_1-\bar {x}m'_1)\nonumber\\
&-x\bar x(m'_1+m''_1)({M'}^{2}-{M''}^{2}-q^2){\bf k}_\bot'^2
-{\bf k'_\bot\cdot q_\bot}{\bf k}_\bot'^2 (x-\bar x)(m'_1+m''_1)
\nonumber\\
&-{\bf k'_\bot\cdot q_\bot}(\bar xm'_1-xm_2)\left((\bar xm''_1+xm_2)(m'_1+m''_1)+\bar x^2{\bf q}_\bot^2+x\bar x({M'}^{2}-{M''}^{2})\right)\Big]\nonumber\\
&-\frac{1}{D''_{ V,{\rm LF}}}\Big[x\bar x(m'_1+m''_1)({M'}^{2}-{M''}^{2}-q^2){\bf k'_\bot\cdot k''_\bot}\nonumber\\
&-(x-\bar x)(m'_1+m''_1){\bf  k''_\bot\cdot q_\bot} {\bf k}_\bot'^2-\bar x(\bar xm'_1+xm_2){\bf  k''_\bot\cdot q_\bot} ({\bf k}_\bot'+{\bf  k}_\bot'')\cdot{\bf q}_\bot\nonumber\\
&-(\bar xm'_1+xm_2)\big((m'_1+m''_1)(xm_2-\bar xm''_1)-x\bar x({M'}^{2}-{M''}^{2})
\big){\bf  k''_\bot\cdot q_\bot}
\Big]\nonumber\\
&-\frac{2\bar x^2}{ D'_{ V,{\rm LF}}D''_{ V,{\rm LF}}}\Big[{\bf k}_\bot'^2{\bf q}_\bot^2-({\bf k'_\bot\cdot q_\bot})^2\Big]\Big[{m'_1}^{2}-{m''_1}^{2}-x (M'^{2}-M''^{2})+ {\bf( k'_\bot+ k''_\bot)\cdot q_\bot}\Big]\bigg\}\,,\\
%------------------------------
\widetilde{A}_{2}^{\rm{ SLF}}=&\frac{2\,q^2}{({M'}^{2}-{M''}^{2})^2}\bigg\{
\frac{{\bf k}_\bot'^2}{x^2\bar x}\Big[2(\bar x-x){\bf k}_\bot'^2+\bar x(2+x^2-5x\bar x){\bf q}_\bot^2+2\bar x^2({m'_1}^{2}+{m''_1}^{2})\nonumber\\
&+2(xm'_1+\bar xm_2)(xm''_1+\bar xm_2)-2(\bar x-x)m_2^2-x^2(x-\bar x) ({M'}^{2}+{M''}^{2})\Big]\nonumber\\
&+\frac{\bf k'_\bot\cdot q_\bot}{x^2}\Big[4(x-\bar x)\kb'^2-2x{\bf k'_\bot\cdot q_\bot}-x\bar x(\bar x-x)q^2+\bar{x}(3x-4)m_1'^2+x\bar x{m''_1}^{2}\nonumber\\
&+4x^2m_2^2-4xm'_1m_2-2x^2(m'_1-m_2)(m''_1-m_2)+(x-\bar x)\cdot x^2({M'}^{2}+{M''}^{2})\Big]\nonumber\\
&-\frac{\bar x m'_1+xm_2}{x^2\bar x} \Big[\bar{x}(x^2 m_2+\bar{x}^2m_1'+\bar{x}m_1')q^2-2m'_1m''_1(\bar x{m''_1}+xm_2)\nonumber\\
&+x^2(\bar x m''_1+xm_2) ({M'}^{2}+{M''}^{2})\Big]\nonumber\\
&+\frac{1}{x\bar xD'_{ V,{\rm LF}}}\Big[\kb'^2(m'_1+m''_1)\big(2{\bf k}_\bot'^2+2m_2^2+2\bar x(m'_1+m_2)(m''_1-m_2)-\bar x x({M'}^{2}+{M''}^{2})\big)\nonumber\\
&+\qb^2 \kb'^2\, \bar x\big((2\bar x-x)m'_1+xm''_1+2(\bar x-x)m_2\big)+\kb'\cdot\qb{\bf k}_\bot'^2\big((x-\bar x)m''_1-(1+4\bar x)m'_1-2\bar xm_2\big)\nonumber\\
&+2\bar x(m'_1-xm''_1+xm_2)(\kb'\cdot\qb)^2+\kb'\cdot\qb(xm_2-\bar xm'_1)\big((m'_1+m''_1)(\bar xm''_1+xm_2)\nonumber\\
&-2m'_1m_2-x\bar x({M'}^{2}+{M''}^{2})\big)+\kb'\cdot\qb\qb^2\bar x^2(xm_2-\bar xm'_1)\Big]\nonumber\\
&+\frac{1}{x\bar xD''_{ V,{\rm LF}}}\Big[\big(({\bf k'_\bot\cdot q_\bot})^2-{\bf k}_\bot'^2 {\bf q}_\bot^2\big)\big(3\bar x^2m'_1+\bar x(x-2\bar x)m_2\big)+2{\bf k}_\bot'^2\kb'\cdot\kb''(m'_1+m''_1)\nonumber\\
&-2{\bf k'_\bot\cdot q_\bot}\kb'\cdot\kb''\bar{x}(m'_1-m_2)+\bar x\big((2\bar x+1)m'_1-xm_1''+(x-2)m_2\big)\kb'\cdot\kb''\qb^2\nonumber\\
&+\kb'\cdot\kb''\big(2m_2^2+2\bar x(m'_1-m_2)(m''_1+m_2)-x\bar x({M'}^{2}+{M''}^{2})\big)(m'_1+m''_1)\nonumber\\
&+\kb''\cdot\qb {\bf k}_\bot'^2\big((x-\bar x)(m'_1+m''_1)-2m'_1\big)+(\kb''\cdot\qb)^2 \bar{x}(\bar xm_1'+xm_2) \nonumber\\
&+\kb''\cdot\qb (\bar xm_1'+xm_2)\big((m'_1+m''_1)(xm_2-\bar xm''_1)-2m'_1m_2+x\bar x({M'}^{2}+{M''}^{2})\big)\Big]\nonumber\\
&+\frac{2}{x D'_{ V,{\rm LF}}D''_{ V,{\rm LF}}}\Big[\big(({\bf k'_\bot\cdot q_\bot})^2-{\bf k}_\bot'^2{\bf q}_\bot^2\big)
\big({\bar x}^2q^2-2{\bf k'_\bot\cdot k''_\bot}-\bar x({m'_1}^{2}+{m''_1}^{2})-2xm_2^2\nonumber\\
&+x\bar x({M'}^{2}+{M''}^{2})\big)\Big]\bigg\}+\widetilde{A}_{1}^{\rm{ SLF}}(x,{\bf k'_\bot},q^2)\,,\\
%------------------------------
\widetilde{A}_{3}^{\rm{ SLF}}=&\frac{4}{x\bar x}\bigg\{\frac{\bf  k_\bot''\cdot q_\bot}{{\bf q}_\bot^2}\Big[(x-\bar{x}){\bf k}_\bot'^2-\bar x^2m_1'^{2}+x^2m_2^2\Big]-{\bf k_\bot'\cdot  k}_\bot''\nonumber\\
&+\frac{1}{D''_{ V,{\rm LF}}}\Big[{\bf k_\bot'\cdot  k_\bot''}\big((x-\bar{x})\bar{x}m'_1+x\bar xm''_1-x\bar xm_2\big)+\bar x^2(\bar xm'_1+xm_2){\bf  k_\bot''\cdot q_\bot}\nonumber\\
&+\Big({\bf k}_\bot'^2+\bar{x}{\bf k'_\bot\cdot q_\bot}-2\frac{({\bf k'_\bot\cdot q_\bot})^2}{{\bf q}_\bot^2}\Big)\left(\bar {x}^2m'_1+x\bar{x}m_2\right)
-\frac{\bf  k_\bot''\cdot q_\bot}{{\bf q}_\bot^2}{\bf k^{\prime2}_\bot}(x-\bar x)(m'_1+m''_1)\nonumber\\
&-\frac{\bf k_\bot''\cdot q_\bot}{{\bf q}_\bot^2}(\bar{x}m'_1+xm_2)(m'_1+m''_1)(xm_2-\bar xm''_1)\Big]\nonumber\\
&+\frac{1}{D'_{ V,{\rm LF}}D''_{ V,{\rm LF}}}\Big[\Big({\bf k}_\bot'^2-\frac{2({\bf k'_\bot\cdot q_\bot})^2}{{\bf q}_\bot^{2}}\Big)\big(\bar {x}{\bf k}_\bot'^2-x\bar{x}^2 q^2\big)+\bar x({\bf k'_\bot\cdot q_\bot}){\bf k}_\bot'^2\nonumber\\
&-{\bf k'_\bot\cdot k''_\bot}(x{\bf k}_\bot'^2-\bar x^2{m'_1}^{2}+x^2m_2^{2})-\frac {\bf k''_\bot\cdot q_\bot }{{\bf q}_\bot^{2}}{\bf k}_\bot'^2(\bar{x}{m'_1}^{2}-\bar{x}{m''_1}^{2}+\bar x^2q^2)\Big]
\bigg\}\,,\label{eq:A3SLF1}\\
%------------------------------
\widetilde{A}_{4}^{\rm{ SLF}}=&\frac{4}{x\bar{x}}\bigg\{\bar{x}{\bf k}_\bot'^2-\frac{\bf k'_\bot\cdot q_\bot}{{\bf q}_\bot^{2}}\Big[(\bar x-x)({\bf k}_\bot'^2-2\bar{x}{\bf k'_\bot\cdot q_\bot})+\bar x^2{m''_1}^{2}-x^2m_2^2-\bar x^3q^2\Big]\nonumber\\
&+\frac{1}{D'_{ V,{\rm LF}}}\Big[2\frac{({\bf k'_\bot\cdot q_\bot})^2}{{\bf q}_\bot^{2}} \left((x\bar{x}-\bar x^2)m'_1+x\bar{x}m''_1+x\bar{x}m_2\right)-x\bar x(m'_1+m''_1){\bf k}_\bot'^2\nonumber\\
&-\bar{x}^2(xm_2-\bar xm'_1){\bf k'_\bot\cdot q_\bot}+\frac{\bf k'_\bot\cdot q_\bot}{{\bf q}_\bot^{2}}{\bf k}_\bot'^2(\bar{x}-x)(m'_1+m''_1)\nonumber\\
&-\frac{\bf k'_\bot\cdot q_\bot}{{\bf q}_\bot^{2}}(xm_2-\bar xm'_1)(m'_1+m''_1)(xm_2+\bar xm''_1)\Big]\nonumber\\
&+\frac{1}{ D'_{ V,{\rm LF}}D''_{ V,{\rm LF}}}\Big[2\bar x\frac{({\bf k'_\bot\cdot q_\bot})^2}{{\bf q}_\bot^2}{\bf k'_\bot\cdot k''_\bot}+{\bf k'_\bot\cdot k''_\bot}\big((x-\bar x){\bf k}_\bot'^2-\bar x^2{m'_1}^{2}+x^2m_2^2\big)\nonumber\\
&+\Big(\bar{x}{\bf k}_\bot'^2+\frac{\bf k'_\bot\cdot q_\bot}{{\bf q}_\bot^2}{\bf k}_\bot'^2-2\bar x\frac{({\bf k'_\bot\cdot q_\bot})^2}{{\bf q}_\bot^2}\Big)\big(
{\bar{x}m'_1}^{2}-\bar{x}{m''_1}^{2}+\bar{x}q^2\big)
\Big]
\bigg\}\,.
\end{align}
%======================
\section{Theoretical results in the  CLF QM}
\subsection{General formalism}
\begin{figure}[t]
\caption{The Feynman diagram for the matrix element $\cal B$.}
\begin{center}
\includegraphics[scale=0.3]{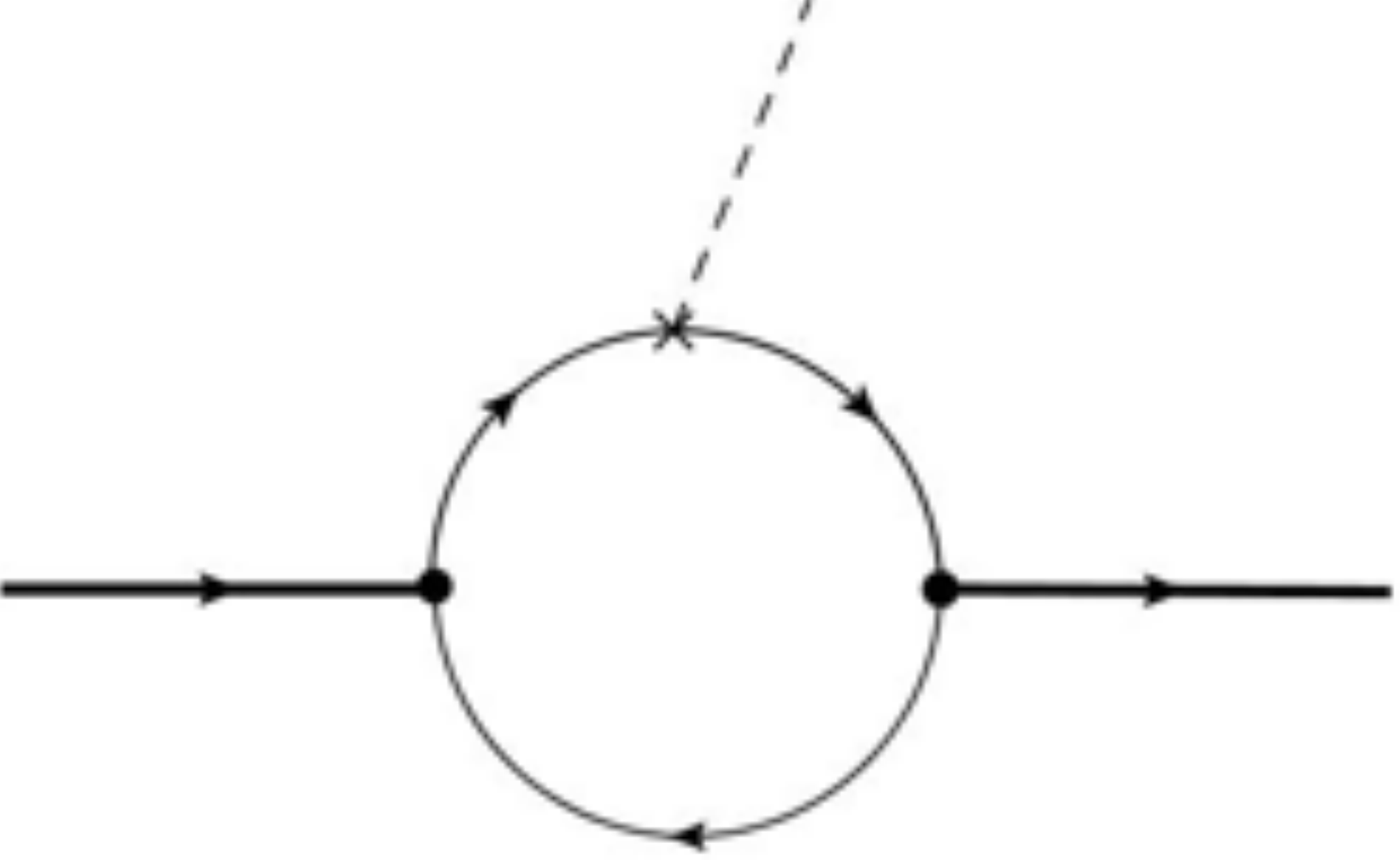}
\end{center}
\label{fig:fayn}
\end{figure}

The theoretical framework of CLF QM has been developed by Jaus with the help of a manifestly covariant BS approach as a guide to the calculation~\cite{Jaus:1999zv,Jaus:2002sv}. One can refer to Refs.~\cite{Jaus:1999zv,Jaus:2002sv,Cheng:2003sm} for the detail. In the CLF QM, the matrix element is obtained by calculating the Feynman diagram shown in Fig.~\ref{fig:fayn}. Using the Feynman rules given in Refs.~\cite{Jaus:1999zv,Cheng:2003sm}, the matrix element of $V'\to V''$ transition can be written as a manifest covariant form,
\begin{eqnarray}\label{eq:Bclf1}
{\cal B}=N_c \int \frac{\d^4 k_1'}{(2\pi)^4} \frac{H_{V'}H_{V''}}{N_1'\,N_1''\,N_2}iS_{\cal B}\,,
\end{eqnarray}
where $\d^4 k_1'=\frac{1}{2} \d k_1'^- \d k_1'^+ \d^2 \mathbf{k}_{\bot}'$, the denominators $N_{1}^{(\prime,\prime\prime)}=k_{1}^{(\prime,\prime\prime)2}-m_1^{(\prime,\prime\prime)2}+i\e$ and $N_{2}=k_{2}^{2}-m_2^{2}+i\e$  come from the fermion propagators, and $H_{V', V''}$ are  vertex functions. The trace term  $S_{\cal B}$  associated with the fermion loop is written as
\begin{eqnarray}
S_{\cal B}={\rm Tr}\left[\Gamma\, (\not\!k'_1+m'_1)\,(i\Gamma_{V'})\,(-\!\not\!k_2+m_2)\,(i\r^0{\Gamma}_{V''}^{\dag}\r^0) (\not\!k_1''+m_1'')\right]\,,
\end{eqnarray}
where $\Gamma_{V^{(\prime,\prime\prime)}}$ is the vertex operator written as~\cite{Cheng:2003sm,Choi:2013mda}
\begin{eqnarray}
i\Gamma_V=i\left[\gamma^\mu-\frac{ (k_1-k_2)^\mu}{D_{ V,{\rm con}}}\right]\,,\quad D_{V,{\rm con}}=M+m_1+m_2\,.
\end{eqnarray}

Integrating out the minus components of the loop momentum, one goes from the covariant calculation to the LF one. By closing the contour in the upper complex $k_1'^-$ plane and assuming that $H_{V', V''}$ are analytic within the contour, the integration picks up a residue at $k_2^2=\hat{k}_2^2=m_2^2$ corresponding to put the spectator antiquark on its mass-shell. Consequently, one has the following replacements~\cite{Jaus:1999zv,Cheng:2003sm}
\begin{eqnarray}
N_1 \to \hat{N}_1=x \left( M^2-M_0^2\right)
\end{eqnarray}
and%~\footnote{The Choi's convention for the vertex function, $\chi_M$, is related to the $H_M$ though $\chi_M=H_M/N$.}
\begin{eqnarray}\label{eq:type1}
\chi_V = H_V/N\to h_V/\hat{N}\,,\qquad  D_{V,{\rm  con}} \to D_{V,{\rm  LF}}\,,\qquad \text{(type-I)}
\end{eqnarray}
where the LF form of vertex function, $h_V$,  is given by
\begin{eqnarray}
h_V/\hat{N}=\frac{1}{\sqrt{2N_c}}\sqrt{\frac{\bar{x}}{x}}\frac{\psi}{\hat{M}_0}\,.
\label{eq:vPV}
\end{eqnarray}
The Eq.~\eqref{eq:type1} shows the correspondence between manifest covariant  and LF approaches.  In Eq.~\eqref{eq:type1},  the correspondence between $\chi$ and $\psi$  can be clearly derived by matching the CLF expressions to the SLF ones via some zero-mode independent  quantities, such as  $f_P$ and  $f_{+}^{P\to P}(q^2)$~\cite{Jaus:1999zv,Cheng:2003sm}, however, the validity of the correspondence for the $D$ factor appearing in the vertex operator, $D_{V{\rm  con}} \to D_{V,{\rm  LF}}$, has not yet been clarified explicitly~\cite{Choi:2013mda}. Instead of the traditional type-I correspondence, a much more generalized correspondence,
\begin{eqnarray}\label{eq:type2}
\chi_M = H_M/N\to h_M/\hat{N}\,,\qquad  M\to  M_0\,,\qquad \text{(type-II)}
\end{eqnarray}
is suggested by Choi {\it et al.} for  the purpose of  self-consistent results for $f_{A,V}$~\cite{Choi:2013mda,Chang:2018zjq}. Our following theoretical results are given within traditional type-I scheme unless otherwise specified. The ones within type-II scheme can be easily obtained by making an additional replacement $M\to  M_0$.

After integrating out $k_1'^-$, the matrix element, Eq.~\eqref{eq:Bclf1}, can be reduced to the LF form
\begin{eqnarray}
\label{eq:Bclf2}
\hat{{\cal B}}=N_c \int \frac{\d x \d^2 \mathbf{k}_{\bot}'}{2(2\pi)^3}\frac{h_{M'}h_{M''}}{\bar{x} \hat{N}_1'\,\hat{N}_1''\,}\hat{S}_{\cal B}\,.
\end{eqnarray}
It should be noted that ${\cal B}$ receives additional spurious contributions proportional to the light-like vector $\omega^\mu=(0,2,\mathbf{0}_\bot)$, and these undesired spurious contributions are expected to be cancelled out by the zero-mode contributions~\cite{Jaus:1999zv,Cheng:2003sm}. The inclusion of the zero-mode contributions in practice amounts to some proper replacements in  $\hat{S}_{\cal B}$ under integration~\cite{Jaus:1999zv}. In this work, we need
\begin{align}
 \label{eq:repFF1}
\hat{k}_1'^{\mu} \to& P^\u A_1^{(1)}+q^\u A_2^{(1)} \,,\\
%--------------------------------
\hat{k}_1'^{\mu}\hat{k}_1'^{\nu} \to &g^{\u\v}A_1^{(2)}+P^\u P^\v A_2^{(2)}+(P^\u q^\v+q^\u P^\v)A_3^{(2)}+q^\u q^\v A_4^{(2)}\nonumber\\
&+\frac{P^\u\omega^\v+\omega^\u P^\v}{\omega\cdot P}B_1^{(2)}\,,\\
%--------------------------------
\hat k_1'^{\mu}\hat k_1'^{\nu}\hat k_1'^{\alpha}\to&\left(g^{\mu \nu}P^\alpha+g^{\mu \alpha}P^\nu+g^{\nu\alpha}P^\mu\right)A_1^{(3)}+\left(g^{\mu \nu}q^\alpha+g^{\mu \alpha}q^\nu+g^{\nu\alpha}q^\mu\right)A_2^{(3)}\nonumber\\
&+P^\mu P^\nu P^\alpha A_3^{(3)}+\left(P^\mu P^\nu q^\alpha+P^\mu q^\nu P^\alpha+q^\mu P^\nu P^\alpha\right)A_4^{(3)}\nonumber\\
&+\left(q^\mu q^\nu P^\alpha+q^\mu P^\nu q^\alpha+P^\mu q^\nu q^\alpha\right)A_5^{(3)}+q_\mu q_\nu q_\alpha A_6^{(3)}\nonumber\\
&+\frac{1}{\omega \cdot P}\left(P^\mu P^\nu\omega^\alpha+P^\mu \omega^\nu P^\alpha+\omega^\mu P^\nu P^\alpha\right)B_1^{(3)}\nonumber\\
&+\frac{1}{\omega\cdot P}\left[\left(P^\mu q^\nu+q^\mu P^\nu\right)\omega^\alpha+\left(P^\mu q^\alpha+q^\mu P^\alpha\right)\omega^\nu+\left(P^\alpha q^\nu+q^\alpha P^\nu\right)\omega^\mu\right]B_2^{(3)}\,,\\
%--------------------------------
k_1'^{\mu}\hat{N}_2\to& q^\u\left(A_2^{(1)}Z_2+\frac{q\cdot P}{q^2}A_1^{(2)} \right) \,,\\
\hat{k}_1'^{\mu}\hat{k}_1'^{\nu}\hat{N}_2\to &g^{\u\v}A_1^{(2)}Z_2+q^\u q^\v\left( A_4^{(2)}Z_2+2\frac{q\cdot P}{q^2}A_2^{(1)}A_1^{(2)}\right)+\frac{P^\u\omega^\v+\omega^\u P^\v}{\omega\cdot P}B_3^{(3)}\,,\\
%--------------------------------
\hat k_1'^{\mu}\hat k_1'^{\nu}\hat k_1'^{\alpha}\hat N_2\to &\left(g^{\mu \nu}q^\alpha+g^{\mu \alpha}q^\nu+g^{\nu \alpha}q^\mu\right)\left(A_2^{(3)}Z_2+\frac{q\cdot P}{q^2}A_1^{(4)}\right)+q^\mu q^\nu q^\alpha \left(A_6^{(3)}Z_2+3\frac{q\cdot P}{q^2}A_4^{(4)}\right)\nonumber\\
&+\frac{1}{\omega\cdot P}\left[\left(P^\mu q^\nu+q^\mu P^\nu\right)\omega^\alpha+\left(P^\mu q^\alpha+q^\mu P^\alpha\right)\omega^\nu+\left(P^\alpha q^\nu+q^\alpha P^\nu\right)\omega^\mu\right]B_5^{(4)}
 \label{eq:repFF}
\end{align}
where $A$ and $ B$ functions are given by
\begin{align}
A_1^{(1)}&=  \frac{x}{2}\,,\qquad
A_2^{(1)}=\frac{x}{2} -\frac{\kb' \cdot \qb }{q^2}\,;\\
A_1^{(2)}&=-\kb'^2 -\frac{(\kb' \cdot \qb)^2}{q^2}\,,\qquad
A_2^{(2)}=(A_1^{(1)})^2\,,\qquad
A_3^{(2)}=A_1^{(1)}A_2^{(1)}\,,\qquad
A_4^{(2)}=(A_2^{(1)})^2\,;\\
A_1^{(3)}&=A_1^{(1)}A_1^{(2)}\,,\qquad
A_2^{(3)}=A_2^{(1)}A_1^{(2)}\,,\qquad
A_3^{(3)}=A_1^{(1)}A_2^{(2)}\,,\qquad
A_4^{(3)}=A_2^{(1)}A_2^{(2)}\,,\nonumber\\
A_5^{(3)}&=A_1^{(1)}A_4^{(2)}\,,\qquad
A_6^{(3)}=A_2^{(1)}A_4^{(2)}-\frac{2}{q^2}A_2^{(1)}A_1^{(2)}\,;\\
A_3^{(4)}&=A_1^{(1}A_2^{(3)}\,;\\
B_1^{(2)}&=\frac{x}{2}Z_2-A_1^{(2)}\,;\\
% B_2^{(3)}&=A_1^{(1)}C_1^{(2)} -A_2^{(1)}A_1^{(2)}
 B_2^{(3)}&=\frac{x}{2}Z_2A_2^{(1)}+A_1^{(1)}A_1^{(2)} \frac{P\cdot q}{q^2}-A_2^{(1)}A_1^{(2)},
~~B_3^{(3)}=B_1^{(2)}Z_2+\left(P^2-\frac{(q\cdot P)^2}{q^2}\right)A_1^{(1)}A_1^{(2)};\\
 B_1^{(4)}&=\frac{x}{2}Z_2A_1^{(2)}-A_1^{(4)}\,,\qquad
B_5^{(4)}=B_2^{(3)}Z_2+\frac{q\cdot P}{q^2}B_1^{(4)}+\left[P^2-\frac{\left(q\cdot P\right)^2}{q^2}\right]A_3^{(4)} \,;\\
Z_2&=\hat{N}_1'+m_1'^2-m_2^2+(\bar{x}-x)M'^2+(q^2+q\cdot P)\frac{\kb' \cdot \qb}{q^2}\,.
\end{align}
In these formulas, the $\w$-dependent terms associated with the  $C$ functions are not shown because they can be eliminated exactly by the inclusion of the zero-mode contributions~\cite{Jaus:1999zv}. It should be noted that there are still some residual $\w$-dependences associated with the  $B$ functions, which can be clearly seen from Eq.~(\ref{eq:repFF1}-\ref{eq:repFF}). As illustrated in Ref.~\cite{Jaus:1999zv}, the $B$ functions  play a special role since, on the one hand, it is combined with $\w^\u$, on the other hand, there is no zero-mode contribution associated with $B$ due to $x\hat{N}_2=0$.  Therefore, a different mechanism is required to neutralize the residual $\w$-dependence .

\subsection{Theoretical results}
Using the formalism introduced in the last subsection, we can obtain  ${\cal B}_{\rm  CLF}^\u$ for the $V'\to V''$ transition.  Then, matching ${\cal B}_{\rm  CLF}^\u(\Gamma=\r^\u)$ and ${\cal B}_{\rm  CLF}^\u(\Gamma=\r^\u\r_5)$  to the definitions of  form factors, Eq.~\eqref{eq:defFV} and Eq.~\eqref{eq:defFA}, respectively, we can extract the CLF results for the form factors of $V'\to V''$ transition directly. They can be written as
\begin{eqnarray}\label{eq:FCLF}
[\mathcal F(q^2)]_{\rm CLF}=N_c\int\frac{\d x\d^2{\bf k'_\bot}}{2(2\pi)^3}\frac{\chi_V'\chi_V''}{\bar x}{\cal \widetilde{\cal F}}^{\rm CLF}(x,{\bf k'_\bot},q^2)
\end{eqnarray}
where, the integrands are
\begin{align}
\label{eq:V1CLF}
\widetilde{V}_1^{\rm{CLF}}=&2\bigg\{x(m'_1+m''_1)m_2+\frac{1}{\bar x}(\kb'^2+x^2m_2^2)-{\bf k'_\bot\cdot q_\bot}+\bar xm'_1m''_1+8A_1^{(3)}\nonumber\\
&-\frac{2}{D'_{ V,{\rm con}}}\left[m'_1\left(-A_1^{(2)}+4A_1^{(3)}\right)+m'_1 A_1^{(2)}+4m_2A^{(3)}_1\right]\nonumber\\
&-\frac{2}{D''_{ V,{\rm con}}}\left[m'_1 A_1^{(2)}+m''_1\left(-A_1^{(2)}+4A_1^{(3)}\right)+4m_2 A^{(3)}_1\right]\nonumber\\
&-\frac{4}{D'_{ V,{\rm con}}D''_{ V,{\rm con}}}A_1^{(2)}\Big[\frac{1}{\bar x}({\bf k'_\bot}^{2}+x^2m_2^2)-{\bf k'_\bot\cdot q_\bot}+\bar xm'_1m''_1-x(m'_1+m''_1)m_2\Big]
\bigg\}\,, \\
%------------------------------
\widetilde{V}_2^{\rm CLF}=&2{M'}^2-2(m_1'-m_2)^2+(m_1'-m_1'')^2-q^2-\hat N'_1+\hat N''_1-2Z_2-16A_2^{(3)}\nonumber\\
&+2\left[2(m_2-m'_1)(m_2-m''_1)-{M'}^{2}-{M''}^{2}+q^2+2Z_2\right]A_2^{(1)}+4\left(2+\frac{{M'}^{2}-{M''}^{2}}{q^2}\right)A_1^{(2)}\nonumber\\
&+\frac{4}{D'_{ V,{\rm con}}}\left[m'_1\left(4A_2^{(3)}-3A_1^{(2)}\right)+m''_1 A_1^{(2)}+2m_2\left(2A_2^{(3)}-A_1^{(2)}\right)\right]\nonumber\\
&+\frac{4}{D''_{ V,{\rm con}}}\left[-m'_1 A_1^{(2)}+m''_1\left(4A_2^{(3)}-A_1^{(2)}\right)+2m_2\left( 2A_2^{(3)}-A_1^{(2)}\right)\right]\nonumber\\
&+\frac{4}{D'_{ V,{\rm con}}D''_{ V,{\rm con}}}\bigg\{\Big[-2{M'}^{2}-(m'_1-m''_1)^2+2(m'_1-m_2)^2+q^2+\hat N'_1-\hat N''_1+2Z_2\nonumber\\
&-\frac{4({M'}^{2}-{M''}^{2})}{3q^2}A_1^{(2)}\Big]A_1^{(2)}+2\left[{M'}^{2}+{M''}^{2}-2(m'_1+m_2)(m''_1+m_2)-q^2-2Z_2\right]A_2^{(3)}\bigg\}\,,\\
%------------------------------
\widetilde{V}_3^{\rm CLF}=&4\left({M'}^{2}-{M''}^{2}\right)\bigg\{4\left(A_3^{(2)}-A_2^{(2)}+A_3^{(3)}-A_5^{(3)}\right)\nonumber\\
&+\frac{1}{D'_{ V,{\rm con}}}\Big[(x-\bar x)m'_1\left(A_1^{(1)}-A_2^{(1)}-A_2^{(2)}+A_4^{(2)}\right)
+m''_1\left(A_2^{(2)}-2A_3^{(2)}+A_4^{(2)}\right)\nonumber\\
&+xm_2\left(A_1^{(1)}-A_2^{(1)}-2A_2^{(2)}+2A_4^{(2)}\right)\Big]\nonumber\\
&+\frac{1}{D''_{ V,{\rm con}}}\Big[m'_1\left(\bar x-2A_2^{(1)}+A_2^{(2)}+2A_3^{(2)}+A_4^{(2)}\right)+(x-\bar x)m''_1\left(A_1^{(1)}-A_2^{(1)}-A_2^{(2)}+A_4^{(2)}\right)\nonumber\\
&+2m_2\left(A_1^{(1)}+A_2^{(2)}-3A_3^{(2)}-2A_3^{(3)}+2A_5^{(3)}\right)\Big]\nonumber\\
&+\frac{2}{D'_{ V,{\rm con}}D''_{ V,{\rm con}}}\left(A_1^{(1)}-A_2^{(1)}-A_2^{(2)}+A_4^{(2)}\right)\Big[\frac{1}{\bar x}({\bf k'_\bot}^{2}+x^2m_2^2)-{\bf k'_\bot\cdot q_\bot}\nonumber\\
&+\bar xm'_1m''_1-x(m'_1+m''_1)m_2\Big]
\bigg\}\,,\\
%------------------------------
\widetilde{V}_4^{\rm CLF}=&-4q^2\bigg\{2\left(-A_1^{(1)}+A_2^{(1)}+A_2^{(2)}+2A_3^{(2)}-3A_4^{(2)}-2A_4^{(3)}+2A_6^{(3)}\right)\nonumber\\
&+\frac{1}{D'_{ V,{\rm con}}}\Big[m'_1\left(3A_1^{(1)}-3A_2^{(1)}-3A_2^{(2)}-4A_3^{(2)}+7A_4^{(2)}+4A_4^{(3)}-4A_6^{(3)}\right)\nonumber\\
&+m''_1\left(A_2^{(2)}-2A_3^{(2)}+A_4^{(2)}\right)-2m_2\left(A_2^{(2)}+A_3^{(2)}-2A_4^{(2)}-2A_4^{(3)}+2A_6^{(3)}\right)\Big]\nonumber\\
&+\frac{1}{D''_{ V,{\rm con}}}\Big[m'_1\left(-\bar x+2A_2^{(1)}-A_2^{(2)}-2A_3^{(2)}-A_4^{(2)}\right)+m''_1\Big(A_1^{(1)}-A_2^{(1)}-A_2^{(2)}-4A_3^{(2)}\nonumber\\
&+5A_4^{(2)}+4A_4^{(3)}-4A_6^{(3)}\Big)+2m_2\left(A_1^{(1)}-2A_2^{(1)}-A_2^{(2)}-A_3^{(2)}+4A_4^{(2)}+2A_4^{(3)}-2A_6^{(3)}\right)\Big]\nonumber\\
&+\frac{1}{D'_{ V,{\rm con}}D''_{ V,{\rm con}}}\Big[2\left({M'}^{2}+{M''}^{2}-2(m'_1+m_2)(m''_1+m_2)-q^2\right)\Big(A_4^{(2)}-A_3^{(2)}+A_4^{(3)}-A_6^{(3)}\Big)\nonumber\\
&+\Big(2{M'}^{2}+(m'_1-m''_1)^2-2(m'_1+m_2)^2-q^2-\hat N'_1+\hat N''_1\Big)\Big(A_1^{(1)}-A_2^{(1)}-A_2^{(2)}+A_4^{(2)}\Big)\nonumber\\
&+\frac{2({M'}^{2}-{M''}^{2})}{q^2}\Big(A_1^{(2)}-6A_2^{(1)}A_1^{(2)}+6A_2^{(1)}A_2^{(3)}-\frac{2}{q^2}(A_1^{(2)})^2\Big)\nonumber\\
&+Z_2\left(2A_2^{(1)}-6A_4^{(2)}+4A_6^{(3)}\right)\Big]\bigg\}
+\widetilde{V}_3^{\rm CLF}(x,{\bf k'_\bot},q^2)\,,\\
%------------------------------
\widetilde{V}_5^{\rm CLF}=&2\bigg\{{M'}^{2}-(m'_1-m_2)^2-\hat N'_1-Z_2-8\left(A_1^{(3)}-A_2^{(3)}\right)\nonumber\\
&+\left[{M'}^{2}-{M''}^{2}+2(m''_1-m'_1)(m''_1-m_2)-q^2
+2\hat N''_1\right]\left(A_1^{(1)}-A_2^{(1)}\right)\nonumber\\
&+\frac{2}{D'_{ V,{\rm con}}}\Big[m'_1\Big(\big({M''}^{2}-{m''_1}^{2}-m_2^{2}-\hat{N}_1''\big)\big(A_1^{(1)}-A_2^{(1)}\big)+4\big(A_1^{(3)}-A_2^{(3)}\big)+Z_2A_2^{(1)}\nonumber\\
&+\frac{{M'}^{2}-{M''}^{2}}{q^2}A_1^{(2)}\Big)-m''_1\Big(\big({M'}^{2}-{m'_1}^{2}-m_2^{2}-\hat N'_1\big)\big(A_1^{(1)}-A_2^{(1)}\big)+Z_2A_2^{(1)}+\frac{{M'}^{2}-{M''}^{2}}{q^2}A_1^{(2)}\Big)\nonumber\\
&-m_2\left(4A_2^{(3)}-4A_1^{(3)}\right)-m_2\left((m'_1-m''_1)^2-q^2+\hat N'_1+\hat N''_1\right)\left(A_1^{(1)}-A_2^{(1)}\right)\Big]\nonumber\\
&-\frac{2}{D''_{ V,{\rm con}}}\Big[2m'_1A_1^{(2)} -4m''_1\left(A_1^{(3)}-A_2^{(3)}\right)
-2m_2\left(A_1^{(2)}+2A_1^{(3)}-2A_2^{(3)}\right)\Big]\nonumber\\
&+\frac{4}{D'_{ V,{\rm con}}D''_{ V,{\rm con}}}\Big[\Big({M'}^{2}+{M''}^{2}-2(m'_1+m_2)(m''_1+m_2)-q^2\Big)\Big(A_1^{(3)}-A_2^{(3)}\Big)\nonumber\\
&+2Z_2A_2^{(3)}+\frac{2({M'}^{2}-{M''}^{2})}{3q^2}\Big(A_1^{(2)}\Big)^2\Big]
\bigg\}\,,\\
%------------------------------
\widetilde{V}_6^{\rm CLF}=&2\bigg\{{M'}^{2}-(m'_1-m''_1)^2-(m'_1-m_2)^2+q^2-2\hat N'_1-\hat N''_1-Z_2+8\left(A_1^{(2)}-A_1^{(3)}-A_2^{(3)}\right)\nonumber\\
&+\Big[{M''}^{2}-{M'}^{2}+2(m'_1-m''_1)(m'_1-m_2)-q^2+2\hat N'_1\Big]\Big(A_1^{(1)}+A_2^{(1)}\Big)\nonumber\\
&+\frac{2}{D'_{ V,{\rm con}}}\Big[4m'_1\left(A_1^{(3)}+A_2^{(3)}-A_1^{(2)}\right)-2m''_1 A_1^{(2)}-2m_2\left(A_1^{(2)}-2A_1^{(3)}-2A_2^{(3)}\right)\Big]\nonumber\\
&+\frac{2}{D''_{ V,{\rm con}}}\Big[m'_1\Big({M''}^{2}-{m''_1}^{2}-m_2^{2}-\hat N''_1-Z_2+Z_2A_2^{(1)}+\frac{M^{'2}-M^{''2}}{q^2}A_1^{(2)}\Big)\nonumber\\
&+m_1'({m''_1}^{2}+m_2^{2}-{M''}^{2}+\hat N''_1)\left(A_1^{(1)}+A_2^{(1)}\right)+m''_1\left({M'}^{2}-{m'_1}^{2}-m_2^{2}-\hat N'_1\right)\left(A_1^{(1)}+A_2^{(1)}\right)\nonumber\\
&+m''_1\left({m'_1}^{2}+m_2^{2}-{M'}^{2}+\hat N'_1+Z_2-4A_1^{(2)}+4A_1^{(3)}+4A_2^{(3)}-Z_2A_2^{(1)}-\frac{({M'}^{2}-{M''}^{2})}{q^2}A_1^{(2)}\right)\nonumber\\
&+m_2\big((m'_1-m''_1)^2-q^2+\hat N'_1+\hat N''_1\big)\big(1-A_1^{(1)}-A_2^{(1)}\big)+4m_2\big(-A_1^{(2)}+A_1^{(3)}+A_2^{(3)}\big)\Big]\nonumber\\
&+\frac{4}{D'_{ V,{\rm con}}D''_{ V,{\rm con}}}\Big[\big(-{M'}^{2}-{M''}^{2}+q^2+2(m'_1+m_2)(m''_1+m_2)\big)\big(A_1^{(2)}-A_1^{(3)}-A_2^{(3)}\big)\nonumber\\
&+2Z_2(A_1^{(2)}-A_2^{(3)})-\frac{2({M'}^{2}-{M''}^{2})}{3q^2}\big(A_1^{(2)}\big)^2\Big]\bigg\}\,,
\end{align}
%--------------
\begin{align}
\widetilde{A}_1^{\rm{ CLF}}=&-(m''_1-m'_1)^2+q^2-\hat N'_1-\hat N''_1+8A_1^{(2)}+2\Big[M'^{2}+M''^{2}+2(m'_1-m_2)(m_2-m''_1)-q^2\Big] A_1^{(1)}\nonumber\\
&-4(m_1'+m_1'')\left(\frac{1}{D_{ V,{\rm con}}'}+\frac{1}{D_{ V,{\rm con}}''}\right)A_1^{(2)}\,,\\
\widetilde{A}_2^{\rm{ CLF}}=&-\frac{q^2}{M'^{2}-{M''}^{2}} \bigg\{(m'_1-m''_1)^2-2(m_2-m'_1)^2+2{M'}^2-q^2-2Z_2-\hat N'_1+\hat N''_1\nonumber\\
&+\frac{4({M'}^2-{M''}^2)}{q^2}A_1^{(2)}+2\left[2(m_2-m'_1)(m_2-m''_1)-{M'}^2-{M''}^2+q^2+2Z_2 \right]A_2^{(1)}\nonumber\\
&-\frac{4}{D'_{ V,{\rm con}}}\left(m'_1-m''_1+2m_2\right)A_1^{(2)}-\frac{4}{D''_{ V,{\rm con}}}\left(m'_1-m''_1-2m_2\right)A_1^{(2)}\bigg\}+\widetilde{A}_1^{\rm{ CLF}}(x,{\bf k'_\bot},q^2)\,,
\\
\widetilde{A}_3^{\rm{CLF}}=&-4({M'}^{2}-{M''}^{2})\bigg\{A_1^{(1)}-A_2^{(1)}-A_2^{(2)}+A_4^{(2)}
\nonumber\\
&+\frac{1}{D_{ V,{\rm con}}''}\Big[ m'_1\left(-\bar x+2A_2^{(1)}-A_2^{(2)}-2A_3^{(2)}-A_4^{(2)}\right)+m''_1\left(-A_1^{(1)}+A_2^{(1)}+A_2^{(2)}-A_4^{(2)}\right)\nonumber\\
&+ 2m_2\left(-A_1^{(1)}+A_2^{(2)}+A_3^{(2)}\right) \Big]+\frac{2}{D'_{ V,{\rm con}}D''_{ V,{\rm con}}}\left(A_1^{(2)}-A_1^{(3)}-A_2^{(3)}\right)
\bigg\}\,,\\
\widetilde{A}_4^{\rm{CLF}}=&-4({M'}^{2}-{M''}^{2})\bigg\{A_1^{(1)}-A_2^{(1)}-A_2^{(2)}+A_4^{(2)}\nonumber\\
&+\frac{1}{D'_{ V,{\rm con}}}\Big[m'_1\left(A_2^{(1)}-A_1^{(1)}+A_2^{(2)}-A_4^{(2)}\right)+m''_1\left(-A_2^{(2)}+2A_3^{(2)}-A_4^{(2)}\right)+2m_2\left(A_2^{(2)}-A_3^{(2)}\right)\Big]\nonumber\\
&+\frac{2}{D'_{ V,{\rm con}}D''_{ V,{\rm con}}}\left(-A_1^{(3)}+A_2^{(3)}\right)
\bigg\}\,. \label{eq:A4CLF}
\end{align}
It should be noted that the contributions related to the $B$ functions are not included in the results given above. These contributions result in the self-consistence and covariance problems, and will be given and analyzed  separately in the next section.
%Using the formulae given above, one can obtain the full result of $\cal B$, and further extract the form factors. For a given quantity,~${\cal Q}$, its full result can be expressed as the sum of the valence and zero-mode contributions,
%\begin{align}
%{\cal Q}^{\rm full}={\cal Q}^{\rm val.}+{\cal Q}^{\rm z.m.}\,.
%\end{align}
%where $\cal Q$ denotes form factors $g(q^2)$, $f(q^2)$ and $a_\pm(q^2)$.
%In order to evaluate the effect of zero-mode, we also need to calculate ${\cal Q}^{\rm val.}$ and/or ${\cal Q}^{\rm z.m.}$.  In this paper, we employ the strategy introduced in Ref.~\cite{Chang:2018zjq} to calculate ${\cal Q}^{\rm val.}$.

%In this paper, we A simple way to calculate ${\cal Q}^{\rm val.}$ is to assume $k_2^+\neq 0$~(or $p^+\neq 0$), which ensures the pole of $N_2$ is safely located in the contour of  $k^-$~($k'^-$) integral~( the pole of $N_2$ is finite) and implies that the zero-mode contributions are discarded. At this moment, the replacement for $\hat{k}_1^{\u}$ given above have to be disregarded. Instead, one just need to directly use the on-mass-shell condition of spectator antiquark, $k_2^2=m_2^2$, and the conservation of four-momentum at each vertex. Then, the zero-mode contributions can be obviously reflected by  ${\cal Q}^{\rm z.m.}={\cal Q}^{\rm full}-{\cal Q}^{\rm val.}$.

%=============================

\section{Numerical results and discussion}

%%%%%%%%%%%%%%%%%%%%%
\begin{table}[t]
\begin{center}
\caption{\label{tab:qm} \small The values of quark masses (in units of MeV) suggested in the previous works and  used in this work, where $q=u,d$. See text for explanation. }
\vspace{0.4cm}
\let\oldarraystretch=\arraystretch
\renewcommand*{\arraystretch}{1.1}
\setlength{\tabcolsep}{5pt}
\begin{tabular}{lcccccccccccc}
\hline\hline
           &Ref.~\cite{ Choi:2015ywa} &Ref.~\cite{Choi:2009ai}&Ref.~\cite{Choi:2009ai}& Ref.~\cite{Hwang:2010hw} 
           &Ref.~\cite{ Cheng:2003sm} & Ref.~\cite{Verma:2011yw}& This work \\\hline     
$m_q$&$205$  &$220$ &$250$ &$251$    &$260$ &$260$ &$230\pm40$\\
$m_s$&$380$  &$450$  &$480$ &$445$   &$370$ &$450$ &$430\pm60$\\
$m_c$&$1750$&$1800$&$1800$&$1380$ &$1400$&$1400$&$1600\pm300$\\
$m_b$&$5150$&$5200$&$5200$&$4780$ &$4640$&$4640$&$4900\pm400$
\\\hline\hline
\end{tabular}
\end{center}
\end{table}

%%%%%%%%%%%%%%%%%%%%% 其他建议与表中雷同
%%%%%%%%%%%%%%%%%%%%%& This work&$250^{+20}_{-50}$&$450^{+40}_{-70}$&$1400^{+400}_{-200}$&$4640^{+600}_{-200}$
%%%%%%%%%%%%%%%%%%%%%

\begin{table}[t]
\begin{center}
\caption{\label{tab:input} \small The values of Gaussian parameters $\beta$ (in units of MeV).}
\vspace{0.2cm}
\let\oldarraystretch=\arraystretch
\renewcommand*{\arraystretch}{1.1}
\setlength{\tabcolsep}{8.8pt}
\begin{tabular}{lcccccccccc}
\hline\hline
  $\beta_{q\bar{q}}$    &$\beta_{s\bar{q}}$   &$\beta_{s\bar{s}}$
  &$\beta_{c\bar{q}}$    &$\beta_{c\bar{s}}$ \\
  \hline
  $312\pm6$ &$313\pm10$ &$348\pm6$ &$429\pm13$ &$530\pm19$ \\\hline\hline
  $\beta_{c\bar{c}}$    &$\beta_{b\bar{q}}$   &$\beta_{b\bar{s}}$   &$\beta_{b\bar{c}}$   &$\beta_{b\bar{b}}$    \\\hline
  $703\pm7$ &$516\pm15$ &$568\pm10$ &$876\pm20$ &$1390\pm12$\\
\hline\hline
\end{tabular}
\end{center}
\end{table}

Based on the theoretical results given above, we then present our numerical results and discussions. The constituent quark masses and  Gaussian parameters $\beta$ are essential inputs for computing the form factors. Thus, firstly, we would like to clarify their values used in our calculation. The values of constituent quark masses suggested in the previous works based on the LFQMs and Gaussian type WF are collected in Table~\ref{tab:qm}, in which, the second column is the result obtained via variational analyses of meson mass spectra  for the Hamiltonian with a smeared-out hyperfine interaction~\cite{Choi:2015ywa}; the third and fourth columns are the values obtained by the variational principle for the linear and harmonic oscillator~(HO) confining potentials, respectively~\cite{Choi:2009ai} (some similar analyses are made also in Refs.~\cite{Choi:2007se,Choi:1997iq}); in the fifth column,  the light quark masses are fitted by using decay constants $f_\pi$, $f_K$ and the mean square radii $\la r^2_{\pi^+} \ra$, $\la r^2_{K^0} \ra$, and the heavy quark masses are determined by the mass of the spin-weighted average of the heavy quarkonium states and its variational principle~\cite{Hwang:2010hw}; the sixth and seventh columns are some commonly used values in the LFQMs~\cite{ Cheng:2003sm,Verma:2011yw}. The quark masses suggested in the other previous work are generally similar  to one of them.  From Table~\ref{tab:qm}, it can be easily found that the quark masses are model dependent, and their values obtained in the previous works are different from each other more or less.  In this work, we take a moderate choice for the default inputs~(central values) of quark masses, which are listed in the last column of Table~\ref{tab:qm}. In addition, we assign  a conservative error to each quark mass, which covers properly the other values listed in Table~\ref{tab:qm} and therefore can reflect roughly the uncertainties induced by the model dependence of quark mass. 
%As regards parameters $\beta$, their values can be determined by fitting to the decay constants 
Then, in order to determine parameters $\beta$, we make fits to the  data of $f_V$ collected in Ref.~\cite{Chang:2018zjq} following the same way as Refs.~\cite{Chang:2018zjq,Chang:2019mmh} but with the default values of quark masses listed in  Table~\ref{tab:qm} as inputs. The fitting results for  $\beta$ are listed  in Table~\ref{tab:input}.  In addition, the type-II correspondence scheme is employed in the fits, while the fitting results do not affect following comparison between type-I and -II schemes.   
 %\footnote{In the fits, the Lattice QCD results for $f_V$ }  

%The quark mass the ones for the power  law and BSW WFs are not listed here. 

%{ can be determined by fitting to the mass spectra of mesons, but the results are  }we take~\cite{Verma:2011yw}
%\begin{equation}\label{eq:qmass}
%m_{q,s,c,b}=(0.25,0.45,1.40,4.64)\,{\rm GeV}\,,
%\end{equation}
%where $q=u$ and $d$, as default inputs, and  assign $10\%$ uncertainties to them which can cover roughly most of the values suggested in the previous works~\cite{Verma:2011yw,Jaus:2002sv,Cheng:2003sm,Geng:2016pyr,Choi:2007yu,Choi:2007se,Choi:2015ywa}. For the later, we use the results obtained by fitting to the data of $f_V$ with the default values of quark masses, Eq.~\eqref{eq:qmass}, as inputs~\cite{Chang:2018zjq,Chang:2019mmh}. Their values are listed in Table~\ref{tab:input}.

As has been mentioned in the last section, the contributions associated with $B$ functions are not included in the CLF results, Eqs.~(\ref{eq:V1CLF}-\ref{eq:A4CLF}). These contributions to the matrix elements can be written as
\begin{eqnarray}
[\mathcal B]_{\rm B}=N_c\int\frac{\d x\d^2{\bf k'_\bot}}{2(2\pi)^3}\frac{\chi_V'\chi_V''}{\bar x}{\cal \widetilde{\cal B}}_{\rm B}
\end{eqnarray}
where,
\begin{align}
{\cal \widetilde{B}}_{\rm B}^\u(\Gamma=\r^\u\r_5)=&-4i\epsilon_{\mu\nu\alpha\beta}\epsilon'^\nu P^\alpha q^\beta\left(\omega\cdot \epsilon''^{*}\right)\frac{B_1^{(2)}}{\omega\cdot P}\left(1+\frac{m_1'-m_1''-2m_2}{D_{V,{\rm con}}''}\right)\nonumber\\
&-4i\epsilon_{\mu\nu\alpha\beta}\epsilon''^{*\nu} P^\alpha q^\beta\left(\omega\cdot \epsilon'\right)\frac{B_1^{(2)}}{\omega\cdot P}\left(1-\frac{m_1'-m_1''+2m_2}{D_{V,{\rm con}}'}\right)\nonumber\\
&+4i\epsilon_{\mu\nu\alpha\beta}\epsilon'^\nu P^\alpha\omega^\beta\left(q\cdot \epsilon''^{*}\right)\frac{B_1^{(2)}}{\omega\cdot P}\left(1-\frac{m_1'+m_1''}{D_{V,{\rm con}}''}\right)\nonumber\\
&+4i\epsilon_{\mu\nu\alpha\beta}\epsilon'^\nu q^\alpha\omega^\beta\left(q\cdot \epsilon''^{*}\right)\frac{B_1^{(2)}}{\omega\cdot P}\left(1+\frac{m_1'-m_1''-2m_2}{D_{V,{\rm con}}''}\right)\nonumber\\
&+4i\epsilon_{\mu\nu\alpha\beta}\epsilon''^{*\nu} P^\alpha\omega^\beta\left(q\cdot \epsilon'\right)\frac{B_1^{(2)}}{\omega\cdot P}\left(1-\frac{m_1'+m_1''}{D_{V,{\rm con}}'}\right)\nonumber\\
&-4i\epsilon_{\mu\nu\alpha\beta}\epsilon''^{*\nu} q^\alpha\omega^\beta\left(q\cdot \epsilon'\right)\frac{B_1^{(2)}}{\omega\cdot P}\left(1-\frac{m_1'-m_1''+2m_2}{D_{V,{\rm con}}'}\right)\nonumber\\
&+8i\epsilon_{\mu\nu\alpha\beta}\omega^\nu P^\alpha q^\beta\left(q\cdot \epsilon'\right)\left(q\cdot \epsilon''^{*}\right)\frac{B_1^{(3)}-B_1^{(2)}}{\omega\cdot P}\cdot\frac 1{D_{V,{\rm con}}'D_{V,{\rm con}}''}\,, \label{eq:BBA}
\end{align}
and 
\begin{align}
{\cal \widetilde{B}}_B^\u(\Gamma=\r^\u)=&\frac{4}{\omega \cdot P}B_1^{(2)}\bigg\{(\epsilon'\cdot q)(\epsilon''^{*}\cdot \omega)P^\mu\Big[\frac{m'_1-m''_1}{D'_{V,{\rm con}}}+\frac{m'_1+m''_1-2m_2}{D''_{V,{\rm con}}}\nonumber\\
&+\frac{1}{D'_{V,{\rm con}}D''_{V,{\rm con}}}\big((m'_1-m''_1)^2-q^2-\hat N'_1+\hat N''_1\big)\Big]\nonumber\\
&+(\epsilon'\cdot \omega)(\epsilon''^{*}\cdot q)P^\mu\Big[3-\frac{3m'_1+m''_1+2m_2}{D'_{V,{\rm con}}}-\frac{5m''_1+4m_2}{D''_{V,{\rm con}}}\nonumber\\
&-\frac{1}{D'_{V,{\rm con}}D''_{V,{\rm con}}}\big(2M'^2+2M''^2+(m'_1-m''_1)^2-4(m'_1+m_2)(m''_1+m_2)+3q^2+\hat N'_1+\hat N''_1\big)\Big]\nonumber\\
&+q^\mu\Big[(\epsilon'\cdot \omega)(\epsilon''^{*}\cdot q)-(\epsilon'\cdot q)(\epsilon''^{*}\cdot \omega)\Big]\Big[2+\frac{-4m'_1+m''_1-2m_2}{D'_{V,{\rm con}}}-\frac{m''_1+2m_2}{D''_{V,{\rm con}}}\nonumber\\
&+\frac{1}{D'_{V,{\rm con}}D''_{V,{\rm con}}}\big(-2M'^2+2m_2^2+m'^2_1-m''^2_1+2m'_1(m''_1+2m_2)+q^2+\hat N'_1-\hat N''_1\big)\Big]\nonumber\\
&+(\epsilon'\cdot q)(\epsilon''^{*}\cdot q)\omega^\mu\Big[-3+\frac{4m'_1+2m''_1+2m_2}{D'_{V,{\rm con}}}+\frac{m'_1+4m''_1-2m_2}{D''_{V,{\rm con}}}\nonumber\\
&+\frac{2}{D'_{V,{\rm con}}D''_{V,{\rm con}}}(M'^2+M''^2-2(m'_1+m_2)(m''_1+m_2)-q^2)\Big]
\bigg\}\nonumber\\
&+\frac{16}{\omega \cdot P}B_1^{(3)}\bigg\{P^\mu\Big[(\epsilon'\cdot \omega)(\epsilon''^{*}\cdot q)-(\epsilon'\cdot q)(\epsilon''^{*}\cdot \omega)\Big]\Big[-1+\frac{m'_1+m_2}{D'_{V,{\rm con}}}+\frac{m''_1+m_2}{D''_{V,{\rm con}}}\nonumber\\
&+\frac{1}{2D'_{V,{\rm con}}D''_{V,{\rm con}}}(M'^2+M''^2-2(m'_1+m_2)(m''_1+m_2)-q^2)\Big]\nonumber\\
&+(\epsilon'\cdot q)(\epsilon''^{*}\cdot q)\omega^\mu\Big[-1+\frac{m'_1+m_2}{D'_{V,{\rm con}}}+\frac{m''_1+m_2}{D''_{V,{\rm con}}}\nonumber\\
&+\frac{1}{2D'_{V,{\rm con}}D''_{V,{\rm con}}}\big(M'^2+M''^2-2(m'_1+m_2)(m''_1+m_2)-q^2\big)\Big]
\bigg\}\nonumber\\
&+\frac{16}{\omega \cdot P}B_2^{(3)}\bigg\{(\epsilon'\cdot \omega)(\epsilon''^{*}\cdot q)P^\mu\Big[-1+\frac{m'_1+m_2}{D'_{V,{\rm con}}}+\frac{m''_1+m_2}{D''_{V,{\rm con}}}\nonumber\\
&+\frac{1}{2D'_{V,{\rm con}}D''_{V,{\rm con}}}\big(M'^2+M''^2-2(m'_1+m_2)(m''_1+m_2)-q^2\big)\Big]\nonumber\\
&+(\epsilon'\cdot \omega)(\epsilon''^{*}\cdot q)q^\mu\Big[-1+\frac{m'_1+m_2}{D'_{V,{\rm con}}}+\frac{m''_1+m_2}{D''_{V,{\rm con}}}\nonumber\\
&+\frac{1}{2D'_{V,{\rm con}}D''_{V,{\rm con}}}\big(M'^2+M''^2-2(m'_1+m_2)(m''_1+m_2)-q^2\big)\Big]
\bigg\}\nonumber\\
&+\frac{16}{\omega \cdot P}\frac{B_3^{(3)}}{D'_{V,{\rm con}}D''_{V,{\rm con}}}\bigg\{(\epsilon'\cdot \omega)(\epsilon''^{*}\cdot q)P^\mu+\frac{1}{2}\Big[(\epsilon'\cdot \omega)(\epsilon''^{*}\cdot q)-(\epsilon'\cdot q)(\epsilon''^{*}\cdot \omega)\Big]q^\mu\nonumber\\
&-(\epsilon'\cdot q)(\epsilon''^{*}\cdot q)\omega^\mu
\bigg\}\nonumber\\
&+\frac{16}{\omega \cdot P}\frac{B_5^{(4)}}{D'_{V,{\rm con}}D''_{V,{\rm con}}}\bigg\{-\Big[(\epsilon'\cdot \omega)(\epsilon''^{*}\cdot q)+(\epsilon'\cdot q)(\epsilon''^{*}\cdot \omega)\Big]P^\mu\nonumber\\
&-\frac{1}{2}\Big[(\epsilon'\cdot \omega)(\epsilon''^{*}\cdot q)-(\epsilon'\cdot q)(\epsilon''^{*}\cdot \omega)\Big]q^\mu
\bigg\}\,.\label{eq:BBV}
\end{align}
They may present nontrivial contributions to the form factors and  lead to the self-consistency and covariance problems of CLF QM. Then, the full results for form factors in the CLF QM can be expressed as
\begin{align}
[{\cal F}]^{\rm full}=[{\cal F}]^{\rm CLF}+[{\cal F}]^{\rm B}\,.
\end{align}

\begin{table}[t]
\begin{center}
\caption{\label{tab:A3} \small Numerical results of form factor $A_3({\bf{q}}_\perp^2)$  at ${\bf q}_{\bot}^2=(0,2,4,9) \,{\rm GeV^2}$ for $B^*\to D^*$ transition. See text for further explanation. }
\vspace{0.2cm}
\let\oldarraystretch=\arraystretch
\renewcommand*{\arraystretch}{1}
\setlength{\tabcolsep}{4pt}
\begin{tabular}{l|cccccccccccc}
\hline\hline
$$    & &$ [A_3]^{\text{SLF}}_{\lambda'=\lambda''=\pm}$
                 &$[A_3]^{\text{SLF}}_{\lambda'=0,\lambda''=\pm}$
                 &$[A_3]^{\rm full}_{\lambda'=\pm,\lambda''=0}$
                 &$[A_3]^{\rm full}_{\lambda'=\lambda''=\pm}$
                 &$[A_3]^{\rm val.}$
                 &$[A_3]^{\rm CLF}$\\\hline
\multirow{2}{*}{${\bf{q}}_\perp^2=0$}
&type-I    &$0.067$&$0.071$&$0.058$&$0.068$&$0.069$&$0.068$
              \\
&type-II    &$0.067$&$0.067$&$0.067$&$0.067$&$0.067$&$0.067$
\\\hline
\multirow{2}{*}{${\bf{q}}_\perp^2=2$}
&type-I     &$0.060$&$0.063$&$0.051$&$0.061$&$0.062$&$0.061$
              \\
&type-II    &$0.060$&$0.060$&$0.060$&$0.060$&$0.060$&$0.060$
\\\hline
\multirow{2}{*}{${\bf{q}}_\perp^2=4$}
&type-I     &$0.054$&$0.057$&$0.046$&$0.055$&$0.056$&$0.055$
              \\
&type-II    &$0.054$&$0.054$&$0.054$&$0.054$&$0.054$&$0.054$
\\\hline
\multirow{2}{*}{${\bf{q}}_\perp^2=9$}
&type-I     &$0.042$&$0.045$&$0.035$&$0.043$&$0.044$&$0.043$
              \\
&type-II    &$0.042$&$0.042$&$0.042$&$0.042$&$0.042$&$0.042$
\\\hline\hline
\end{tabular}
\end{center}
\end{table}

\begin{figure}[t]
\caption{The dependences of $\Delta^{A_3}_{\rm full}(x)$, $\Delta^{A_3}_{\rm SLF}(x)$ and $\d [A_3]_{\rm z.m.}/\d x$ on $x$ for  $B^*\to D^*$ transition at ${\bf q}_{\bot}^2=(0,9)\,{\rm GeV^2}$ and  for $D^*\to \rho$ transition at ${\bf q}_{\bot}^2=(0,4)\,{\rm GeV^2}$. See text for the detailed explanations and discussions.}
\begin{center}
\subfigure[]{\includegraphics[scale=0.4]{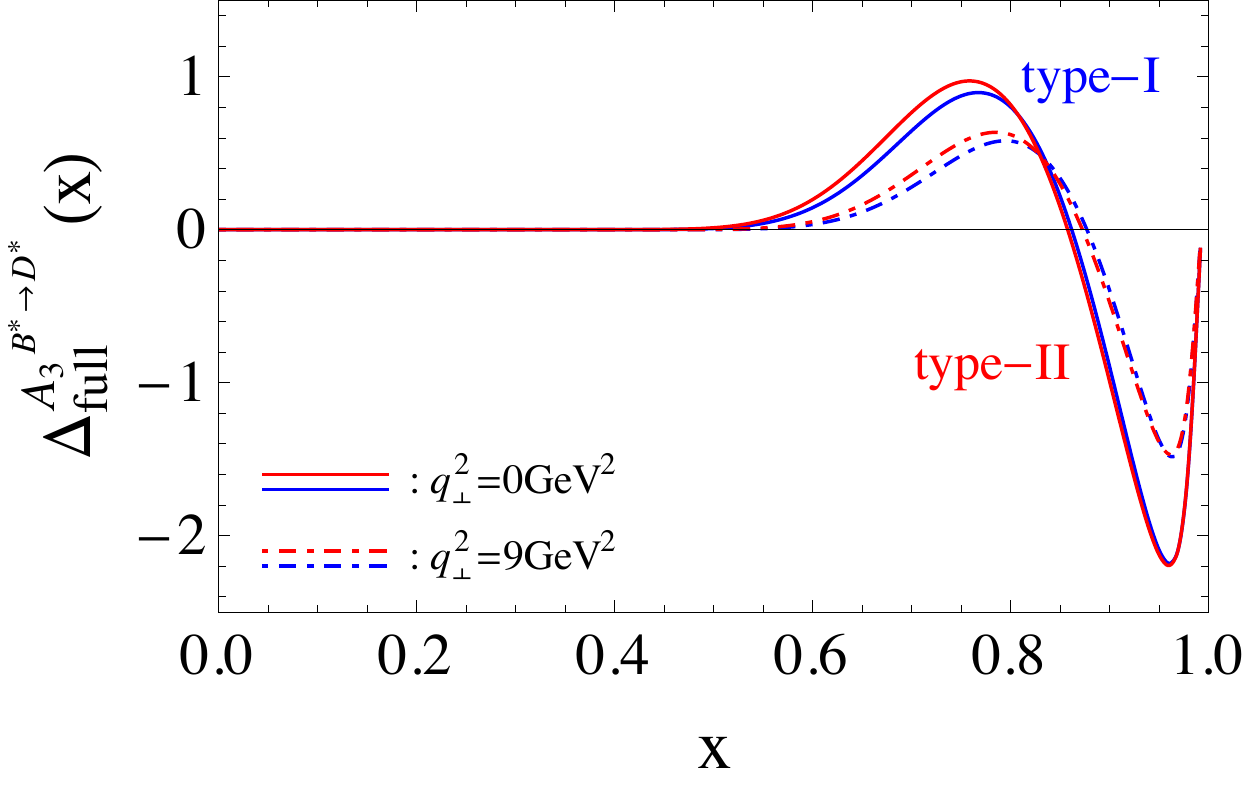}}\quad
\subfigure[]{\includegraphics[scale=0.42]{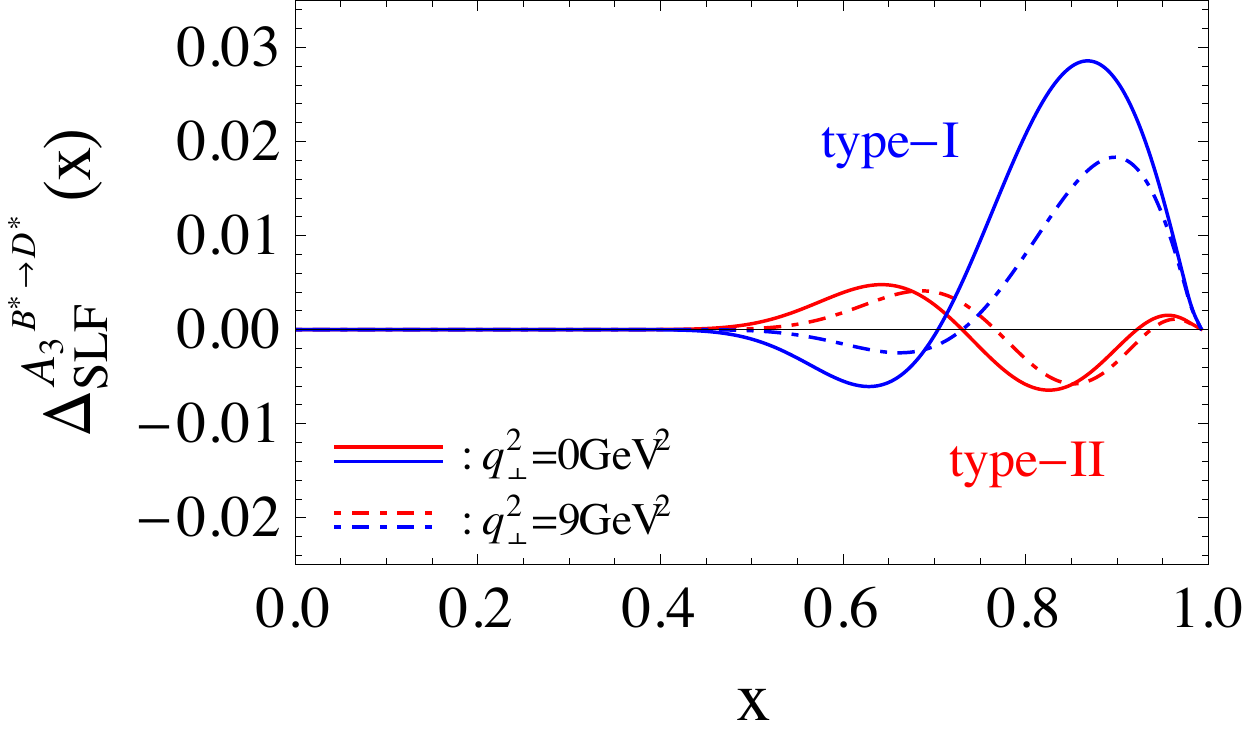}}\quad
\subfigure[]{\includegraphics[scale=0.4]{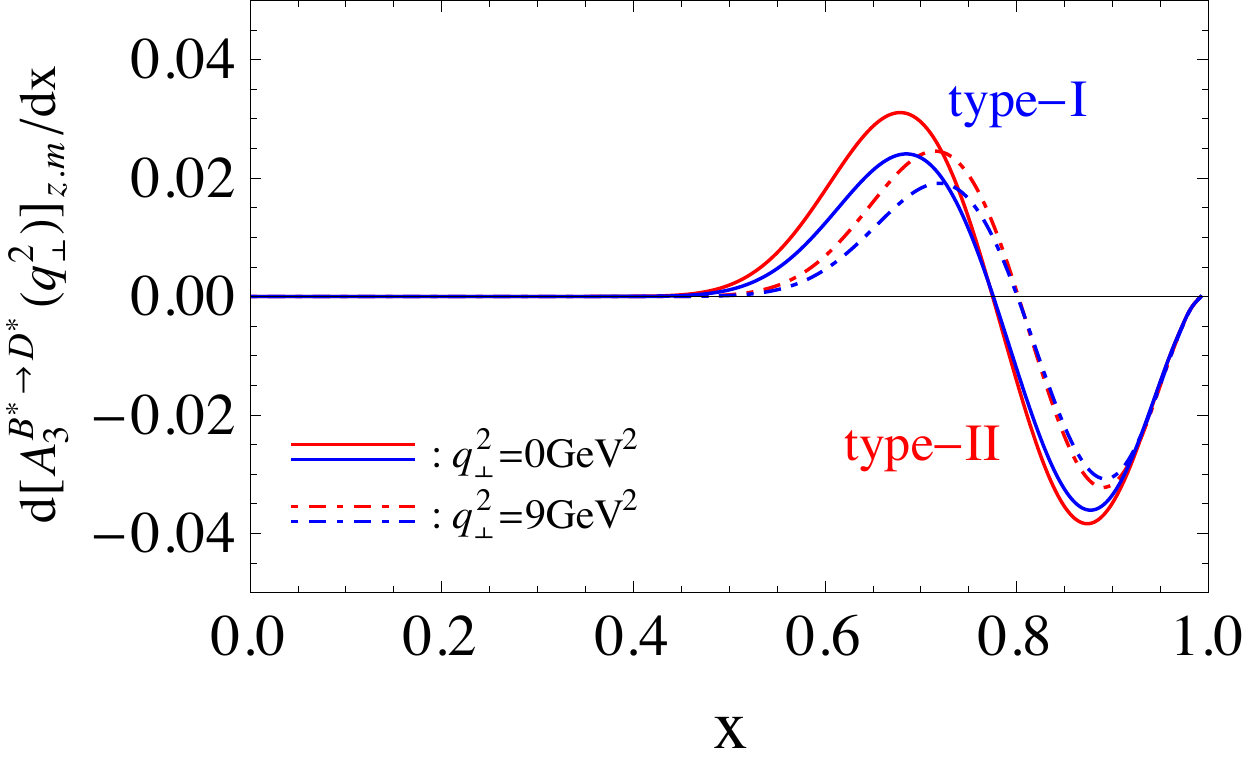}}\\
\subfigure[]{\includegraphics[scale=0.4]{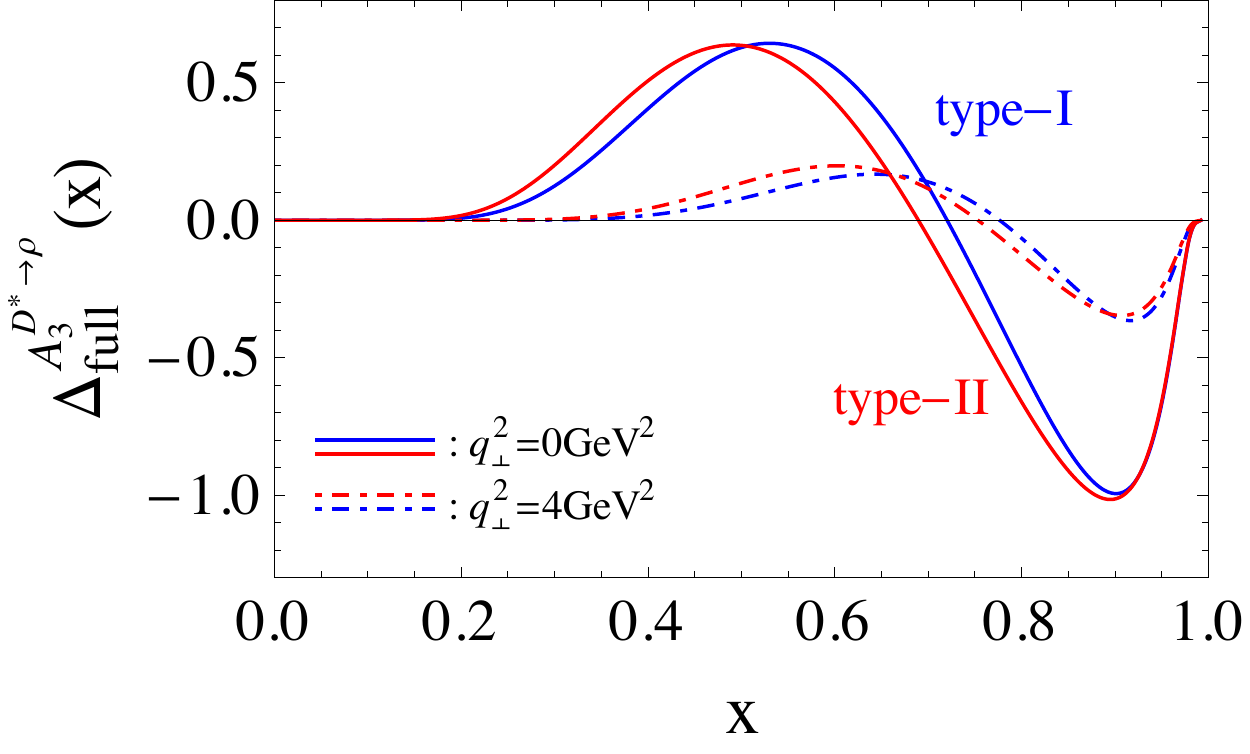}}\qquad
\subfigure[]{\includegraphics[scale=0.4]{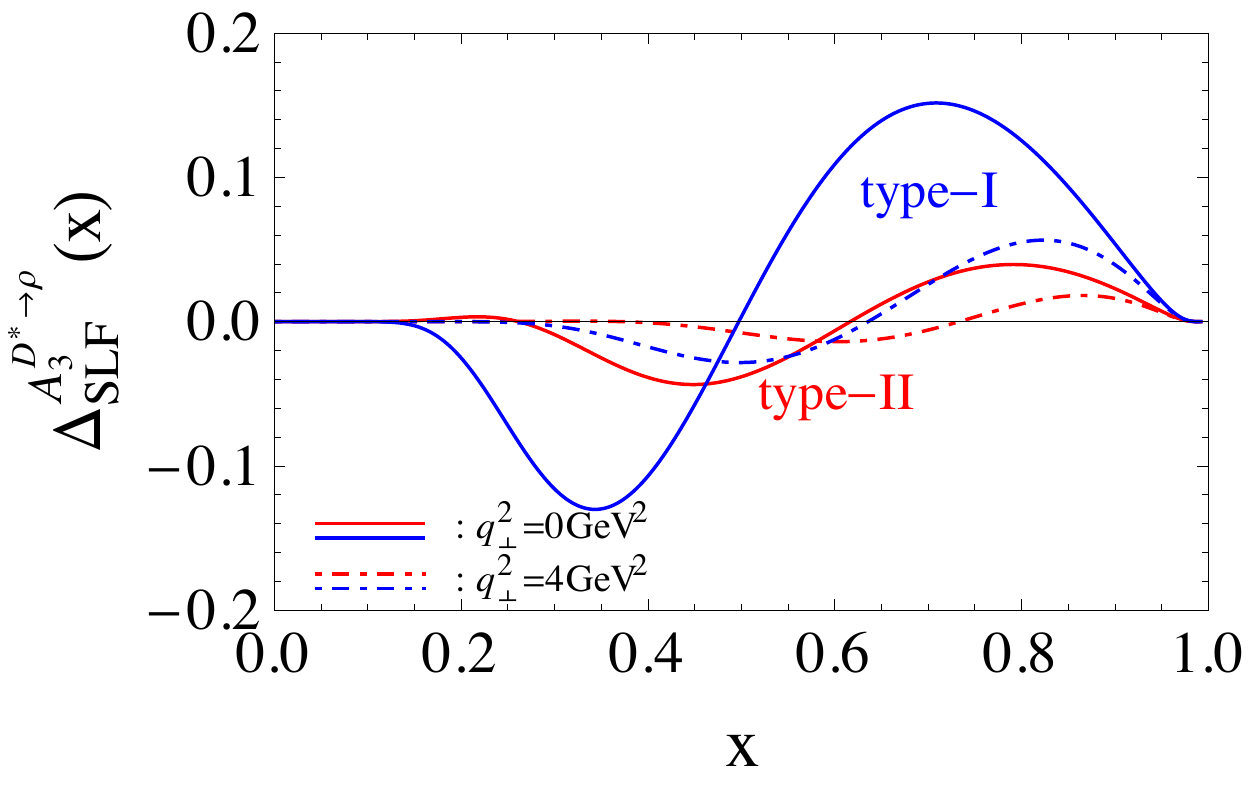}}\qquad
\subfigure[]{\includegraphics[scale=0.4]{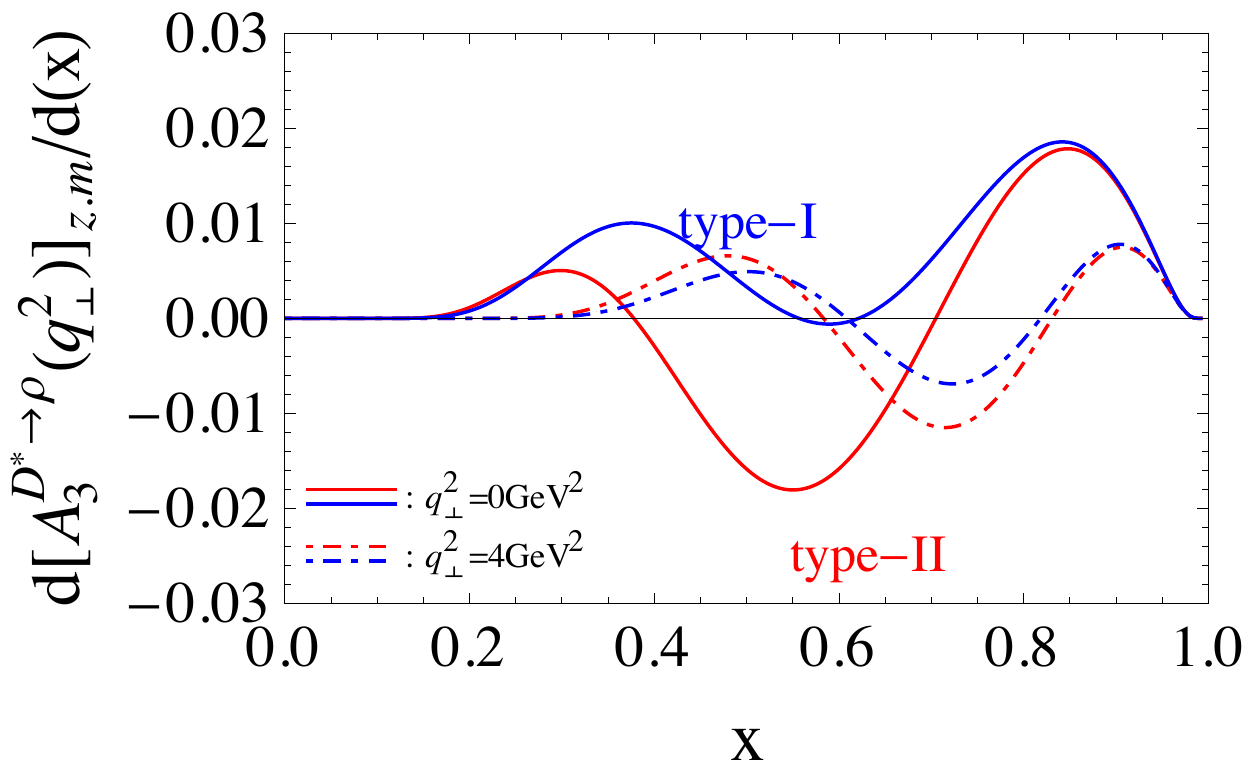}}\\
\end{center}
\label{fig:A3}
\end{figure}

Based on these formulas, we have following discussions and findings:
\begin{itemize}
\item  In Eq.~\eqref{eq:BBA}, the first and the second term presents contribution to $A_3$ and $A_4$, respectively; the other terms correspond to the unphysical form factors. For convenience of discussion, we take the first term as an example and name it as $[{\cal \widetilde{B}}_B^\u(\Gamma=\r^\u\r_5)]_{\rm term1}$.  It can be easily found  that $A_3$ could receive the contribution from $[{\cal \widetilde{B}}_B^\u(\Gamma=\r^\u\r_5)]_{\rm term1}$ written as
\begin{align}
{\widetilde{A}_3}^{\rm B}=4\frac{M'^2-M''^2}{ \epsilon''^{*}\cdot q}\,\frac{\omega\cdot \epsilon''^{*} }{\omega\cdot P}B_1^{(2)}\left(1+\frac{m_1'-m_1''-2m_2}{D_{V,{\rm con}}''}\right)\,,
\end{align}
which  is dependent on the choice of $\lambda''$, {\it i.e.},
\begin{equation}
{ \widetilde{A}_3}^{\rm B}=\left\{
             \begin{array}{lr}
           4\frac{M'^2-M''^2}{M'^2-M''^2+q_\bot^2}B_1^{(2)}\left(1+\frac{m_1'-m_1''-2m_2}{D_{V,{\rm con}}''}\right)\,, &  \qquad \lambda''=0 \\
             0\,.& \qquad \lambda''=\pm  \\
             \end{array}
\right.
\end{equation}
Further considering the fact that ${\cal \widetilde{F}}^{\rm CLF}$ is independent of the choices of $\lambda^{\prime\,,\prime\prime}$, it can be found that $A_3$ in the CLF QM suffers from the problem of self-consistence, $[A_3]^{\rm full}_{\lambda''=0}\neq [A_3]^{\rm full}_{\lambda''=\pm}$, except that $A_3^{\rm B}$ vanishes at least numerically which is equivalent to the condition
\begin{equation}
\int\frac{\d x\d^2{\bf k'_\bot}}{2(2\pi)^3}\frac{\chi_V'\chi_V''}{\bar x} B_1^{(2)}=0 \,,\qquad \int\frac{\d x\d^2{\bf k'_\bot}}{2(2\pi)^3}\frac{\chi_V'\chi_V''}{\bar x} \frac{B_1^{(2)}}{D_V''} =0 \,.
\end{equation}

In order to show clearly the performance of type-I and -II correspondence schemes, we take $B^*\to D^*$ transition as an example, and list the numerical results of $[A_3]^{\rm full}_{\lambda''=0}$, $[A_3]^{\rm full}_{\lambda''=\pm}$ and $[A_3]^{\rm CLF}$ at $\qb^2=(0,2,4,9)\,{\rm GeV^2}$ in Table~\ref{tab:A3}. In addition, we define the difference,
\begin{align}
\Delta^{\cal F}_{\rm full}(x) \equiv \frac{\d [{\cal F}]^{\rm full}_{\lbd''=0}}{ \d x}-\frac{\d [{\cal F}]^{\rm full}_{\lbd''=\pm}}{ \d x}\,,
\end{align}
which is equal to $N_c\int\frac{\d^2{\bf k'_\bot}}{2(2\pi)^3}\frac{\chi_V'\chi_V''}{\bar x} {\widetilde{A}_3}^{\rm B}$ for $A_3$, and show $\Delta^{A_3}_{\rm full}(x)$ for   $B^*\to D^*$ and  $D^*\to \rho$ transitions in Fig.~\ref{fig:A3}~(a) and (d). From these results, it can be easily found that such self-consistence condition is violated in the  traditional type-I  scheme, but can be satisfied by using the type-II scheme. Moreover, we have checked that all of the contributions of $B$ functions in Eqs.~\eqref{eq:BBA} and \eqref{eq:BBV} vanish numerically within  type-II scheme.

Therefore, it can be concluded that the CLF results for the form factors of $V'\to V''$ transition have the self-consistency problem within the  type-I correspondence scheme, but it can be resolved by employing  type-II  scheme and moreover the unphysical form factors~(for instance, the one corresponds to the third term in Eqs.~\eqref{eq:BBA}) vanish.

\item  In fact, the way to deal with the  contributions of $B$ functions is ambiguous. For instance,  instead of the treatment on $[{\cal \widetilde{B}}_B^\u(\Gamma=\r^\u\r_5)]_{\rm term1}$ in the last item, we can also decompose $[{\cal \widetilde{B}}_B^\u(\Gamma=\r^\u\r_5)]_{\rm term1}$ by using the identity
 \begin{align}\label{eq:dcomp}
P^\u  \frac{\e\cdot \w}{\w\cdot P}=&\e^\u-{\frac{q^\u}{q^2}}\left({\e\cdot q}-q\cdot P \frac{\w\cdot\e}{\w\cdot P}\right)-\frac{\w^\u}{\w\cdot P} \Big[ \e\cdot P-\e\cdot q \frac{q\cdot P}{q^2}-\e\cdot\w \frac{P^2}{\w\cdot P} \nonumber\\
&+\e\cdot\w\frac{(q\cdot P)^2}{q^2\w\cdot P} \Big]-\frac{i\lbd \,}{\,\w\cdot P}\frac{\e\cdot q}{q^2}\ve^{\u\a\b\v}\w_\a q_\b P_\v\,,
 \end{align}
where, $\e=\e''^*$. In Eq.~\eqref{eq:dcomp}, the second term vanishes in $[{\cal \widetilde{B}}_B^\u(\Gamma=\r^\u\r_5)]_{\rm term1}$, and the last two terms would introduce more unphysical form factors.  While, in this way, $A_3$ does not receive the contribution from $[{\cal \widetilde{B}}_B^\u(\Gamma=\r^\u\r_5)]_{\rm term1}$ anymore; such contribution, as well as the corresponding self-consistence problem,  transfers from $A_3$ to $A_2$ via the first term in Eq.~\eqref{eq:dcomp}.

Therefore, it is hard to determine which form factor the $[{\cal \widetilde{B}}_B^\u(\Gamma=\r^\u\r_5)]_{\rm term1}$  contributes to. This ambiguity results in significant uncertainty of CLF prediction, and thus is unacceptable. Fortunately, this problem exists only in the type-I scheme, and becomes trivial  in the type-II scheme because all of the contributions related to $B$ functions vanish numerically.

\item Besides of the self-consistency, the contributions of $B$ functions also result in the covariance problem because many terms in ${\cal \widetilde{B}}_B^\u(\Gamma=\r^\u\r_5)$ and ${\cal \widetilde{B}}_B^\u(\Gamma=\r^\u)$ are dependent on $\w$, which violates the Lorentz covariance of ${\cal {B}}^\u$. It can be clearly seen from Eqs.~\eqref{eq:BBA} and \eqref{eq:BBV}. The covariance problem caused by $B$ function contributions can not be avoided in the type-I scheme, but does not exist in  the type-II scheme since, as has mentioned above, $B$ function contributions exist only in form and vanish numerically.
\end{itemize}

Besides of the CLF QM, the SLF QM also suffers from  the problem of self-consistency. Again, we take $A_3$ as an example. In the section 2.2,  $\lbd'=\lbd''=\pm$ is taken in the calculation of $A_3$, and the result of $[{A}_{3}]^{\rm SLF}_{\lbd'=\lbd''=\pm}$ given by Eq.~\eqref{eq:A3SLF1} is obtained.  Instead of such choice, one can also take $\lbd'=0$ and $\lbd''=\pm$. In this way, we can obtain
 \begin{align}
\widetilde{A}_{3}^{{\rm SLF}}\big |_{\lbd'=0, \lbd''=\pm}&=\frac{-2(M'^2-M''^2)}{x\bar{x}M'M'_0 q^2}\bigg\{\Big[{\bf k}_\bot'^2+\bar{x}{\bf k'_\bot\cdot q_\bot}-2\frac{({\bf k'_\bot\cdot q_\bot})^2}{{\bf q}_\bot^2}\Big]\left({\bf k}_\bot'^2+\bar{x}m_1'^2+xm_2^2+x\bar{x}M_0'^2\right)\nonumber\\
&+\frac{1}{D_{V,{\rm LF}}''}\Big[\Big(2\frac{({\bf k'_\bot\cdot q_\bot})^2}{{\bf q}_\bot^2}-{\bf k}_\bot'^2-\bar{x}{\bf k'_\bot\cdot q_\bot}\Big)\Big(\left(m_1'+m_1''\right)\left({\bf k}_\bot'^2+xm_2^2+\bar{x}m_1'm_1''+x\bar{x}M_0'^2\right)\nonumber\\
&+\bar{x}m_2m_1'^2-\bar{x}m_2m_1''^2+\bar{x}^2\left(m_1'-m_2\right){\bf q}_\bot^2\Big)+x\bar{x}{\bf k_\bot'\cdot k_\bot''}{\bf q}_\bot^2\left(m_1'-m_2\right)\nonumber\\
&-{\bf k_\bot''\cdot q_\bot}\left[m_1'{\bf k}_\bot'^2+\left(\bar{x}m_1'+xm_2\right)\left(m_1'm_2+x\bar{x}M_0'^2\right)\right]\nonumber\\
&-\bar{x}\left(m_1'-m_2\right)\Big(4\frac{\left({\bf k'_\bot\cdot q_\bot}\right)^3}{{\bf q}_\bot^2}-{\bf k''_\bot\cdot q_\bot}{\bf k}_\bot'^2-2{\bf k'_\bot\cdot q_\bot}{\bf k}_\bot'^2-2\bar{x}\left({\bf k'_\bot\cdot q_\bot}\right)^2\Big)\Big]\nonumber\\
&-\frac{1}{2x\bar{x}D_{V,{\rm LF}}'D_{V,{\rm LF}}''}\left[\left(x-\bar{x}\right){\bf k}_\bot'^2-\bar{x}m_1'^2+xm_2^2-x\bar{x}\left(\bar{x}-x\right)M_0'^2
\right]\Big[{\bf k_\bot''\cdot q_\bot}\Big(x{\bf k}_\bot'^2\nonumber\\
&-\bar{x}^2m_1'^2+x^2m_2^2\Big)+4\bar{x}\frac{\left({\bf k'_\bot\cdot q_\bot}\right)^3}{{\bf q}_\bot^2}-2\bar{x}^2\left({\bf k'_\bot\cdot q_\bot}\right)^2 -x\bar{x}{\bf k_\bot'\cdot k_\bot''}{\bf q}_\bot^2-\bar{x}{\bf k''_\bot\cdot q_\bot}{\bf k}_\bot'^2\nonumber\\
&-2\bar{x}{\bf k'_\bot\cdot q_\bot}{\bf k}_\bot'^2+\left(2\frac{({\bf k'_\bot\cdot q_\bot})^2}{{\bf q}_\bot^2}-{\bf k}_\bot'^2-\bar{x}{\bf k'_\bot\cdot q_\bot}\right)\left(\bar{x}m_1'^2-\bar{x}m_1''^2-\bar{x}^2{\bf q}_\bot^2\right)
\Big]
\bigg\}\,.\label{eq:A3SLF2}
\end{align}
Comparing Eq.~\eqref{eq:A3SLF1} with Eq.~\eqref{eq:A3SLF2}, it can be found that $[{A}_{3}]^{\rm SLF}_{\lbd'=\lbd''=\pm}$ and $[{A}_{3}]^{\rm SLF}_{\lbd'=0, \lbd''=\pm}$ are  different from each other, which implies that the problem of self-consistency exists possibly also in the traditional SLF QM. In order to verify that, we take $B^*\to D^*$ transition as an example and list the numerical results of  $[{A}_{3}]^{\rm SLF}_{\lbd'=\lbd''=\pm}$ and $[{A}_{3}]^{\rm SLF}_{\lbd'=0, \lbd''=\pm}$  in Table~\ref{tab:A3}; meanwhile,  we also show the difference, $\Delta^{A_3}_{\rm SLF}(x)$, defined as
\begin{align}
\Delta^{ A_3}_{\rm SLF}(x) \equiv \frac{\d [{A_3}]^{\rm SLF}_{\lbd'=0, \lbd''=\pm}}{ \d x}-\frac{\d [{A_3}]^{\rm SLF}_{\lbd'=\lbd''=\pm}}{ \d x}\,,
\end{align}
for $B^*\to D^*$ and  $D^*\to \rho$ transitions in  Fig.~\ref{fig:A3}~(b)and (e).  Form these results,  it can be easily found that   $[{A}_{3}]^{\rm SLF}_{\lbd'=\lbd''=\pm}\neq [{A}_{3}]^{\rm SLF}_{\lbd'=0, \lbd''=\pm}$ in the traditional SLF QM~(named as type-I SLF QM for convince of discussion ); and meanwhile, it is interesting that the self-consistence can be recovered, because $[{A}_{3}]^{\rm SLF}_{\lbd'=\lbd''=\pm}\dot = [{A}_{3}]^{\rm SLF}_{\lbd'=0, \lbd''=\pm}$ numerically,  when an additional replacement $M\to M_0$~(named as type-II SLF QM) is taken. Such replacement is also the main difference between type-I and -II correspondence schemes in the CLF QM.

% Therefore, we can conclude that the replacement $M\to M_0$ is required by not only the CLF QM but also the SLF QM due to the self-consistency.

Combining the findings mentioned above, we can conclude that the replacement $M\to M_0$ is necessary for the strict self-consistency and covariance of the CLF QM, as well as for the  self-consistency of the SLF QM.  This implies possibly that the effect of interaction has not  yet been taken into account in a proper way at least in the SLF QM,
which is easy to be understood since $\hat{M}^2=M_0^2+\hat{I}$~($\hat{M}$ and $\hat{I}$ denote mass and interaction operators, respectively) in the LF dynamics.  Therefore, the formulas for form factors with $M\to M_0$ should be treated  as the results only at ``leading-order'' approximation or in the zero-binding-energy limit with ``dressed'' constituents~\cite{Choi:2013mda,Choi:2014ifm,Choi:2017uos,Choi:2017zxn}. Further, mapping the CLF result to the corresponding  SLF one, the type-II correspondence scheme is expected to be obtained\footnote{The CLF vertex  obtained by mapping to the SLF QM is not the only choice for the CLF QM, while, if such  vertex  is used, the other correspondences should be applied simultaneously for consistence. At this moment, the CLF QM can be treated as a covariant expression for the SLF QM but with the zero-mode contributions taken into account.   }, which will be checked in the following.  In the mapping, in order to obtain the complete correspondence and avoid the effects of zero-mode  contribution, one should use the valence contribution, $[{\cal F}]^{\rm val.}$, in the CLF QM instead  of  choosing only some special zero-mode independent quantities\footnote{The traditional type-I correspondence is obtained via zero-mode independent $f_P$ and/or $f_+^{P\to P}$. The LF results of these quantities are very simple and their integrands are irrelevant to $M$; as a result,  the traditional type-I scheme limits $M\to M_0$ only in the $D$ factor, {\it i.e.} $D_{V,{\rm  con}} \to D_{V,{\rm  LF}}$, which is possibly incomplete.     }.  Here, we take $A_3$ as an example again. Its valence result can be written as
\begin{align}
\widetilde{A}_3^{\rm{ val.}}=&4\Bigg\{\frac{{\bf k}_\bot''\cdot{\bf q}_\bot}{{\bf q}_\bot^2}\left(\frac{{\bf k}_\bot'^2+m_2^2}{\bar{x}}-\bar{x}M'^2\right)-{\bf k}_\bot'\cdot{\bf k}_\bot''+\frac{1}{D_{V,{\rm con}}{''}}\bigg[{\bf k}_\bot'\cdot{\bf k}_\bot''\left(m_1'+m_1''\right)\nonumber\\
&-\frac{{\bf k}_\bot''\cdot{\bf q}_\bot}{{\bf q}_\bot^2}\left(\frac{{\bf k}_\bot'^2+m_2^2}{\bar{x}}\left(m_1'+m_1''\right)+\left(xm_2-\bar{x}m_1''\right)M'^2-\left(\bar{x}m_1'+xm_2\right)M''^2\right)
\bigg]\nonumber\\
&-\frac{1}{D_{V,{\rm con}}'D_{V,{\rm con}}''}\bigg[{\bf k}_\bot'\cdot{\bf k}_\bot''\left(\frac{{\bf k}_\bot'^2+m_2^2}{\bar{x}}-\bar{x}M'^2\right)+\frac{{\bf k}_\bot''\cdot{\bf q}_\bot}{{\bf q}_\bot^2}{\bf k}_\bot'^2\left(M'^2-M''^2\right)\nonumber\\
&+2\bar{x}\left({\bf k}_\bot'\cdot{\bf q}_\bot\right)^2-\bar{x}{\bf k}_\bot'^2{\bf q}_\bot^2-{\bf k}_\bot'\cdot{\bf q}_\bot{\bf k}_\bot'^2
\bigg]
\Bigg\}\,.\label{eq:A3CLFV}
\end{align}
Then, comparing $[{A}_3]^{\rm SLF}$ with $[{A}_3]^{\rm val.}$~({\it i.e.}, Eqs.~(\ref{eq:FSLF},\ref{eq:A3SLF1}) with  Eqs.~(\ref{eq:FCLF},\ref{eq:A3CLFV})),  the type-II correspondence can be easily obtained. In other words, one can find that $[{A}_3]^{\rm SLF}=[{A}_3]^{\rm  val.}$ in form within type-II correspondence scheme. This confirms again the finding,
\begin{align}
[{\cal O}]^{\rm SLF}=[{\cal O}]^{\rm  val.}\,,
\end{align}
obtained via $f_{V,A}$ and form factors of $P\to V$ transition in our previous works~\cite{Chang:2018zjq,Chang:2019mmh}.

The zero-mode contributions to a form factor can be obtained via $[{\cal F}]^{\rm CLF}=[{\cal F}]^{\rm val.}+[{\cal F}]^{\rm z.m.}$. In order to clearly show the effect of zero-mode contribution, we take $A_3^{B^*\to D^*\,,D^*\to \rho}$ as examples and plot the dependence of $\d [{\cal F}]^{\rm z.m.}/\d x$ on $x$ in Fig.~\ref{fig:A3}~(c) and (f).  It can be seen from these Figs that  zero-mode presents nonzero contributions within the traditional type-I correspondence scheme; while, in the  type-II correspondence scheme, these contributions, although existing formally, vanish numerically, {\it i.e.},  $[A_3(q^2)]_{\rm z.m.}\dot{=} 0$~(type-II), because the contribution with small $x$ and the one with large $x$  cancel each other out exactly at each ${\qb^2}$ point. From such finding, one can further conclude that
 \begin{equation}\label{eq:CLFval}
[{\cal O}]^{\rm  val.}\,\dot{=}\,[{\cal O}]^{\rm  CLF}\,,
\end{equation}
which can also be found from the numerical example given in Table~\ref{tab:A3}. This confirms  Eq.~\eqref{eq:sc} mentioned in the introduction.
%obtained via $f_{V,A}$ and form factors of $P\to V$ transition in our previous works~\cite{Chang:2018zjq,Chang:2019mmh}.,

 \begin{figure}[t]
\caption{ $q^2$~(in unit of $\rm GeV^2$) dependence of form factors of $c\to (s,q)$ induced $D^*\to (K^*\,,\rho)$, $D^*_s\to (\phi\,,K^*)$, $J/\Psi\to (D^*_s\,,D^*)$,  $B^*_c\to (B^*_s\,,B^*)$ transitions with the parameterization scheme given by Eq.~\eqref{eq:para2}. The dots in the space-like region are the results obtained directly via LFQMs, and the lines are fitting results.}
\begin{center}
\subfigure{\includegraphics[scale=0.24]{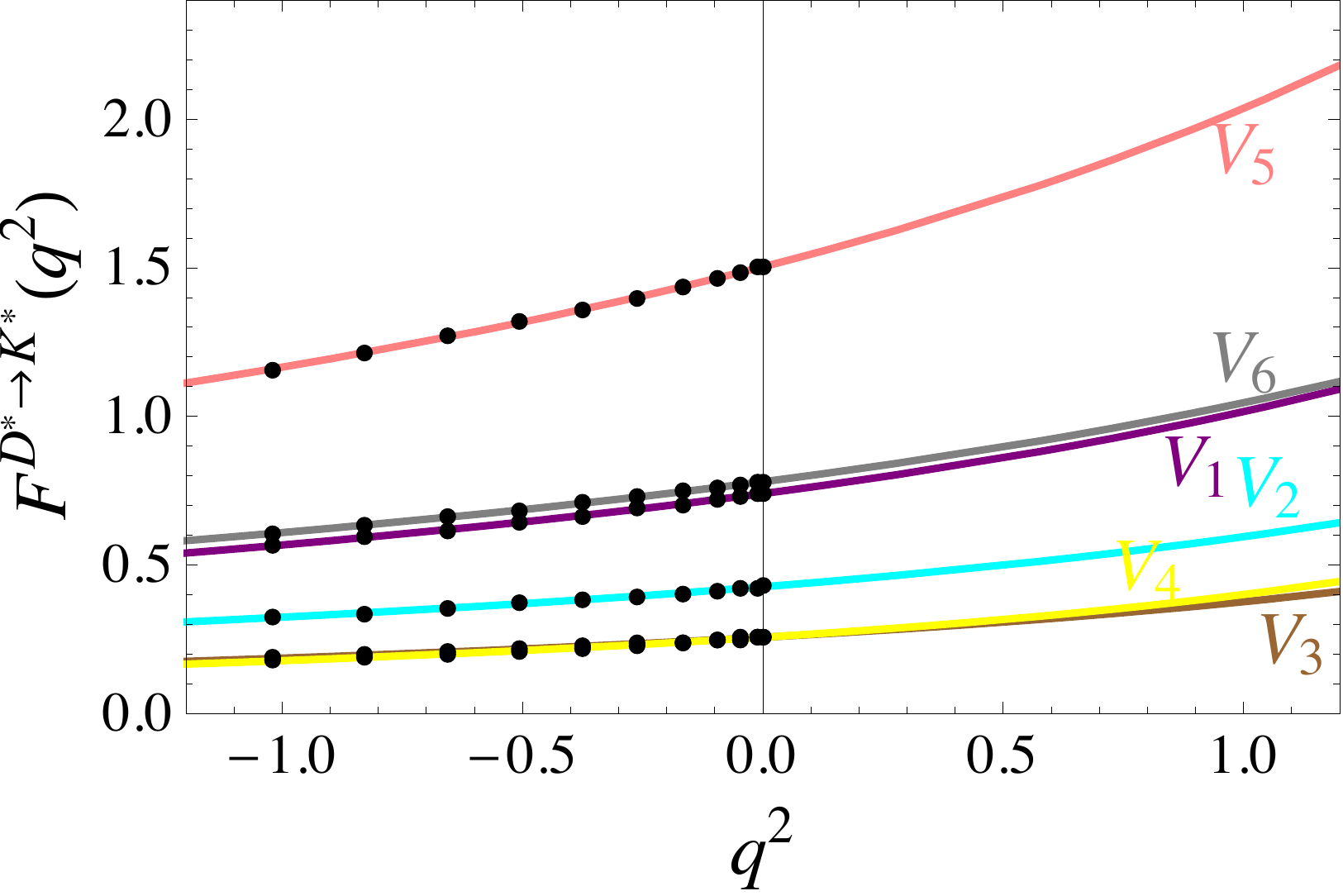}}\,
\subfigure{\includegraphics[scale=0.24]{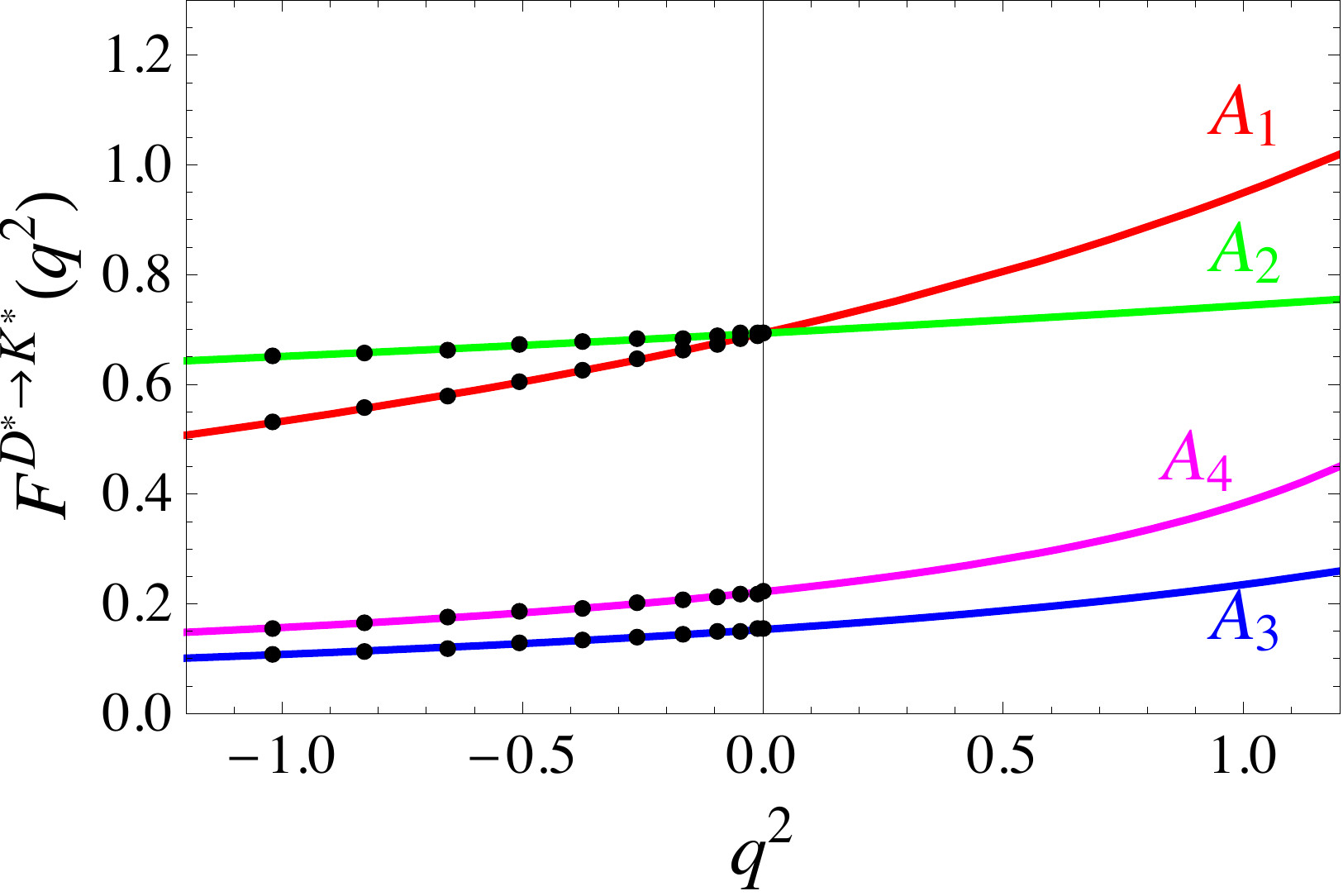}}\,
\subfigure{\includegraphics[scale=0.24]{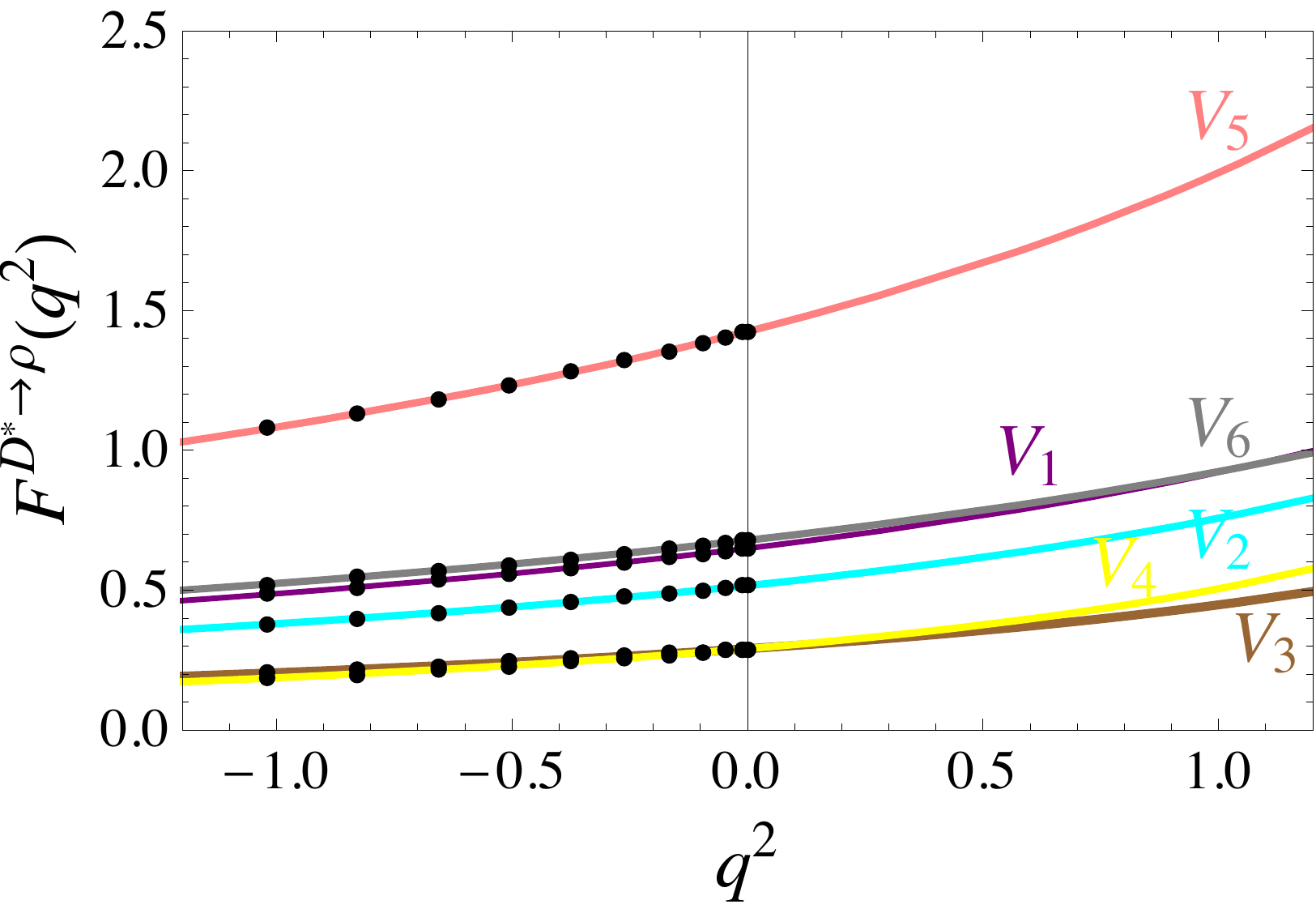}}\,
\subfigure{\includegraphics[scale=0.24]{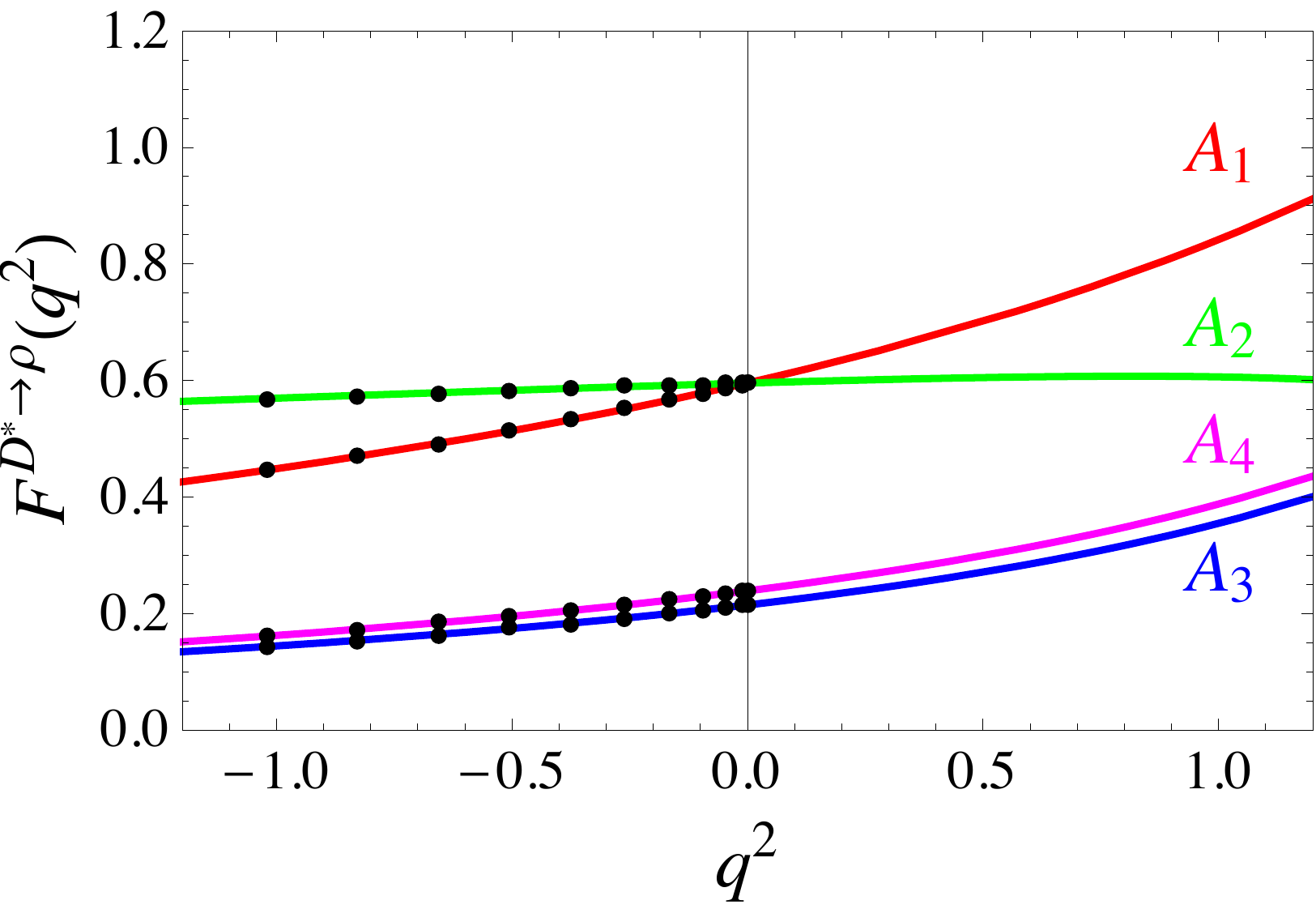}}\\
\subfigure{\includegraphics[scale=0.245]{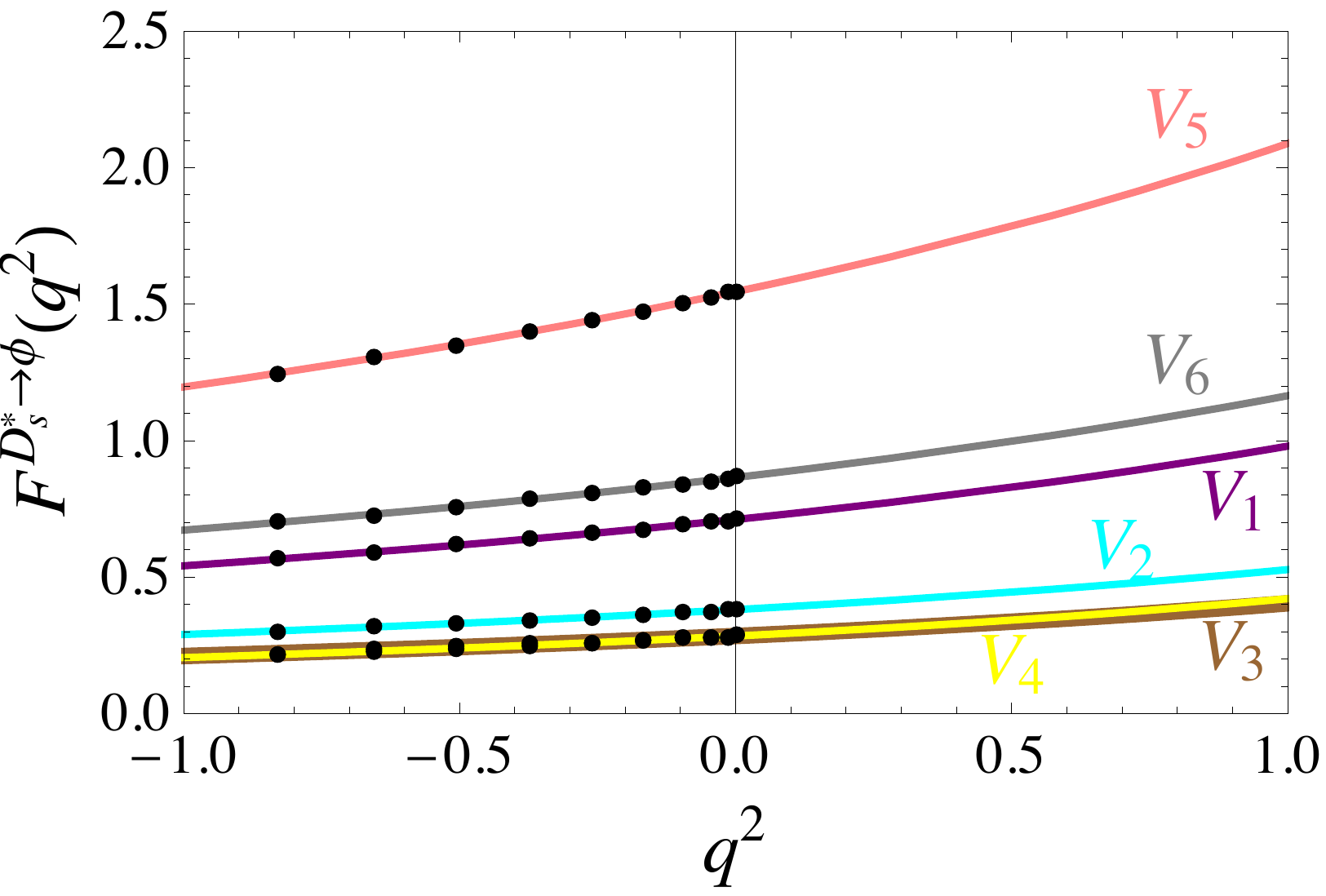}}\,
\subfigure{\includegraphics[scale=0.245]{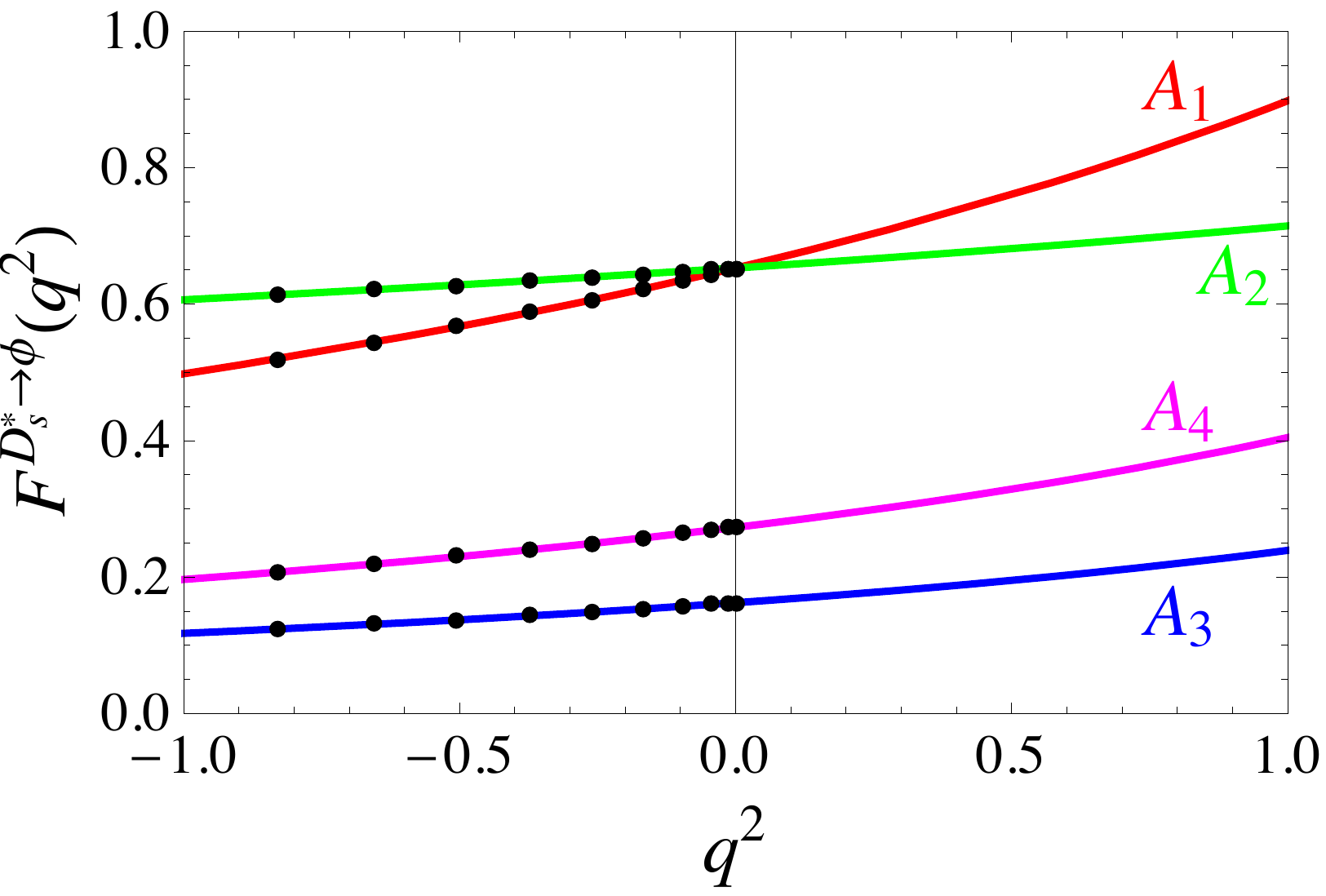}}\,
\subfigure{\includegraphics[scale=0.26]{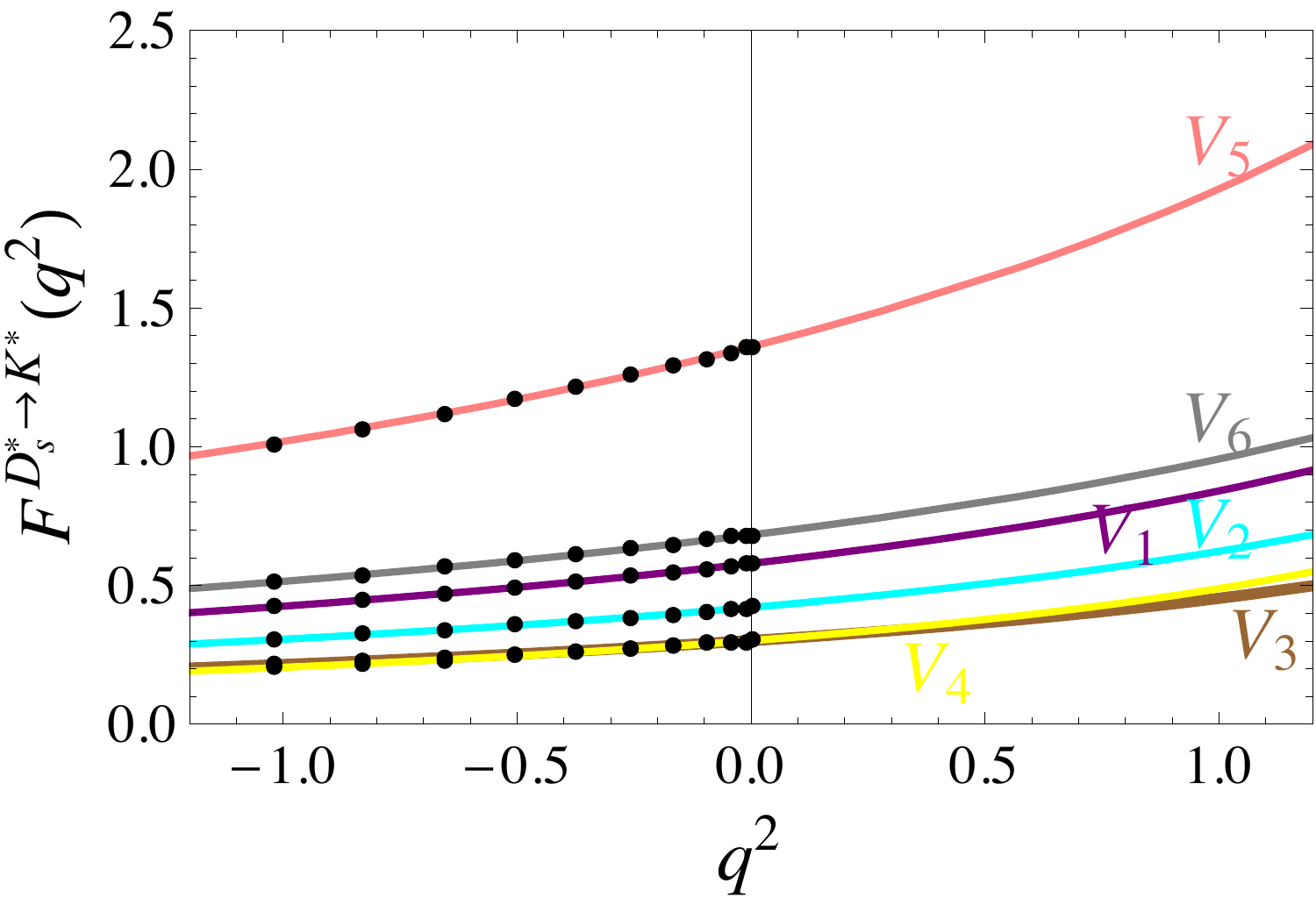}}\,
\subfigure{\includegraphics[scale=0.24]{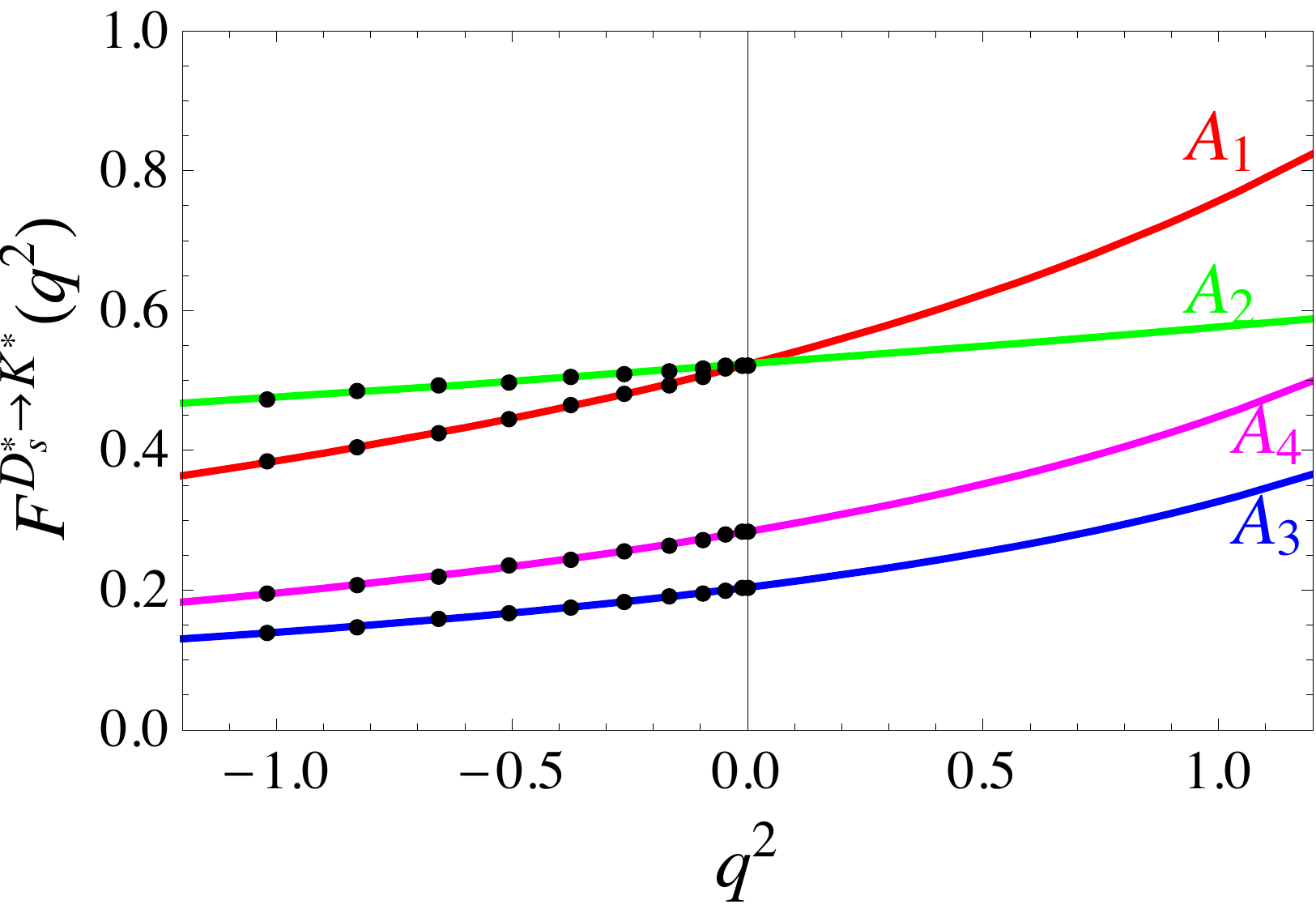}}\\
\subfigure{\includegraphics[scale=0.24]{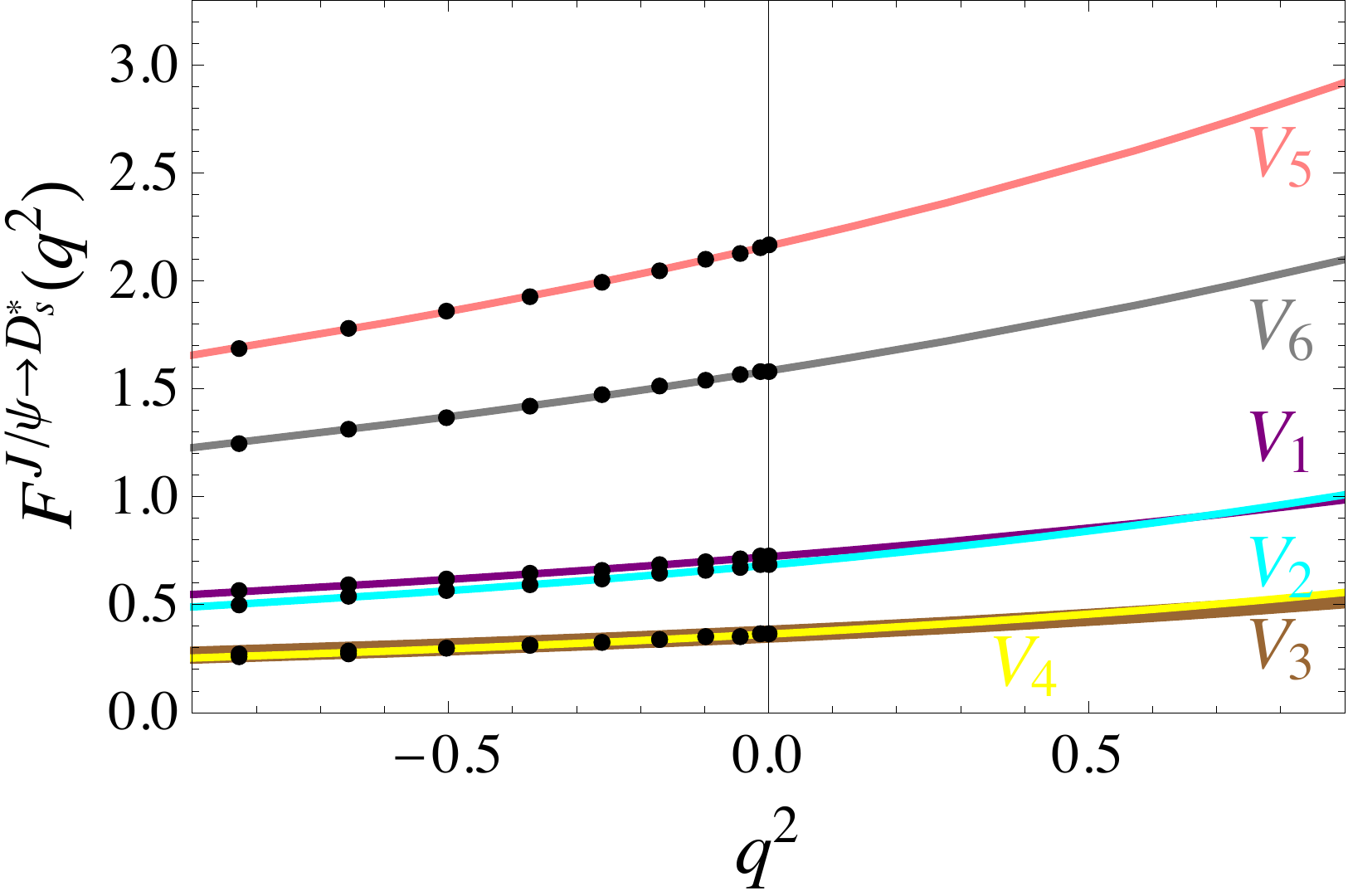}}\,
\subfigure{\includegraphics[scale=0.24]{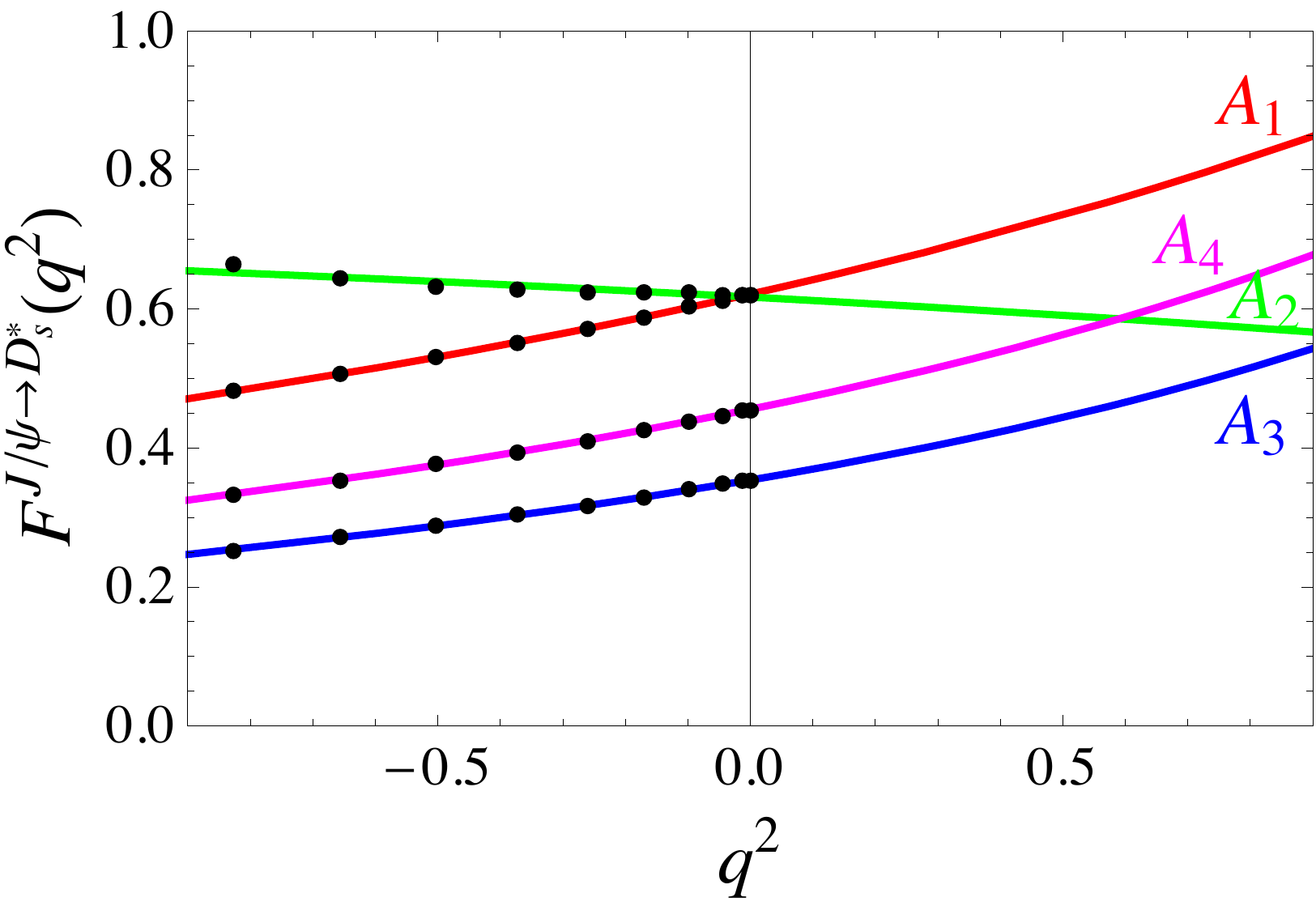}}\,
\subfigure{\includegraphics[scale=0.25]{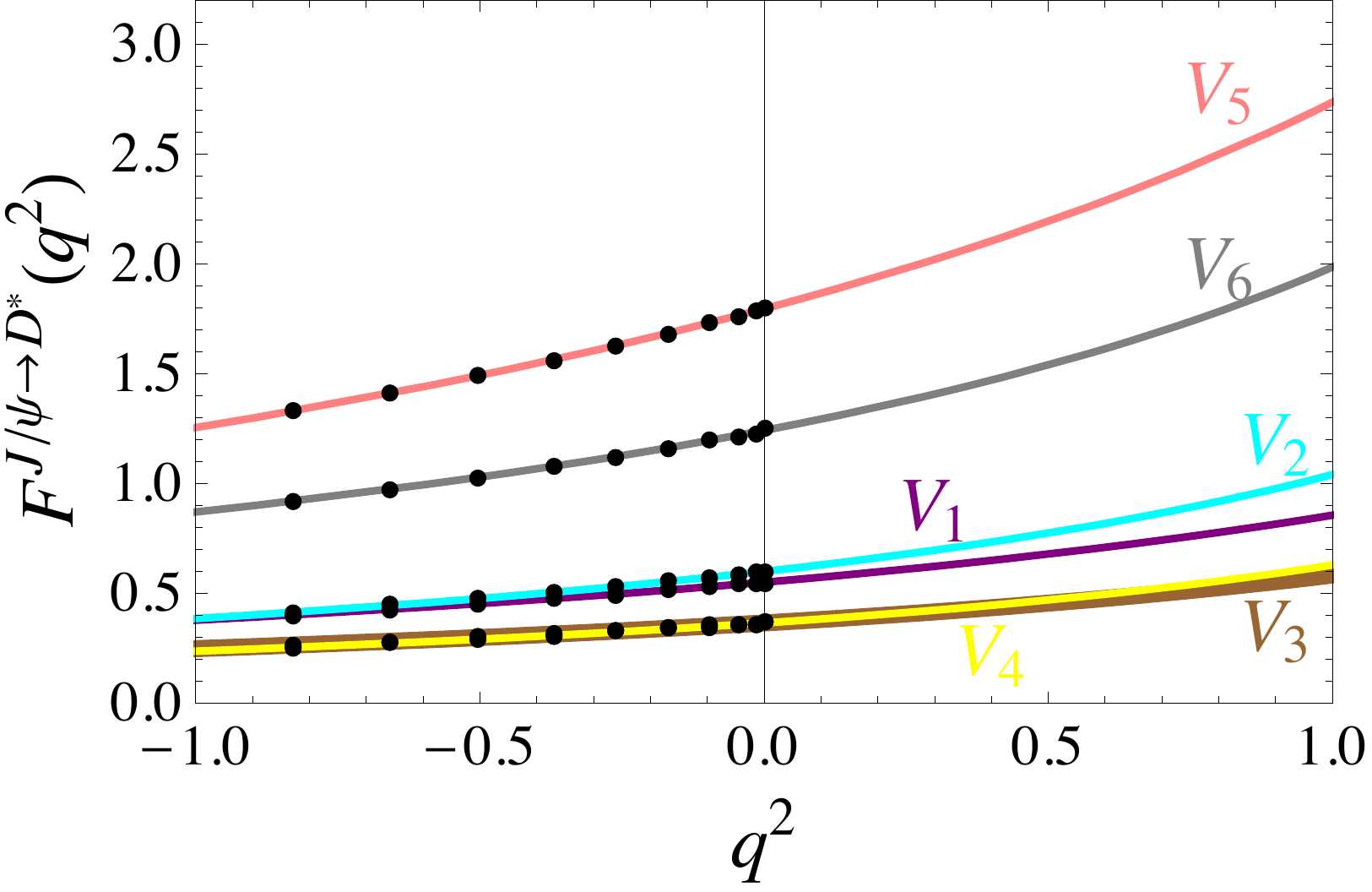}}\,
\subfigure{\includegraphics[scale=0.25]{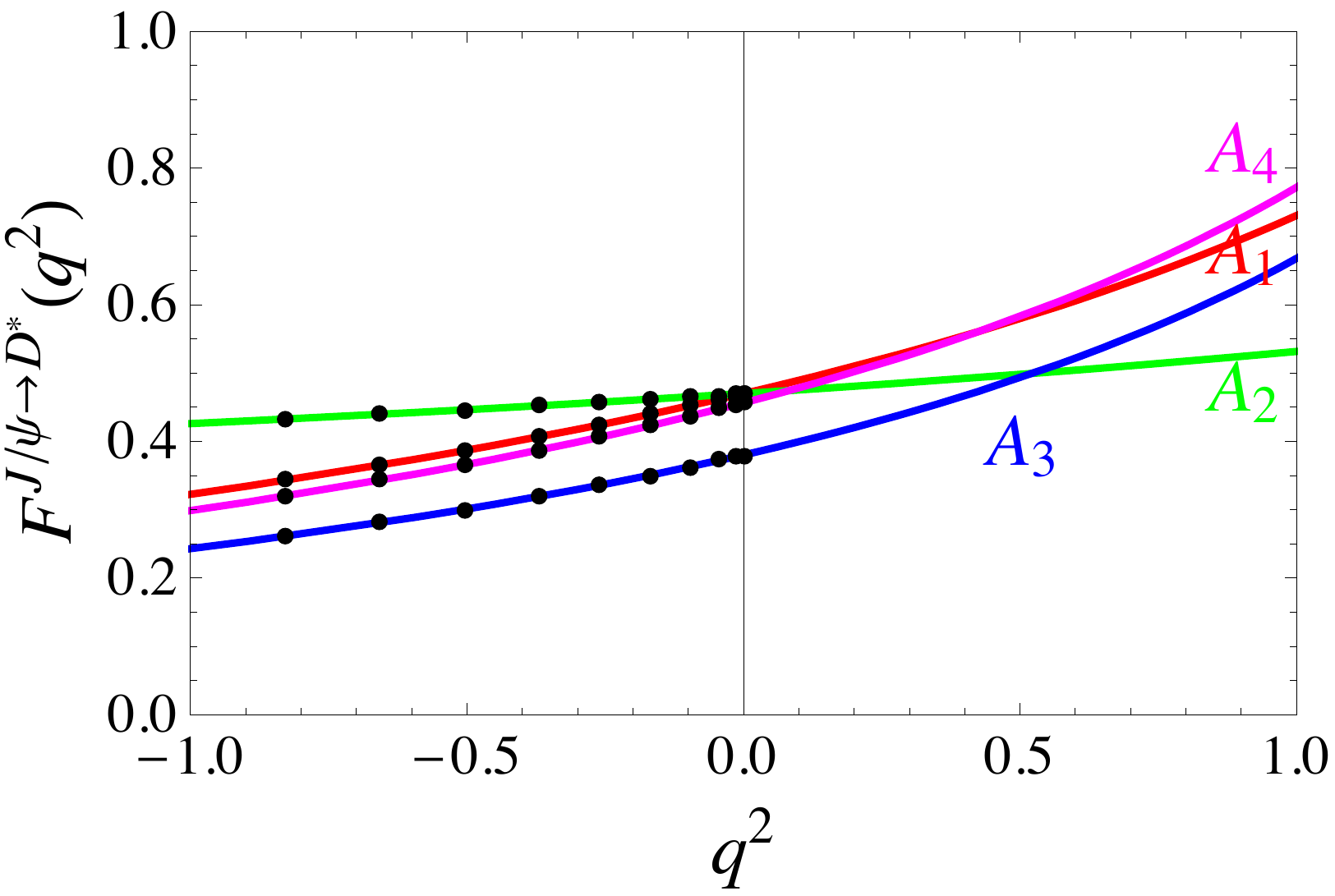}}\\
\subfigure{\includegraphics[scale=0.26]{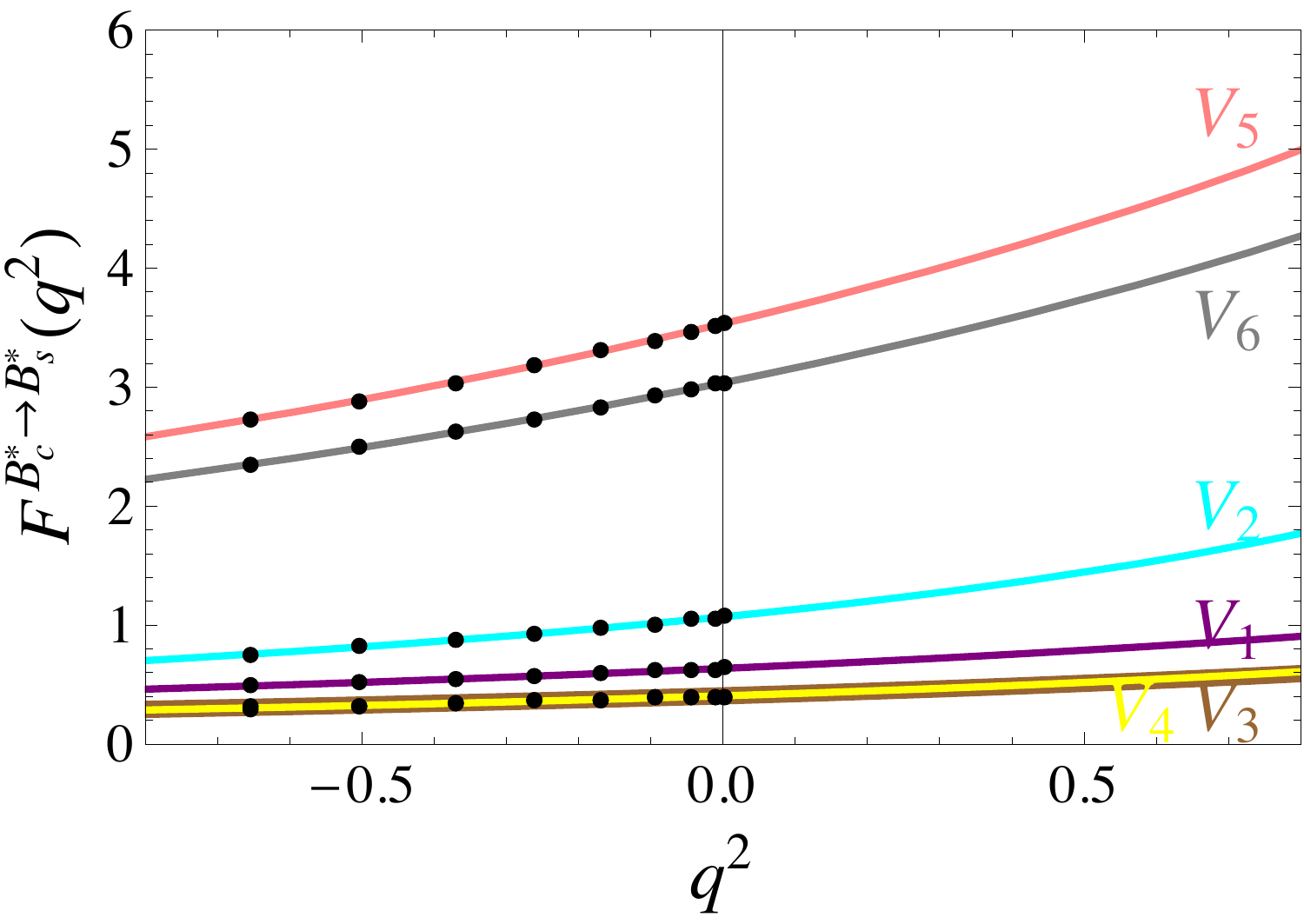}}\,
\subfigure{\includegraphics[scale=0.25]{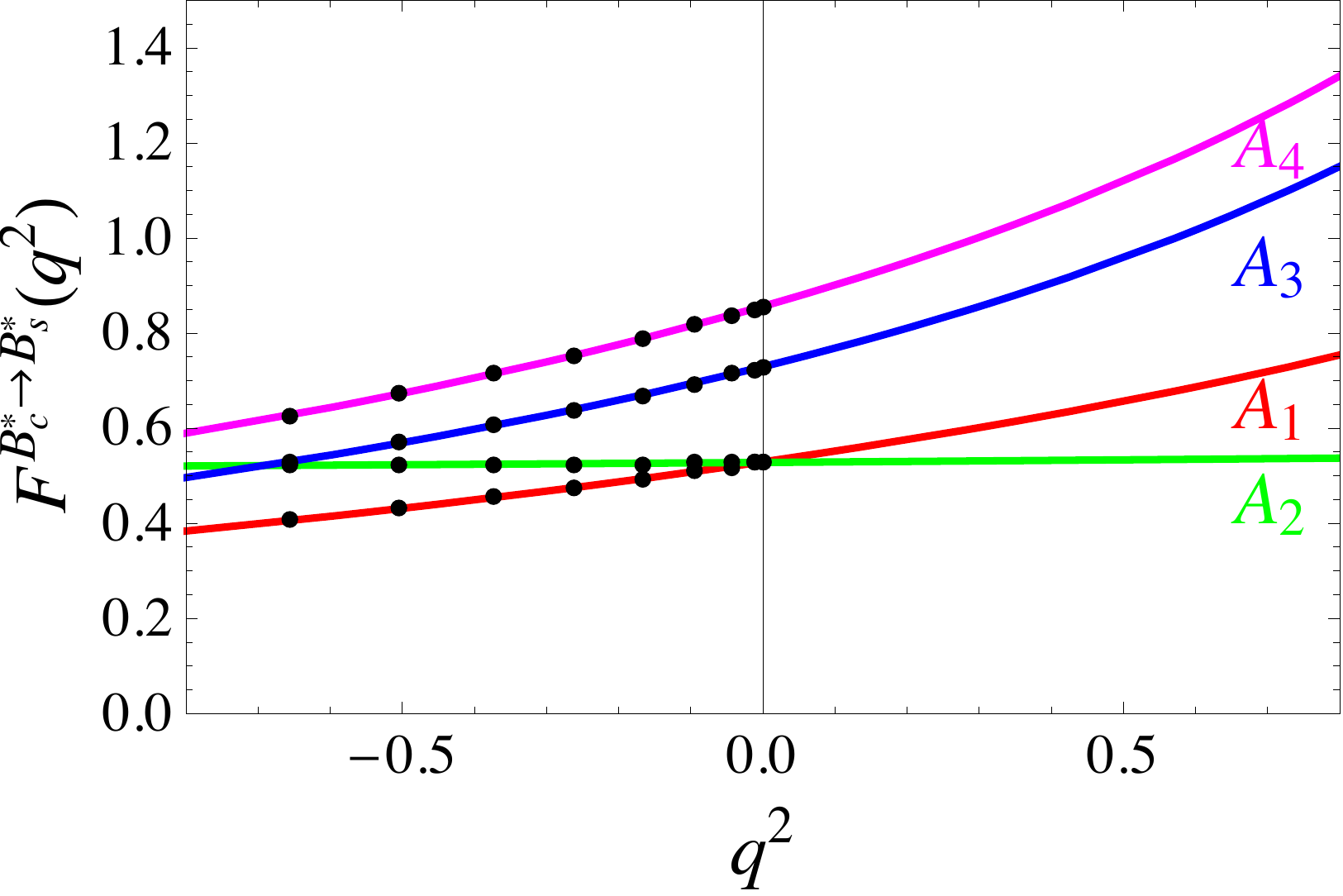}}\,
\subfigure{\includegraphics[scale=0.28]{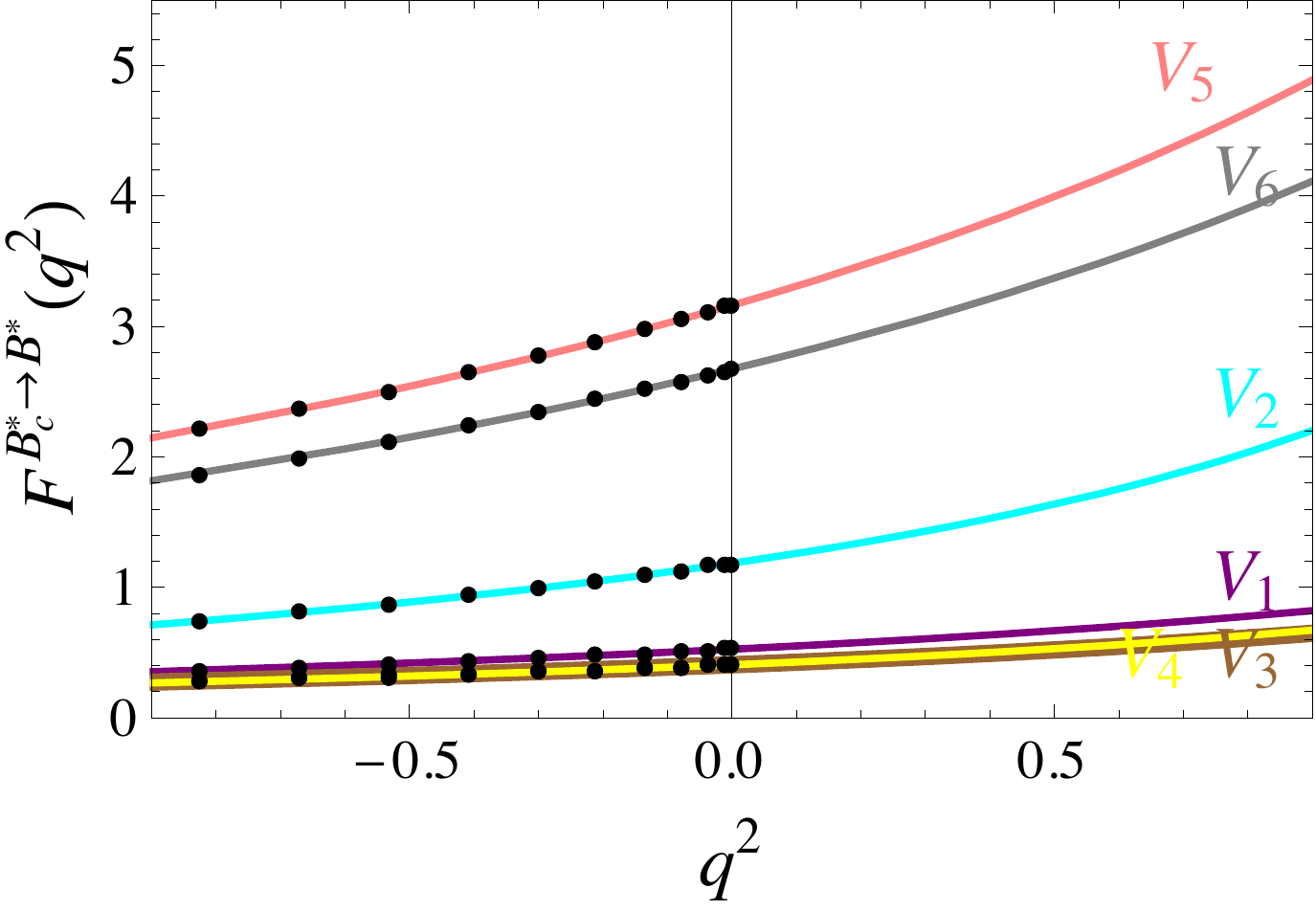}}\,
\subfigure{\includegraphics[scale=0.29]{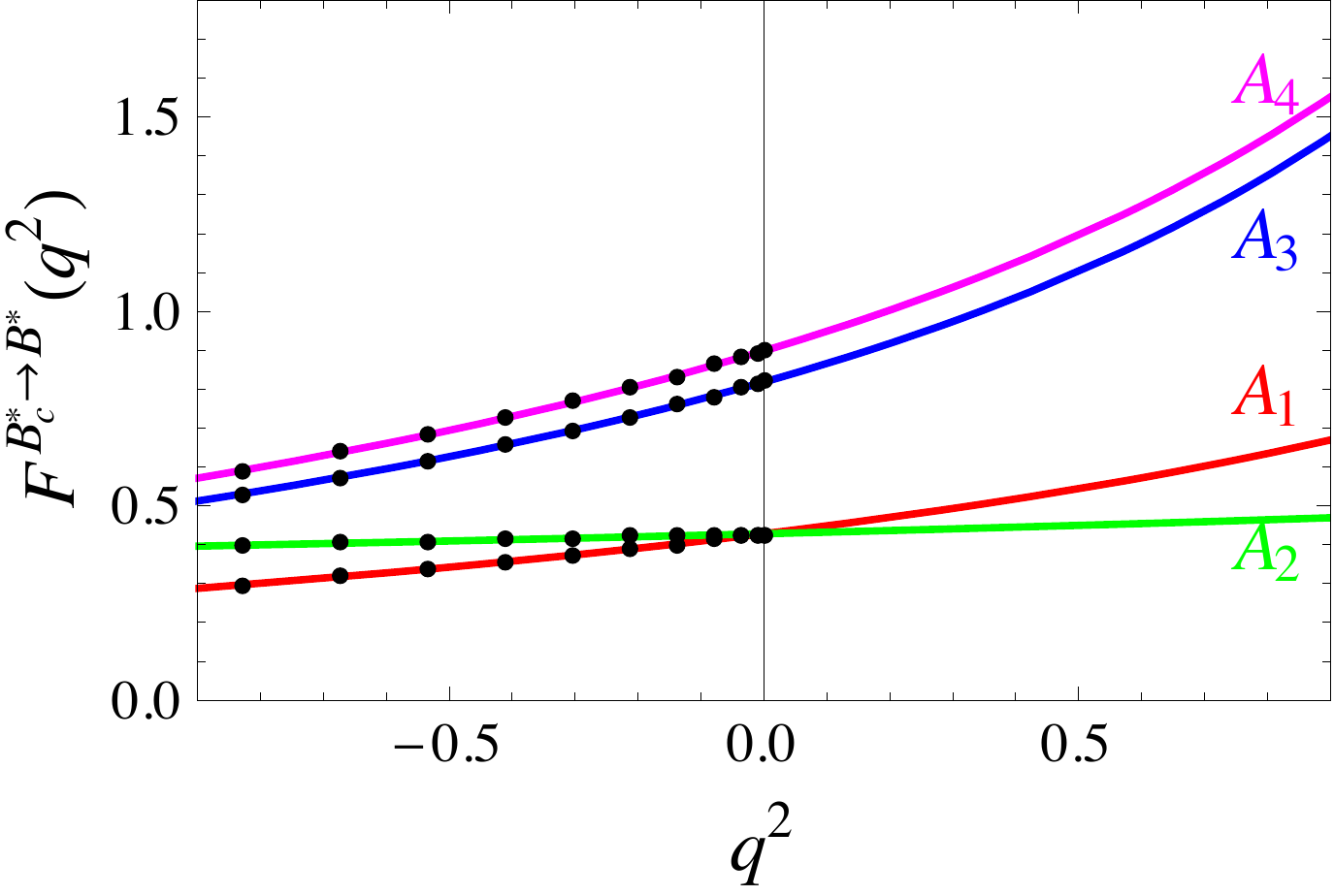}}\\
\end{center}
\label{fig:sche2c}
\end{figure}

 \begin{figure}[t]
\caption{ $q^2$~(in unit of $\rm GeV^2$) dependence of form factors of $b\to (c,s,q)$ induced $B^*\to (D^*\,,K^*\,,\rho)$, $B^*_s\to (D^*_s\,,\phi\,,K^*)$, $B^*_c\to (J/\Psi\,,D^*_s\,,D^*)$, $\Upsilon(1S)\to (B_c^*\,,B_s^*\,,B^*)$ transitions with the parameterization scheme given by Eq.~\eqref{eq:para2}. The other captions are the same as in Fig.~\ref{fig:sche2c}.}
\begin{center}
\subfigure{\includegraphics[scale=0.25]{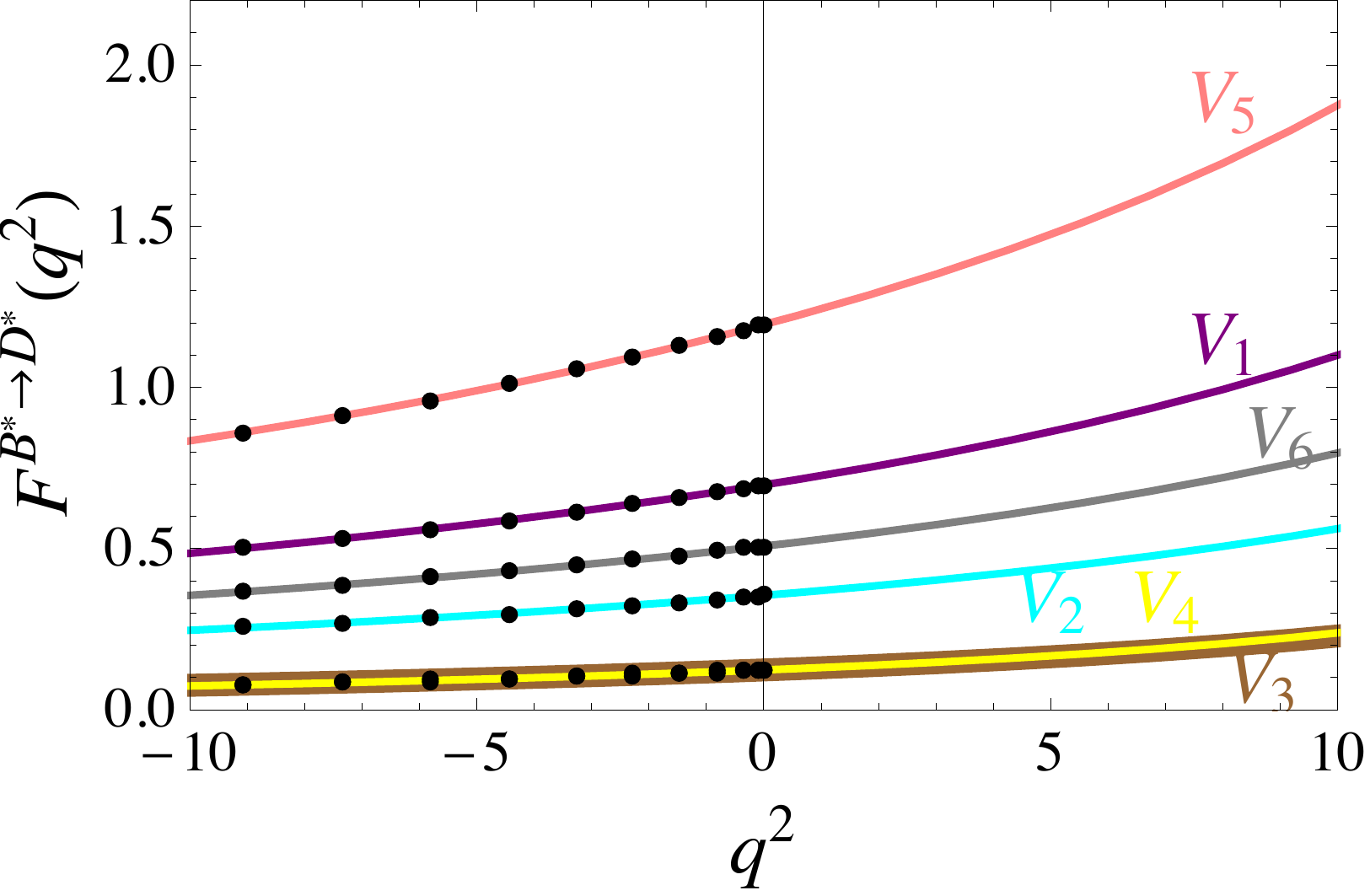}}\,
\subfigure{\includegraphics[scale=0.25]{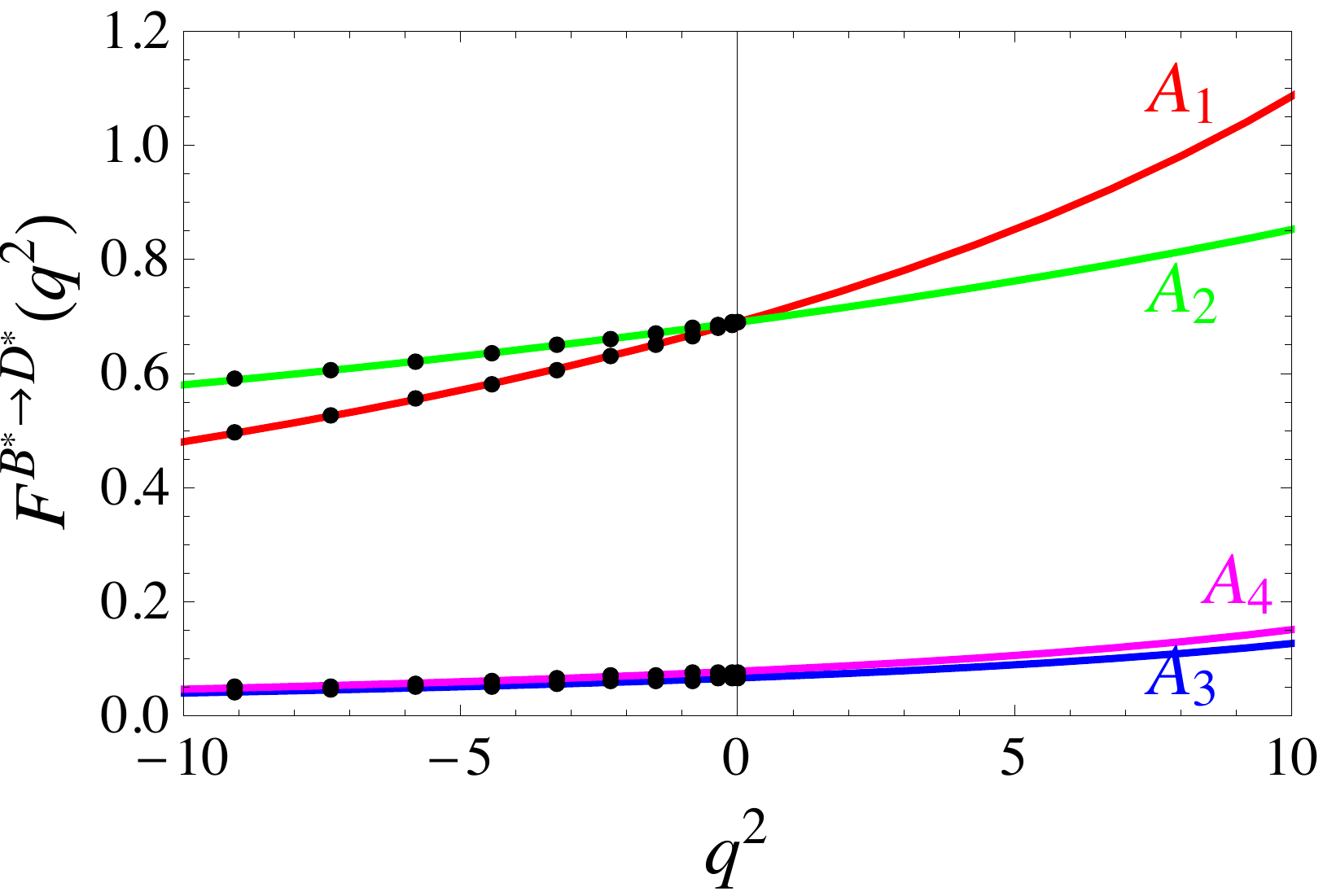}}\,
\subfigure{\includegraphics[scale=0.24]{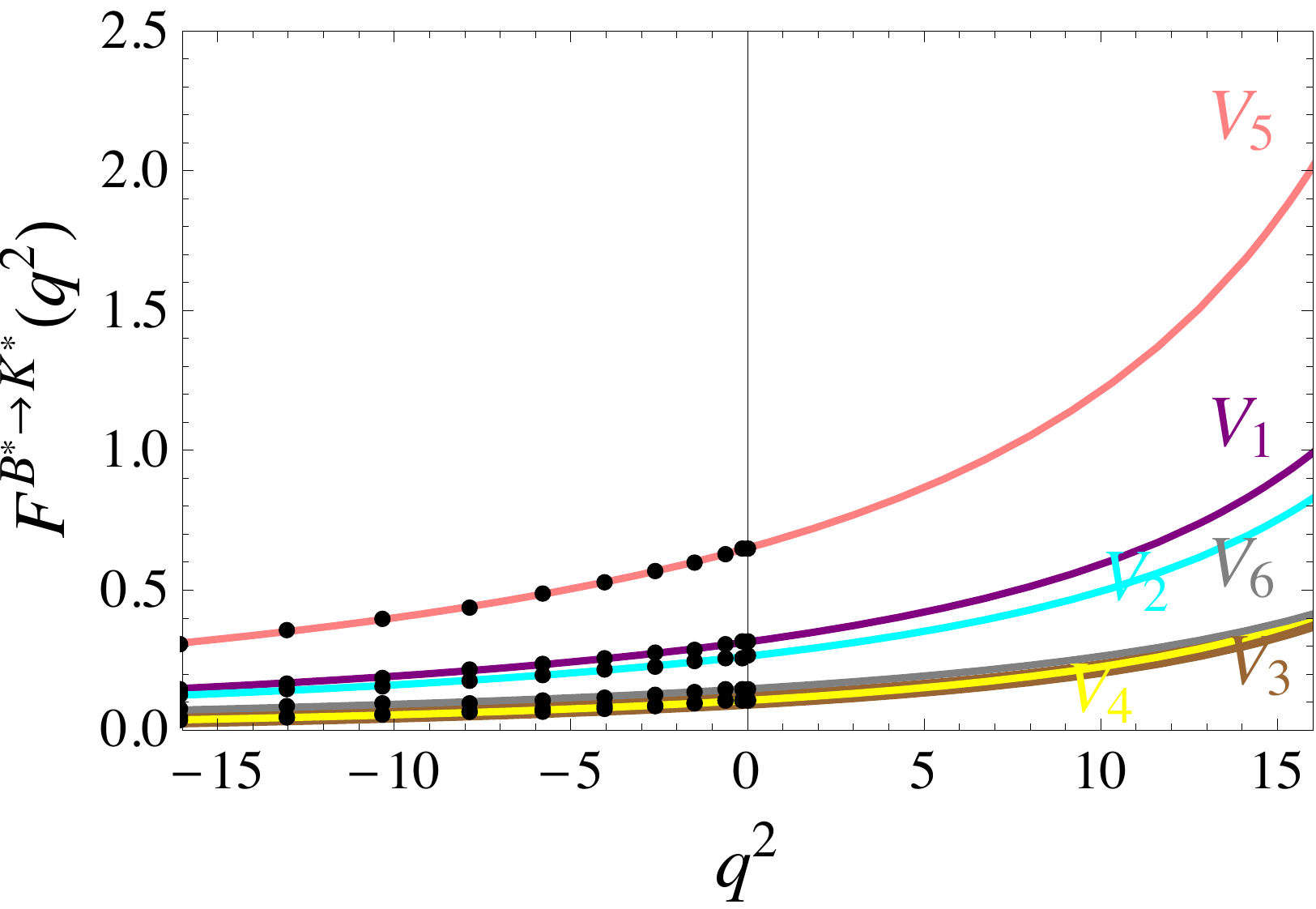}}\,
\subfigure{\includegraphics[scale=0.24]{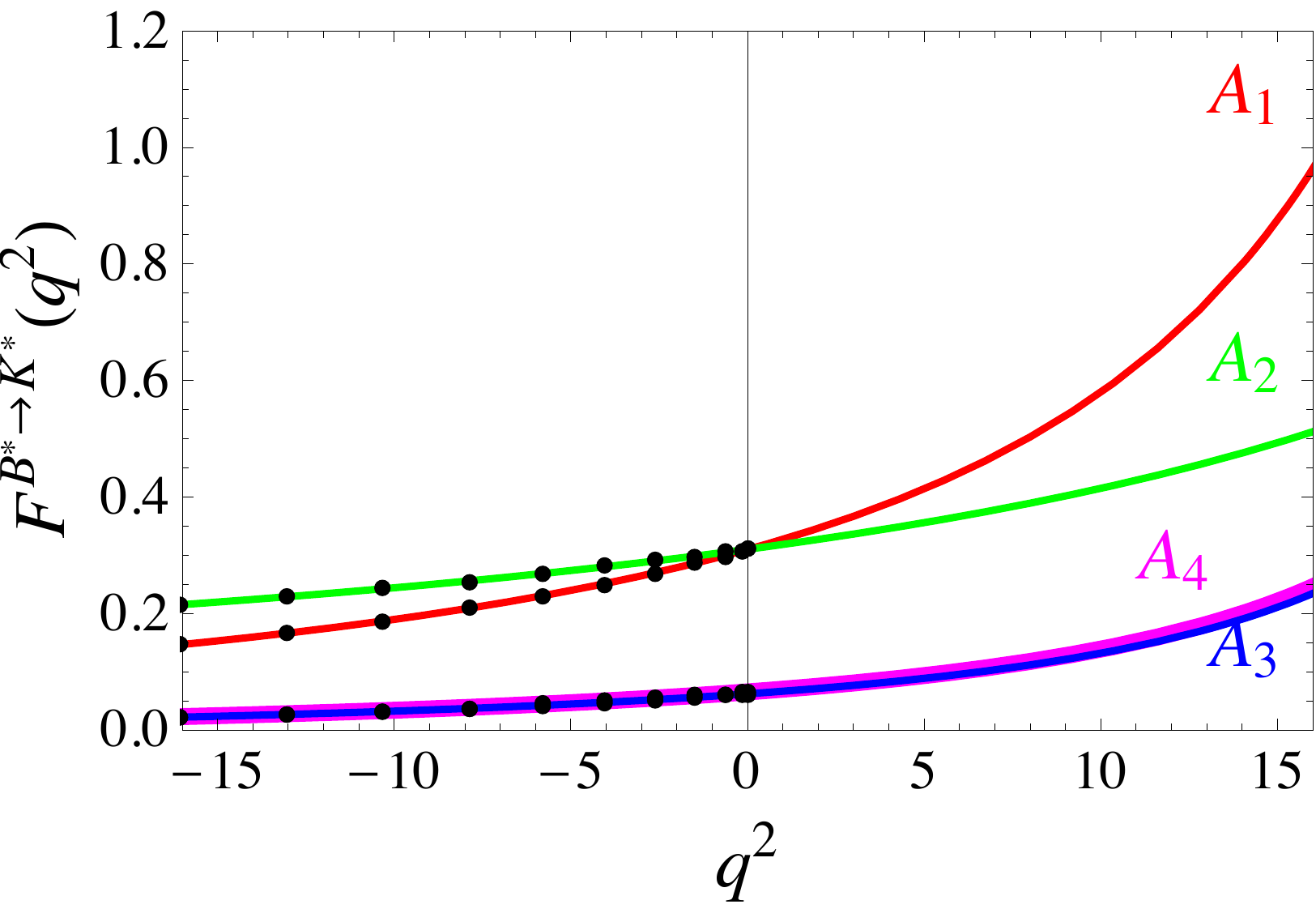}}\\
\subfigure{\includegraphics[scale=0.25]{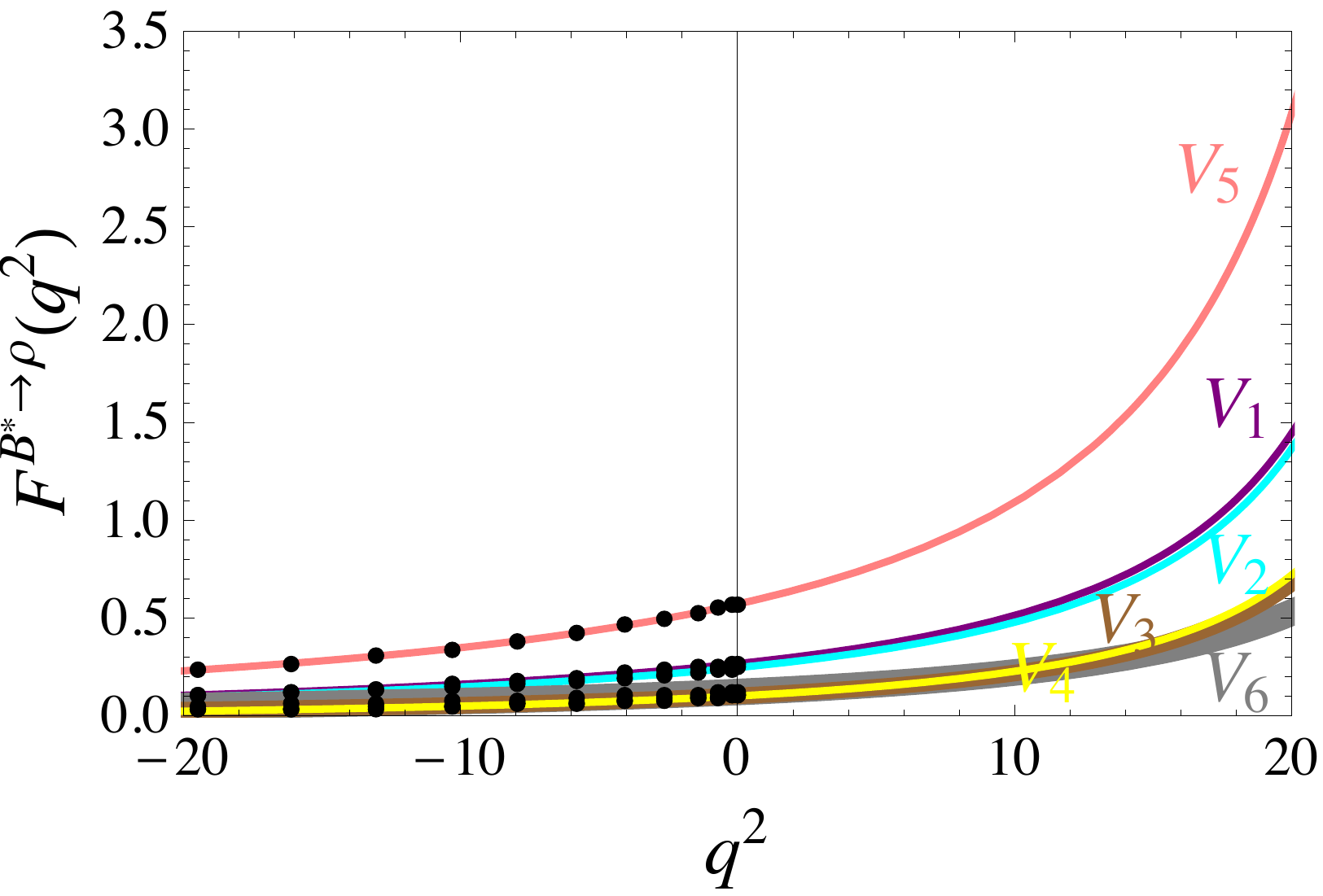}}\,
\subfigure{\includegraphics[scale=0.25]{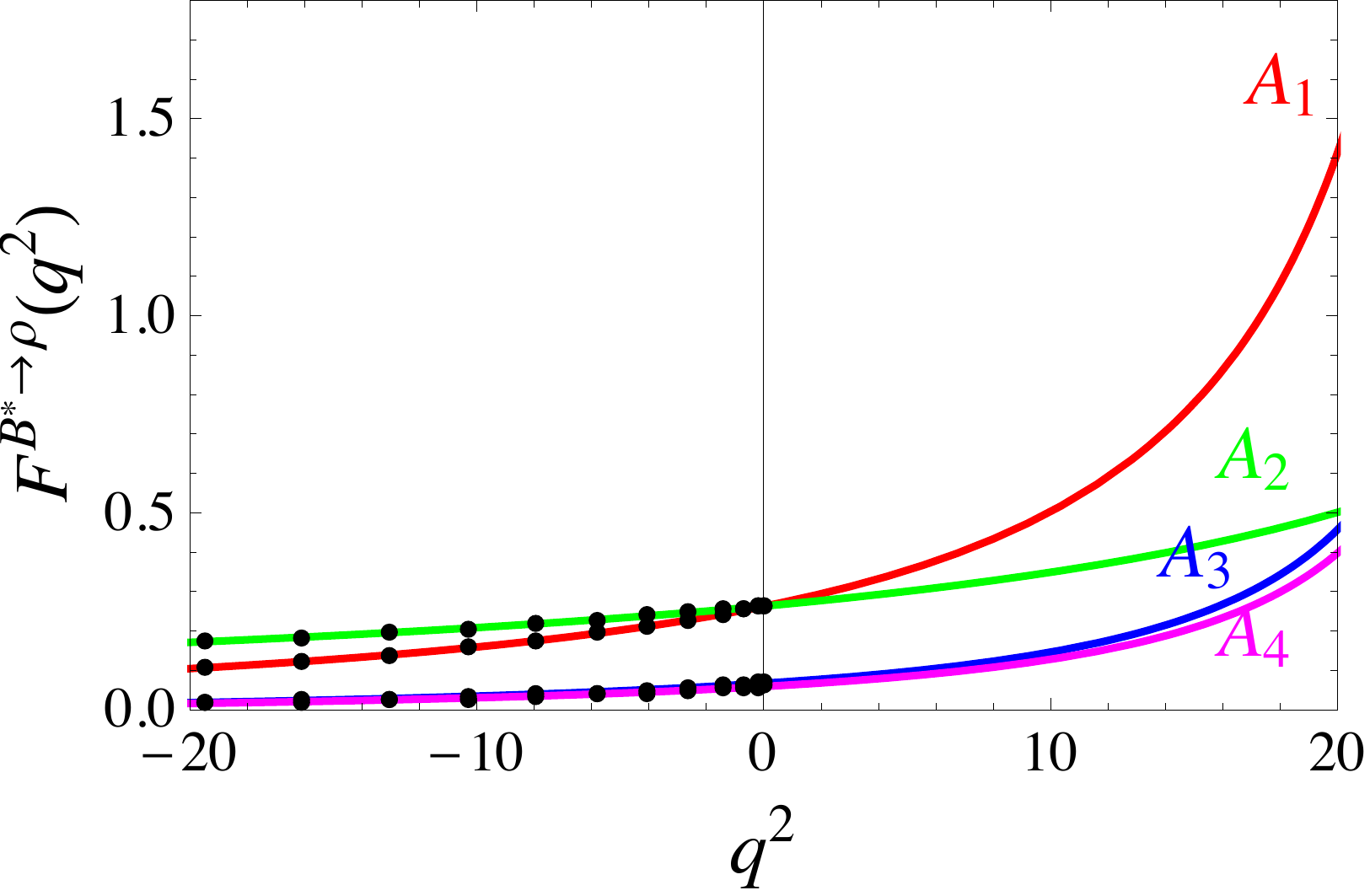}}\,
\subfigure{\includegraphics[scale=0.24]{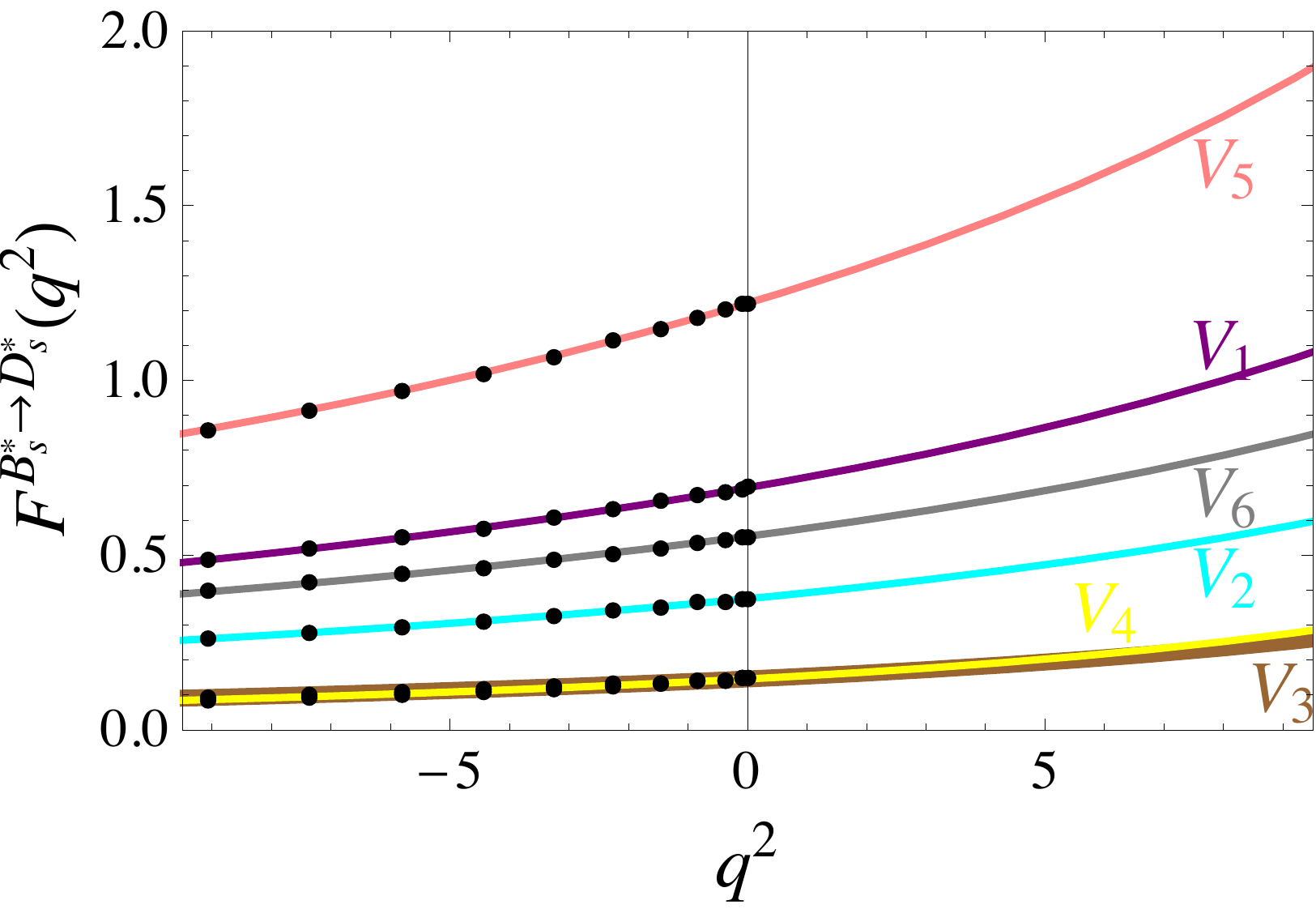}}\,
\subfigure{\includegraphics[scale=0.24]{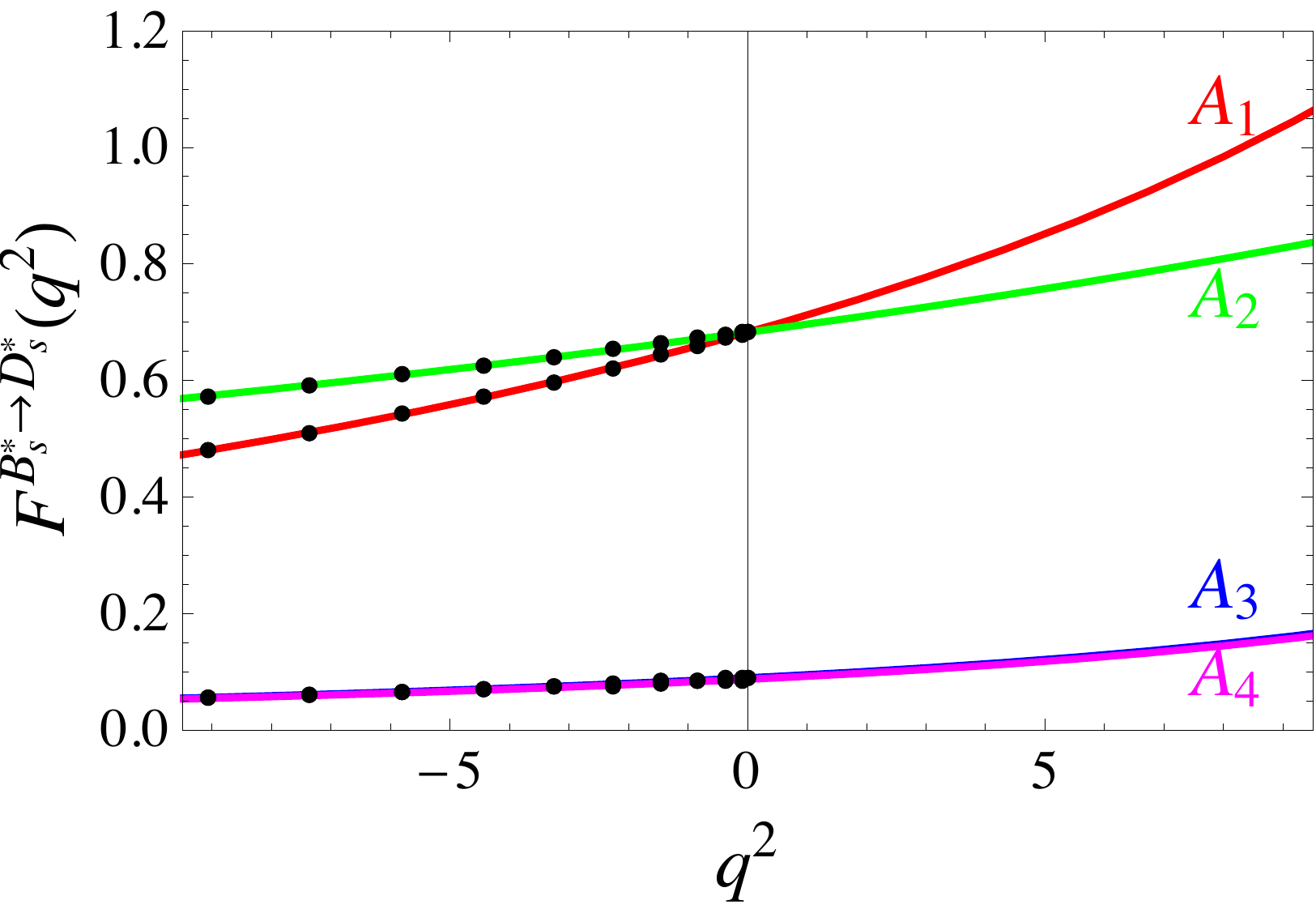}}\\
\subfigure{\includegraphics[scale=0.25]{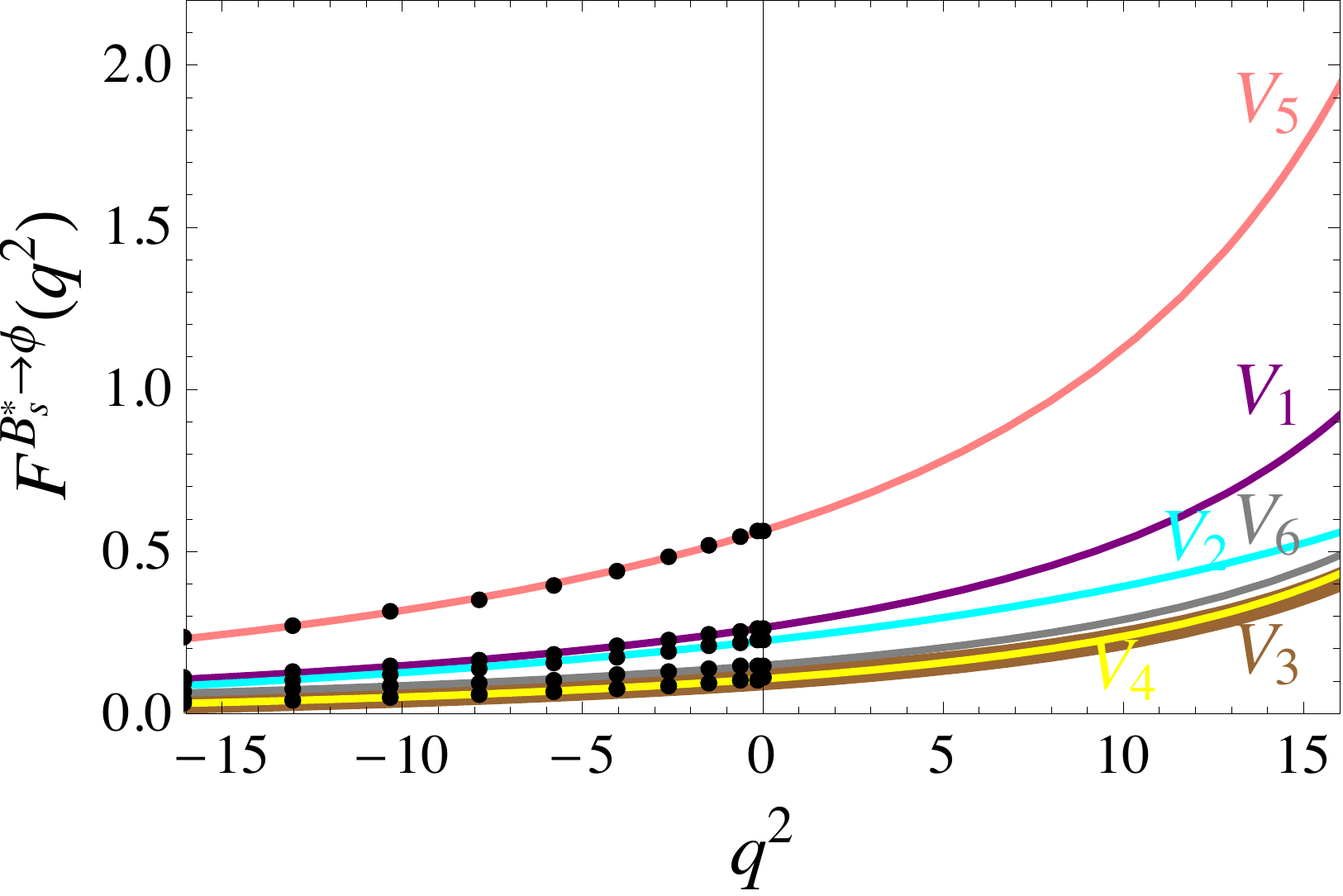}}\,
\subfigure{\includegraphics[scale=0.24]{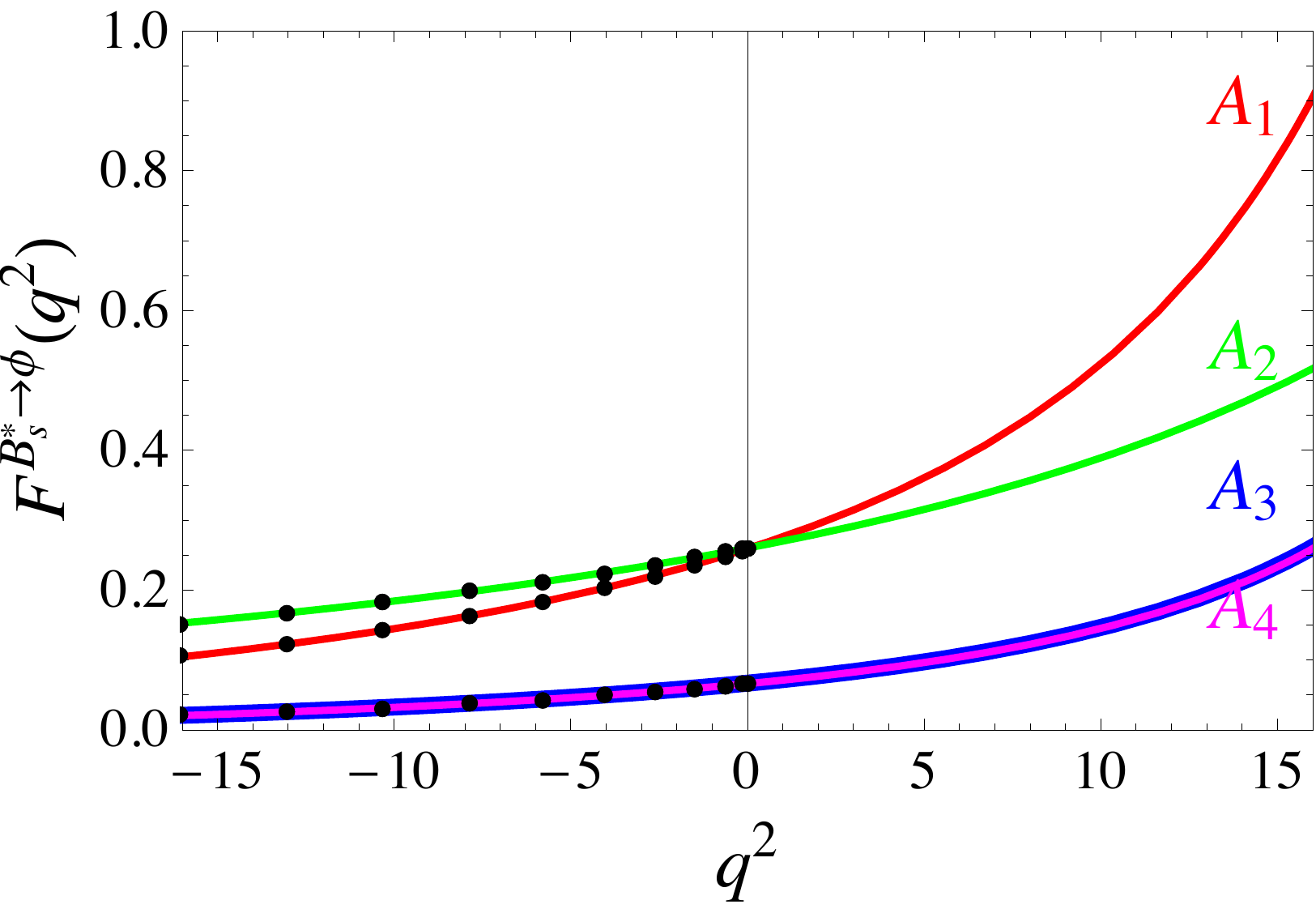}}\,
\subfigure{\includegraphics[scale=0.25]{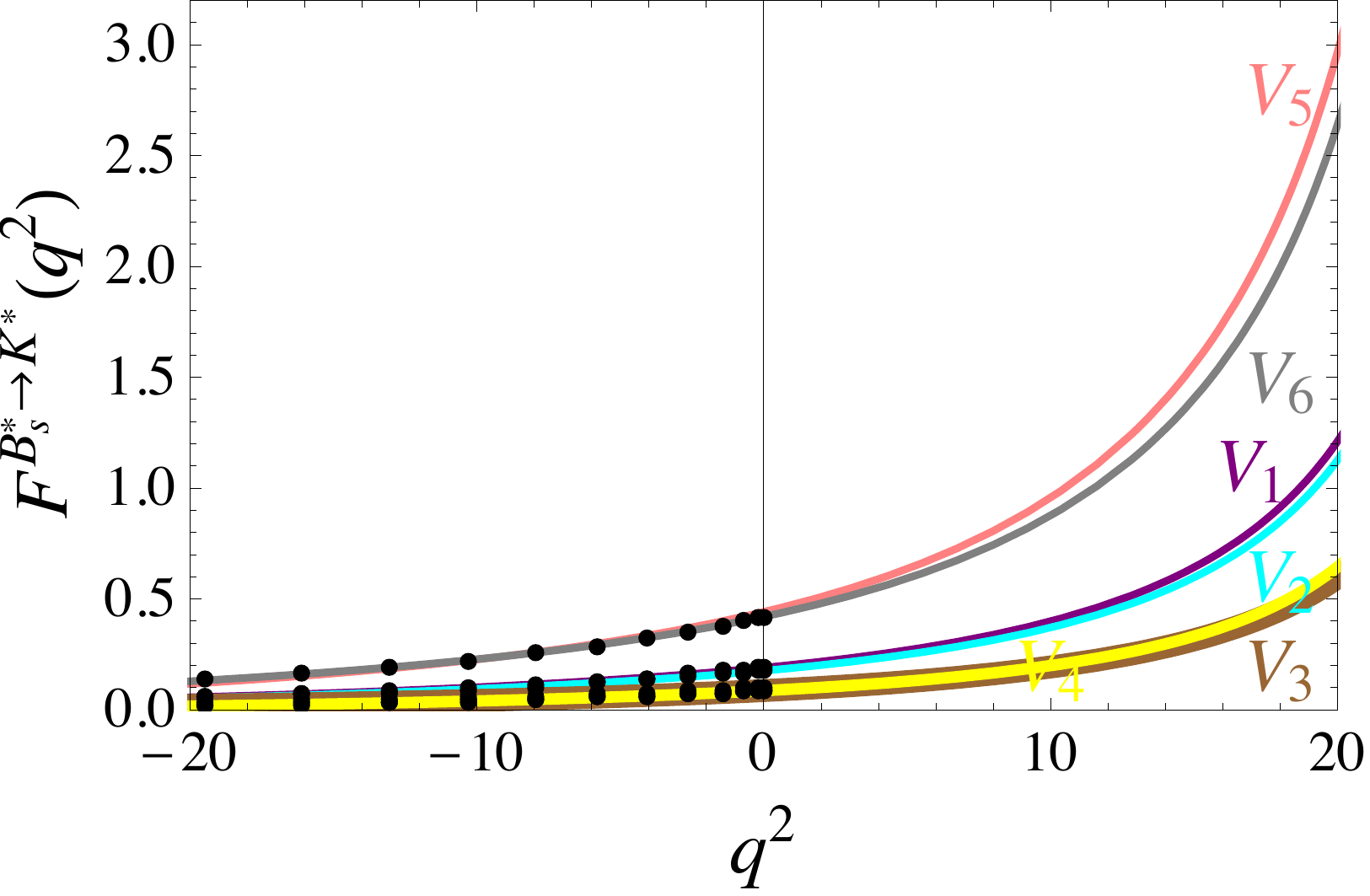}}\,
\subfigure{\includegraphics[scale=0.23]{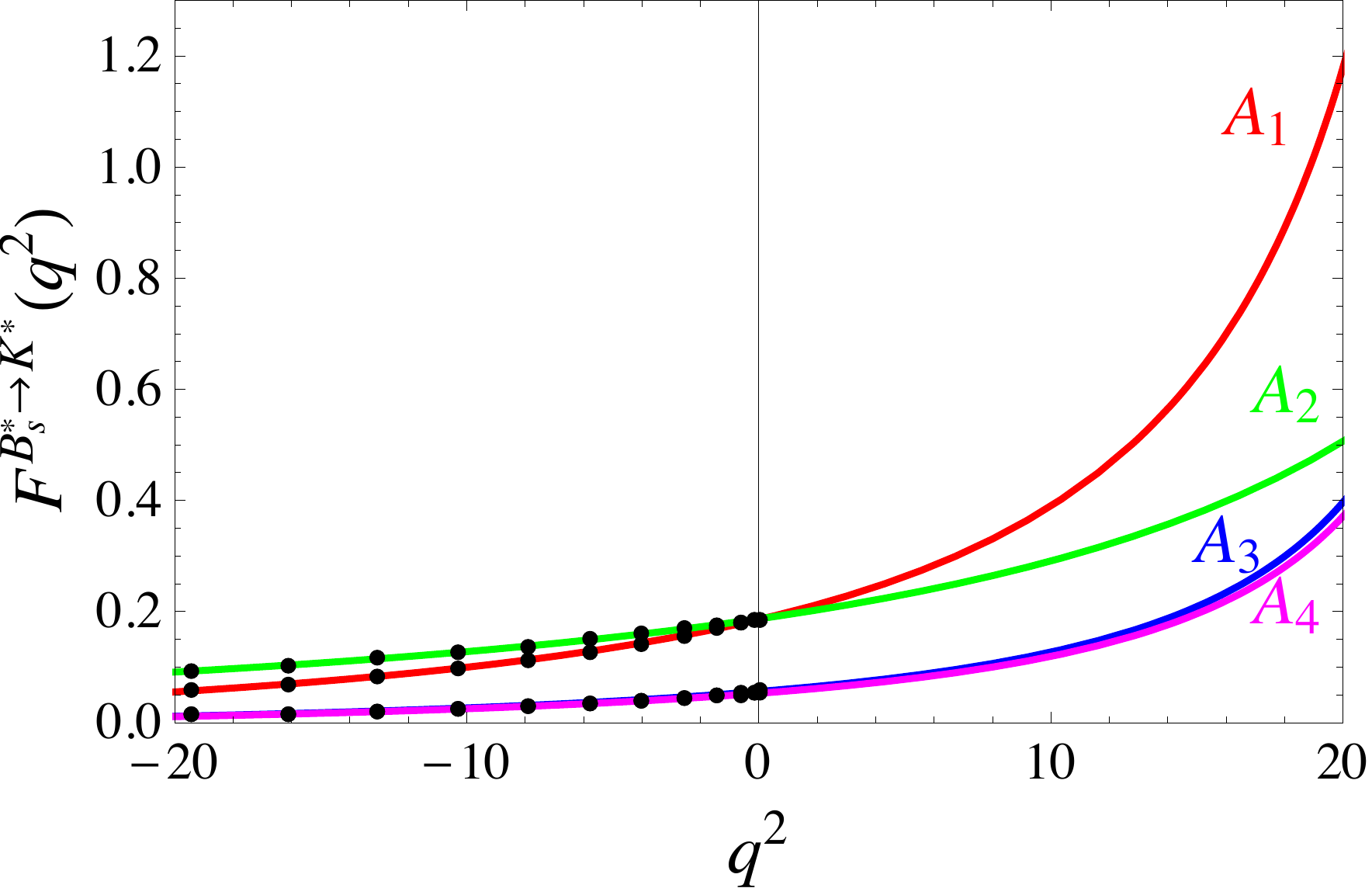}}\\
\subfigure{\includegraphics[scale=0.245]{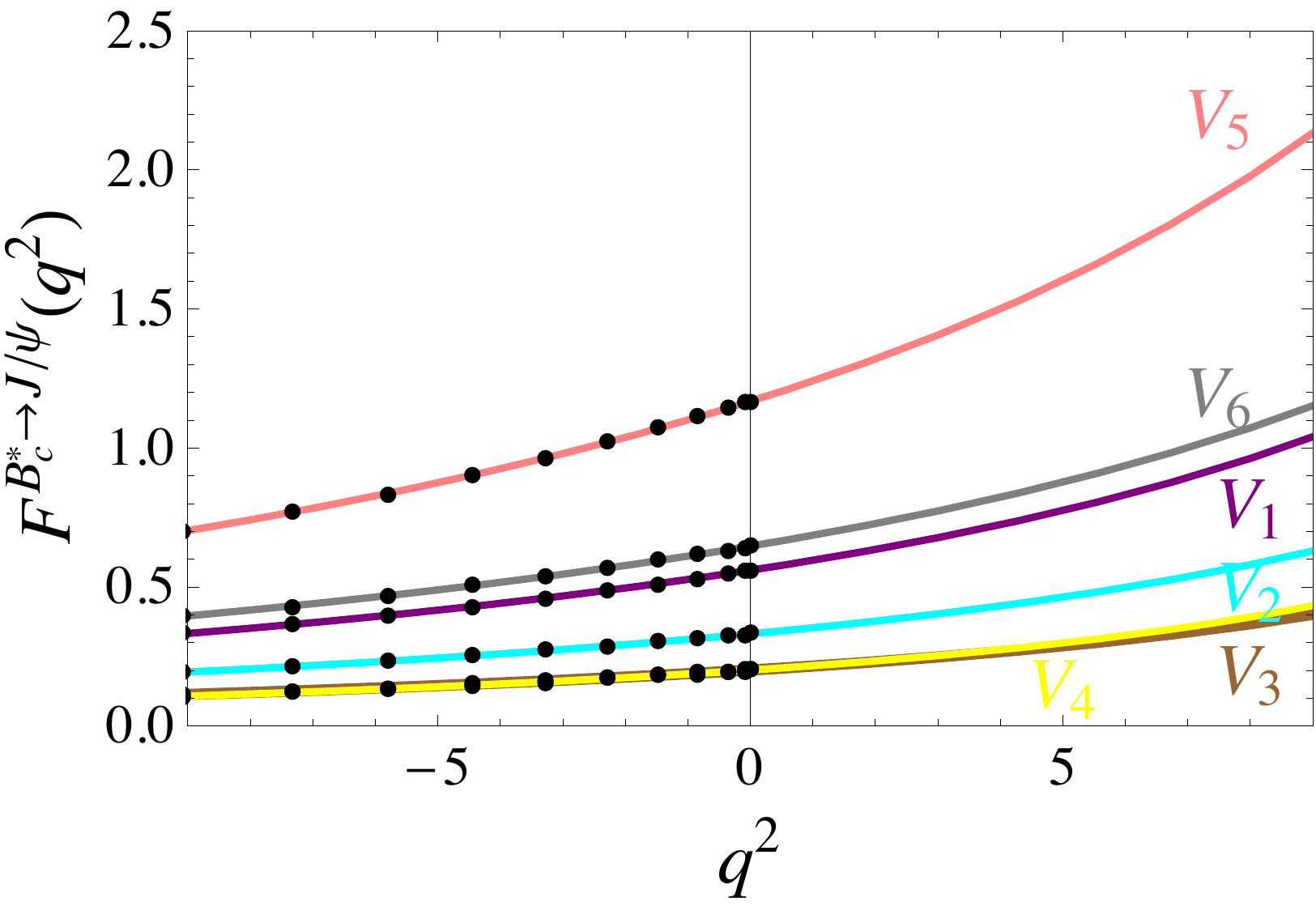}}\,
\subfigure{\includegraphics[scale=0.245]{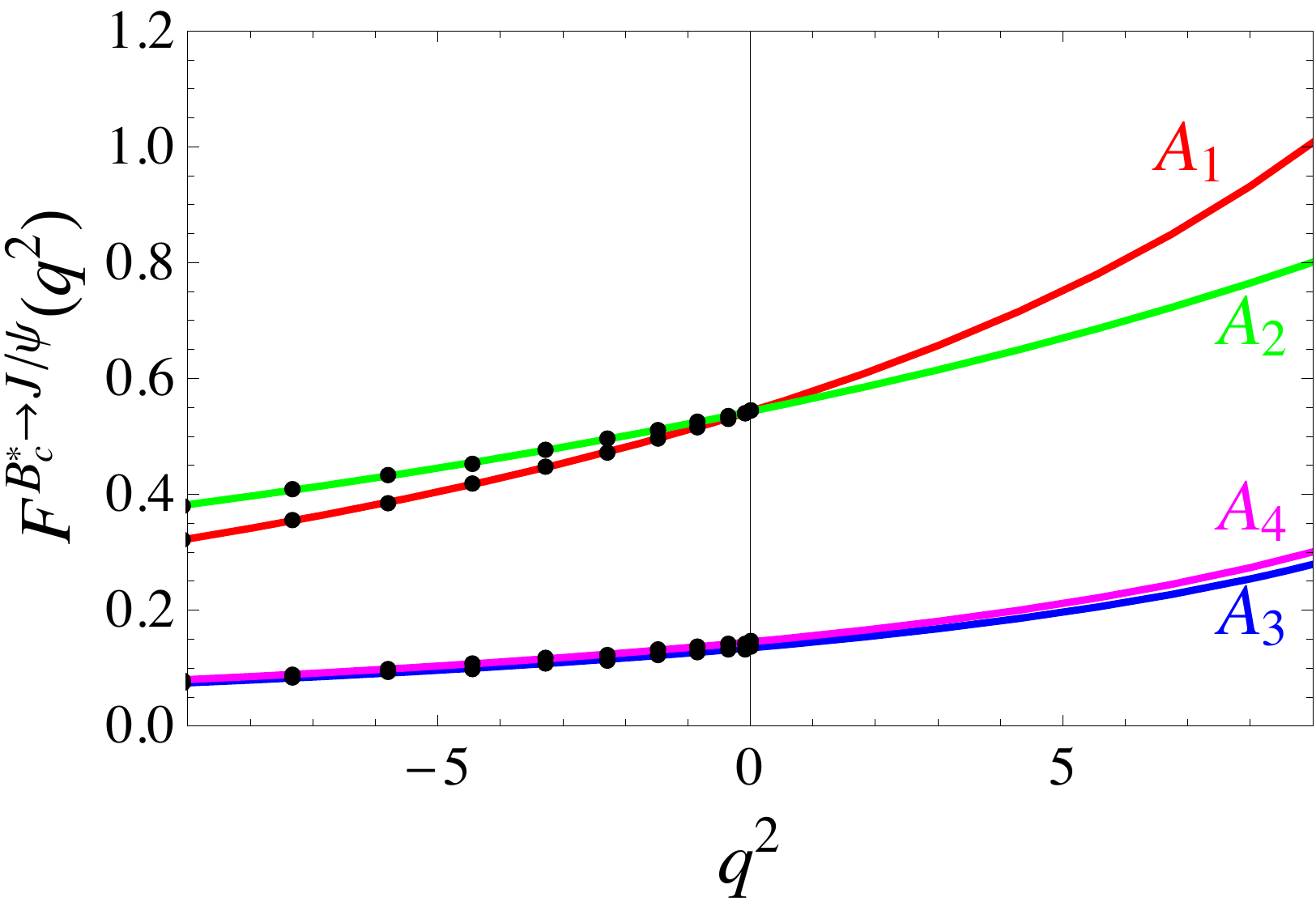}}\,
\subfigure{\includegraphics[scale=0.245]{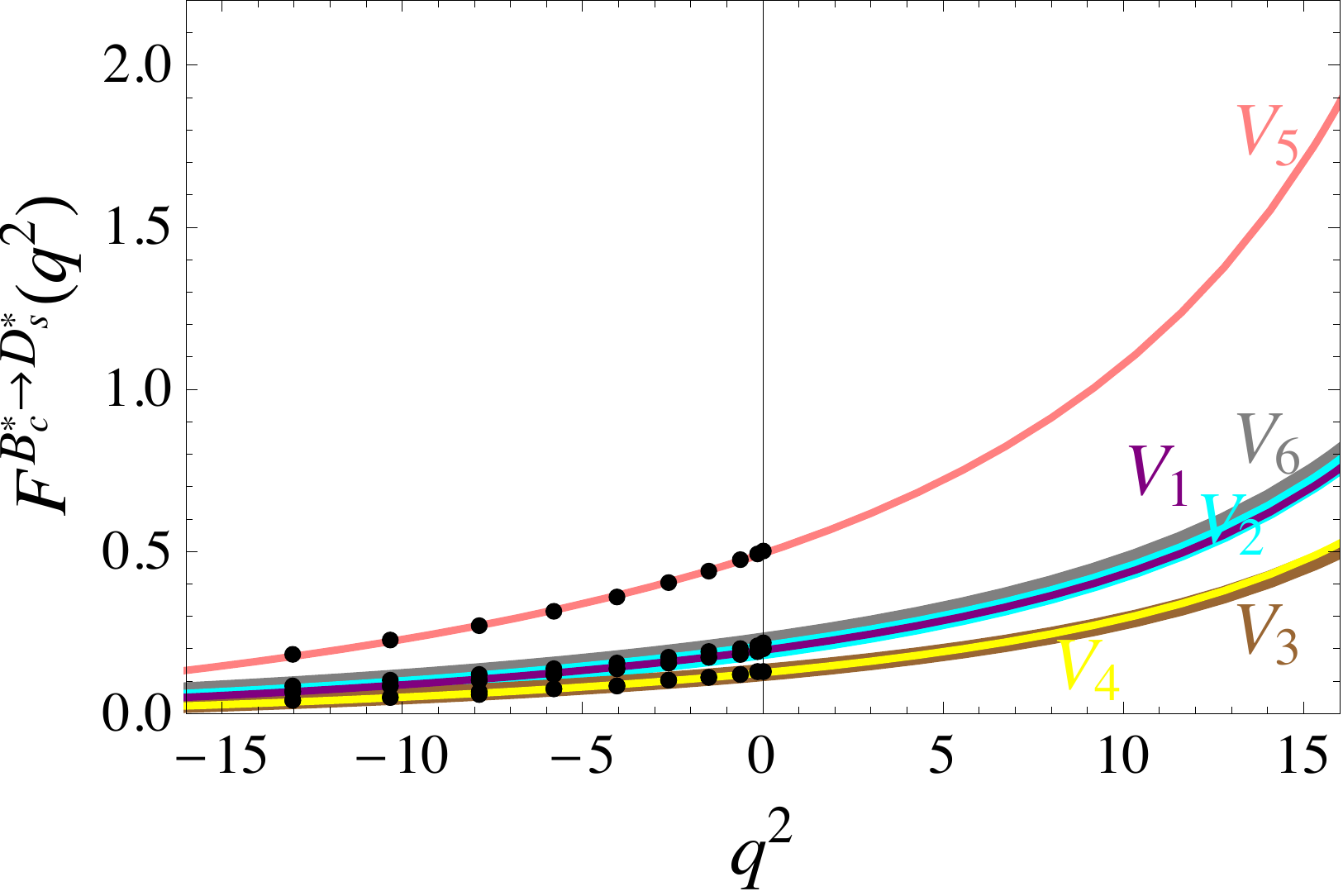}}\,
\subfigure{\includegraphics[scale=0.245]{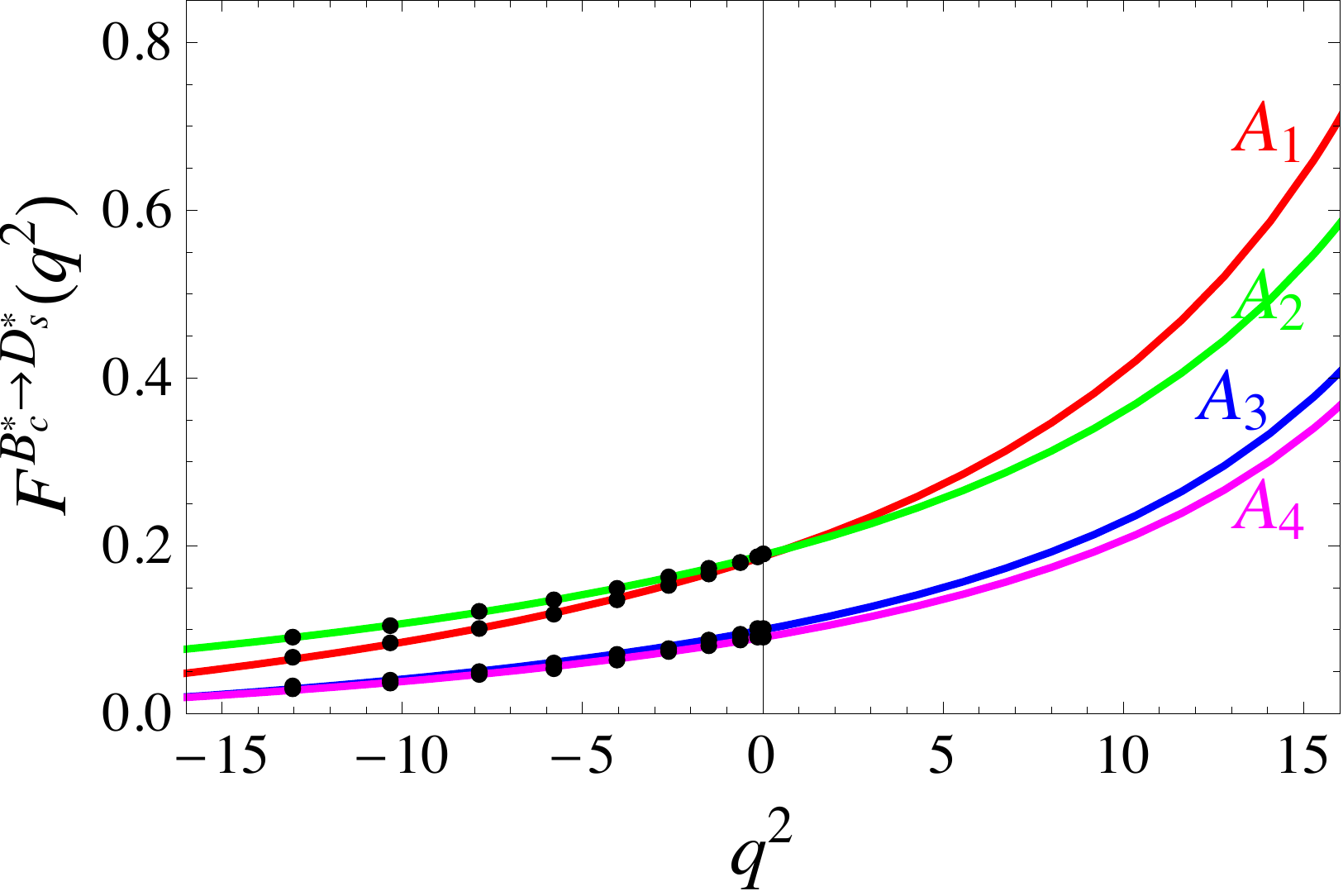}}\\
\subfigure{\includegraphics[scale=0.25]{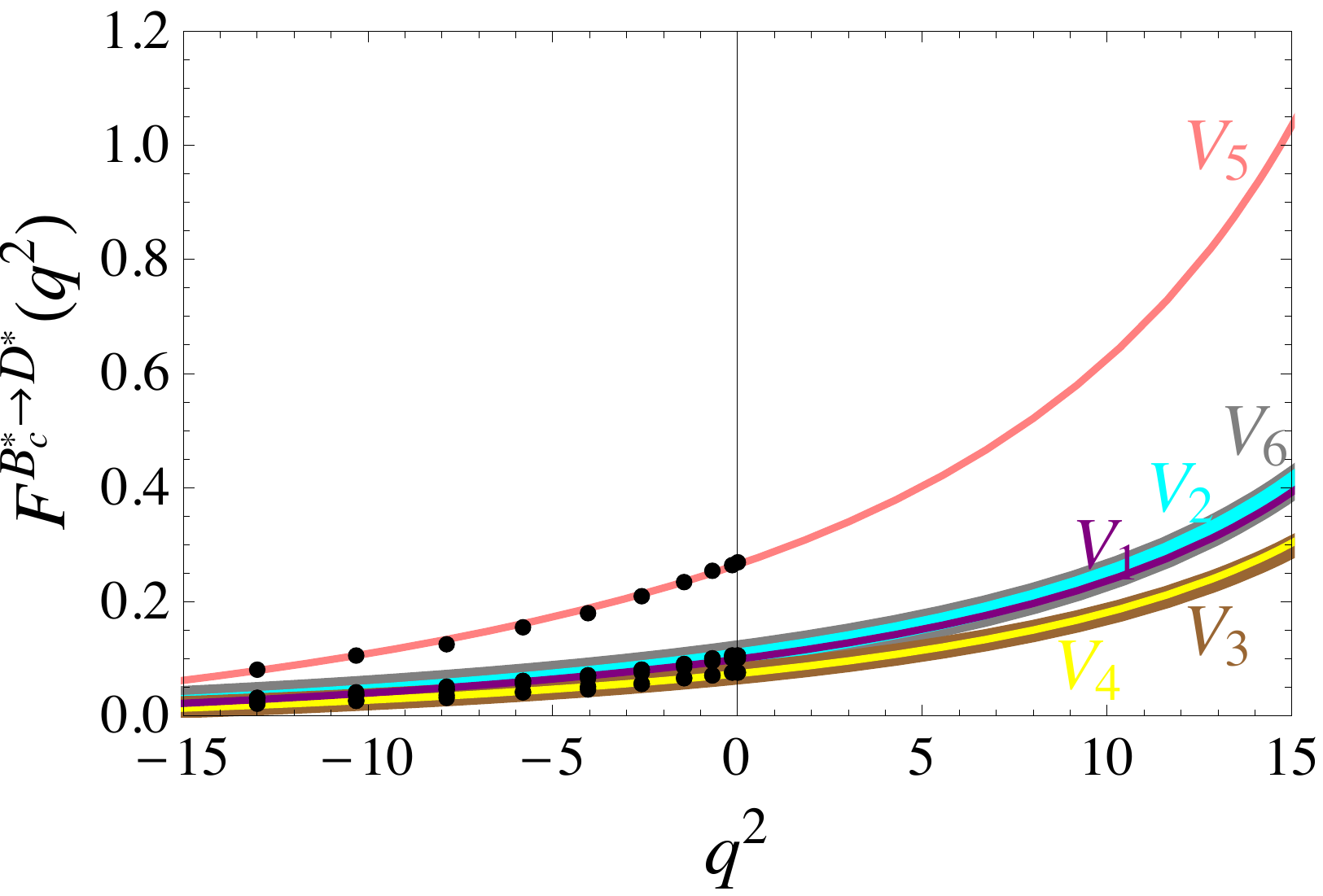}}\,
\subfigure{\includegraphics[scale=0.25]{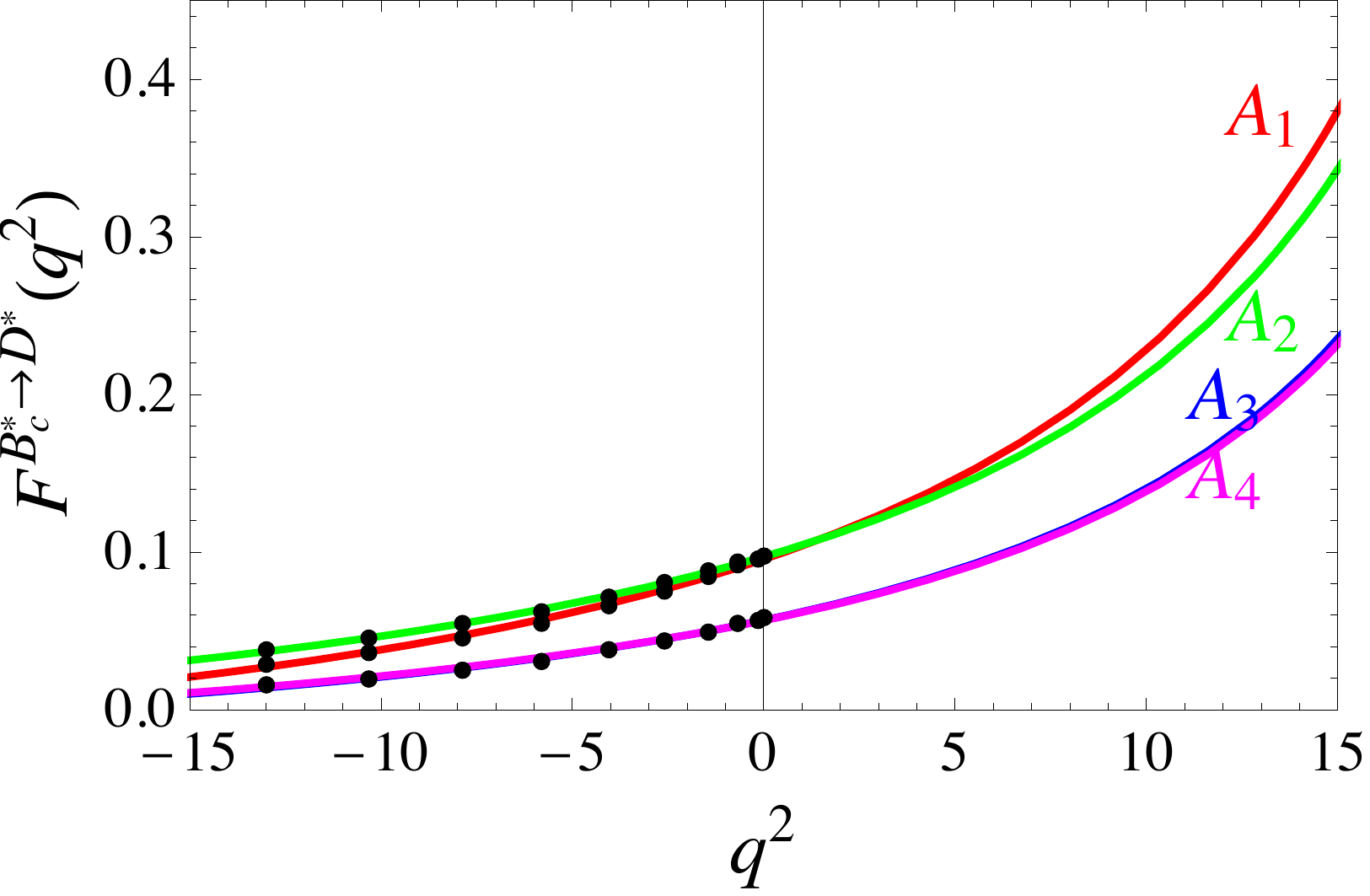}}\,
%\subfigure{\includegraphics[scale=0.26]{BcDV}}\,
%\subfigure{\includegraphics[scale=0.26]{BcDA}}\\
%\subfigure{\includegraphics[scale=0.29]{BcJpsiV}}\,
%\subfigure{\includegraphics[scale=0.28]{BcJpsiA}}\,
\subfigure{\includegraphics[scale=0.265]{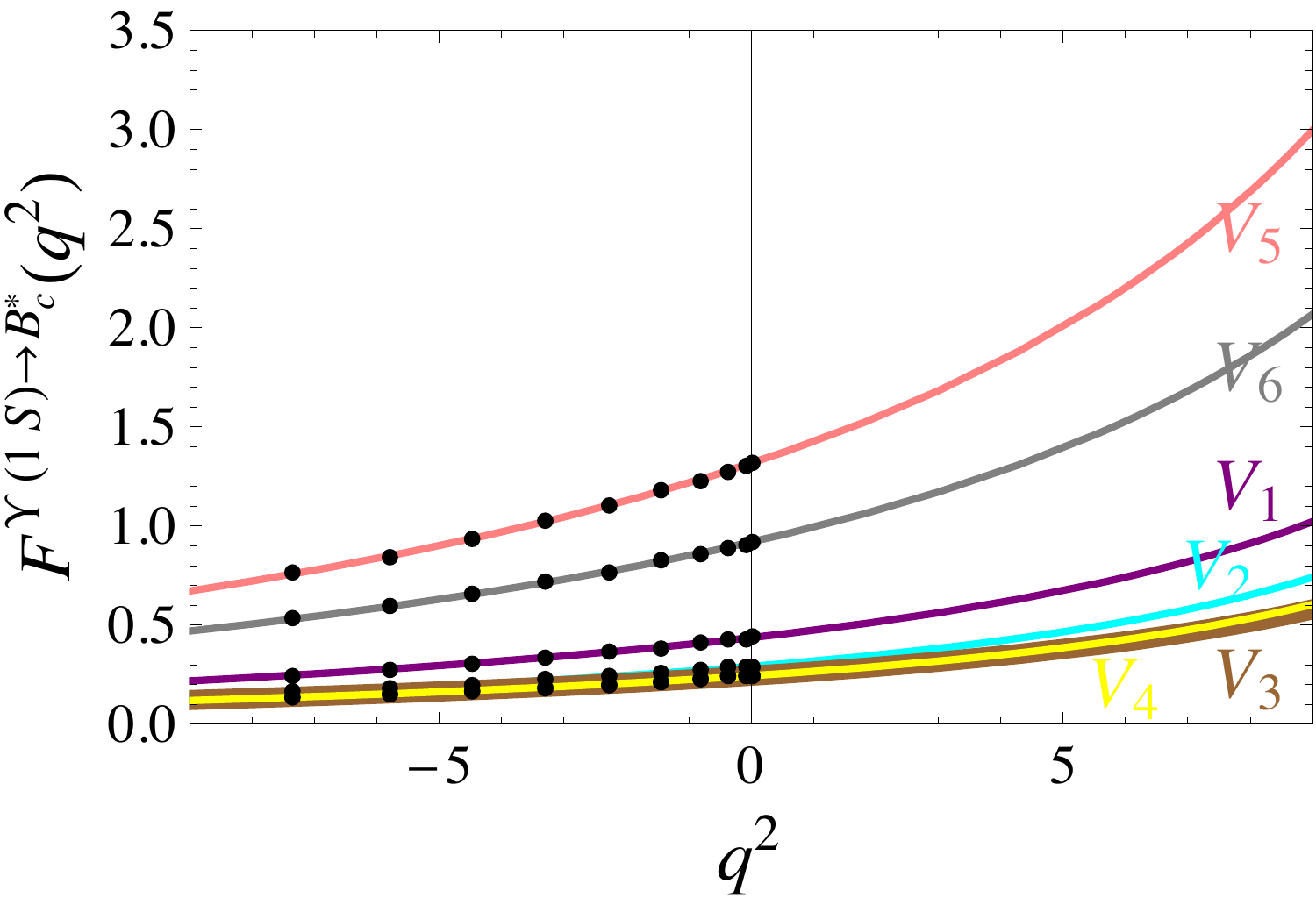}}\,
\subfigure{\includegraphics[scale=0.265]{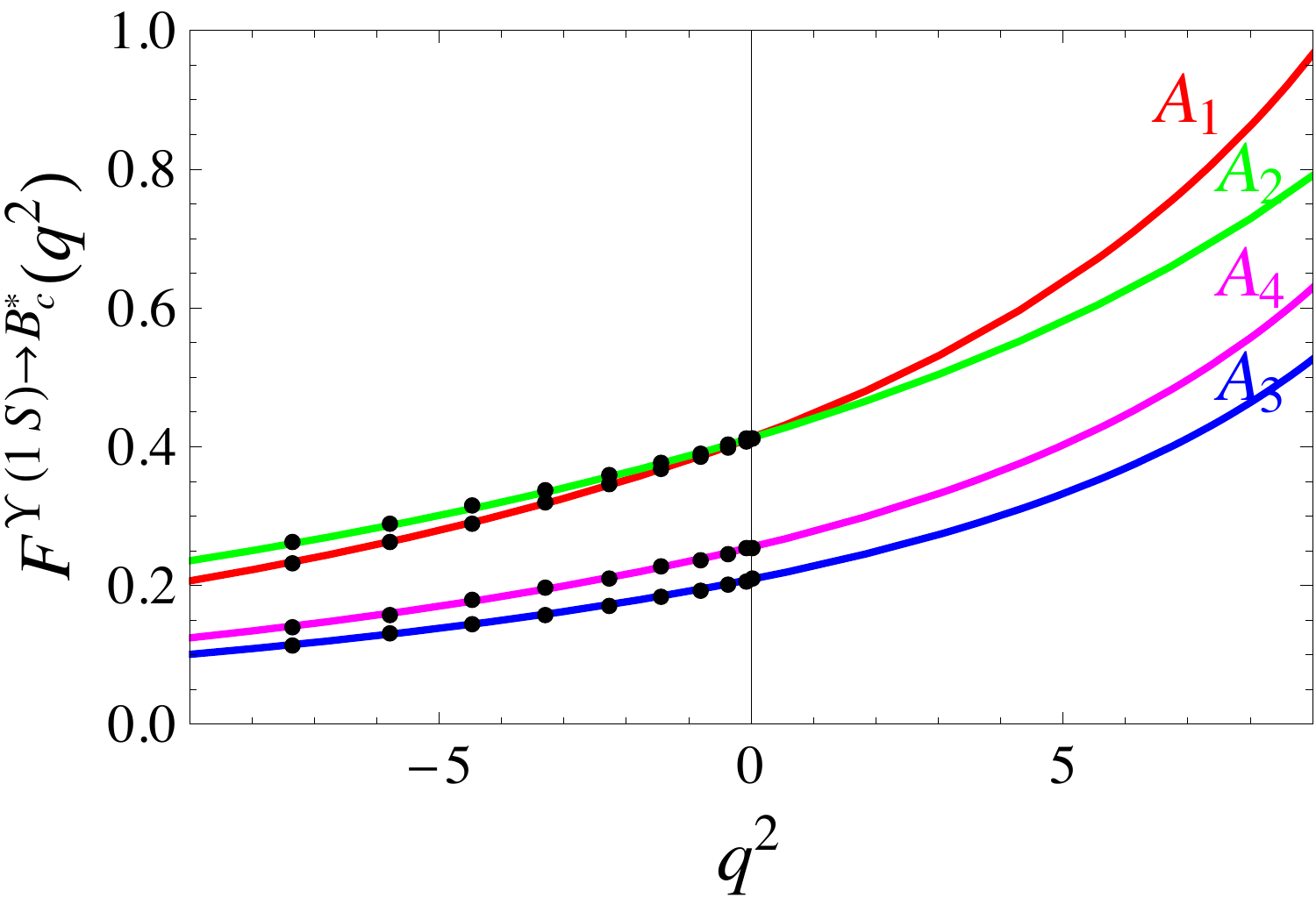}}\\
\subfigure{\includegraphics[scale=0.265]{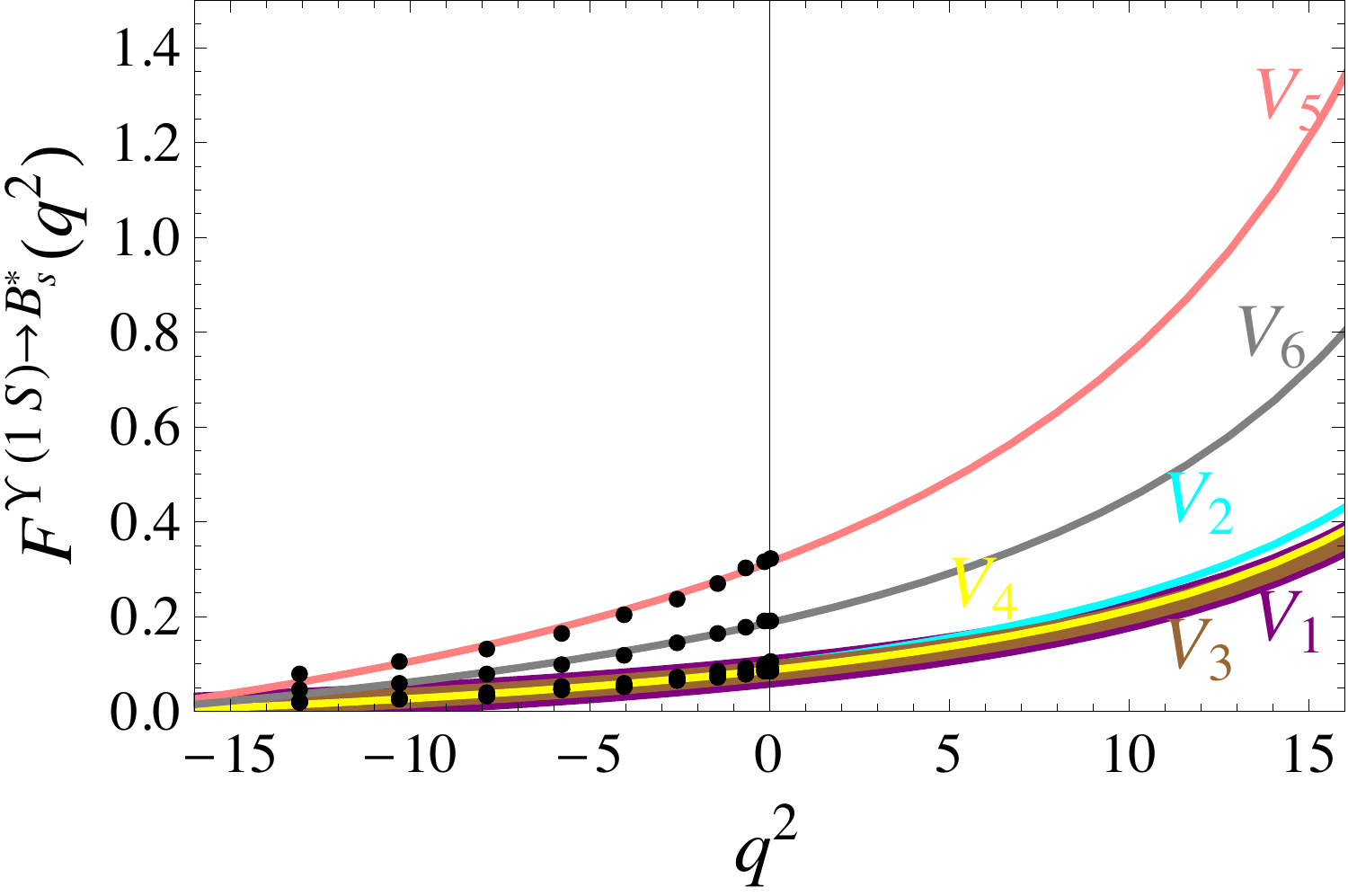}}\,
\subfigure{\includegraphics[scale=0.265]{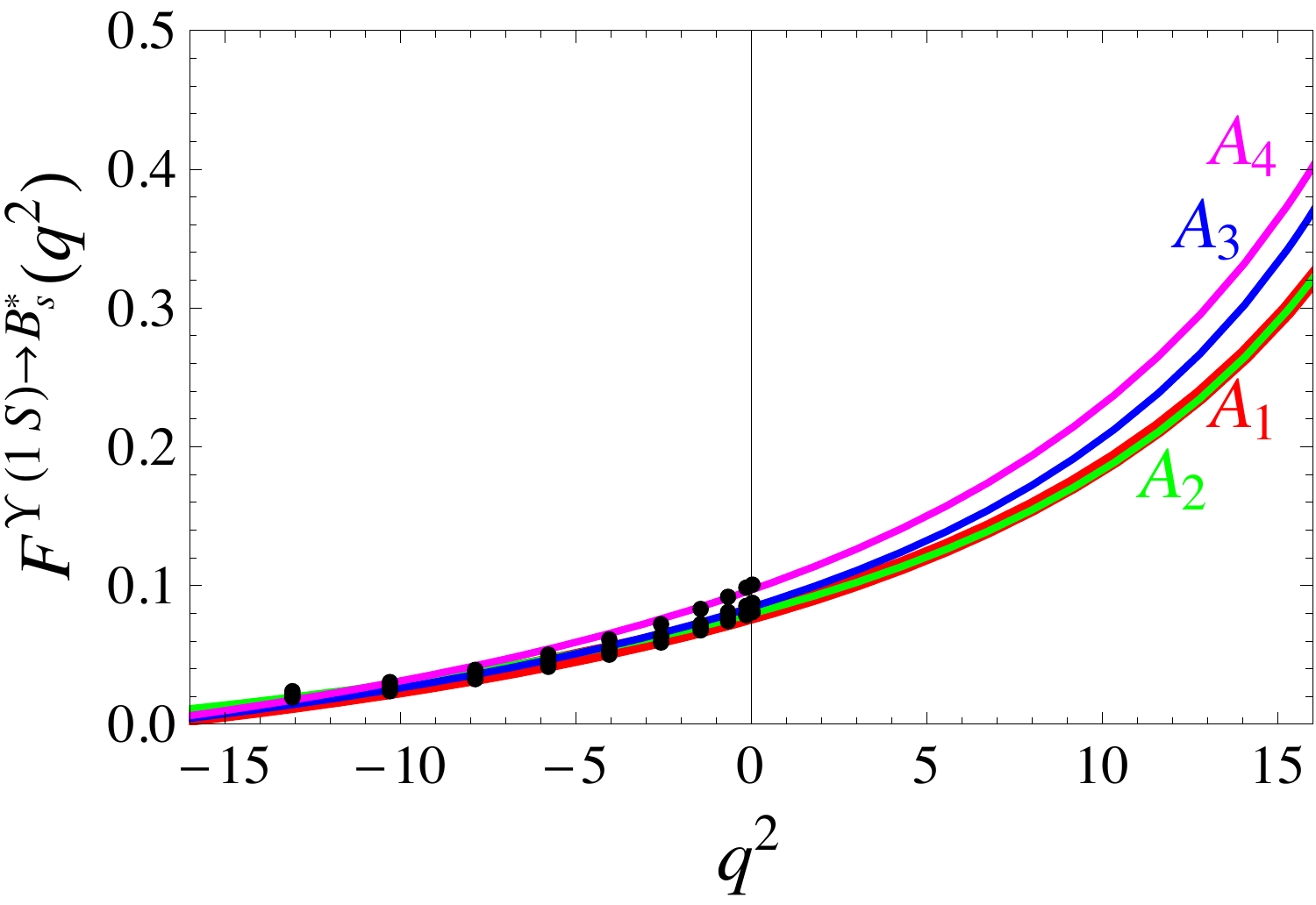}}\,
\subfigure{\includegraphics[scale=0.245]{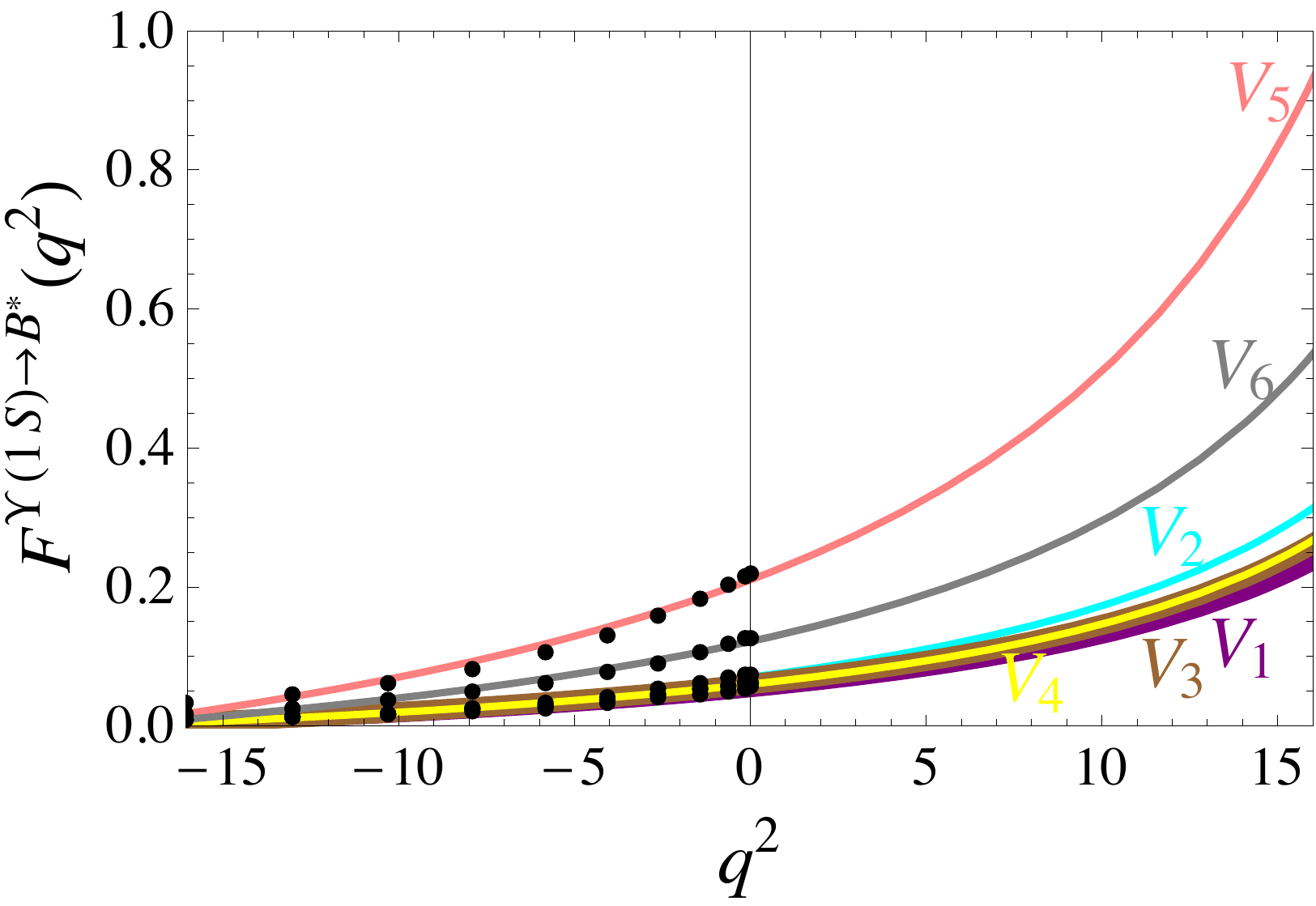}}\,
\subfigure{\includegraphics[scale=0.25]{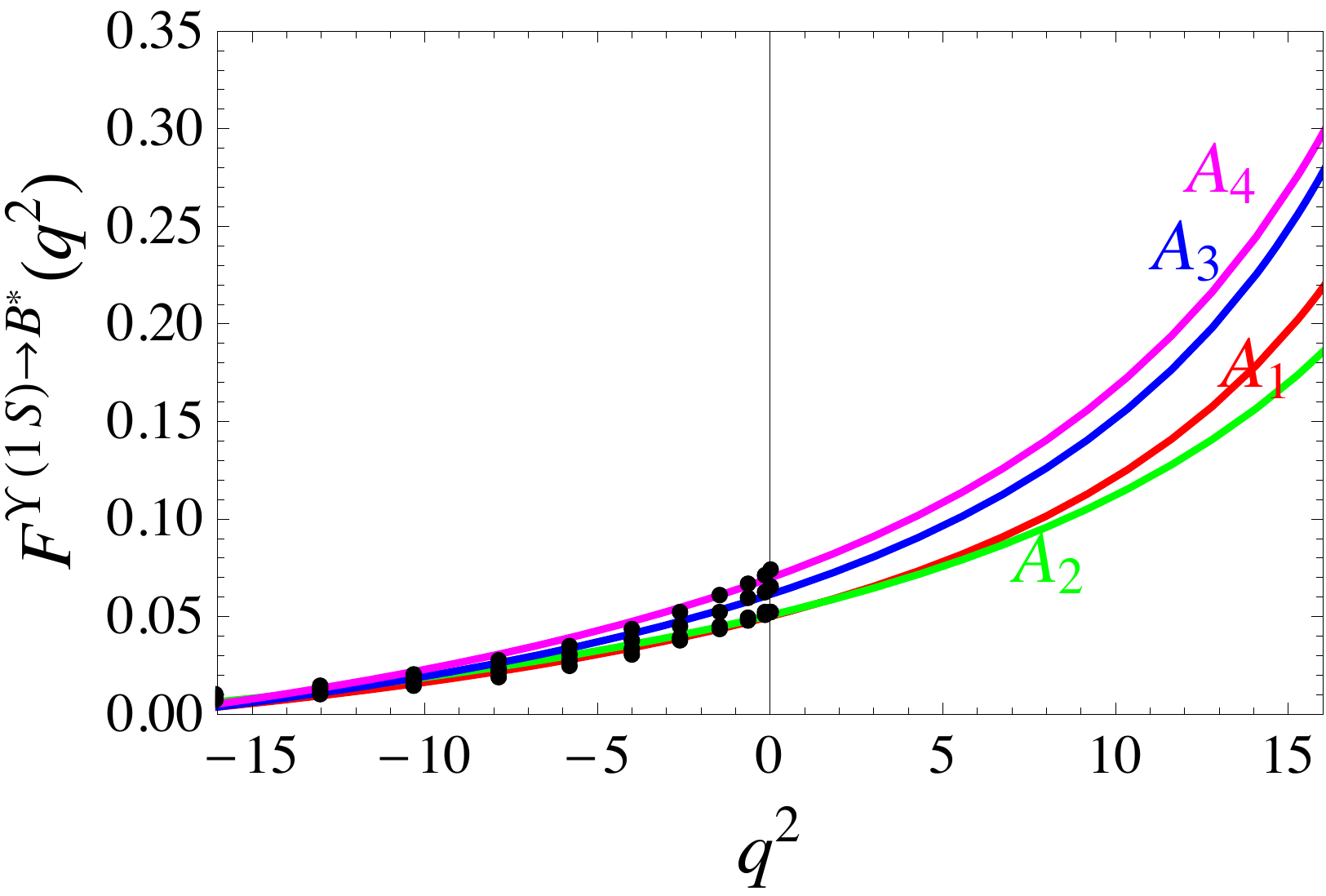}}
\end{center}
\label{fig:sche2b}
\end{figure}

\renewcommand{\baselinestretch}{1.2}
\begin{table}[!t]
\scriptsize
\begin{center}
\caption{\label{tab:sche2c} \small  Fitting results for the form factors of $c\to (q, s)$ induced $D^*\to (\rho\,,K^*)$, $D^*_s\to (K^*\,,\phi)$, $J/\Psi\to (D^*\,,D^*_s)$,  $B^*_c\to (B^*\,,B^*_s)$ transitions with the parameterization scheme given by~Eq.~\eqref{eq:para2}. The errors are due to parameters $\beta$ and quark masses only (cf. section 4 for a more complete discussion). }
\vspace{0.05cm}
\let\oldarraystretch=\arraystretch
\renewcommand*{\arraystretch}{1.12}
\setlength{\tabcolsep}{3pt}
\begin{tabular}{lccc|lccccccc}
\hline\hline
 &${\cal F}(0)$  &$a$   &$b_1$ & &${\cal F}(0)$  &$a$   &$b_1$
\\\hline
$V_1^{D^*\to\rho}$
&$0.65_{-0.01-0.09}^{+0.01+0.09}$ &$0.96_{-0.02-0.23}^{+0.01+0.28}$ &$-2.40_{-0.03-0.03}^{+0.03+0.02}$
&$V_1^{D^*\to K^*}$
&$0.74_{-0.02-0.10}^{+0.01+0.09}$ &$0.87_{-0.01-0.23}^{+0.01+0.35}$
&$-2.48_{-0.01-0.54}^{+0.01+0.42}$
\\
$V_2^{D^*\to\rho}$
&$0.51_{-0.02-0.04}^{+0.02+0.03}$ &$1.07_{-0.01-0.30}^{+0.01+0.37}$ &$-2.31_{-0.04-0.06}^{+0.04+0.11}$
&$V_2^{D^*\to K^*}$
&$0.43_{-0.03-0.08}^{+0.03+0.03}$ &$0.93_{-0.03-0.20}^{+0.02+0.21}$ &$-2.33_{-0.07-0.32}^{+0.05+0.33}$
\\
$V_3^{D^*\to\rho}$
&$0.29_{-0.00-0.03}^{+0.00+0.03}$ &$1.11_{-0.00-0.20}^{+0.01+0.31}$ &$-3.35_{-0.02-0.17}^{+0.02+0.30}$
&$V_3^{D^*\to K^*}$
&$0.26_{-0.01-0.02}^{+0.01+0.02}$ &$0.98_{-0.01-0.26}^{+0.02+0.20}$ &$-3.46_{-0.14-0.45}^{+0.12+0.41}$
\\
$V_4^{D^*\to\rho}$
&$0.29_{-0.00-0.03}^{+0.00+0.03}$ &$1.32_{-0.01-0.20}^{+0.01+0.31}$ &$-4.94_{-0.17-0.56}^{+0.17+0.61}$
&$V_4^{D^*\to K^*}$
&$0.26_{-0.01-0.02}^{+0.01+0.02}$ &$1.11_{-0.02-0.39}^{+0.02+0.47}$ &$-4.14_{-0.10-0.74}^{+0.09+0.62}$
\\
$V_5^{D^*\to\rho}$
&$1.42_{-0.02-0.17}^{+0.02+0.17}$ &$0.94_{-0.00-0.22}^{+0.01+0.26}$ &$-2.08_{-0.04-0.12}^{+0.06+0.18}$
&$V_5^{D^*\to K^*}$
&$1.50_{-0.03-0.16}^{+0.02+0.12}$ &$0.85_{-0.00-0.03}^{+0.01+0.05}$ &$-2.19_{-0.03-0.12}^{+0.03+0.15}$
\\
$V_6^{D^*\to\rho}$
&$0.68_{-0.01-0.17}^{+0.01+0.17}$ &$0.86_{-0.00-0.22}^{+0.01+0.31}$ &$-2.10_{-0.04-0.31}^{+0.05+0.26}$
&$V_6^{D^*\to K^*}$
&$0.78_{-0.02-0.17}^{+0.02+0.23}$ &$0.82_{-0.02-0.18}^{+0.02+0.25}$ &$-2.19_{-0.02-0.35}^{+0.02+0.29}$
\\
$A_1^{D^*\to\rho}$
&$0.59_{-0.01-0.07}^{+0.01+0.07}$ &$0.95_{-0.01-0.22}^{+0.01+0.23}$ &$-2.30_{-0.03-0.06}^{+0.03+0.06}$
&$A_1^{D^*\to K^*}$
&$0.69_{-0.01-0.09}^{+0.01+0.08}$ &$0.87_{-0.01-0.11}^{+0.01+0.10}$ &$-2.40_{-0.02-0.07}^{+0.02+0.07}$
\\
$A_2^{D^*\to\rho}$
&$0.59_{-0.01-0.07}^{+0.01+0.07}$ &$1.72_{-0.05-0.07}^{+0.05+0.08}$ &$12.08_{-0.21-0.49}^{+0.25+0.52}$
&$A_2^{D^*\to K^*}$
&$0.69_{-0.01-0.09}^{+0.01+0.08}$ &$1.04_{-0.03-0.06}^{+0.05+0.07}$ &$6.47_{-0.05-0.50}^{+0.03+0.53}$
\\
$A_3^{D^*\to\rho}$
&$0.22_{-0.01-0.03}^{+0.01+0.03}$ &$1.27_{-0.01-0.35}^{+0.01+0.43}$ &$-3.84_{-0.12-0.39}^{+0.11+0.50}$
&$A_3^{D^*\to K^*}$
&$0.15_{-0.01-0.04}^{+0.01+0.02}$ &$1.09_{-0.04-0.19}^{+0.03+0.22}$ &$-3.78_{-0.13-0.24}^{+0.11+0.25}$
\\
$A_4^{D^*\to\rho}$
&$0.24_{-0.01-0.04}^{+0.01+0.06}$ &$1.23_{-0.01-0.34}^{+0.02+0.41}$ &$-3.78_{-0.05-0.32}^{+0.03+0.23}$
&$A_4^{D^*\to K^*}$
&$0.22_{-0.01-0.03}^{+0.01+0.05}$ &$1.07_{-0.02-0.15}^{+0.02+0.21}$ &$-3.93_{-0.12-0.42}^{+0.12+0.51}$
\\
\hline
%%%%%%%%%%%%%%%%%%%%%%%%%%%%%%%%%%%%%%%%%%
$V_1^{D_s^*\to K^*}$  &$0.58_{-0.02-0.09}^{+0.02+0.10}$ &$0.98_{-0.01-0.21}^{+0.01+0.29}$ &$-3.35_{-0.04-0.09}^{+0.03+0.15}$
&$V_1^{D_s^*\to \phi}$ &$0.71_{-0.01-0.10}^{+0.01+0.09}$ &$0.94_{-0.01-0.11}^{+0.01+0.11}$ &$-3.25_{-0.07-0.09}^{+0.07+0.14}$
\\
$V_2^{D_s^*\to K^*}$  &$0.42_{-0.04-0.03}^{+0.04+0.02}$ &$1.06_{-0.03-0.27}^{+0.02+0.33}$ &$-3.12_{-0.11-0.22}^{+0.09+0.30}$
&$V_2^{D_s^*\to \phi}$ &$0.38_{-0.03-0.11}^{+0.03+0.04}$ &$0.98_{-0.05-0.06}^{+0.04+0.04}$ &$-3.00_{-0.14-0.25}^{+0.12+0.28}$
\\
$V_3^{D_s^*\to K^*}$  &$0.30_{-0.01-0.03}^{+0.00+0.03}$ &$1.06_{-0.02-0.27}^{+0.02+0.34}$ &$-3.77_{-0.10-0.25}^{+0.12+0.33}$
&$V_3^{D_s^*\to \phi}$ &$0.28_{-0.01-0.02}^{+0.01+0.02}$ &$1.00_{-0.04-0.25}^{+0.04+0.30}$ &$-3.76_{-0.14-0.34}^{+0.14+0.43}$
\\
$V_4^{D_s^*\to K^*}$  &$0.30_{-0.01-0.03}^{+0.00+0.03}$ &$1.24_{-0.04-0.34}^{+0.03+0.30}$ &$-4.26_{-0.51-0.37}^{+0.45+0.49}$
&$V_4^{D_s^*\to \phi}$ &$0.28_{-0.01-0.02}^{+0.01+0.02}$ &$1.13_{-0.06-0.30}^{+0.04+0.39}$ &$-3.98_{-0.48-0.36}^{+0.35+0.47}$
\\
$V_5^{D_s^*\to K^*}$  &$1.36_{-0.04-0.20}^{+0.04+0.19}$ &$0.95_{-0.01-0.20}^{+0.01+0.27}$ &$-2.87_{-0.05-0.22}^{+0.05+0.28}$
&$V_5^{D_s^*\to \phi}$ &$1.54_{-0.03-0.16}^{+0.02+0.16}$ &$0.91_{-0.01-0.18}^{+0.01+0.23}$ &$-2.82_{-0.08-0.49}^{+0.09+0.43}$
\\
$V_6^{D_s^*\to K^*}$  &$0.68_{-0.03-0.15}^{+0.03+0.21}$ &$0.91_{-0.02-0.23}^{+0.02+0.30}$ &$-2.89_{-0.05-0.46}^{+0.03+0.45}$
&$V_6^{D_s^*\to \phi}$ &$0.86_{-0.02-0.19}^{+0.02+0.25}$ &$0.91_{-0.03-0.23}^{+0.03+0.25}$ &$-2.70_{-0.18-0.35}^{+0.18+0.33}$
\\
$A_1^{D_s^*\to K^*}$  &$0.52_{-0.02-0.08}^{+0.02+0.07}$ &$0.98_{-0.01-0.21}^{+0.01+0.28}$ &$-3.25_{-0.04-0.12}^{+0.04+0.20}$
&$A_1^{D_s^*\to \phi}$ &$0.65_{-0.01-0.09}^{+0.01+0.08}$ &$0.94_{-0.01-0.16}^{+0.01+0.17}$ &$-3.17_{-0.07-0.11}^{+0.07+0.17}$
\\
$A_2^{D_s^*\to K^*}$  &$0.52_{-0.02-0.08}^{+0.02+0.08}$ &$0.05_{-0.01-0.04}^{+0.01+0.03}$ &$-3.07_{-0.11-0.70}^{+0.19+0.87}$
&$A_2^{D_s^*\to \phi}$ &$0.65_{-0.01-0.09}^{+0.01+0.08}$ &$1.06_{-0.05-0.16}^{+0.04+0.14}$ &$6.17_{-0.12-0.17}^{+0.09+0.15}$
\\
$A_3^{D_s^*\to K^*}$  &$0.20_{-0.02-0.03}^{+0.02+0.03}$ &$1.21_{-0.04-0.34}^{+0.03+0.36}$ &$-4.22_{-0.17-0.44}^{+0.15+0.40}$
&$A_3^{D_s^*\to \phi}$ &$0.16_{-0.01-0.05}^{+0.01+0.02}$ &$1.10_{-0.07-0.07}^{+0.05+0.07}$ &$-4.04_{-0.21-0.16}^{+0.14+0.20}$
\\
$A_4^{D_s^*\to K^*}$  &$0.28_{-0.01-0.05}^{+0.01+0.04}$ &$1.16_{-0.03-0.24}^{+0.03+0.30}$ &$-4.28_{-0.24-0.25}^{+0.23+0.27}$
&$A_4^{D_s^*\to \phi}$ &$0.27_{-0.01-0.04}^{+0.01+0.05}$ &$1.12_{-0.06-0.31}^{+0.05+0.39}$ &$-4.10_{-0.10-0.58}^{+0.07+0.40}$
\\
\hline
%%%%%%%%%%%%%%%%%%%%%%%%%%%%%%%%%%%%%%%%%%
$V_1^{J/\psi\to D^*}$   &$0.55_{-0.02-0.12}^{+0.02+0.12}$ &$1.09_{-0.02-0.15}^{+0.02+0.13}$ &$-13.71_{-0.25-0.66}^{+0.26+0.53}$
&$V_1^{J/\psi\to D_s^*}$ &$0.72_{-0.02-0.13}^{+0.01+0.12}$ &$0.97_{-0.03-0.14}^{+0.03+0.18}$ &$-11.44_{-0.58-1.30}^{+0.54+1.26}$
\\
$V_2^{J/\psi\to D^*}$   &$0.60_{-0.05-0.10}^{+0.05+0.15}$ &$1.37_{-0.03-0.41}^{+0.02+0.47}$ &$-14.80_{-0.08-3.17}^{+0.05+2.59}$
&$V_2^{J/\psi\to D_s^*}$ &$0.68_{-0.06-0.11}^{+0.06+0.14}$ &$1.23_{-0.03-0.36}^{+0.03+0.35}$ &$-12.59_{-0.53-3.17}^{+0.51+2.97}$
\\
$V_3^{J/\psi\to D^*}$   &$0.36_{-0.01-0.03}^{+0.01+0.03}$ &$1.20_{-0.02-0.32}^{+0.02+0.34}$ &$-12.66_{-0.48-2.33}^{+0.52+2.22}$
&$V_3^{J/\psi\to D_s^*}$ &$0.36_{-0.01-0.02}^{+0.01+0.01}$ &$1.17_{-0.02-0.32}^{+0.02+0.40}$ &$-11.59_{-0.35-1.48}^{+0.35+2.15}$
\\
$V_4^{J/\psi\to D^*}$   &$0.36_{-0.01-0.03}^{+0.01+0.03}$ &$1.37_{-0.02-0.36}^{+0.02+0.42}$ &$-14.02_{-0.34-2.40}^{+0.31+2.31}$
&$V_4^{J/\psi\to D_s^*}$ &$0.36_{-0.01-0.02}^{+0.01+0.01}$ &$1.33_{-0.02-0.36}^{+0.02+0.36}$ &$-14.33_{-0.04-1.82}^{+0.06+2.57}$
\\
$V_5^{J/\psi\to D^*}$   &$1.79_{-0.05-0.34}^{+0.05+0.52}$ &$1.06_{-0.02-0.27}^{+0.02+0.27}$ &$-12.48_{-0.22-1.94}^{+0.21+1.75}$
&$V_5^{J/\psi\to D_s^*}$ &$2.16_{-0.04-0.36}^{+0.04+0.51}$ &$0.96_{-0.02-0.22}^{+0.03+0.30}$ &$-10.41_{-0.27-2.19}^{+0.26+2.28}$
\\
$V_6^{J/\psi\to D^*}$   &$1.23_{-0.05-0.34}^{+0.04+0.53}$ &$1.33_{-0.05-0.30}^{+0.06+0.32}$ &$-7.22_{-0.30-1.72}^{+0.35+1.43}$
&$V_6^{J/\psi\to D_s^*}$ &$1.58_{-0.04-0.40}^{+0.04+0.58}$ &$0.89_{-0.04-0.22}^{+0.03+0.31}$ &$-10.50_{-0.32-2.37}^{+0.30+2.40}$
\\
$A_1^{J/\psi\to D^*}$   &$0.47_{-0.01-0.09}^{+0.01+0.09}$ &$1.10_{-0.03-0.06}^{+0.03+0.08}$ &$-13.60_{-0.26-0.70}^{+0.26+0.52}$
&$A_1^{J/\psi\to D_s^*}$ &$0.62_{-0.01-0.10}^{+0.01+0.08}$ &$0.98_{-0.03-0.15}^{+0.03+0.17}$ &$-11.35_{-0.25-0.73}^{+0.31+1.12}$
\\
$A_2^{J/\psi\to D^*}$   &$0.47_{-0.01-0.09}^{+0.01+0.09}$ &$0.72_{-0.02-0.17}^{+0.02+0.18}$ &$7.17_{-0.19-1.18}^{+0.23+1.30}$
&$A_2^{J/\psi\to D_s^*}$ &$0.62_{-0.01-0.10}^{+0.01+0.08}$ &$0.50_{-0.09-0.21}^{+0.09+0.27}$
&$20.69_{-1.27-4.02}^{+1.19+5.33}$
\\
$A_3^{J/\psi\to D^*}$   &$0.38_{-0.01-0.08}^{+0.01+0.11}$ &$1.41_{-0.01-0.30}^{+0.01+0.34}$ &$-14.69_{-0.08-2.21}^{+0.05+2.84}$
&$A_3^{J/\psi\to D_s^*}$ &$0.35_{-0.01-0.06}^{+0.01+0.07}$ &$1.35_{-0.01-0.27}^{+0.01+0.35}$ &$-13.69_{-0.30-2.84}^{+0.21+3.12}$
\\
$A_4^{J/\psi\to D^*}$   &$0.45_{-0.01-0.11}^{+0.01+0.15}$ &$1.32_{-0.03-0.27}^{+0.02+0.26}$ &$-14.00_{-0.33-2.58}^{+0.28+2.55}$
&$A_4^{J/\psi\to D_s^*}$ &$0.45_{-0.01-0.09}^{+0.01+0.13}$ &$1.26_{-0.03-0.35}^{+0.02+0.45}$ &$-12.91_{-0.36-2.79}^{+0.20+2.49}$
\\
%%%%%%%%%%%%%%%%%%%%%%%%%%%%%%%%%%%%%%%%%%
\hline
$V_1^{B_c^*\to B^*}$   &$0.52_{-0.02-0.11}^{+0.02+0.11}$   &$1.18_{-0.03-0.16}^{+0.03+0.24}$ &$-92.66_{-0.25-1.25}^{+0.23+2.13}$
&$V_1^{B_c^*\to B_s^*}$ &$0.63_{-0.01-0.12}^{+0.01+0.12}$   &$1.17_{-0.02-0.17}^{+0.01+0.23}$ &$-86.23_{-0.61-1.48}^{+0.58+1.31}$
\\
$V_2^{B_c^*\to B^*}$   &$1.18_{-0.13-0.28}^{+0.13+0.30}$   &$1.63_{-0.03-0.38}^{+0.03+0.46}$ &$-109.58_{-0.10-14.01}^{+0.14+14.65}$
&$V_2^{B_c^*\to B_s^*}$ &$1.06_{-0.11-0.40}^{+0.11+0.20}$   &$1.74_{-0.03-0.44}^{+0.03+0.55}$ &$-96.20_{-1.02-1.52}^{+0.98+1.92}$
\\
$V_3^{B_c^*\to B^*}$   &$0.40_{-0.01-0.04}^{+0.01+0.03}$   &$1.30_{-0.04-0.26}^{+0.04+0.34}$ &$-81.00_{-0.49-2.30}^{+0.53+2.47}$
&$V_3^{B_c^*\to B_s^*}$ &$0.40_{-0.01-0.03}^{+0.01+0.01}$   &$1.33_{-0.03-0.27}^{+0.03+0.33}$ &$-77.08_{-0.73-3.07}^{+0.68+3.56}$
\\
$V_4^{B_c^*\to B^*}$   &$0.40_{-0.01-0.04}^{+0.01+0.03}$   &$1.40_{-0.04-0.37}^{+0.03+0.41}$ &$-84.44_{-0.58-3.01}^{-0.61+3.25}$
&$V_4^{B_c^*\to B_s^*}$ &$0.40_{-0.01-0.03}^{+0.01+0.01}$   &$1.42_{-0.02-0.26}^{+0.02+0.34}$ &$-80.47_{-1.57-5.81}^{+1.49+6.00}$
\\
$V_5^{B_c^*\to B^*}$   &$3.15_{-0.11-0.82}^{+0.10+1.12}$   &$1.16_{-0.02-0.21}^{+0.02+0.29}$ &$-88.43_{-0.04-5.82}^{+0.04+8.03}$
&$V_5^{B_c^*\to B_s^*}$ &$3.52_{-0.06-0.87}^{+0.06+1.12}$   &$1.16_{-0.02+0.21}^{+0.01+0.28}$ &$-82.15_{-0.85-5.10}^{+0.85+8.23}$
\\
$V_6^{B_c^*\to B^*}$   &$2.66_{-0.10-0.79}^{+0.10+1.11}$   &$1.14_{-0.03-0.21}^{+0.03+0.29}$ &$-89.98_{-0.31-6.80}^{+0.24+7.69}$
&$V_6^{B_c^*\to B_s^*}$ &$3.02_{-0.06-0.85}^{+0.06+1.15}$   &$1.13_{-0.02-0.21}^{+0.01+0.29}$ &$-83.43_{-0.66-6.58}^{+0.64+8.16}$
\\
$A_1^{B_c^*\to B^*}$   &$0.43_{-0.02-0.08}^{+0.01+0.07}$   &$1.18_{-0.03-0.17}^{+0.02+0.16}$ &$-92.23_{-0.29-2.02}^{+0.25+1.99}$
&$A_1^{B_c^*\to B_s^*}$ &$0.53_{-0.01-0.09}^{+0.01+0.08}$   &$1.17_{-0.01-0.17}^{+0.01+0.25}$ &$-85.77_{-0.70-1.03}^{+0.66+1.36}$
\\
$A_2^{B_c^*\to B^*}$   &$0.43_{-0.02-0.08}^{+0.01+0.07}$   &$0.66_{-0.01-0.24}^{+0.01+0.25}$ &$37.14_{-0.07-7.22}^{+0.06+7.01}$
&$A_2^{B_c^*\to B_s^*}$ &$0.53_{-0.01-0.09}^{+0.01+0.08}$   &$0.67_{-0.08-0.17}^{+0.09+0.25}$ &$71.55_{-1.80-7.33}^{+2.42+8.44}$
\\
$A_3^{B_c^*\to B^*}$   &$0.81_{-0.03-0.20}^{+0.03+0.25}$   &$1.56_{-0.01-0.35}^{+0.01+0.40}$ &$-94.78_{-0.60-5.14}^{+0.73+7.38}$
&$A_3^{B_c^*\to B_s^*}$ &$0.73_{-0.02-0.15}^{+0.02+0.16}$   &$1.59_{-0.01-0.37}^{+0.01+0.41}$ &$-88.56_{-1.59-5.53}^{+1.90+6.86}$
\\
$A_4^{B_c^*\to B^*}$   &$0.89_{-0.02-0.24}^{+0.02+0.32}$   &$1.49_{-0.04-0.33}^{+0.04+0.45}$ &$-93.00_{-0.94-5.58}^{+0.91+7.77}$
&$A_4^{B_c^*\to B_s^*}$ &$0.85_{-0.01-0.20}^{+0.01+0.25}$   &$1.60_{-0.12-0.33}^{+0.10+0.32}$ &$-80.22_{-1.29-8.29}^{+1.57+9.02}$
\\
\hline\hline
\end{tabular}
\end{center}
\end{table}

%%%%%%%%%%%%%%%%%%%%%%%%%%%%%%%%%%%%%%%%%%%%%%%%%%%%%%%%%%%%%%%%%%%%%%%%%%%%%%%%%
\renewcommand{\baselinestretch}{1.2}
\begin{table}[!t]
\scriptsize
\begin{center}
\caption{\label{tab:sche2b1}  Same as Table \ref{tab:sche2c} except for  $b\to (q,s,c)$ induced $B^*\to (\rho\,,K^*\,,D^*)$ and $B_s\to (K^*\,,\phi\,,D^*_s)$ transitions. The errors are due to parameters $\beta$ and quark masses only (cf. section 4 for a more complete discussion).}
%Fitting results for the form factors of  $b\to (q,s,c)$ induced $B^*\to (D^*\,,K^*\,,\rho)$, $B^*_s\to (D^*_s\,,\phi\,,K^*)$, $B^*_c\to (J/\Psi\,,D^*_s\,,D^*)$, $\Upsilon(1S)\to (B_c^*\,,B_s^*\,,B^*)$ transitions  with the parameterization scheme given by~Eq.~\eqref{eq:para2}. 
\vspace{0.2cm}
\let\oldarraystretch=\arraystretch
\renewcommand*{\arraystretch}{1.3}
\setlength{\tabcolsep}{3pt}
\begin{tabular}{lccc|lccc}
\hline\hline
 &${\cal F}(0)$  &$a$   &$b_1$& &${\cal F}(0)$  &$a$   &$b_1$
\\\hline
$V_1^{B^*\to\rho}$&$0.27_{-0.01-0.04}^{+0.01+0.05}$ &$0.99_{-0.01-0.05}^{+0.01+0.04}$ &$-3.25_{-0.15-0.24}^{+0.14+0.12}$
&$V_1^{B^*\to K^*}$&$0.31_{-0.01-0.06}^{+0.01+0.06}$ &$1.00_{-0.01-0.08}^{+0.01+0.07}$ &$-3.28_{-0.06-0.12}^{+0.05+0.10}$
\\
$V_2^{B^*\to\rho}$
&$0.25_{-0.01-0.03}^{+0.01+0.03}$ &$1.00_{-0.04-0.05}^{+0.03+0.05}$ &$-3.34_{-0.06-0.14}^{+0.06+0.22}$
&$V_2^{B^*\to K^*}$
&$0.26_{-0.01-0.04}^{+0.01+0.03}$ &$1.00_{-0.05-0.06}^{+0.03+0.05}$ &$-3.27_{-0.16-0.17}^{+0.12+0.09}$
\\
$V_3^{B^*\to\rho}$
&$0.10_{-0.00-0.01}^{-0.00+0.02}$ &$1.00_{-0.01-0.03}^{+0.00+0.02}$ &$-5.91_{-0.13-0.50}^{+0.09+0.38}$
&$V_3^{B^*\to K^*}$
&$0.11_{-0.00-0.02}^{-0.00+0.02}$ &$0.98_{-0.00-0.02}^{+0.01+0.01}$ &$-5.18_{-0.08-0.41}^{+0.16+0.34}$
\\
$V_4^{B^*\to\rho}$
&$0.10_{-0.00-0.01}^{+0.00+0.02}$ &$0.99_{-0.00-0.01}^{+0.01+0.01}$ &$-5.91_{-0.09-0.43}^{+0.14+0.36}$
&$V_4^{B^*\to K^*}$
&$0.11_{-0.00-0.02}^{+0.00+0.02}$ &$0.99_{-0.00-0.04}^{+0.00+0.04}$ &$-5.69_{-0.09-0.32}^{+0.09+0.27}$
\\
$V_5^{B^*\to\rho}$
&$0.57_{-0.01-0.09}^{+0.01+0.10}$ &$1.00_{-0.01-0.08}^{+0.01+0.06}$ &$-3.10_{-0.07-0.18}^{+0.05+0.13}$
&$V_5^{B^*\to K^*}$
&$0.65_{-0.02-0.11}^{+0.02+0.11}$ &$0.99_{-0.01-0.05}^{+0.01+0.04}$ &$-3.17_{-0.13-0.14}^{+0.13+0.09}$
\\
$V_6^{B^*\to\rho}$
&$0.12_{-0.00-0.02}^{+0.00+0.03}$ &$0.89_{-0.01-0.01}^{+0.00+0.02}$ &$-3.24_{-0.11-0.35}^{+0.11+0.31}$
&$V_6^{B^*\to K^*}$
&$0.15_{-0.01-0.03}^{+0.01+0.04}$ &$0.90_{-0.01-0.04}^{+0.00+0.09}$ &$-3.34_{-0.15-0.20}^{+0.15+0.19}$
\\
$A_1^{B^*\to\rho}$
&$0.26_{-0.01-0.04}^{+0.01+0.05}$ &$0.99_{-0.02-0.04}^{+0.01+0.04}$ &$-3.21_{-0.17-0.23}^{+0.14+0.11}$
&$A_1^{B^*\to K^*}$
&$0.31_{-0.01-0.06}^{+0.01+0.06}$ &$0.98_{-0.01-0.05}^{+0.01+0.04}$ &$-3.28_{-0.03-0.22}^{+0.01+0.19}$
\\
$A_2^{B^*\to\rho}$
&$0.26_{-0.01-0.04}^{+0.01+0.05}$ &$0.51_{-0.01-0.06}^{+0.01+0.07}$ &$-1.14_{-0.01-0.13}^{+0.02+0.15}$
&$A_2^{B^*\to K^*}$
&$0.31_{-0.01-0.06}^{+0.01+0.06}$ &$0.55_{-0.06-0.15}^{+0.03+0.14}$ &$-1.21_{-0.20-0.28}^{+0.11+0.24}$
\\
$A_3^{B^*\to\rho}$
&$0.07_{-0.00-0.01}^{+0.00+0.01}$ &$1.00_{-0.00-0.01}^{+0.01+0.01}$ &$-5.39_{-0.03-0.59}^{+0.07+0.56}$
&$A_3^{B^*\to K^*}$
&$0.06_{-0.00-0.01}^{+0.00+0.01}$ &$1.00_{-0.02-0.01}^{+0.00+0.01}$ &$-5.48_{-0.03-0.64}^{+0.01+0.51}$
\\
$A_4^{B^*\to\rho}$
&$0.06_{-0.00-0.01}^{+0.00+0.01}$ &$0.99_{-0.00-0.01}^{+0.01+0.02}$ &$-5.38_{-0.12-0.61}^{+0.21+0.48}$
&$A_4^{B^*\to K^*}$
&$0.07_{-0.00-0.01}^{+0.00+0.01}$ &$0.98_{-0.01-0.02}^{+0.01+0.01}$ &$-5.56_{-0.09-0.55}^{+0.05+0.41}$
\\
\hline
%%%%%%%%%%%%%%%%%%%%%%%%%%%%%%%%%%%%%%%%%%%%%%%%%%%%%%%%%%%%%%%%%%%%%%%%%%%%%%%%%%%%%%%%%%%
$V_1^{B^*\to D^*}$   &$0.70_{-0.01-0.11}^{+0.01+0.10}$ &$1.13_{-0.01-0.07}^{+0.01+0.07}$ &$-2.54_{-0.01-0.20}^{+0.01+0.15}$
&$V_1^{B_s^*\to K^*}$ &$0.19_{-0.01-0.05}^{+0.01+0.06}$ &$1.00_{-0.02-0.02}^{+0.01+0.02}$ &$-5.12_{-0.24-0.20}^{+0.24+0.17}$
\\
$V_2^{B^*\to D^*}$   &$0.35_{-0.01-0.04}^{+0.01+0.02}$ &$1.15_{-0.01-0.07}^{+0.01+0.05}$ &$-2.51_{-0.02-0.16}^{+0.03+0.23}$
&$V_2^{B_s^*\to K^*}$ &$0.18_{-0.01-0.04}^{+0.01+0.04}$ &$1.00_{-0.01-0.02}^{+0.01+0.02}$ &$-5.14_{-0.23-0.18}^{+0.21+0.17}$
\\
$V_3^{B^*\to D^*}$   &$0.12_{-0.00-0.01}^{+0.00+0.02}$ &$1.39_{-0.02-0.25}^{+0.01+0.13}$ &$-4.05_{-0.06-1.02}^{+0.06+0.47}$
&$V_3^{B_s^*\to K^*}$ &$0.09_{-0.01-0.02}^{+0.01+0.02}$ &$0.98_{-0.01-0.02}^{+0.02+0.02}$ &$-6.37_{-0.20-0.33}^{+0.23+0.26}$
\\
$V_4^{B^*\to D^*}$   &$0.12_{-0.00-0.01}^{+0.00+0.02}$ &$1.47_{-0.01-0.40}^{+0.01+0.39}$ &$-4.27_{-0.03-0.99}^{+0.03+0.78}$
&$V_4^{B_s^*\to K^*}$ &$0.09_{-0.01-0.02}^{+0.01+0.02}$ &$1.00_{-0.02-0.03}^{+0.01+0.02}$ &$-7.00_{-0.18-0.30}^{+0.17+0.17}$
\\
$V_5^{B^*\to D^*}$   &$1.19_{-0.01-0.14}^{+0.01+0.09}$ &$1.13_{-0.02-0.08}^{+0.01+0.06}$ &$-2.48_{-0.06-0.21}^{+0.05+0.13}$
&$V_5^{B_s^*\to K^*}$ &$0.42_{-0.03-0.10}^{+0.02+0.12}$ &$1.00_{-0.03-0.03}^{+0.02+0.04}$ &$-4.95_{-0.30-0.19}^{+0.27+0.21}$
\\
$V_6^{B^*\to D^*}$   &$0.51_{-0.01-0.13}^{+0.01+0.15}$ &$1.13_{-0.02-0.09}^{+0.01+0.07}$ &$-2.43_{-0.05-0.21}^{+0.02+0.18}$
&$V_6^{B_s^*\to K^*}$ &$0.10_{-0.01-0.03}^{+0.01+0.03}$ &$1.00_{-0.01-0.02}^{+0.02+0.01}$ &$-4.62_{-0.17-0.17}^{+0.19+0.15}$
\\
$A_1^{B^*\to D^*}$   &$0.69_{-0.01-0.11}^{+0.01+0.10}$ &$1.13_{-0.01-0.07}^{+0.01+0.07}$ &$-2.52_{-0.01-0.21}^{+0.01+0.24}$
&$A_1^{B_s^*\to K^*}$ &$0.19_{-0.01-0.05}^{+0.01+0.05}$ &$0.99_{-0.02-0.04}^{+0.02+0.03}$ &$-5.09_{-0.13-0.06}^{+0.07+0.05}$
\\
$A_2^{B^*\to D^*}$   &$0.69_{-0.01-0.11}^{+0.01+0.10}$ &$0.80_{-0.04-0.15}^{+0.04+0.14}$ &$0.24_{-0.14-0.13}^{+0.08+0.17}$
&$A_2^{B_s^*\to K^*}$ &$0.19_{-0.01-0.05}^{+0.01+0.05}$ &$0.61_{-0.03-0.02}^{+0.03+0.03}$ &$-3.12_{-0.17-0.32}^{+0.13+0.38}$
\\
$A_3^{B^*\to D^*}$   &$0.07_{-0.00-0.01}^{+0.00+0.01}$ &$1.45_{-0.03-0.17}^{+0.02+0.27}$ &$-4.18_{-0.02-0.66}^{+0.05+0.54}$
&$A_3^{B_s^*\to K^*}$ &$0.06_{-0.00-0.01}^{+0.00+0.01}$ &$0.98_{-0.00-0.01}^{+0.01+0.01}$ &$-6.74_{-0.16-0.27}^{+0.20+0.30}$
\\
$A_4^{B^*\to D^*}$   &$0.08_{-0.00-0.01}^{+0.00+0.01}$ &$1.47_{-0.00-0.21}^{+0.01+0.30}$ &$-4.25_{-0.01-0.27}^{+0.01+0.33}$
&$A_4^{B_s^*\to K^*}$ &$0.05_{-0.00-0.01}^{+0.00+0.02}$ &$0.98_{-0.00-0.03}^{+0.01+0.01}$ &$-6.70_{-0.16-0.33}^{+0.17+0.23}$
\\
\hline
%%%%%%%%%%%%%%%%%%%%%%%%%%%%%%%%%%%%%%%%%%%%%%%%%%%%%%%%%%%%%%%%%%%%%%%%%%%%%%%%%%%%%%%%%%%
$V_1^{B_s^*\to \phi}$  &$0.27_{-0.01-0.06}^{+0.01+0.07}$ &$0.99_{-0.01-0.02}^{+0.01+0.02}$ &$-4.84_{-0.17-0.14}^{+0.16+0.10}$
&$V_1^{B_s^*\to D_s^*}$ &$0.69_{-0.01-0.12}^{+0.01+0.11}$ &$1.10_{-0.02-0.04}^{+0.02+0.05}$ &$-3.34_{-0.10-0.29}^{+0.09+0.30}$
\\
$V_2^{B_s^*\to \phi}$  &$0.23_{-0.01-0.05}^{+0.01+0.05}$ &$0.53_{-0.03-0.27}^{+0.03+0.34}$ &$-6.39_{-0.34-0.76}^{+0.32+1.08}$
&$V_2^{B_s^*\to D_s^*}$ &$0.38_{-0.01-0.03}^{+0.01+0.03}$ &$1.16_{-0.02-0.20}^{+0.02+0.25}$ &$-3.34_{-0.08-0.03}^{+0.05+0.03}$
\\
$V_3^{B_s^*\to \phi}$  &$0.11_{-0.00-0.02}^{+0.00+0.02}$ &$1.00_{-0.02-0.01}^{+0.02+0.02}$ &$-6.23_{-0.14-0.28}^{+0.08+0.27}$
&$V_3^{B_s^*\to D_s^*}$ &$0.15_{-0.00-0.02}^{+0.00+0.02}$ &$1.36_{-0.02-0.25}^{+0.02+0.34}$ &$-4.51_{-0.09-0.80}^{+0.08+0.63}$
\\
$V_4^{B_s^*\to \phi}$  &$0.11_{-0.00-0.02}^{+0.00+0.02}$ &$1.00_{-0.02-0.03}^{+0.02+0.03}$ &$-6.90_{-0.17-0.20}^{+0.19+0.15}$
&$V_4^{B_s^*\to D_s^*}$ &$0.15_{-0.00-0.02}^{+0.00+0.02}$ &$1.52_{-0.01-0.53}^{+0.01+0.30}$ &$-5.43_{-0.03-1.43}^{+0.03+0.91}$
\\
$V_5^{B_s^*\to \phi}$  &$0.57_{-0.02-0.12}^{+0.02+0.14}$ &$0.99_{-0.02-0.01}^{+0.03+0.01}$ &$-4.67_{-0.17-0.19}^{+0.24+0.10}$
&$V_5^{B_s^*\to D_s^*}$ &$1.22_{-0.02-0.16}^{+0.01+0.11}$ &$1.09_{-0.02-0.04}^{+0.03+0.05}$ &$-3.36_{-0.11-0.30}^{+0.10+0.30}$
\\
$V_6^{B_s^*\to \phi}$  &$0.15_{-0.01-0.04}^{+0.01+0.05}$ &$0.97_{-0.01-0.02}^{+0.01+0.02}$ &$-4.40_{-0.14-0.27}^{+0.14+0.23}$
&$V_6^{B_s^*\to D_s^*}$ &$0.55_{-0.01-0.14}^{+0.01+0.15}$ &$1.06_{-0.03-0.07}^{+0.02+0.06}$ &$-3.15_{-0.13-0.27}^{+0.07+0.23}$
\\
$A_1^{B_s^*\to \phi}$  &$0.26_{-0.01-0.06}^{+0.01+0.07}$ &$0.99_{-0.01-0.02}^{+0.01+0.02}$ &$-4.70_{-0.16-0.12}^{+0.17+0.09}$
&$A_1^{B_s^*\to D_s^*}$ &$0.68_{-0.01-0.12}^{+0.01+0.11}$ &$1.10_{-0.02-0.05}^{+0.02+0.05}$
&$-3.33_{-0.10-0.28}^{+0.10+0.30}$
\\
$A_2^{B_s^*\to \phi}$  &$0.26_{-0.01-0.06}^{+0.01+0.07}$ &$0.58_{-0.01-0.03}^{+0.02+0.05}$ &$-2.81_{-0.08-0.20}^{+0.13+0.44}$
&$A_2^{B_s^*\to D_s^*}$ &$0.68_{-0.01-0.12}^{+0.01+0.11}$ &$0.51_{-0.04-0.05}^{+0.04+0.07}$ &$-1.68_{-0.08-0.58}^{+0.05+0.76}$
\\
$A_3^{B_s^*\to \phi}$  &$0.07_{-0.00-0.01}^{+0.00+0.01}$ &$0.97_{-0.01-0.02}^{+0.00+0.01}$ &$-6.46_{-0.01-0.60}^{+0.01+0.42}$
&$A_3^{B_s^*\to D_s^*}$ &$0.09_{-0.00-0.01}^{+0.00+0.02}$ &$1.45_{-0.01-0.28}^{+0.01+0.37}$ &$-4.62_{-0.01-0.79}^{+0.01+0.66}$
\\
$A_4^{B_s^*\to \phi}$  &$0.07_{-0.00-0.02}^{+0.00+0.02}$ &$0.97_{-0.01-0.02}^{+0.00+0.01}$ &$-6.50_{-0.12-0.44}^{+0.11+0.32}$
&$A_4^{B_s^*\to D_s^*}$ &$0.09_{-0.00-0.01}^{+0.00+0.01}$ &$1.46_{-0.04-0.29}^{+0.04+0.33}$ &$-4.62_{-0.07-0.95}^{+0.07+0.56}$
\\
\hline\hline
\end{tabular}
\end{center}
\end{table}

%%%%%%%%%%%%%%%%%%%%%%%%%%%%%%%%%%%%%%%%%%%%%%%%%%%%%%%%%%%%%%%%%%%%%%%%%%%%%%%%%
\renewcommand{\baselinestretch}{1.2}
\begin{table}[!t]
\scriptsize
\begin{center}
\caption{\label{tab:sche2b2}  Same as Table \ref{tab:sche2c} except for  $b\to (q,s,c)$ induced $B^*_c\to (D^*\,,D^*_s\,,J/\Psi)$ and $\Upsilon(1S)\to (B^*\,,B_s^*\,,B_c^*)$ transitions. The errors are due to parameters $\beta$ and quark masses only (cf. section 4 for a more complete discussion). }
%Fitting results for the form factors of  $b\to (q,s,c)$ induced $B^*\to (D^*\,,K^*\,,\rho)$, $B^*_s\to (D^*_s\,,\phi\,,K^*)$, $B^*_c\to (J/\Psi\,,D^*_s\,,D^*)$, $\Upsilon(1S)\to (B_c^*\,,B_s^*\,,B^*)$ transitions  with the parameterization scheme given by~Eq.~\eqref{eq:para2}. 
\vspace{0.2cm}
\let\oldarraystretch=\arraystretch
\renewcommand*{\arraystretch}{1.3}
\setlength{\tabcolsep}{3pt}
\begin{tabular}{lccc|lccc}
\hline\hline
 &${\cal F}(0)$  &$a$   &$b_1$& &${\cal F}(0)$  &$a$   &$b_1$
\\\hline
$V_1^{B_c^*\to D^*}$   &$0.10_{-0.01-0.04}^{+0.01+0.06}$   &$0.99_{-0.01-0.01}^{+0.01+0.02}$ &$-13.44_{-0.56-0.72}^{+0.56+0.92}$
&$V_1^{B_c^*\to D_s^*}$ &$0.20_{-0.02-0.07}^{+0.02+0.10}$   &$1.00_{-0.02-0.01}^{+0.02+0.01}$ &$-12.03_{-0.76-0.68}^{+0.65+0.80}$
\\
$V_2^{B_c^*\to D^*}$   &$0.11_{-0.01-0.04}^{+0.01+0.06}$   &$0.99_{-0.01-0.01}^{+0.01+0.01}$ &$-13.71_{-0.43-0.70}^{+0.43+0.91}$
&$V_2^{B_c^*\to D_s^*}$ &$0.20_{-0.02-0.06}^{+0.02+0.07}$   &$1.00_{-0.01-0.02}^{+0.01+0.02}$ &$-12.45_{-0.40-0.69}^{+0.57+0.74}$
\\
$V_3^{B_c^*\to D^*}$   &$0.08_{-0.01-0.03}^{+0.01+0.04}$   &$0.98_{-0.02-0.01}^{+0.02+0.01}$ &$-13.94_{-0.55-0.63}^{+0.56+0.78}$
&$V_3^{B_c^*\to D_s^*}$ &$0.13_{-0.01-0.04}^{+0.01+0.04}$   &$0.99_{-0.01-0.04}^{+0.01+0.03}$ &$-12.94_{-0.63-0.70}^{+0.61+0.53}$
\\
$V_4^{B_c^*\to D^*}$   &$0.08_{-0.01-0.03}^{+0.01+0.04}$   &$0.99_{-0.01-0.01}^{+0.01+0.01}$ &$-14.61_{-0.48-0.57}^{+0.48+0.72}$
&$V_4^{B_c^*\to D_s^*}$ &$0.13_{-0.01-0.04}^{+0.01+0.04}$   &$1.00_{-0.01-0.04}^{+0.01+0.04}$ &$-14.20_{-0.59-0.59}^{+0.51+0.52}$
\\
$V_5^{B_c^*\to D^*}$   &$0.27_{-0.03-0.11}^{+0.03+0.15}$   &$1.00_{-0.01-0.02}^{+0.01+0.02}$ &$-13.04_{-0.58-0.80}^{+0.56+0.98}$
&$V_5^{B_c^*\to D_s^*}$ &$0.50_{-0.05-0.16}^{+0.05+0.20}$   &$1.00_{-0.02-0.01}^{+0.02+0.01}$ &$-11.63_{-0.69-0.72}^{+0.66+0.88}$
\\
$V_6^{B_c^*\to D^*}$   &$0.11_{-0.01-0.04}^{+0.01+0.06}$   &$1.00_{-0.01-0.02}^{+0.01+0.02}$ &$-12.82_{-0.60-0.73}^{+0.56+0.97}$
&$V_6^{B_c^*\to D_s^*}$ &$0.22_{-0.02-0.07}^{+0.02+0.08}$   &$0.98_{-0.02-0.01}^{+0.02+0.01}$ &$-11.21_{-0.65-0.71}^{+0.75+0.96}$
\\
$A_1^{B_c^*\to D^*}$   &$0.10_{-0.01-0.04}^{+0.01+0.06}$   &$1.00_{-0.01-0.02}^{+0.01+0.01}$ &$-13.61_{-0.57-0.69}^{+0.56+0.92}$
&$A_1^{B_c^*\to D_s^*}$ &$0.19_{-0.02-0.06}^{+0.02+0.08}$   &$0.98_{-0.02-0.02}^{+0.02+0.02}$ &$-12.07_{-0.61-0.58}^{+0.67+0.89}$
\\
$A_2^{B_c^*\to D^*}$   &$0.10_{-0.01-0.04}^{+0.01+0.06}$   &$0.99_{-0.02-0.01}^{+0.01+0.01}$ &$-10.30_{-0.76-1.31}^{+0.61+1.68}$
&$A_2^{B_c^*\to D_s^*}$ &$0.19_{-0.02-0.06}^{+0.02+0.08}$   &$0.95_{-0.05-0.06}^{+0.03+0.03}$ &$-7.61_{-1.03-1.15}^{+0.52+1.28}$
\\
$A_3^{B_c^*\to D^*}$   &$0.06_{-0.01-0.02}^{+0.01+0.03}$   &$1.00_{-0.01-0.01}^{+0.01+0.01}$ &$-14.87_{-0.52-0.51}^{+0.47+0.67}$
&$A_3^{B_c^*\to D_s^*}$ &$0.10_{-0.01-0.03}^{+0.01+0.03}$   &$0.98_{-0.01-0.01}^{+0.02+0.02}$ &$-13.83_{-0.60-0.43}^{+0.55+0.55}$
\\
$A_4^{B_c^*\to D^*}$   &$0.06_{-0.01-0.02}^{+0.01+0.03}$   &$1.00_{-0.01-0.01}^{+0.01+0.01}$ &$-14.47_{-0.44-0.55}^{+0.51+0.64}$
&$A_4^{B_c^*\to D_s^*}$ &$0.09_{-0.01-0.03}^{+0.01+0.03}$   &$0.99_{-0.02-0.01}^{+0.01+0.01}$ &$-13.44_{-0.61-0.37}^{+0.59+0.37}$
\\
\hline
%%%%%%%%%%%%%%%%%%%%%%%%%%%%%%%%%%%%%%%%%%%%%%%%%%%%%%%%%%%%%%%%%%%%%%%%%%%%%%%%%%%%%%%%%%%%
$V_1^{B_c^*\to J/\psi}$       &$0.56_{-0.01-0.17}^{+0.01+0.17}$  &$1.51_{-0.03-0.08}^{+0.03+0.07}$  &$-8.50_{-0.21-0.89}^{+0.21+1.05}$
&$V_1^{\Upsilon(1S)\to B^*}$ &$0.06_{-0.01-0.02}^{+0.01+0.03}$  &$0.98_{-0.02-0.02}^{+0.01+0.02}$ &$-50.22_{-1.20-0.36}^{+1.10+0.36}$
\\
$V_2^{B_c^*\to J/\psi}$       &$0.33_{-0.01-0.04}^{+0.01+0.05}$  &$1.57_{-0.02-0.06}^{+0.01+0.04}$  &$-8.62_{-0.09-0.57}^{+0.08+0.72}$
&$V_2^{\Upsilon(1S)\to B^*}$ &$0.08_{-0.01-0.03}^{+0.01+0.03}$  &$0.98_{-0.02-0.02}^{+0.02+0.01}$ &$-52.09_{-1.05-0.50}^{+1.06+0.51}$
\\
$V_3^{B_c^*\to J/\psi}$       &$0.20_{-0.00-0.02}^{+0.00+0.02}$  &$1.67_{-0.04-0.07}^{+0.04+0.10}$  &$-9.27_{-0.23-0.42}^{+0.23+0.51}$
&$V_3^{\Upsilon(1S)\to B^*}$ &$0.06_{-0.01-0.02}^{+0.01+0.03}$  &$0.98_{-0.02-0.02}^{+0.02+0.02}$ &$-48.95_{-1.32-0.17}^{+1.12+0.18}$
\\
$V_4^{B_c^*\to J/\psi}$       &$0.20_{-0.00-0.02}^{+0.00+0.02}$  &$1.82_{-0.04-0.11}^{+0.05+0.14}$ &$-10.14_{-0.21-0.17}^{+0.19+0.11}$
&$V_4^{\Upsilon(1S)\to B^*}$ &$0.06_{-0.01-0.02}^{+0.01+0.03}$  &$1.00_{-0.03-0.02}^{+0.02+0.02}$ &$-49.50_{-1.22-0.11}^{+1.11+0.10}$
\\
$V_5^{B_c^*\to J/\psi}$       &$1.17_{-0.02-0.29}^{+0.02+0.23}$  &$1.47_{-0.03-0.12}^{+0.03+0.09}$ &$-8.27_{-0.21-0.90}^{+0.22+1.19}$
&$V_5^{\Upsilon(1S)\to B^*}$ &$0.22_{-0.02-0.08}^{+0.03+0.11}$  &$1.00_{-0.02-0.02}^{+0.02+0.02}$ &$-48.96_{-1.23-0.13}^{+1.21+0.07}$
\\
$V_6^{B_c^*\to J/\psi}$       &$0.65_{-0.01-0.19}^{+0.01+0.20}$  &$1.41_{-0.03-0.09}^{+0.03+0.06}$ &$-8.05_{-0.20-0.86}^{+0.20+1.11}$
&$V_6^{\Upsilon(1S)\to B^*}$ &$0.13_{-0.02-0.05}^{+0.01+0.06}$  &$0.99_{-0.02-0.02}^{+0.02+0.02}$ &$-49.46_{-1.27-0.20}^{+1.28+0.11}$
\\
$A_1^{B_c^*\to J/\psi}$       &$0.54_{-0.01-0.17}^{+0.01+0.16}$  &$1.51_{-0.03-0.12}^{+0.03+0.08}$ &$-8.48_{-0.20-0.89}^{+0.21+1.16}$
&$A_1^{\Upsilon(1S)\to B^*}$ &$0.05_{-0.01-0.02}^{+0.01+0.03}$  &$0.97_{-0.02-0.02}^{+0.02+0.01}$ &$-50.24_{-1.09-0.18}^{+1.21+0.16}$
\\
$A_2^{B_c^*\to J/\psi}$       &$0.54_{-0.01-0.17}^{+0.01+0.16}$  &$0.94_{-0.03-0.12}^{+0.02+0.06}$ &$-6.16_{-0.19-1.07}^{+0.20+1.56}$
&$A_2^{\Upsilon(1S)\to B^*}$ &$0.05_{-0.01-0.02}^{+0.01+0.03}$  &$0.86_{-0.10-0.09}^{+0.12+0.10}$ &$-46.15_{-1.48-0.70}^{+2.49+0.59}$
\\
$A_3^{B_c^*\to J/\psi}$       &$0.13_{-0.00-0.02}^{+0.00+0.03}$  &$1.75_{-0.02-0.45}^{+0.02+0.58}$ &$-9.56_{-0.11-2.25}^{+0.13+2.05}$
&$A_3^{\Upsilon(1S)\to B^*}$ &$0.06_{-0.01-0.02}^{+0.01+0.03}$  &$1.00_{-0.14-0.02}^{+0.11+0.02}$ &$-51.25_{-1.17-0.55}^{+0.94+0.37}$
\\
$A_4^{B_c^*\to J/\psi}$       &$0.14_{-0.00-0.02}^{+0.00+0.02}$  &$1.75_{-0.04-0.10}^{+0.04+0.13}$ &$-9.59_{-0.22-0.36}^{+0.22+0.41}$
&$A_4^{\Upsilon(1S)\to B^*}$ &$0.07_{-0.01-0.03}^{+0.01+0.04}$  &$0.96_{-0.02-0.03}^{+0.02+0.03}$ &$-50.14_{-0.96-0.32}^{+1.09+0.34}$
\\
\hline
%%%%%%%%%%%%%%%%%%%%%%%%%%%%%%%%%%%%%%%%%%%%%%%%%%%%%%%%%%%%%%%%%%%%%%%%%%%%%%%%%%%%%%%%%%%%
$V_1^{\Upsilon(1S)\to B_s^*}$
&$0.09_{-0.01-0.03}^{+0.01+0.04}$ &$1.00_{-0.02-0.05}^{+0.03+0.05}$ &$-51.38_{-0.86-0.59}^{+0.98+0.31}$
&$V_1^{\Upsilon(1S)\to B_c^*}$
&$0.44_{-0.01-0.13}^{+0.01+0.15}$ &$2.01_{-0.05-0.05}^{+0.05+0.03}$  &$-30.90_{-0.89-0.35}^{+0.85+0.89}$
\\
$V_2^{\Upsilon(1S)\to B_s^*}$
&$0.11_{-0.01-0.03}^{+0.01+0.04}$ &$0.98_{-0.03-0.04}^{+0.03+0.03}$ &$-53.90_{-0.26-0.88}^{+0.27+0.62}$
&$V_2^{\Upsilon(1S)\to B_c^*}$
&$0.29_{-0.03-0.05}^{+0.03+0.08}$ &$2.18_{-0.04-0.39}^{+0.03+0.45}$  &$-32.25_{-0.26-3.57}^{+0.28+2.20}$
\\
$V_3^{\Upsilon(1S)\to B_s^*}$
&$0.09_{-0.01-0.03}^{+0.01+0.04}$ &$1.00_{-0.02-0.04}^{+0.02+0.06}$ &$-50.08_{-1.01-0.54}^{+0.86+0.35}$
&$V_3^{\Upsilon(1S)\to B_c^*}$
&$0.24_{-0.01-0.04}^{+0.01+0.03}$ &$2.04_{-0.04-0.09}^{+0.05+0.05}$  &$-30.63_{-0.80-0.87}^{+0.78+1.63}$
\\
$V_4^{\Upsilon(1S)\to B_s^*}$
&$0.09_{-0.01-0.03}^{+0.01+0.04}$ &$0.99_{-0.05-0.09}^{+0.05+0.09}$ &$-53.00_{-1.53-1.29}^{+2.88+1.32}$
&$V_4^{\Upsilon(1S)\to B_c^*}$
&$0.24_{-0.01-0.04}^{+0.01+0.03}$ &$2.13_{-0.04-0.09}^{+0.04+0.03}$  &$-31.13_{-0.84-1.42}^{+0.69+2.47}$
\\
$V_5^{\Upsilon(1S)\to B_s^*}$
&$0.32_{-0.02-0.11}^{+0.02+0.14}$ &$1.00_{-0.02-0.06}^{+0.02+0.06}$ &$-50.01_{-1.03-0.25}^{+1.02+0.25}$
&$V_5^{\Upsilon(1S)\to B_c^*}$
&$1.31_{-0.04-0.33}^{+0.04+0.33}$ &$1.96_{-0.06-0.05}^{+0.05+0.04}$  &$-29.98_{-0.88-0.60}^{+0.74+0.87}$
\\
$V_6^{\Upsilon(1S)\to B_s^*}$
&$0.19_{-0.01-0.06}^{+0.01+0.09}$ &$1.00_{-0.02-0.06}^{+0.02+0.06}$ &$-50.40_{-1.04-0.69}^{+0.88+0.65}$
&$V_6^{\Upsilon(1S)\to B_c^*}$
&$0.92_{-0.03-0.25}^{+0.03+0.35}$ &$1.92_{-0.04-0.23}^{+0.06+0.07}$  &$-30.17_{-0.76-5.28}^{+0.99+3.11}$
\\
$A_1^{\Upsilon(1S)\to B_s^*}$
&$0.08_{-0.01-0.03}^{+0.01+0.04}$ &$0.95_{-0.02-0.04}^{+0.02+0.03}$ &$-51.88_{-0.91-0.42}^{+0.96+0.78}$
&$A_1^{\Upsilon(1S)\to B_c^*}$
&$0.41_{-0.01-0.12}^{+0.01+0.14}$ &$2.01_{-0.06-0.06}^{+0.06+0.03}$  &$-30.88_{-0.88-0.45}^{+0.85+0.80}$
\\
$A_2^{\Upsilon(1S)\to B_s^*}$
&$0.08_{-0.01-0.03}^{+0.01+0.04}$ &$1.00_{-0.06-0.10}^{+0.05+0.09}$ &$-44.97_{-1.10-0.93}^{+0.60+0.97}$
&$A_2^{\Upsilon(1S)\to B_c^*}$
&$0.41_{-0.01-0.12}^{+0.01+0.14}$ &$1.48_{-0.07-0.08}^{+0.07+0.05}$ &$-25.53_{-1.12-0.69}^{+1.10+0.89}$
\\
$A_3^{\Upsilon(1S)\to B_s^*}$
&$0.09_{-0.01-0.03}^{+0.01+0.03}$ &$1.00_{-0.04-0.07}^{+0.04+0.05}$ &$-52.91_{-0.98-0.33}^{+0.60+0.57}$
&$A_3^{\Upsilon(1S)\to B_c^*}$
&$0.21_{-0.01-0.04}^{+0.01+0.05}$ &$2.17_{-0.02-0.37}^{+0.02+0.42}$  &$-31.97_{-0.26-3.93}^{+0.16+2.80}$
\\
$A_4^{\Upsilon(1S)\to B_s^*}$
&$0.10_{-0.01-0.03}^{+0.01+0.04}$ &$0.96_{-0.04-0.02}^{+0.03+0.02}$ &$-51.82_{-1.07-0.68}^{+0.73+0.60}$
&$A_4^{\Upsilon(1S)\to B_c^*}$
&$0.25_{-0.01-0.04}^{+0.01+0.03}$ &$2.12_{-0.05-0.20}^{+0.02+0.15}$  &$-31.59_{-0.80-1.99}^{+0.76+2.51}$
\\
\hline\hline
\end{tabular}
\end{center}
\end{table}

Using the values of input parameters collected in Tables \ref{tab:qm} and \ref{tab:input} and employing the type-II scheme, we then present our  numerical predictions for the  form factors of $D^*\to (K^*\,,\rho)$, $D^*_s\to (\phi\,,K^*)$, $J/\Psi\to (D^*_s\,,D^*)$,  $B^*_c\to (B^*_s\,,B^*)$ transitions induced by $c\to (q,s)$, where $q=u$ and $d$,  and  $B^*\to (D^*\,,K^*\,,\rho)$, $B^*_s\to (D^*_s\,,\phi\,,K^*)$, $B^*_c\to (J/\Psi\,,D^*_s\,,D^*)$, $\Upsilon(1S)\to (B_c^*\,,B_s^*\,,B^*)$ transitions induced by $b\to (q,s,c)$. It should be noted that the theoretical  results given in the last sections are obtained in the $q^+=0$ frame, which implies that the form factors are known only for space-like momentum transfer, $q^2=-\mathbf{q}_\bot^2\leqslant 0$, and the results in the time-like region need  an additional $q^2$ extrapolation. To achieve this purpose, the three parameters form~\cite{Ball:1998kk}
\begin{align}\label{eq:para1}
{\cal F}(q^2)=\frac{{\cal F}(0)}{1-a(q^2/M_{B,D}^2)+b(q^2/M_{B,D}^2)^2}\,,
\end{align}
is usually employed by the LFQMs.  Here, $M_{B,D}$ is the mass of the relevant $B$ and $D$ mesons, {\it i.e.}, $M_{B_{q,s,c}}$ and $M_{D_{q,s}}$ for $b\to (q,s,c)$ and $c\to (q,s)$ transitions respectively; $a$ and $b$ are parameters obtained by fitting to the results computed  directly within LFMQs. Our results for the form factors based on Eq.~\eqref{eq:para1} are collected in appendix. From these results, it is found that the LFQMs' results obtained in the space-like region~(dots in Figs.~\ref{fig:sche1c} and \ref{fig:sche1b} ) can be well reproduced via  Eq.~\eqref{eq:para1}; however, for the case of $b\to$ light-quark transition with a heavy spectator quark~(for instance, $B_c^*\to D^*$), the fitting results for $b$ are very large, $b\sim 10$,  which result in the  non-monotonic $q^2$ dependences of some form factors in the time-like region. It can be clearly seen from Fig.~\ref{fig:sche1b}.

Besides of the three parameters form given by Eq.~\eqref{eq:para1}, in order to avoid the abnormal $q^2$ dependence mentioned above, we also employ  the $z$-series parameterization scheme~\cite{Bourrely:1980gp}. For the phenomenological application, we adopt
\begin{align}\label{eq:para2}
{\cal F}(q^2)=\frac{{\cal F}(0)}{1-a q^2/M_{B^*,D^*}}\left\{1+\sum_{k=1}^N b_k \left[ z(q^2, t_0)^k- z(0, t_0)^k\right]   \right\}\,,
\end{align}
where, $z(q^2, t_0)=\frac{\sqrt{t_+-q^2}-\sqrt{t_+-t_0}}{\sqrt{t_+-q^2}+\sqrt{t_+-t_0}}$, $t_\pm=(M'\pm M'')^2$. It is similar to the BCL version of the $z$-series expansion~\cite{Bourrely:2008za,Gao:2019lta}, but an additional parameter $a\sim1$ is introduced to improve the performance of Eq.~\eqref{eq:para2}. In the practice, we will truncate the expansion at $N=1$. In addition, since $M_{B_c^{*}}$ hasn't been measured yet, we take $ M_{B_c^{*}}-M_{B_c}=54\,{\rm MeV}$~predicted by lattice QCD~\cite{Dowdall:2012ab}. Then, we collect our numerical results for ${\cal F}(0)$, $a$ and $b_1$ in Tables~\ref{tab:sche2c}, \ref{tab:sche2b1} and \ref{tab:sche2b2}.
%where the theoretical  errors for ${\cal F}(0)$ are caused by $\beta$ and quark masses.  
The $q^2$ dependences of form factors are shown in Figs.~\ref{fig:sche2c} and \ref{fig:sche2b}. From these results, it can be found that the LFQMs' results obtained in the space-like region can be well reproduced via the Eq.~\eqref{eq:para2}, and the $q^2$ dependences are monotonic in the whole allowed $q^2$ region. In addition, our results for all of  transitions respect the relations
\begin{align}
V_3(0)=V_4(0)\,,\qquad A_1(0)=A_2(0)\,,
\end{align}
 which are essential to assure that the hadronic matrix element of $V'\to V''$ is divergence free at $q^2 = 0$. These results can be applied further in the relevant phenomenological studies of meson decays.

%%%%%%%%%%%%%%%%%%%%%

\renewcommand{\baselinestretch}{1.2}
\begin{table}[!t]
\footnotesize
\begin{center}
\caption{\label{tab:comp} \small Theoretical predictions for the form factors of $J/\Psi\to D^*$ and $J/\Psi\to D^*_s$ transitions at $q^2=0$ in this work, QCD SR~\cite{ Wang:2007ys},  CCQM~\cite{Ivanov:2015woa} and BS method~\cite{Wang:2016dkd}. }
\vspace{0.2cm}
\let\oldarraystretch=\arraystretch
\renewcommand*{\arraystretch}{1.3}
\setlength{\tabcolsep}{4pt}
\begin{tabular}{l|cccc|cccccccc}
\hline\hline
    &\multicolumn{4}{c|}{$J/\Psi\to D^*$}  &\multicolumn{4}{c}{$J/\Psi\to D^*_s$}\\\hline
      &this work  &QCD SR~\cite{ Wang:2007ys} & CCQM~\cite{Ivanov:2015woa}& BS~\cite{Wang:2016dkd}
      & this work  &QCD SR~\cite{ Ivanov:2015woa} & CCQM~\cite{Ivanov:2015woa}& BS~\cite{Wang:2016dkd} \\
 \hline
$V_1'$&$0.55^{+0.02+0.12}_{-0.02-0.12}$&$0.41^{+0.01}_{-0.01}$&$0.51$&$0.58$
            &$0.72^{+0.01+0.12}_{-0.02-0.13}$&$0.54^{+0.01}_{-0.01}$&$0.60$&$0.71$\\
$V_2'$&$0.60^{+0.05+0.15}_{-0.05-0.10}$&$0.63^{+0.01}_{-0.04}$&$0.39$&$0.29$
            &$0.68^{+0.06+0.14}_{-0.06-0.11}$&$0.69^{+0.05}_{-0.06}$&$0.34$&$0.26$\\
$V_3'$&$0.36^{+0.01+0.03}_{-0.01-0.03}$&$0.22^{+0.03}_{-0.01}$&$0.11$&$0.35$
            &$0.36^{+0.01+0.01}_{-0.01-0.02}$&$0.24^{+0.03}_{-0.01}$&$0.10$&$0.36$\\
$V_4'$&$0.36^{+0.01+0.03}_{-0.01-0.03}$&$0.26^{+0.03}_{-0.05}$&$0.11$&$0.35$
            &$0.36^{+0.01+0.01}_{-0.01-0.02}$&$0.26^{+0.03}_{-0.03}$&$0.10$&$0.36$\\
$V_5'$&$1.79^{+0.05+0.52}_{-0.05-0.34}$&$1.37^{+0.08}_{-0.03}$&$1.68$&$1.66$
            &$2.16^{+0.04+0.51}_{-0.04-0.36}$&$1.69^{+0.10}_{-0.03}$&$1.84$&$1.83$\\
$V_6'$&$1.23^{+0.04+0.53}_{-0.05-0.34}$&$0.87^{+0.05}_{-0.01}$&$1.05$&$1.23$
            &$1.58^{+0.04+0.58}_{-0.04-0.40}$&$1.14^{+0.08}_{-0.01}$&$1.23$&$1.38$\\
\hline
$A_1'$&$0.47^{+0.01+0.09}_{-0.01-0.09}$&$0.40^{+0.03}_{-0.01}$&$0.42$&$-0.43$
           &$0.62^{+0.01+0.08}_{-0.01-0.10}$&$0.53^{+0.03}_{-0.01}$&$0.51$&$-0.56$\\
$A_2'$&$0.47^{+0.01+0.09}_{-0.01-0.09}$&$0.44^{+0.10}_{-0.04}$&$0.42$&$-0.43$
           &$0.62^{+0.01+0.08}_{-0.01-0.10}$&$0.53^{+0.05}_{-0.01}$&$0.51$&$-0.56$\\
$A_3'$&$0.76^{+0.02+0.22}_{-0.02-0.16}$&$0.86^{+0.05}_{-0.01}$&$0.41$&$0.14$
            &$0.70^{+0.02+0.14}_{-0.02-0.12}$&$0.91^{+0.05}_{-0.01}$&$0.37$&$0.25$\\
$A_4'$&$0.90^{+0.02+0.30}_{-0.02-0.22}$&$0.91^{+0.06}_{-0.04}$&$0.41$&$-0.14$
           &$0.90^{+0.02+0.26}_{-0.02-0.18}$&$0.91^{+0.06}_{-0.01}$&$0.37$&$-0.25$
\\\hline\hline
\end{tabular}
\end{center}
\end{table}

Some semileptonic decays induced by $B_{u,s,c}^*\to V$ transitions are studied within the BS method~\cite{Wang:2018dtb}, but the relevant form factors are not given.  The form factors of  $J/\Psi\to (D^*_s\,,D^*)$ transition have also been evaluated by other approaches, for instance, the QCD sum rules~(QCD SR)~\cite{Wang:2007ys}, a covariant constituent quark model~(CCQM)~\cite{Ivanov:2015woa} and the BS method~\cite{Wang:2016dkd}. These theoretical predictions are collected in Table~ \ref{tab:comp}, in which the convention for the definitions of form factors in Refs.~\cite{Wang:2007ys,Ivanov:2015woa} is used. The LF form factors~($V_{1-6}$ and $A_{1-4}$) defined by Eqs.~\eqref{eq:defFV} and \eqref{eq:defFA} are related to the ones ($V'_{1-6}$ and $A'_{1-4}$) defined in Ref.~\cite{Wang:2007ys,Ivanov:2015woa} via
\begin{align}
V'_{1-6}=V_{1-6}\,,  \quad A_{1,2}'=A_{1,2}\,, \quad A_{3,4}'=2A_{3,4}\,;
\end{align}
and the form factors~($t_{1-6}$ and $h_{3-6}$)  in the BS method~\cite{Wang:2016dkd} are related to $V'_{1-6}$ and $A'_{1-4}$ via
\begin{align}
&V_1'=\frac{t_5+t_6}{2}\,,\quad V_2'=\frac{t_6-t_5}{2}\,,\quad V_3'=(t_1+t_2)\frac{M'^2-M''^2}{2M'^2}\,,\quad V_4'=V_3'+(t_1-t_2)\frac{q^2}{2M'^2}\,,\nonumber\\
& V'_{5,6}=t_{4,3}\,, \quad  A_1'=\frac{h_6-h_5}{2}\,,\quad A_2'=A_1'-\frac{q^2(h_6+h_5)}{2(M'^2-M''^2)}\,,\quad  A_{3}'=h_{3}\frac{M'^2-M''^2}{M'^2}\,, \nonumber\\
& A_{4}'=-h_{4}\frac{M'^2-M''^2}{M'^2}\,.
\end{align}
Through comparison of these results listed in Table \ref{tab:comp}, it can be found that they are different from each other more or less but are still  in rough consistence  within theoretical uncertainties.

In the most of previous works based on the LFQMs, the theoretical errors are not given since the error analysis is hard to be made in a systematical way. 
% is an inherent problem of quark models. 
 Finally, we would like to discuss  briefly the possible sources of errors/uncertainties of this work as follows:
\begin{itemize}
\item  An obvious source of error is the input parameters $\beta$ and quark masses. Especially, the  quark masses are model dependent~(for instance, their values are dependent on the assumptions for the potential),  which has been explained at the beginning of this section. The errors of this part have been evaluated and given above. Numerically, these input parameters result in about $[10,30]\%$ errors for  most of transitions\footnote{For a few form factors of heavy-heavy meson $\to$ light-heavy meson transitions~(for instance, $V_1^{\Upsilon(1S)\to B^*}(0)=0.06^{+0.01+0.03}_{-0.01-0.02}$), because the central values of their results are very small, the errors caused by input parameters can reach up to about $60\%$.  \label{fn:d} }. More explicitly, taking $D^*\to \rho $ transition as an example, one can find from Table~\ref{tab:sche2c} that the errors are at the  level of  about $(15,12,10,10,13,26)\%$ for $V_{1-6}(0)$ and $(14,14,18,21)\%$ for $A_{1-4}(0)$.   Besides the errors induced by the quark masses and $\beta$, there are  some other sources of errors/uncertainties, which  are not fully considered in this work and will be briefly discussed  in the following. 
%Besides the errors induced by the inputs,  other sources of errors/uncertainties will be briefly discussed  in the following.
 %}
%---------
\item   The WF is also an important input in the  LFQMs.
%The WF is also an important  input in the calculation, but it can not been obtained by LFQMs directly. 
In principle, it can be obtained by solving the light-front QCD bound state equation. However, except in some simplified cases, the full solution has remained a challenge, therefore we would have to be contented with some phenomenological WFs.  In this work, we have employed a  Gaussian type WF given by Eq.~\eqref{eq:RWFs}, which is  commonly used in the LFQMs. In addition to Eq.~\eqref{eq:RWFs}, there are several other popular alternatives, for instance, BSW WF~\cite{Wirbel:1985ji}, BHL WF~(another Gaussian type WF)~\cite{Lepage:1983}, power-law type WF~\cite{Cheung:1995ub} and so on, which have been employed in some of previous works~\cite{Cheng:1996if,Cheng:1997au,Hwang:2001hj,Geng:2001de,Choi:2017uos,Choi:1997qh}. The different choices for the WF may result in some uncertainties of the theoretical prediction more or less. For instance, the difference between the predictions for $f_{D^*}$ with Gaussian~(G) and power-law~(PL) type WFs   can reach up to $|f_{D^*}^{\rm PL }-f_{D^*}^{\rm G }|/f_{D^*}^{\rm G }\sim 15\%$~\cite{Hwang:2010hw}~(one can refer to Ref.~\cite{Hwang:2010hw} for more examples). For the form factor concerned in this paper,  taking $D^*\to \rho $ transition as an example again, we find that $|{\cal F}^{\rm PL }-{\cal F}^{\rm G }|/{\cal F}^{\rm G }\sim (5,4,1,1,2,7)\%$ for ${\cal F}=V_{1-6}$ and $ \sim (3,3,18,1)\%$ for ${\cal F}=A_{1-4}$ at $q^2=0$~\footnote{It should be noted that the values of input parameters are different for different WFs. In the calculation of ${\cal F}^{\rm PL }(0)$, the PL type WF given by Eq.~(4.4) in Ref.~\cite{Hwang:2010hw} is employed; and correspondingly, the quark masses $m_{q,c}^{\rm PL}=(0.172,1.36)\,{\rm GeV}$ obtained with such PL type WF~\cite{Hwang:2010hw} and  $\beta_{q\bar{q},c\bar{q}}^{\rm PL}=(0.292,0.497)\,{\rm GeV}$ updated by fitting to the decay constants are used. }.
%we find that $|{\cal F}^{\rm PL }(0)-{\cal F}^{\rm G }(0)|/{\cal F}^{\rm G }(0)\sim (14,10,14,14,13,24)\%$ for ${\cal F}=V_{1-6}$ and $ \sim (10,10,23,29)\%$ for ${\cal F}=A_{1-4}$
%---------
\item  The predictions of LFQMs may be affected also by the radiative corrections.
%, which are not considered in this paper. 
The quark model is an approximation and effective picture of  QCD, the soft QCD effects are expected to be encoded in  the  WF~(vertex function) and relevant parameters, while the short-distance corrections are hard to be calculated precisely from the first principle of QCD in a quark model because the bridge between QCD and quark model is not fully clear for now.  For the SLF QM, the treatment of explicit gluon exchange between constituent quarks goes beyond the valence quark picture\footnote{As has been shown in Ref.~\cite{Jaus:1996np},  it is expected that the calculation of the form factor in the light-front formalism automatically takes into account a subset of higher-order gluon exchange diagrams. }, which is a basic assumption of the SLF QM.  For the CLF QM~\cite{Jaus:1999zv}, the short-distance corrections can not be calculated directly either when the LF (valence) vertex function obtained by matching to the SLF QM at 1-loop approximation is used. Interestingly, it is claimed in  Ref.~\cite{Jaus:1999dw} that an alternative and more general covariant approach to treat hadronic matrix elements in the light-front formalism can be established by combining with the methods developed in Ref.~\cite{Carbonell:1998rj}. This modified CLF QM is valid for a general vertex function, and is employed to calculate radiative corrections of ${\cal O}(\alpha)$ for pion beta decay,  while the gluon exchange diagrams are not calculated directly but approximated by means of $\rho$ exchange~(vector meson dominance ansatz), and the correction to pion beta  decay rate is very small and can be safely neglected~\cite{Jaus:1999dw}.  In addition, it is expected that the spurious $\omega$-dependent contributions associated with $B$ functions~(for instance,  $R_{\mu\nu}$ term in Ref.~\cite{Jaus:1999dw} ) can be canceled by higher order gluon exchange contributions, but it has not been demonstrated. More efforts are needed still for evaluating the radiative corrections to form factors in a systematical way within the framework of LFQMs.  
% and Ref.~\cite{Jaus:1999zv} has been applied
%---------
\item  As has been mentioned in the last section, the $q^+=0$ frame is employed in this paper for convenience of calculation. The form factors are known only for space-like momentum transfer and the results in the time-like region are obtained by  an additional $q^2$ extrapolation.  Thus, the predictions in the time-like region may have some uncertainties caused by the extrapolation scheme, which  can be clearly seen by comparing the results in two parameterization schemes given above.   Although ones can take $q^+>0$ frame instead of $q^+=0$ frame to evaluate directly the form factors in the time-like region, the theoretical uncertainties exist still because the nonvalence contributions from so-called $Z$ graph should be fully considered in such case, and it is unavoidable to introduce some additional assumptions and/or help from other models  for calculating these contributions. 
%---------
\item Besides, the effect of isospin symmetry breaking is not considered in this work. Such effect in the LFQMs is  generated mainly by quark mass difference $\Delta m=m_d-m_u$, which is small enough and can be the neglected except for the case that the quantity considered is proportional to $\Delta m$~(for instance, the decay constant of scalar meson). 
\end{itemize}
From above discussions and taking $D^*\to \rho$ as an example, we can find that the combined errors caused by the input parameters and WFs are about  $(20,16,11,11,15,34)\%$ for $V_{1-6}(0)$ and $ (17,17,36,21)\%$ for $A_{1-4}(0)$. Finally, further considering that the effect of isospin symmetry breaking is trivial, and supposing that the radiative corrections are not as large as the errors caused by the input parameters and WFs, we may roughly estimate that the total errors are at the level of about $[20,50]\%$ for most of ${\cal F}(0)$ with a few exceptions\textsuperscript{\ref{fn:d}}.

 \section{Summary}
In this paper,  the matrix elements and  relevant form factors of $V'\to V''$ transition are  calculated within the SLF and the CLF approach.  The self-consistency and Lorentz covariance of the CLF QM are analyzed in detail. It is found that both of them are violated within the traditional correspondence scheme~(type-I) between  the  manifest covariant BS  approach and the LF approach given by Eq.~\eqref{eq:type1}, while they can be recovered by employing the type-II correspondence scheme given by Eq.~\eqref{eq:type2} which requires an additional replacement $M\to M_0$  relative to type-I scheme. Such replacement is also favored by the self-consistency  of the SLF QM. Within the type-II correspondence scheme, the zero-mode contributions to the form factors exist only in form but vanish numerically, and the valence contributions are exactly the same as the SLF results, which can be concluded as  the relation $[{\cal Q}]_{\rm SLF}=[{\cal Q}]_{\rm val.}\doteq[{\cal Q}]_{\rm CLF}$. The findings mentioned above confirm again the conclusions obtained via $f_{V,A}$ and $P\to V$  transition in the previous works~\cite{Choi:2013mda,Chang:2018zjq,Chang:2019mmh}.  Finally, we present  our  numerical predictions for the form factors of $c\to (q,s)$~($q=u,d$) induced $D^*\to (K^*\,,\rho)$, $D^*_s\to (\phi\,,K^*)$, $J/\Psi\to (D^*_s\,,D^*)$,  $B^*_c\to (B^*_s\,,B^*)$ transitions and $b\to (q,s,c)$ induced $B^*\to (D^*\,,K^*\,,\rho)$, $B^*_s\to (D^*_s\,,\phi\,,K^*)$, $B^*_c\to (J/\Psi\,,D^*_s\,,D^*)$, $\Upsilon(1S)\to (B_c^*\,,B_s^*\,,B^*)$ transitions, which are collected in Tables \ref{tab:sche2c}, {\ref{tab:sche2b1} and \ref{tab:sche2b2}. These results can be applied further to the relevant phenomenological studies of meson decays.

\newpage
\begin{appendix}
\section*{Appendix: Results for the form factors with dipole approximation given by Eq.~\eqref{eq:para1}.}
Using the values of input parameters collected in Tables \ref{tab:qm} and \ref{tab:input}  and employing the parameterization scheme given by Eq.~\eqref{eq:para1}, we  present our numerical results for ${\cal F}(0)$, $a$ and $b$ in Tables~\ref{tab:sche1c}, \ref{tab:sche1b1} and \ref{tab:sche1b2}.  Besides, the $q^2$ dependences of form factors are shown in Figs.~\ref{fig:sche1c} and \ref{fig:sche1b}.

 \begin{figure}[h]
\caption{Same as Fig.~\ref{fig:sche2c} except  with the parameterization scheme given by Eq.~\eqref{eq:para1}. }
\begin{center}
\subfigure{\includegraphics[scale=0.285]{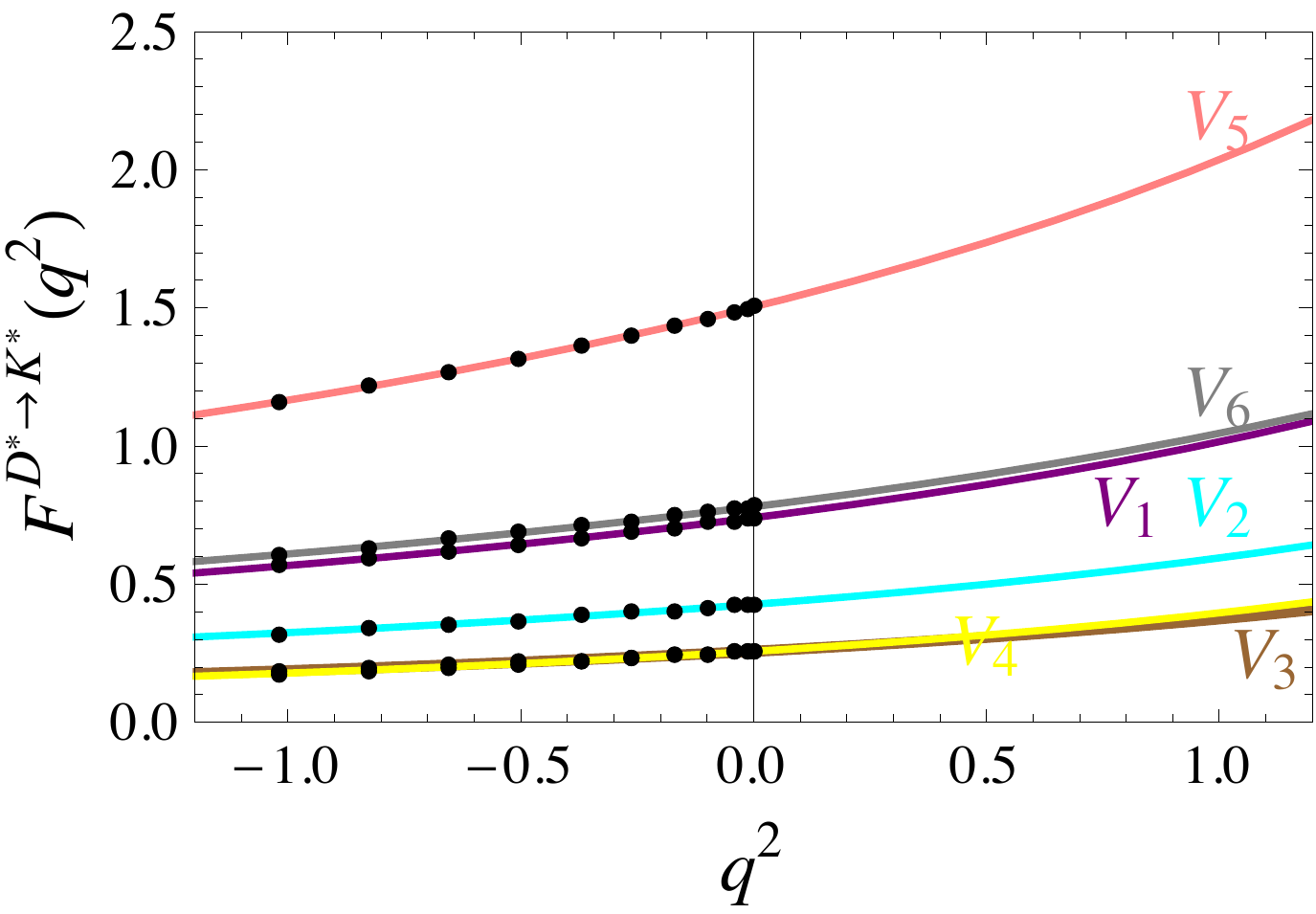}}\,
\subfigure{\includegraphics[scale=0.24]{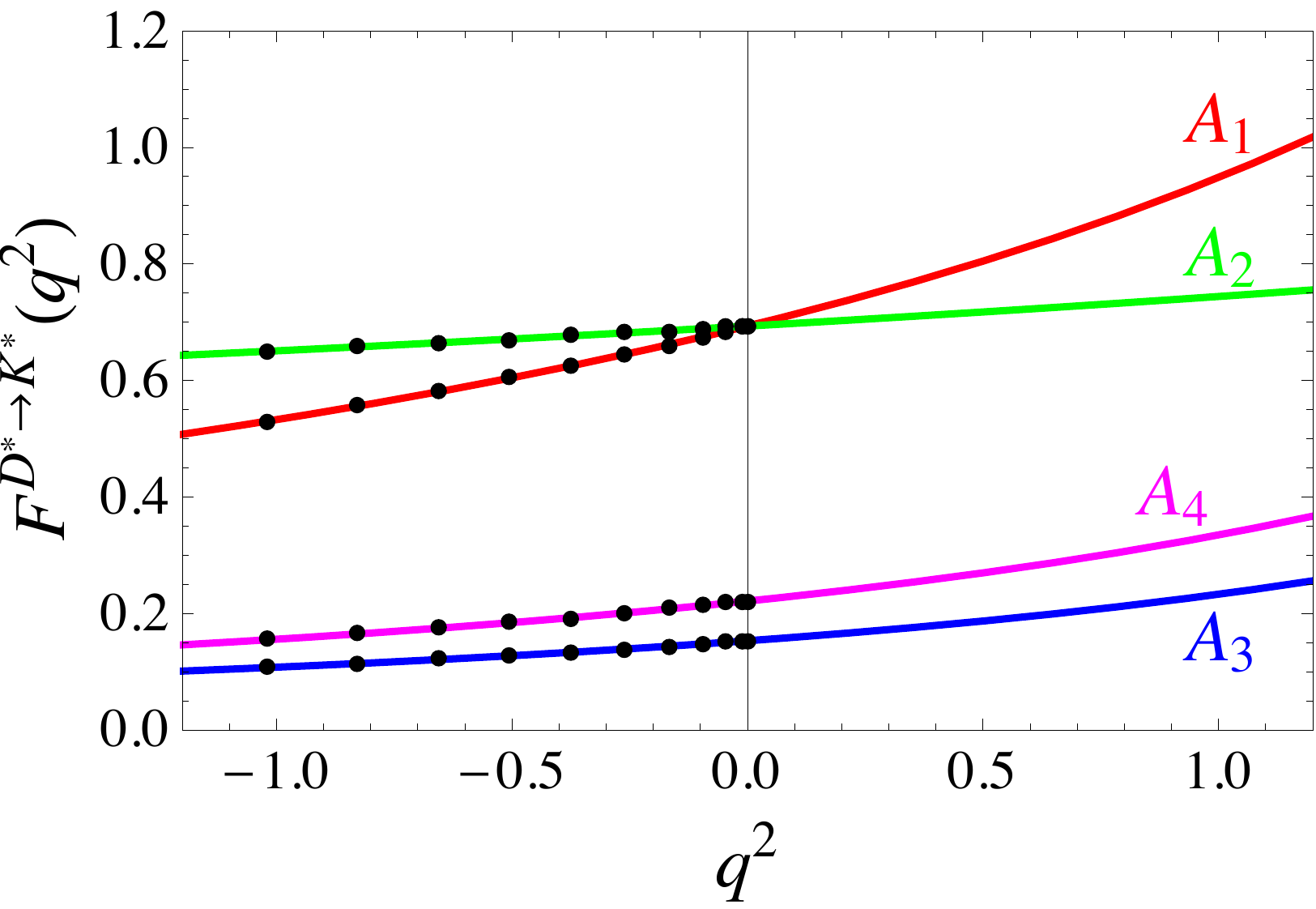}}\,
\subfigure{\includegraphics[scale=0.245]{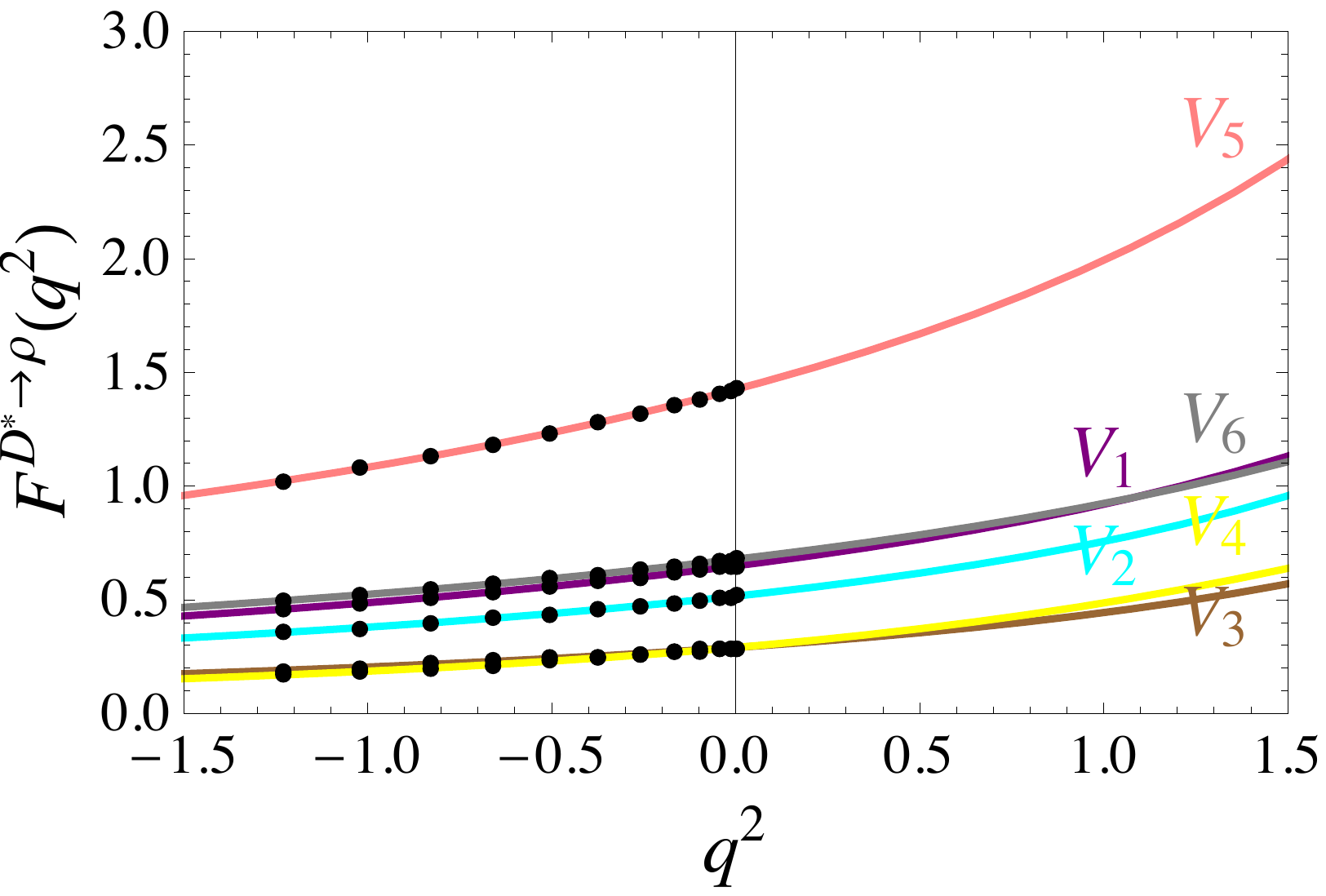}}\,
\subfigure{\includegraphics[scale=0.245]{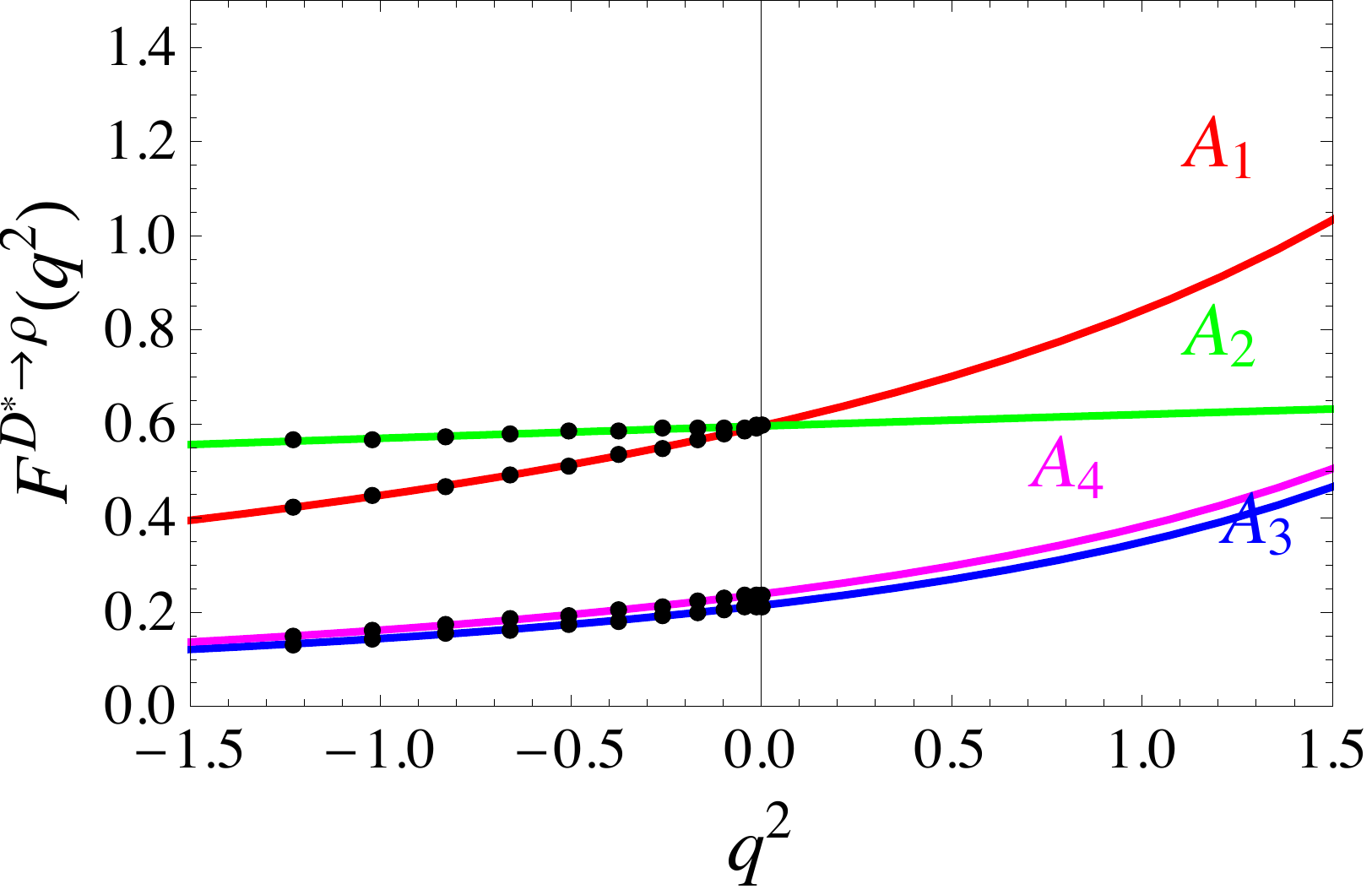}}\\
\subfigure{\includegraphics[scale=0.265]{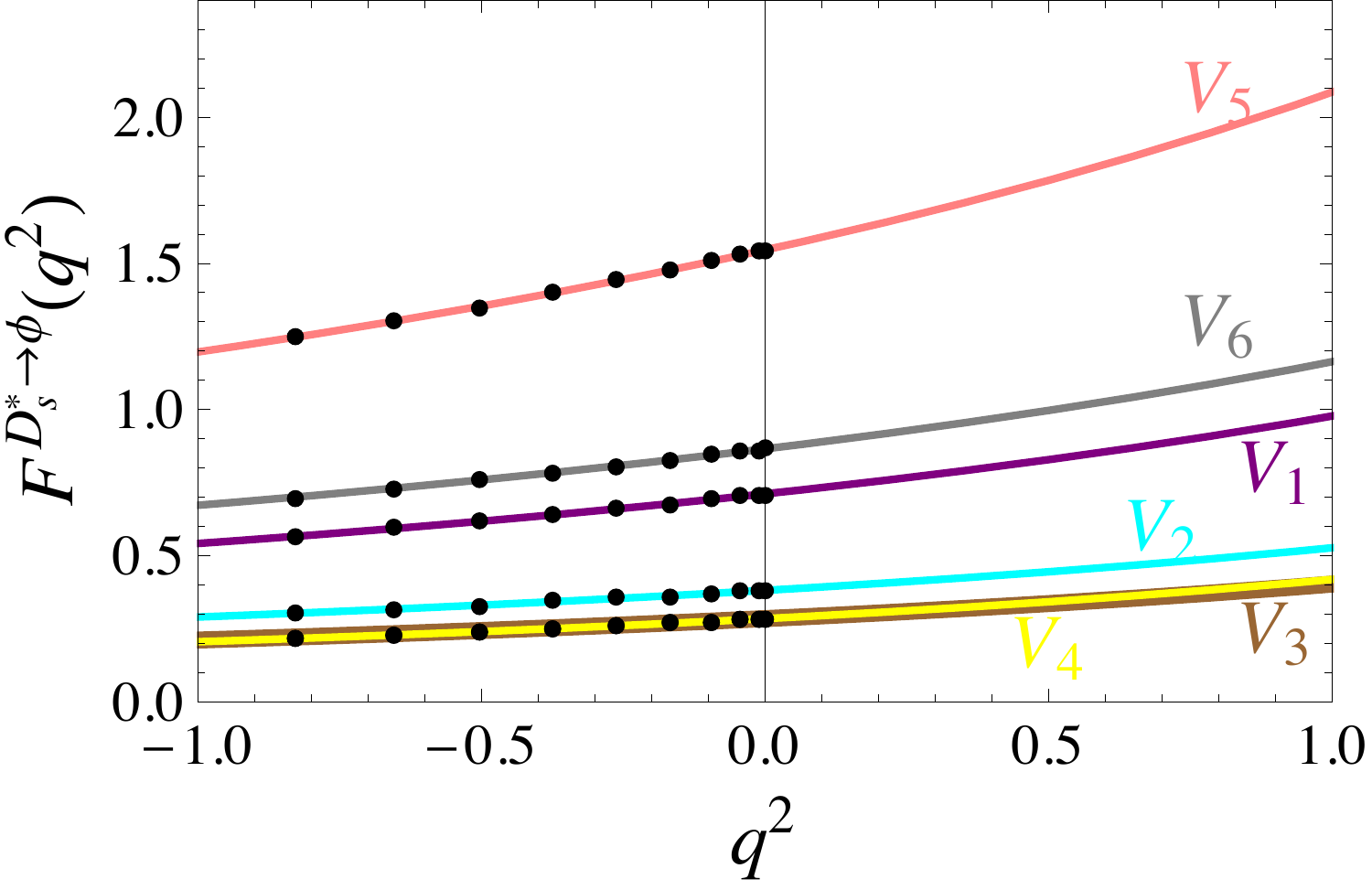}}\,
\subfigure{\includegraphics[scale=0.265]{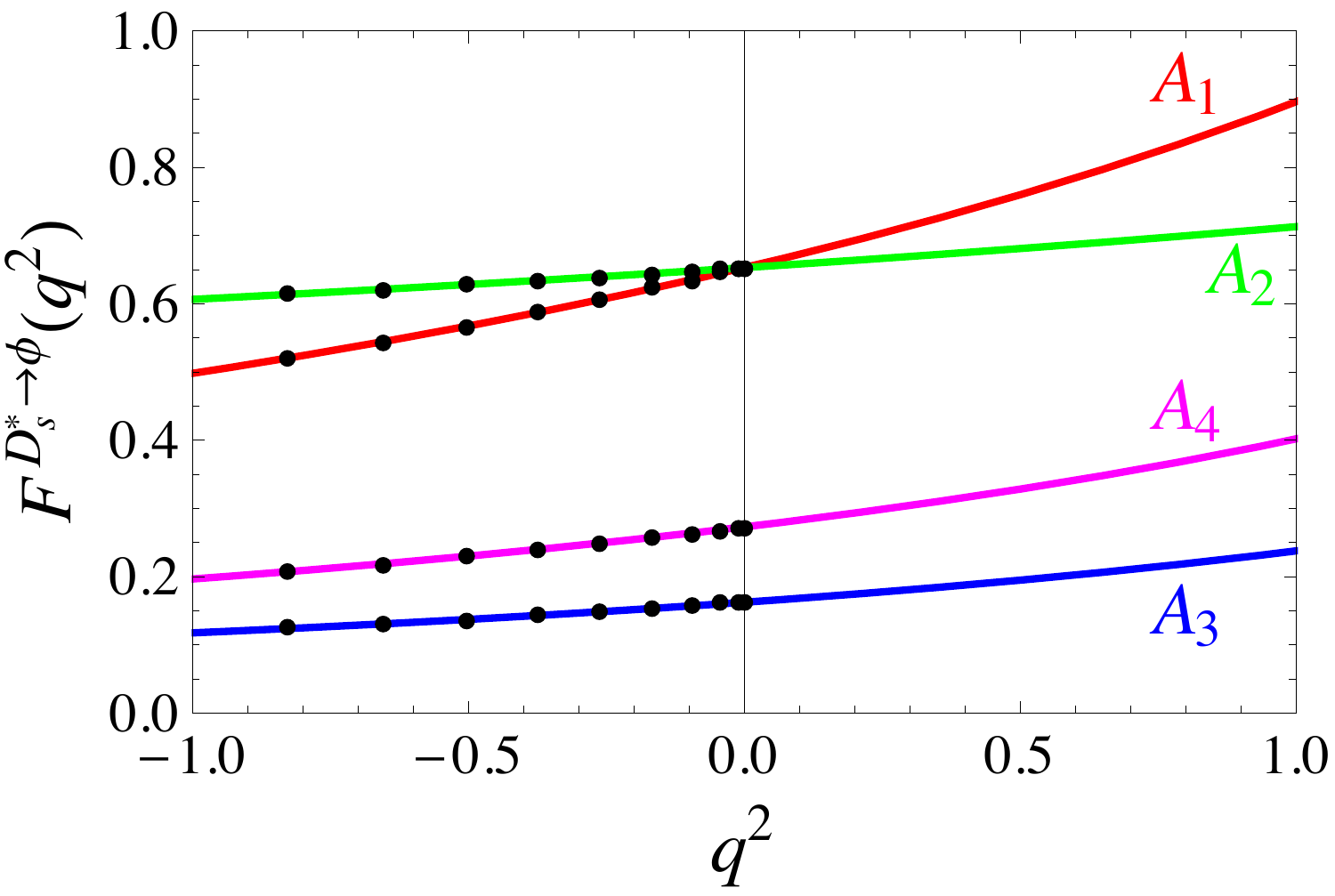}}\,
\subfigure{\includegraphics[scale=0.24]{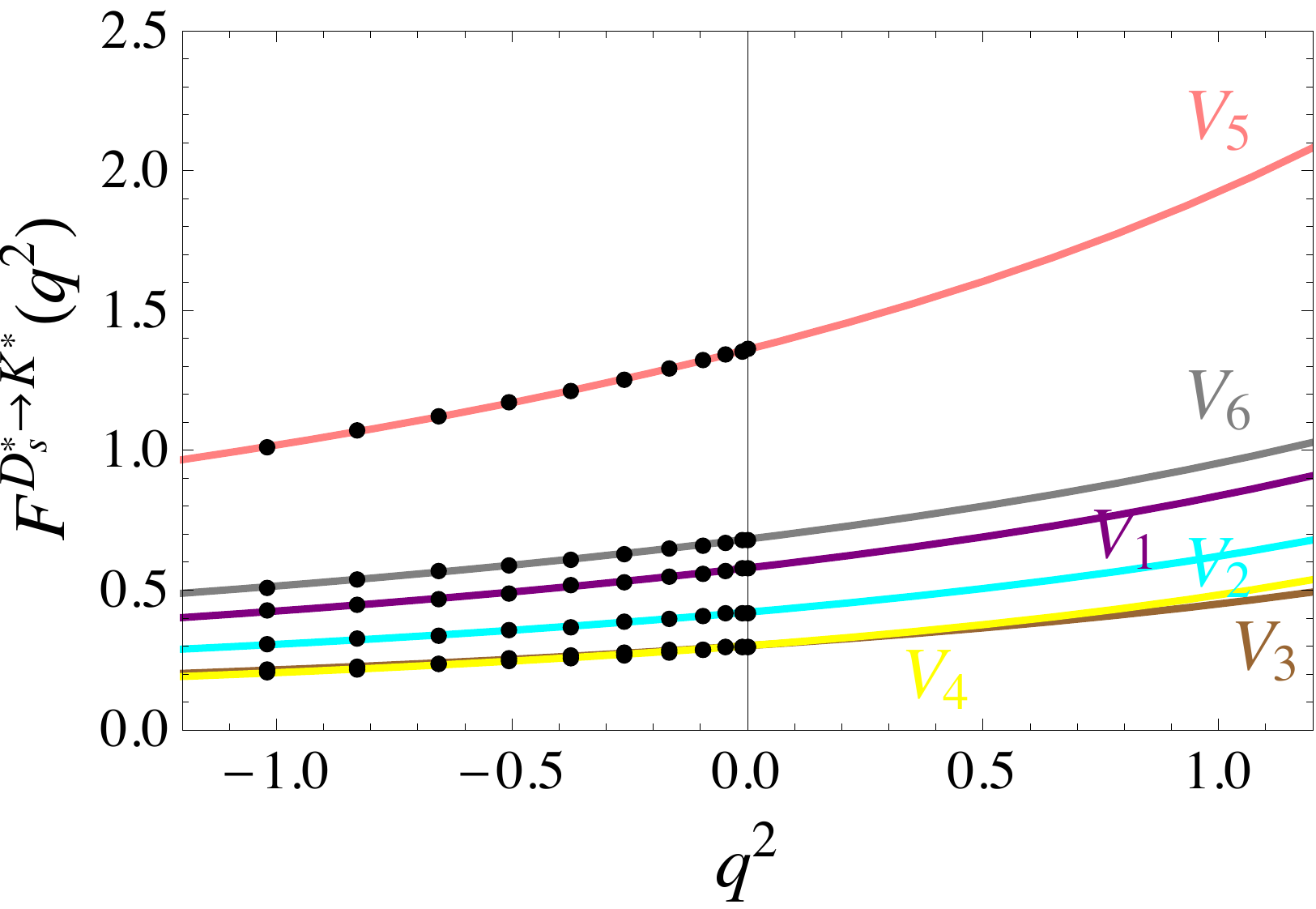}}\,
\subfigure{\includegraphics[scale=0.24]{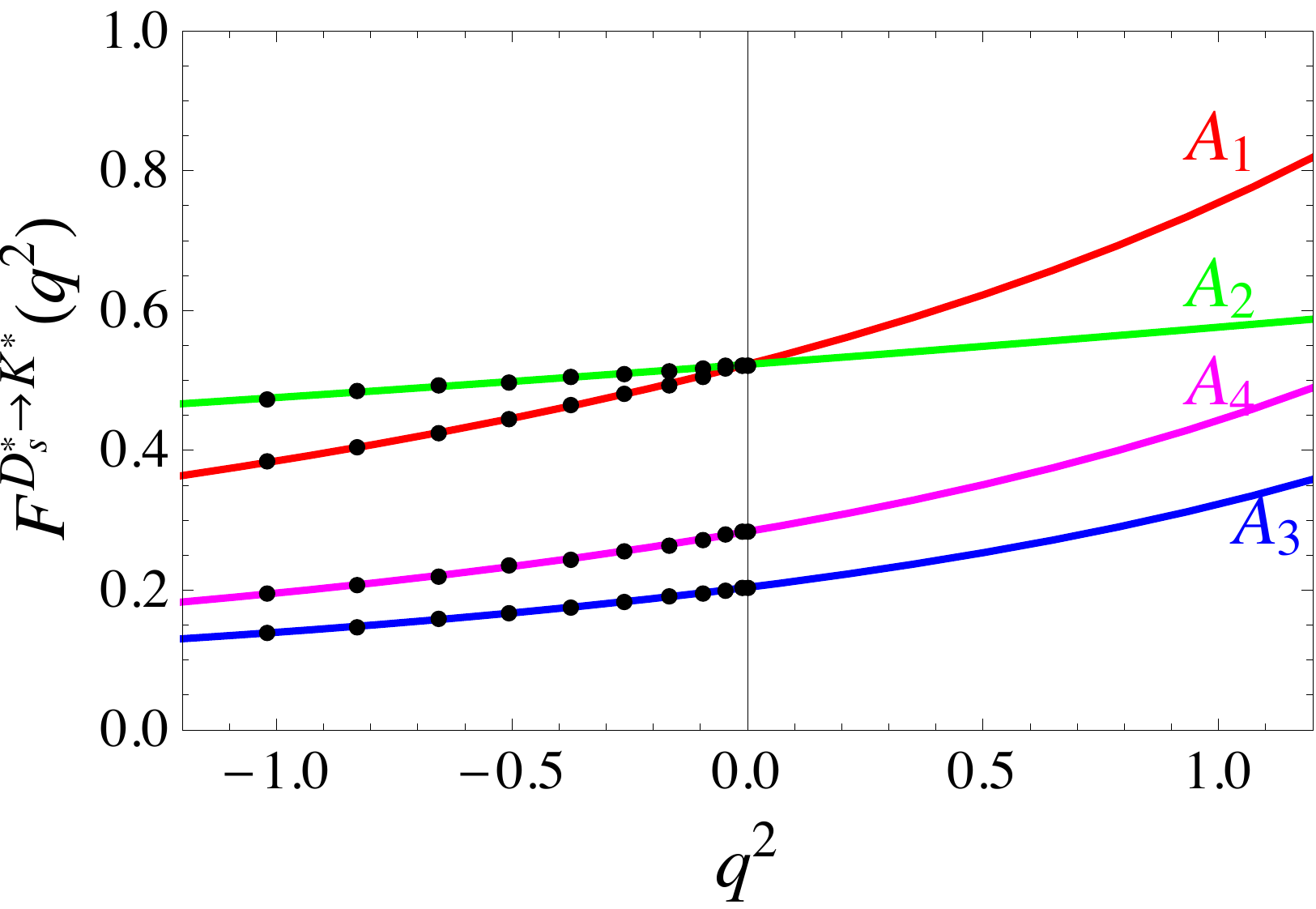}}\\
\subfigure{\includegraphics[scale=0.32]{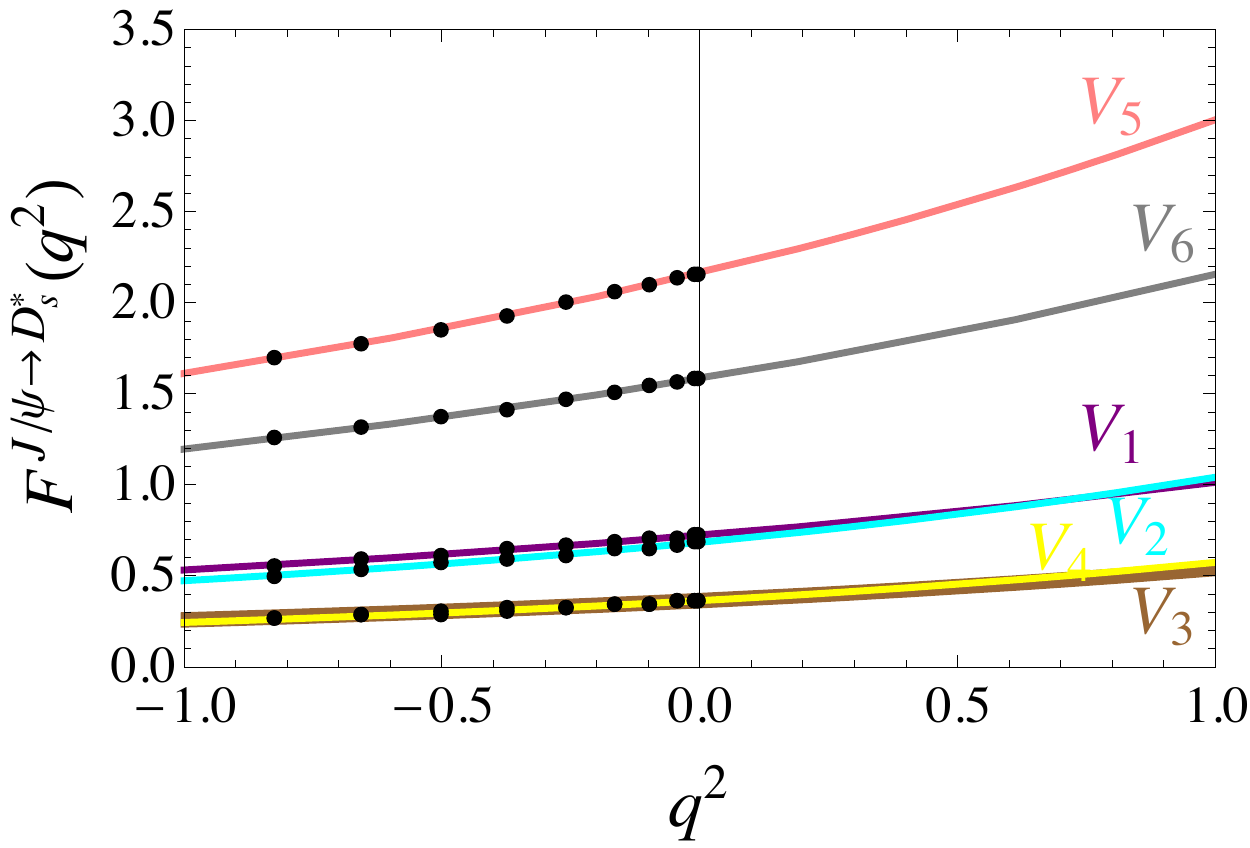}}\,
\subfigure{\includegraphics[scale=0.32]{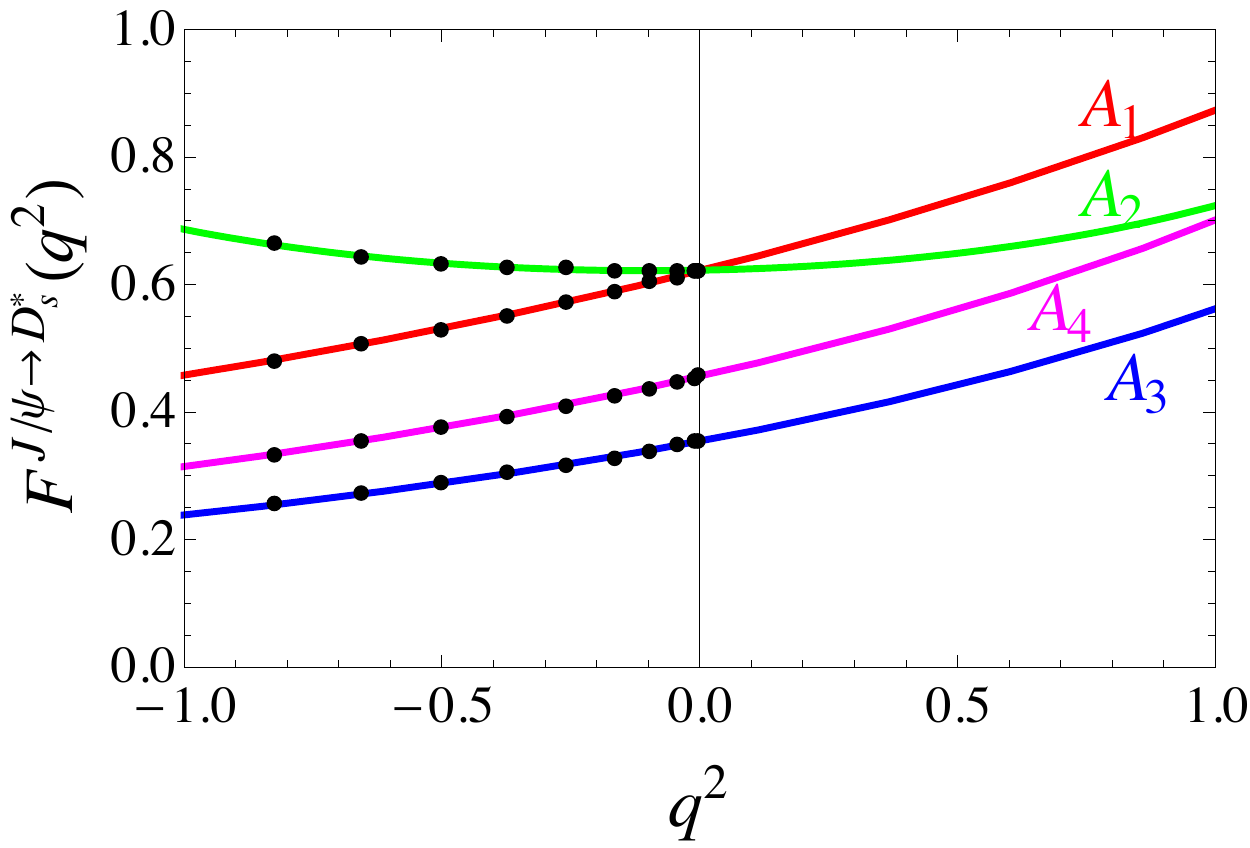}}\,
\subfigure{\includegraphics[scale=0.25]{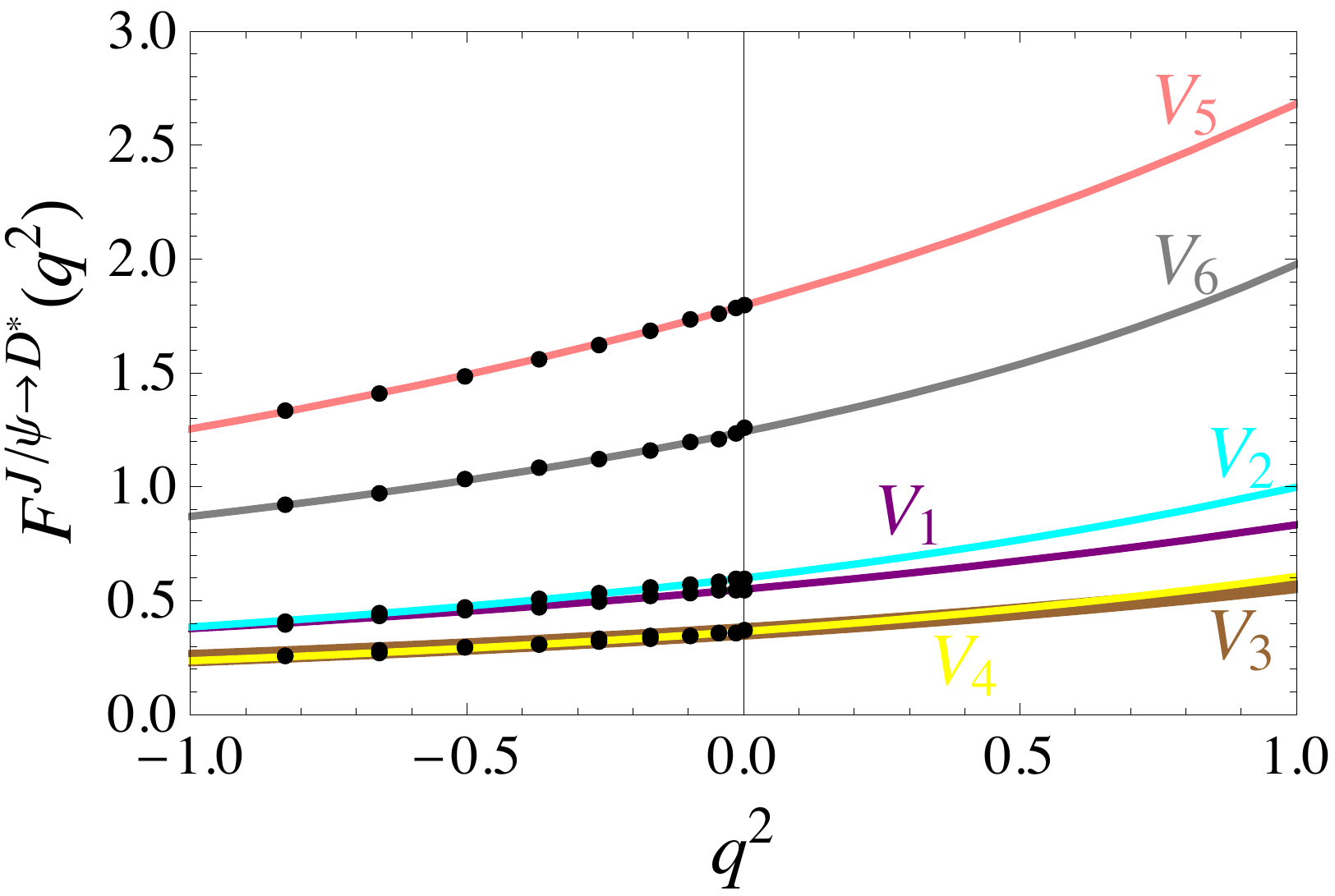}}\,
\subfigure{\includegraphics[scale=0.265]{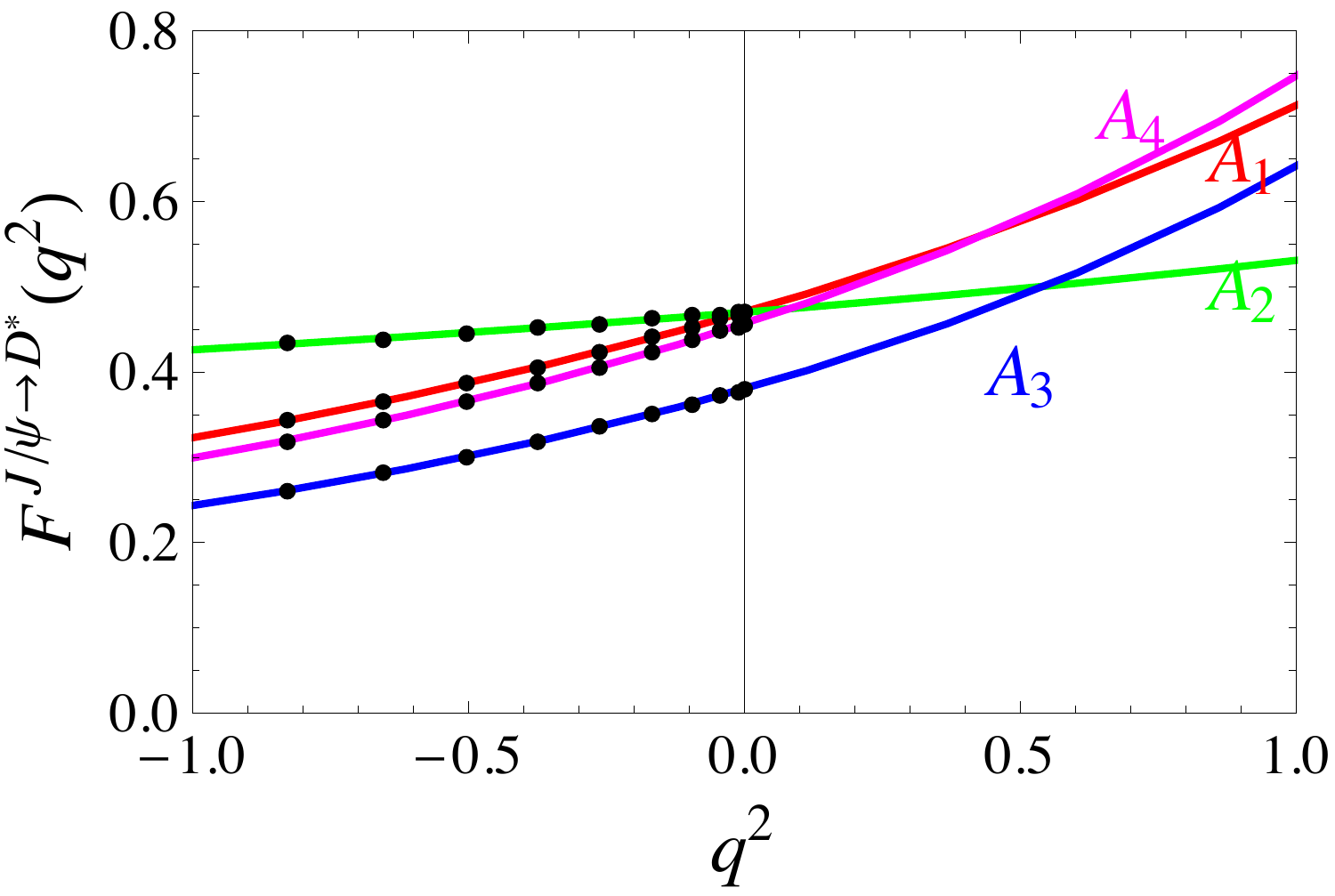}}\\
\subfigure{\includegraphics[scale=0.26]{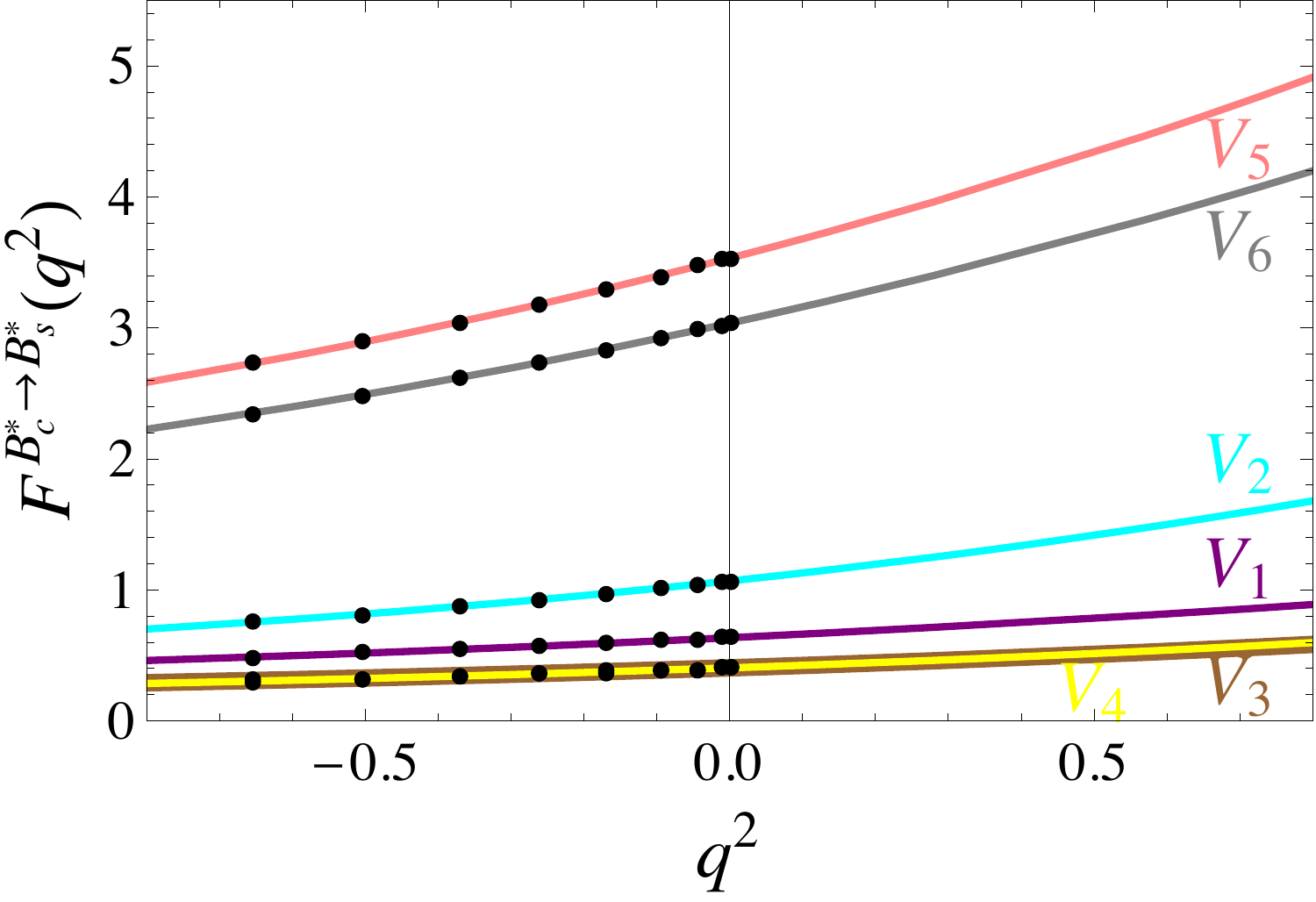}}\,
\subfigure{\includegraphics[scale=0.27]{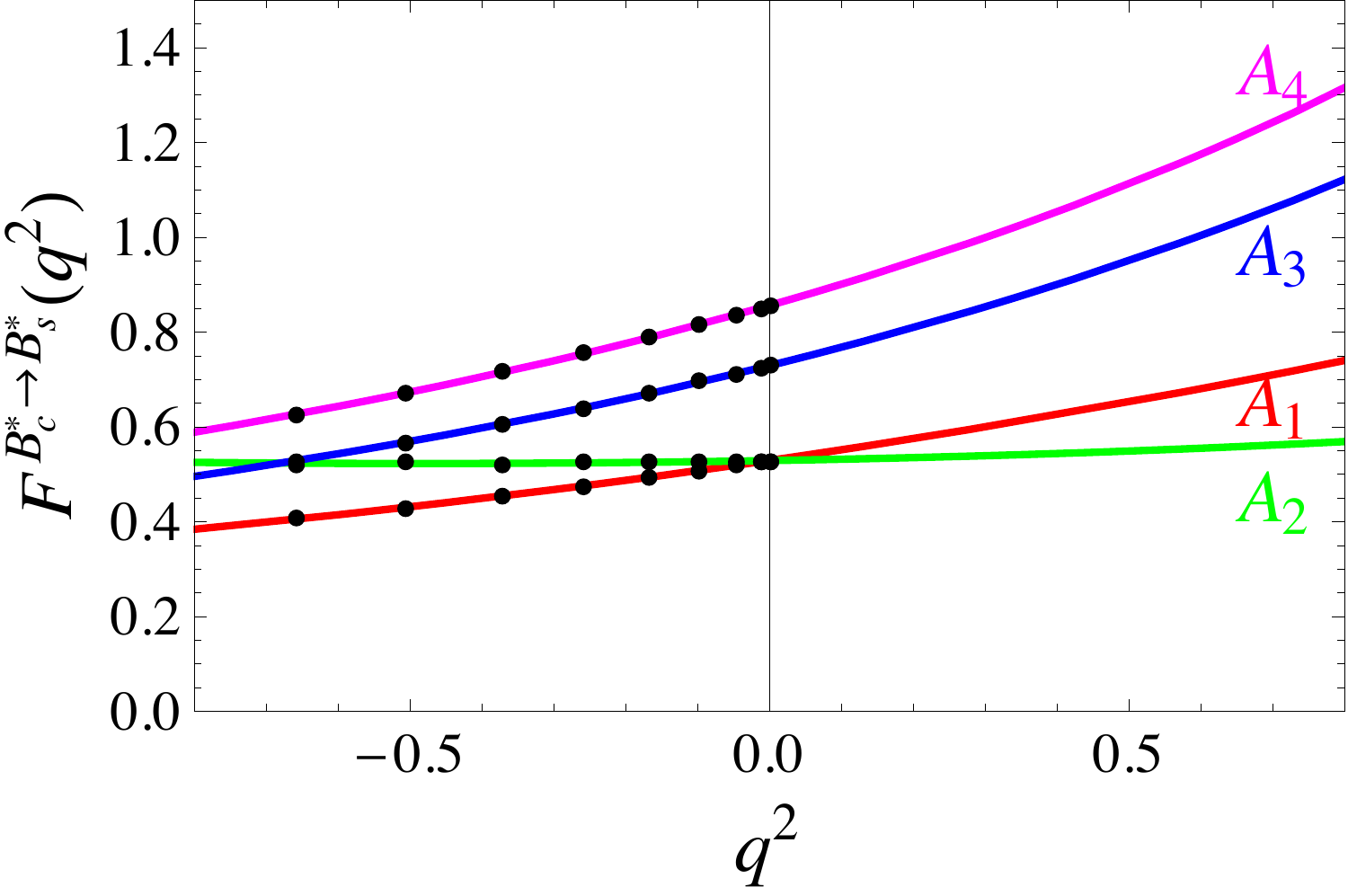}}\,
\subfigure{\includegraphics[scale=0.26]{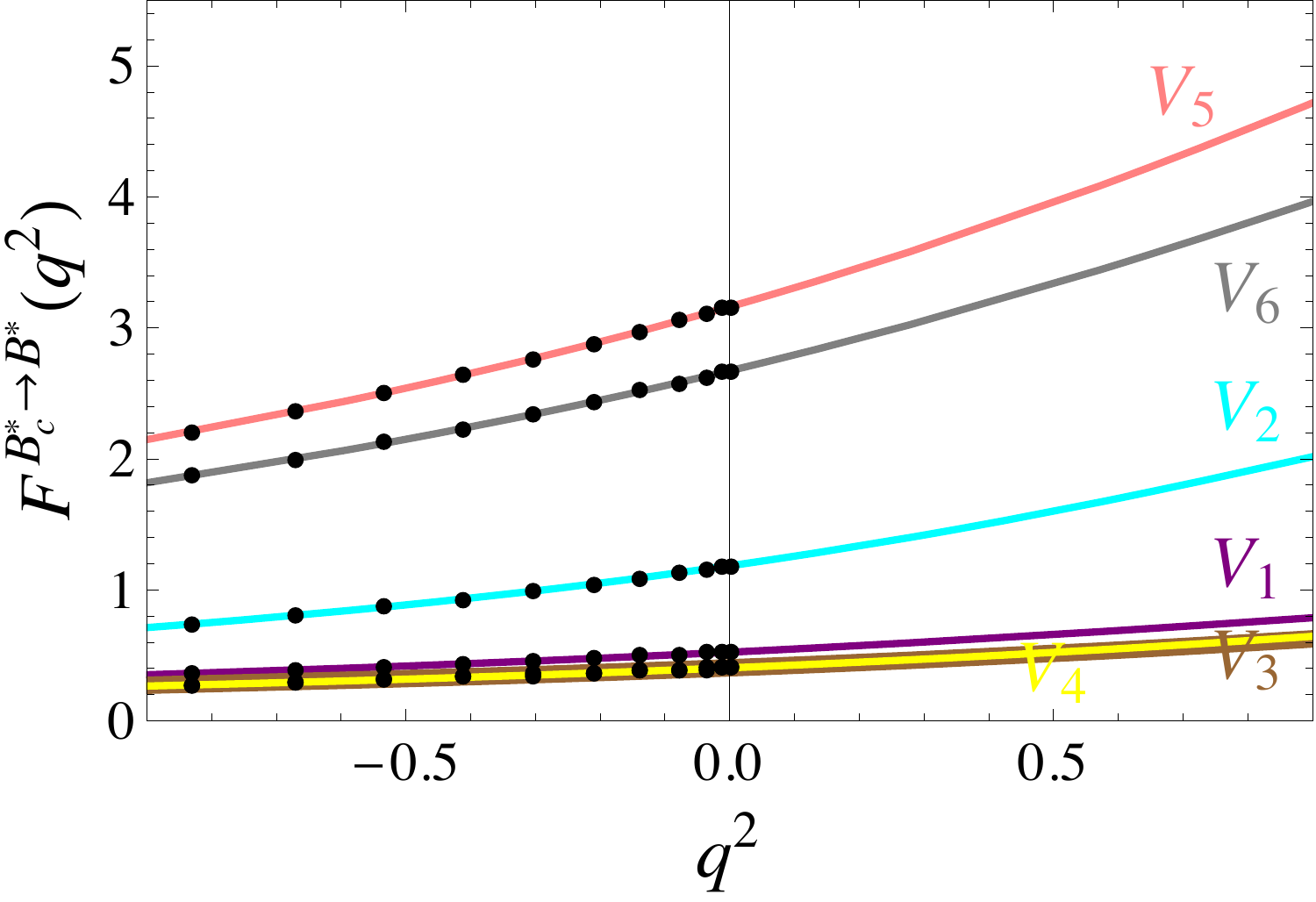}}\,
\subfigure{\includegraphics[scale=0.26]{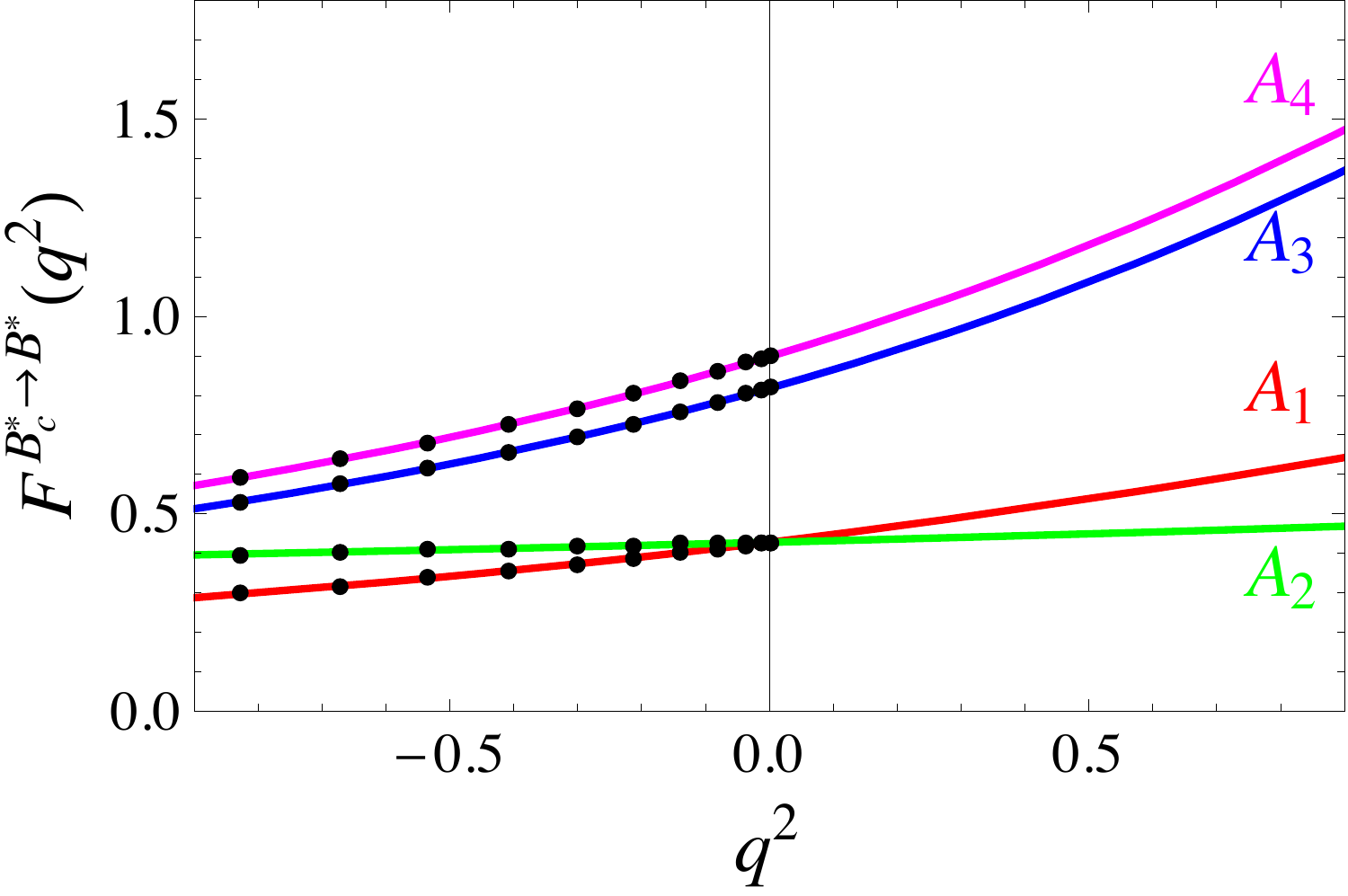}}\\
\end{center}
\label{fig:sche1c}
\end{figure}

 \begin{figure}[]
\caption{ Same as Fig.~\ref{fig:sche2b} except  with the parameterization scheme given by Eq.~\eqref{eq:para1}. }
\begin{center}
\subfigure{\includegraphics[scale=0.24]{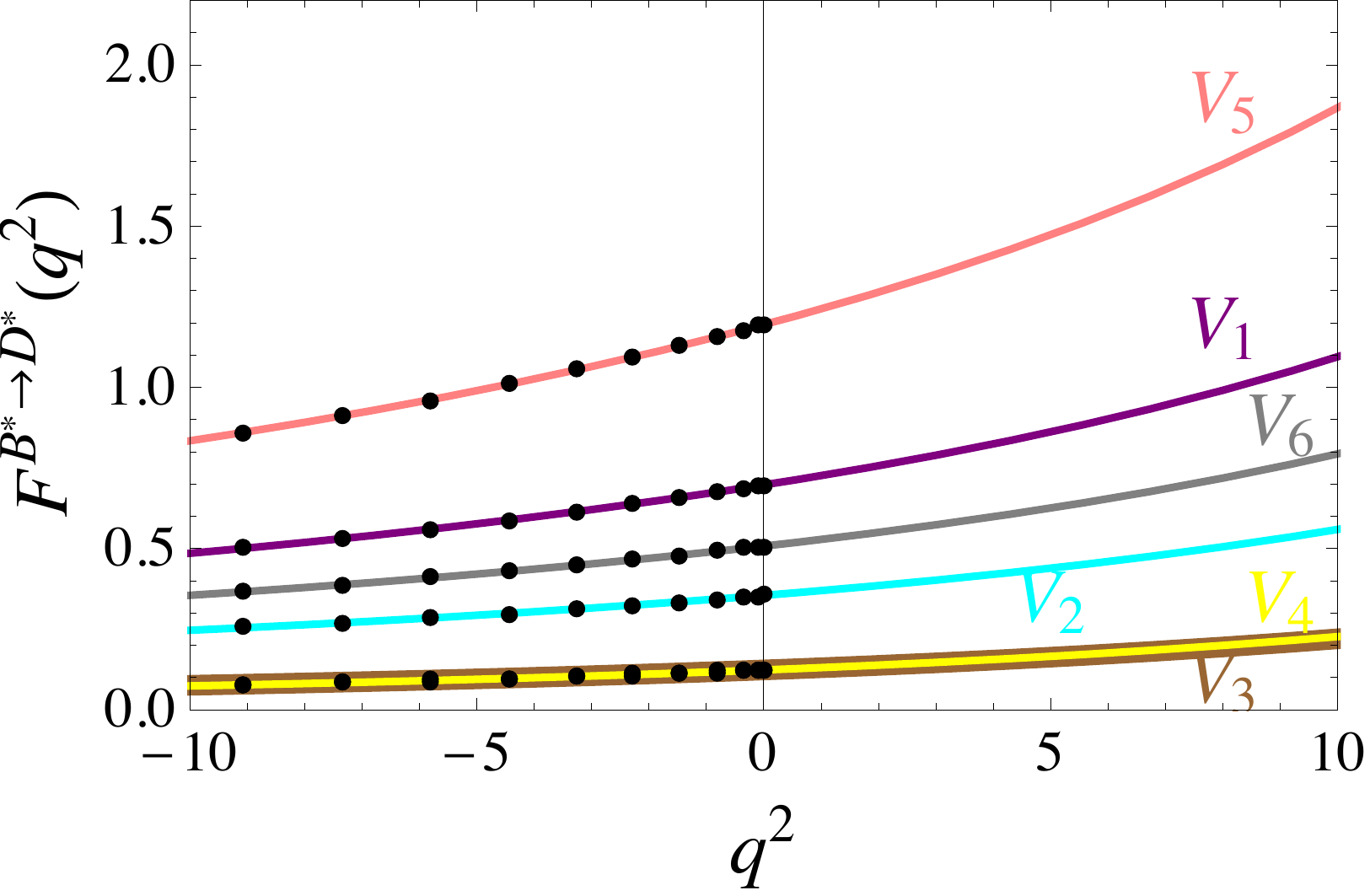}}\,
\subfigure{\includegraphics[scale=0.24]{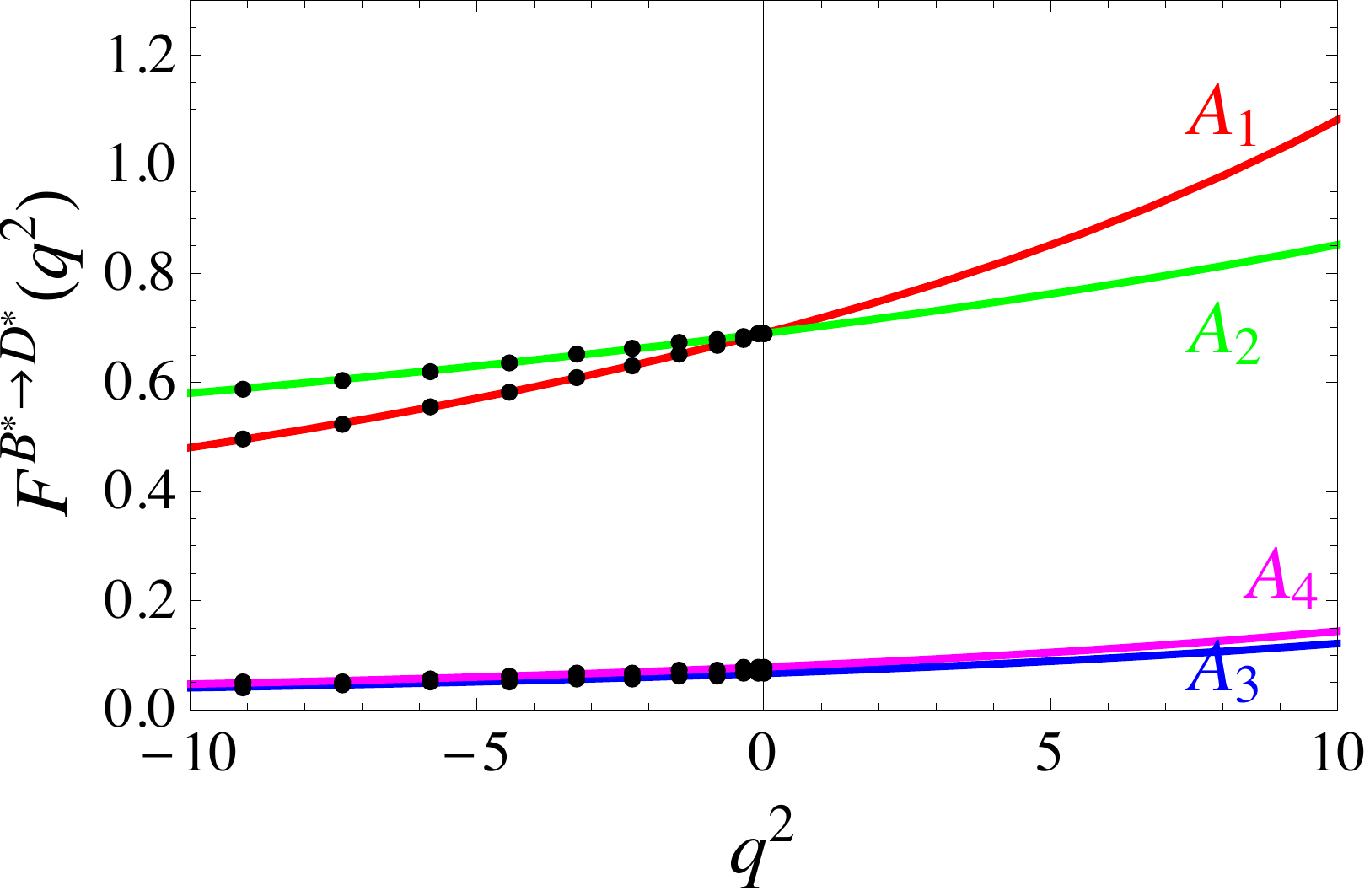}}\,
\subfigure{\includegraphics[scale=0.25]{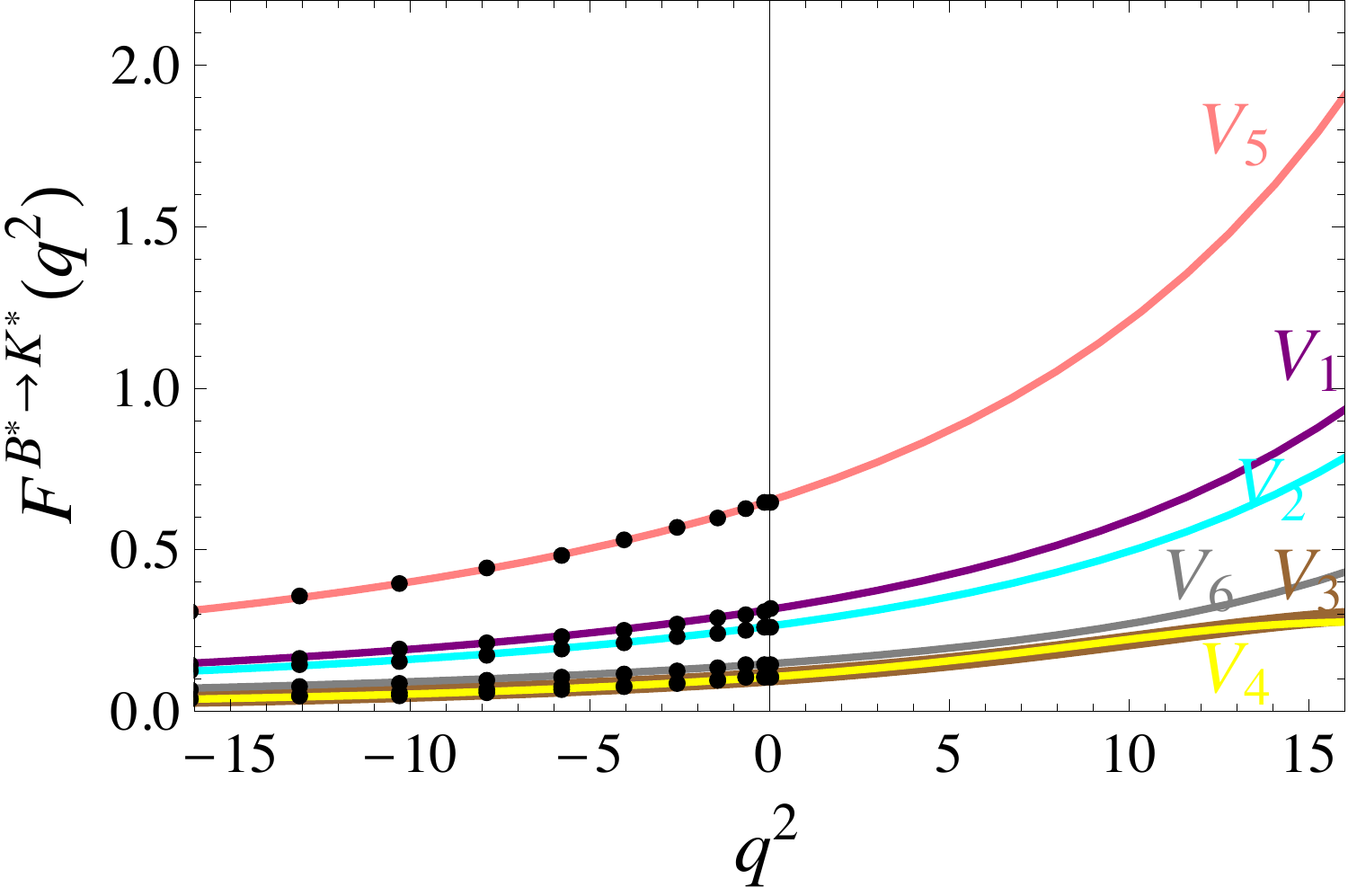}}\,
\subfigure{\includegraphics[scale=0.25]{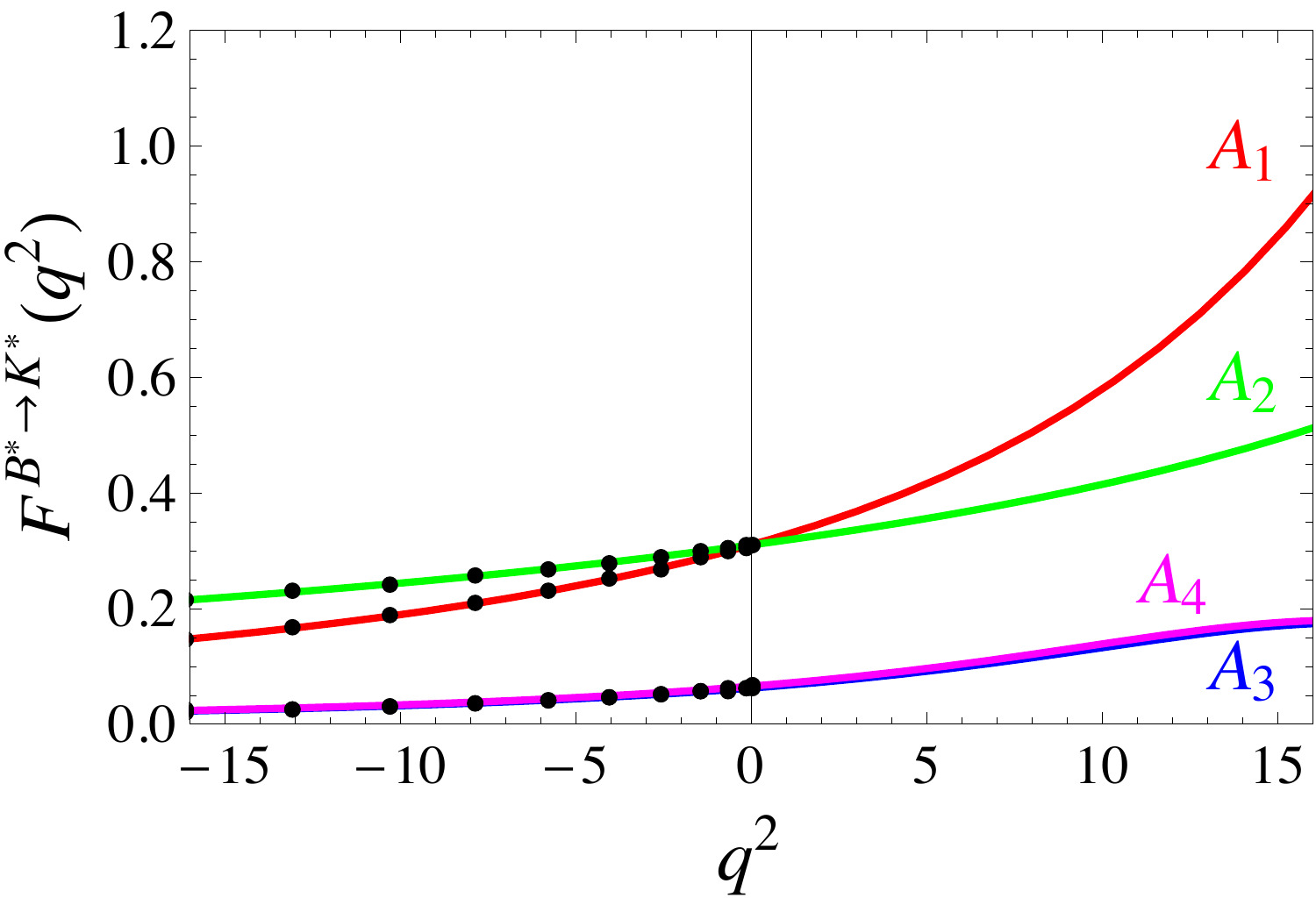}}\\
\subfigure{\includegraphics[scale=0.26]{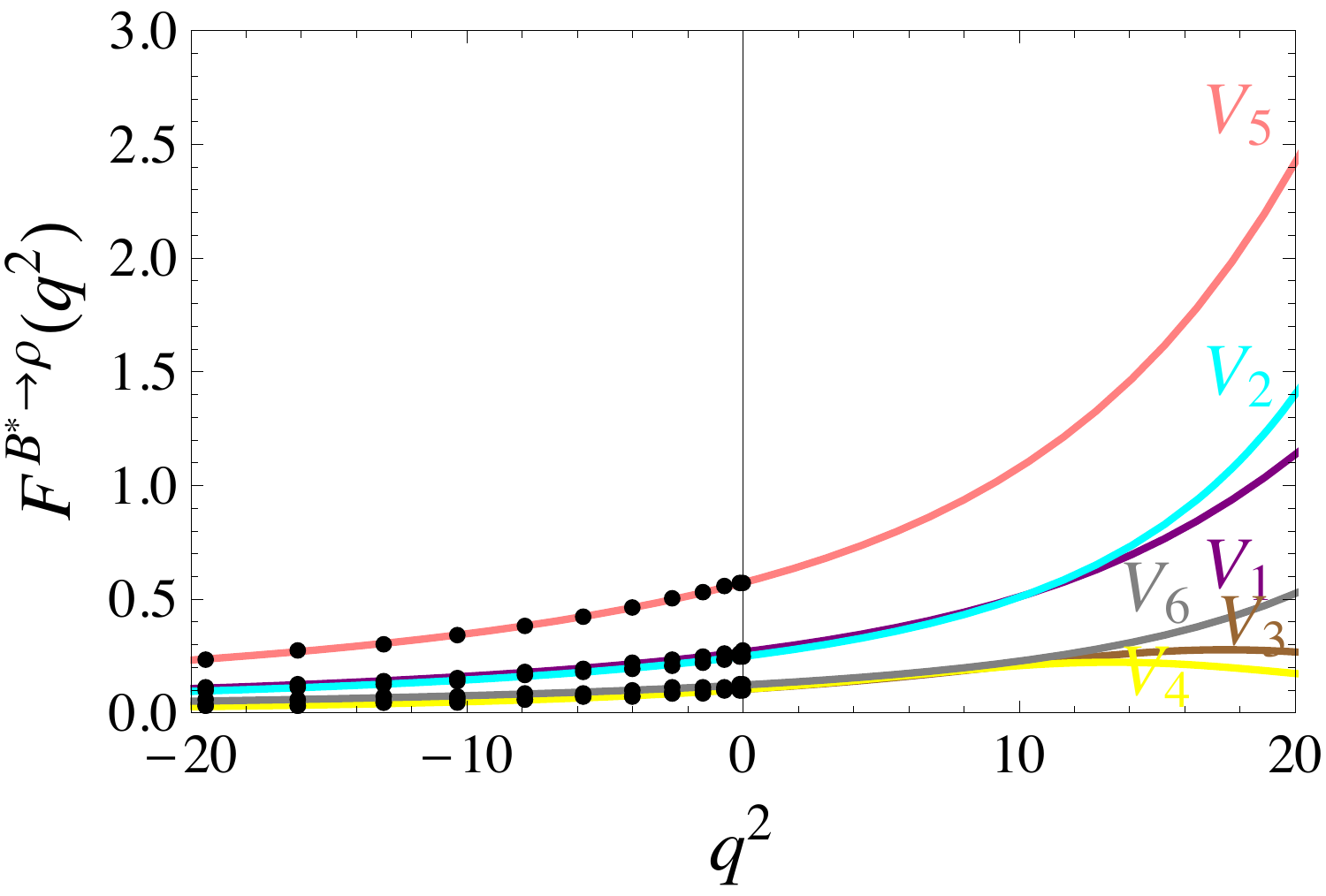}}\,
\subfigure{\includegraphics[scale=0.26]{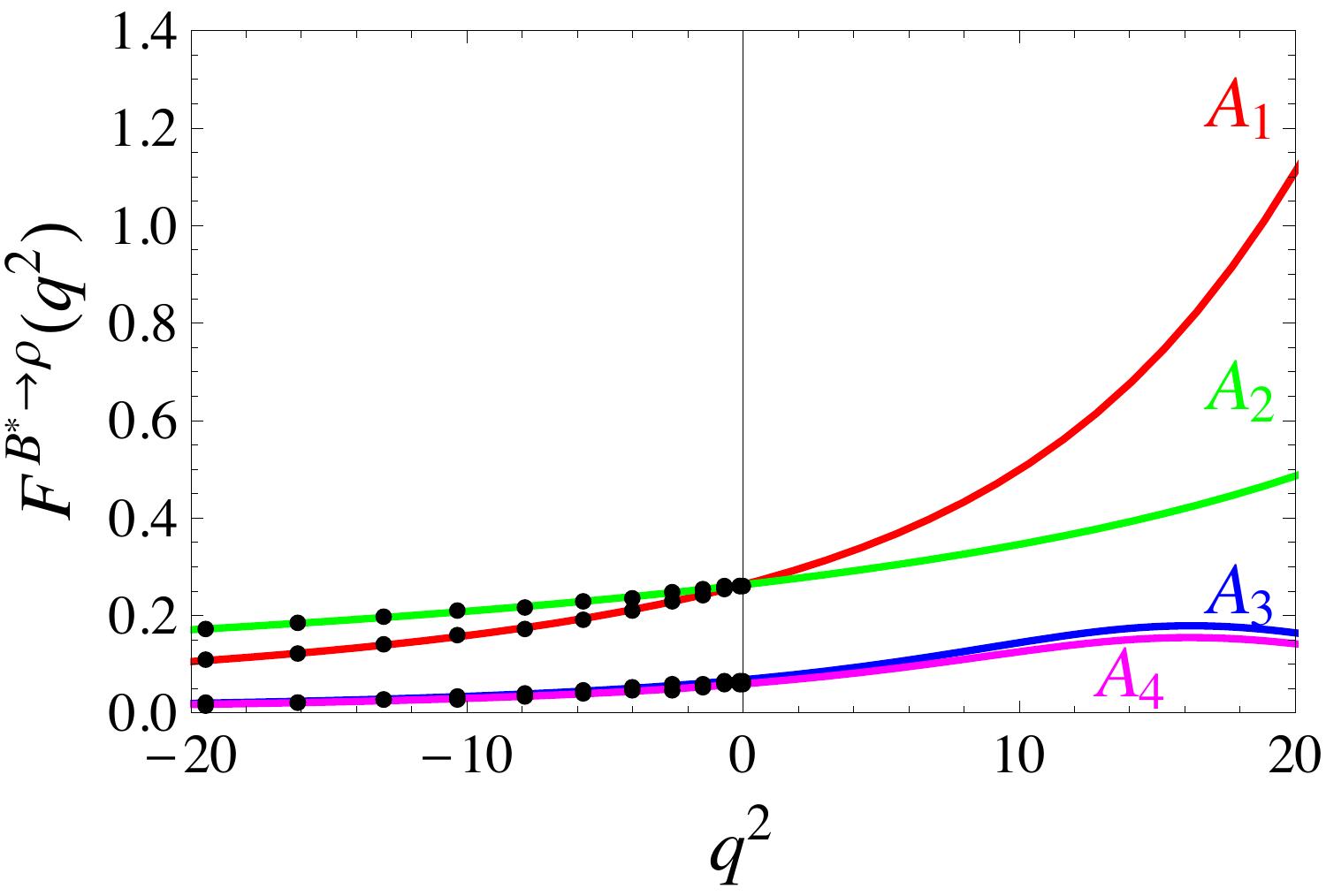}}\,
\subfigure{\includegraphics[scale=0.26]{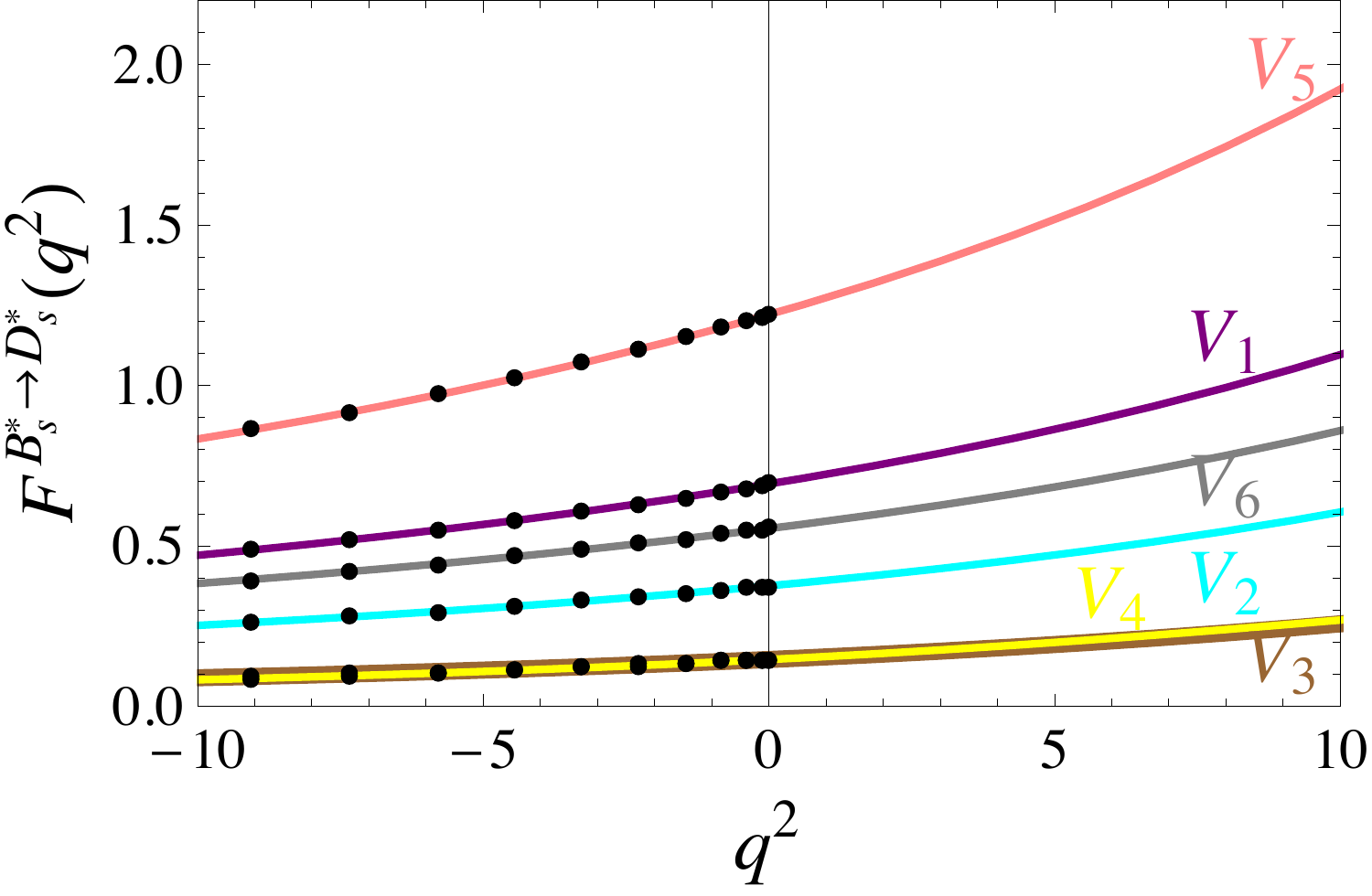}}\,
\subfigure{\includegraphics[scale=0.24]{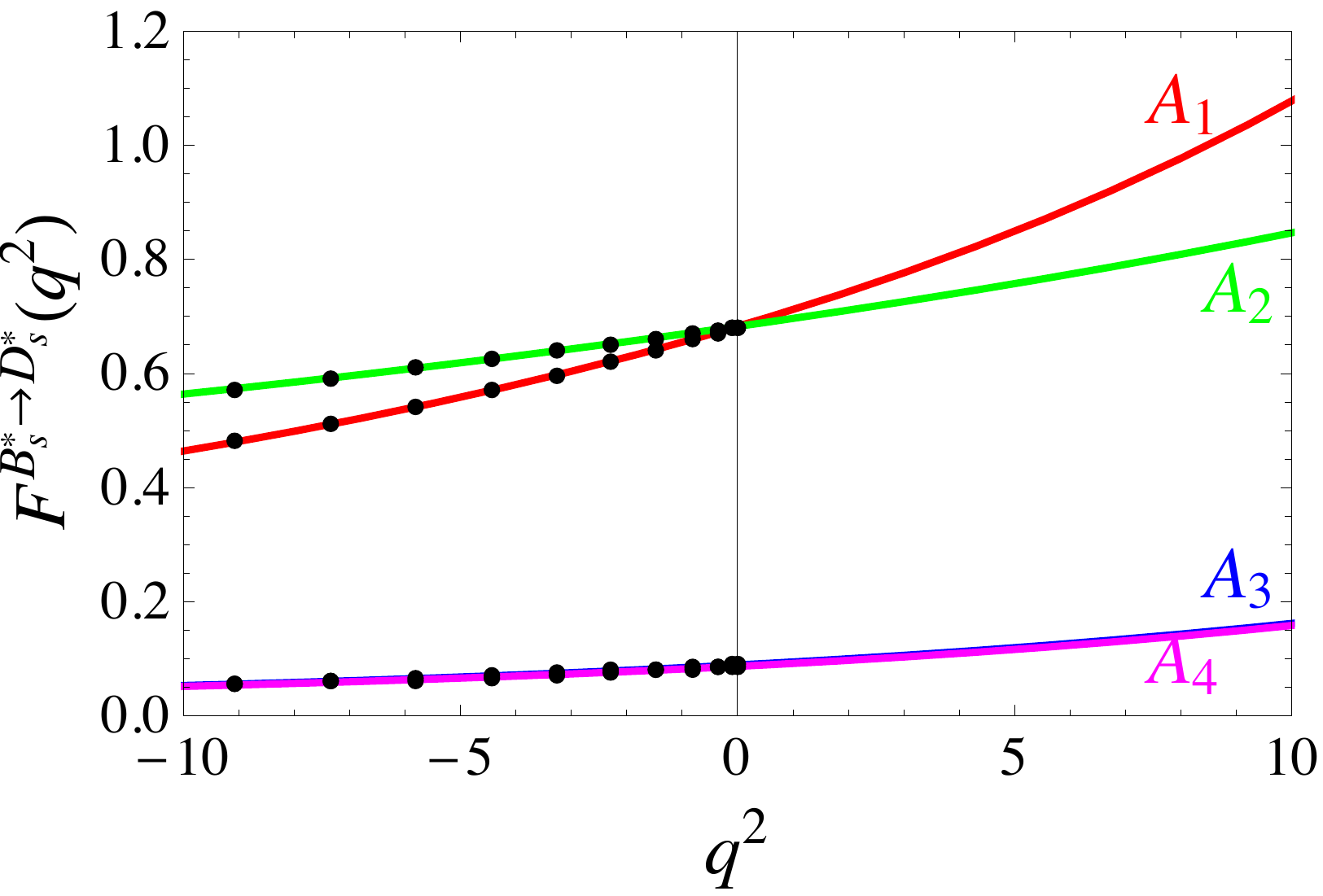}}\\
\subfigure{\includegraphics[scale=0.26]{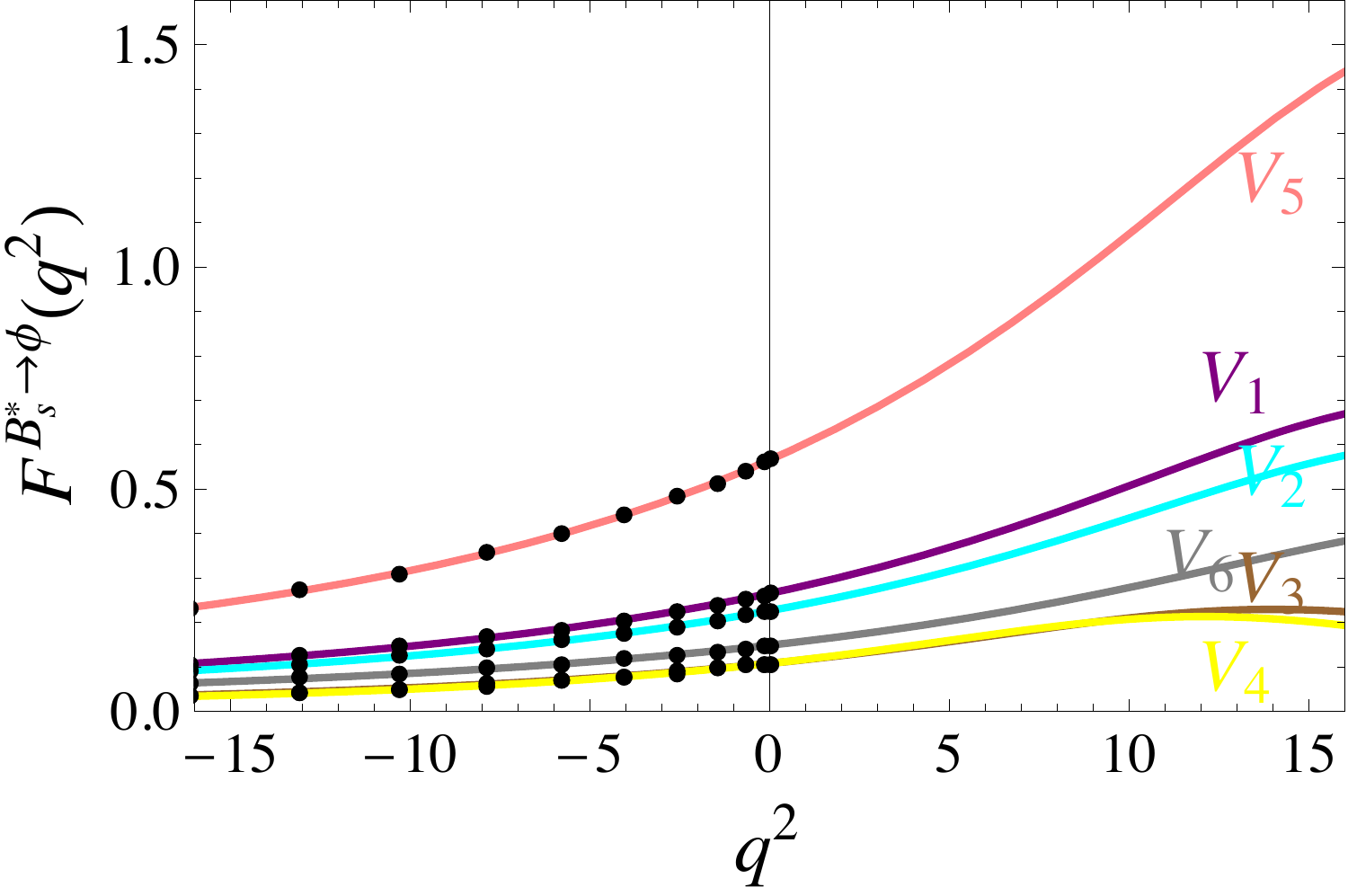}}\,
\subfigure{\includegraphics[scale=0.26]{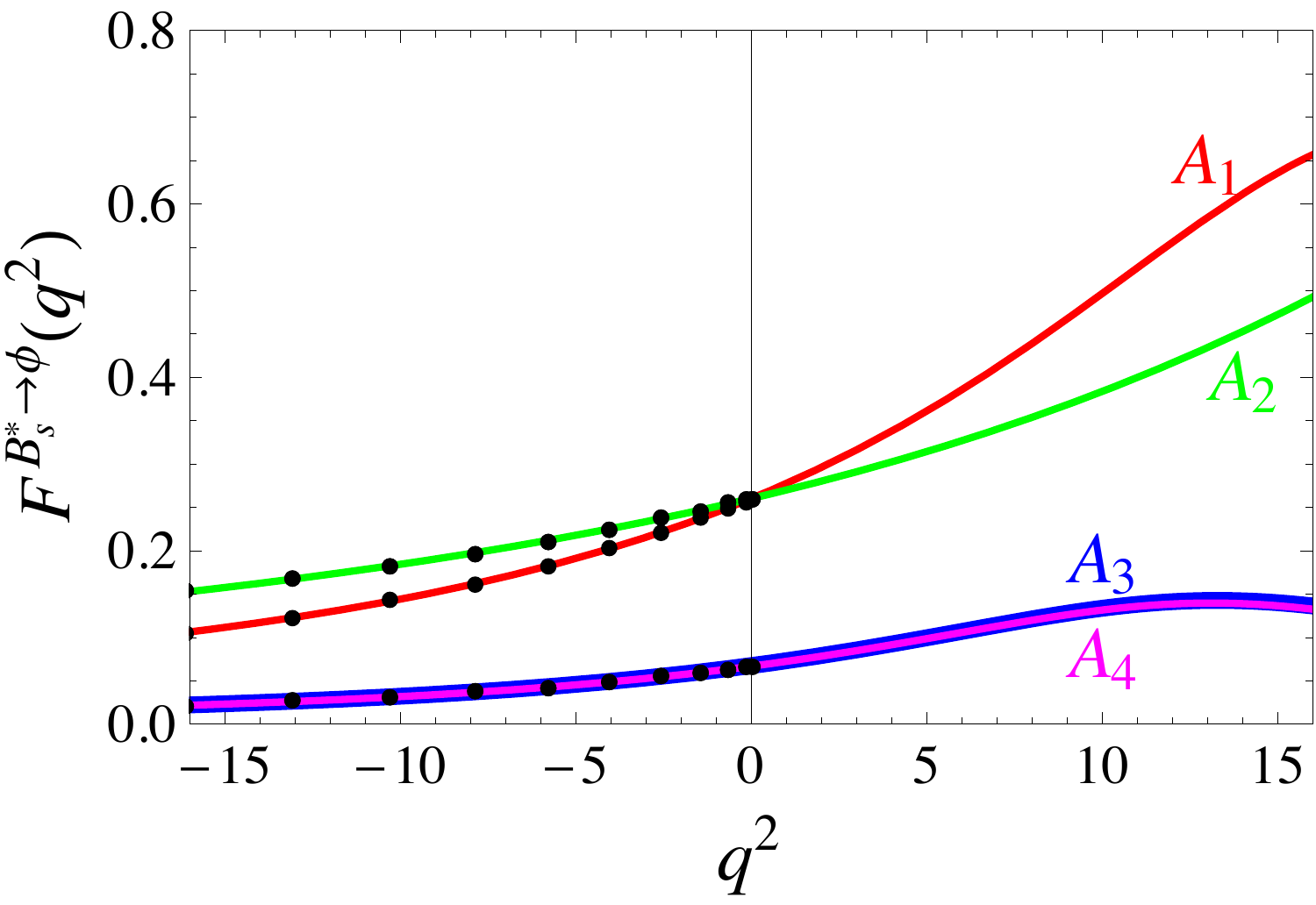}}\,
\subfigure{\includegraphics[scale=0.24]{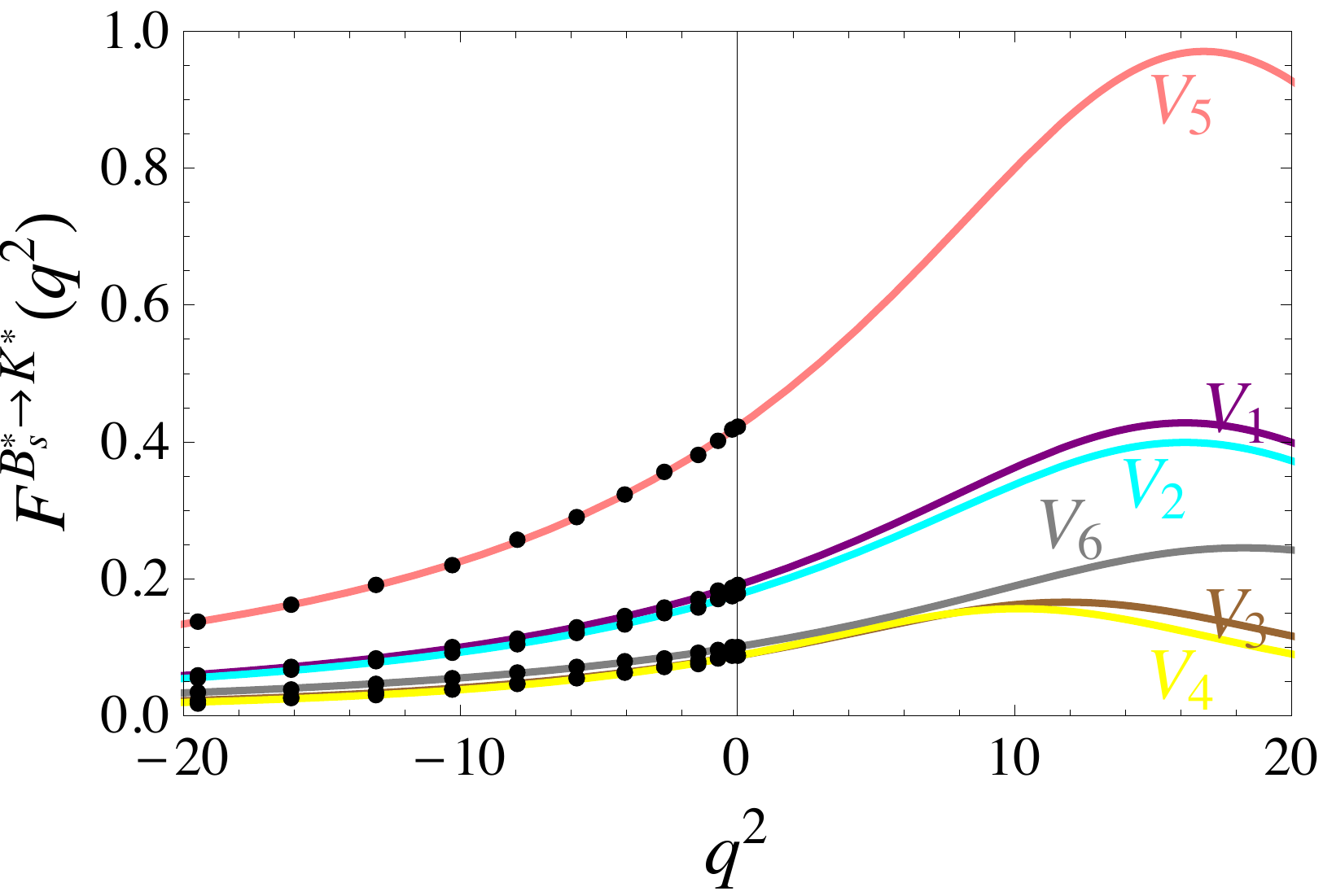}}\,
\subfigure{\includegraphics[scale=0.24]{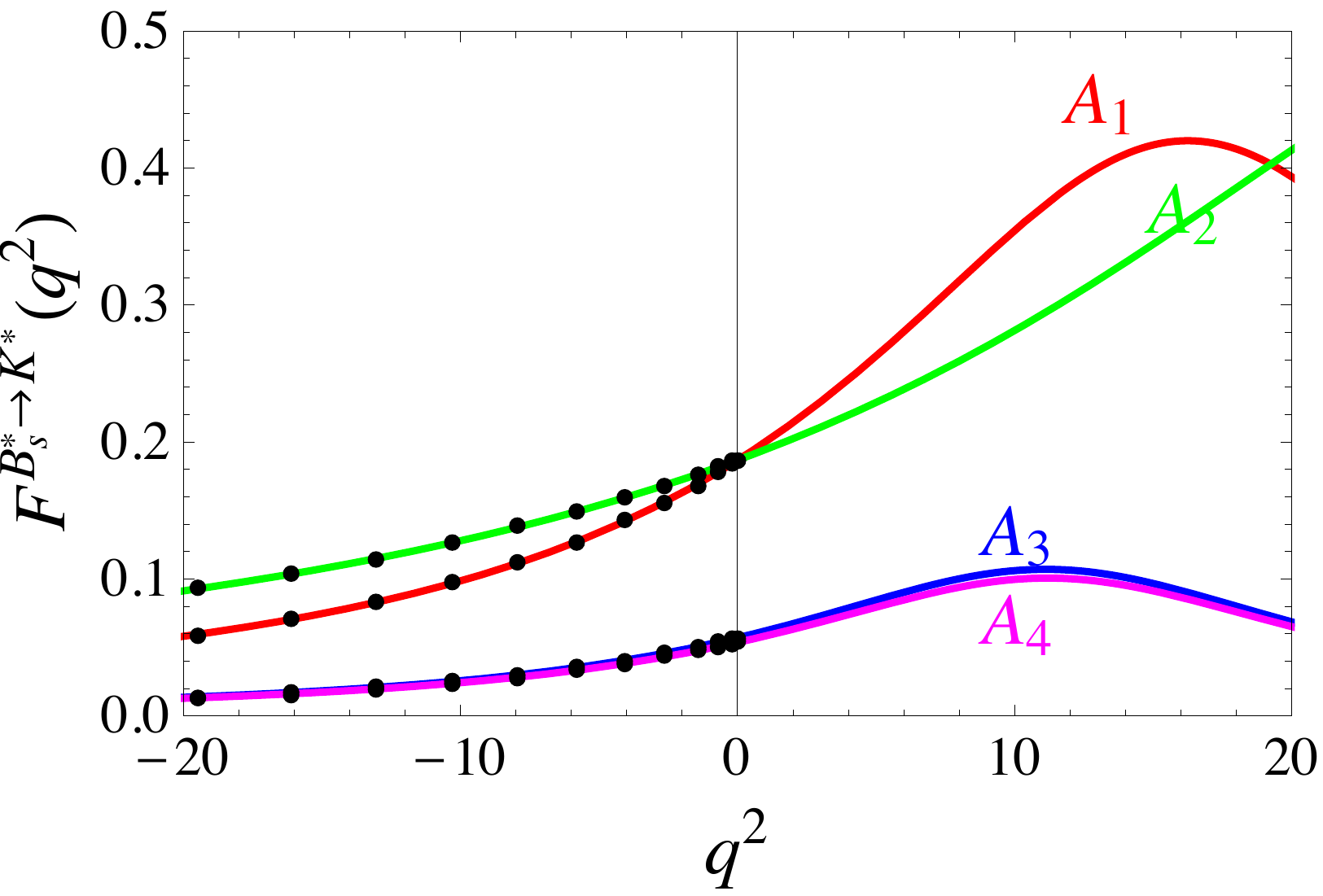}}\\
\subfigure{\includegraphics[scale=0.26]{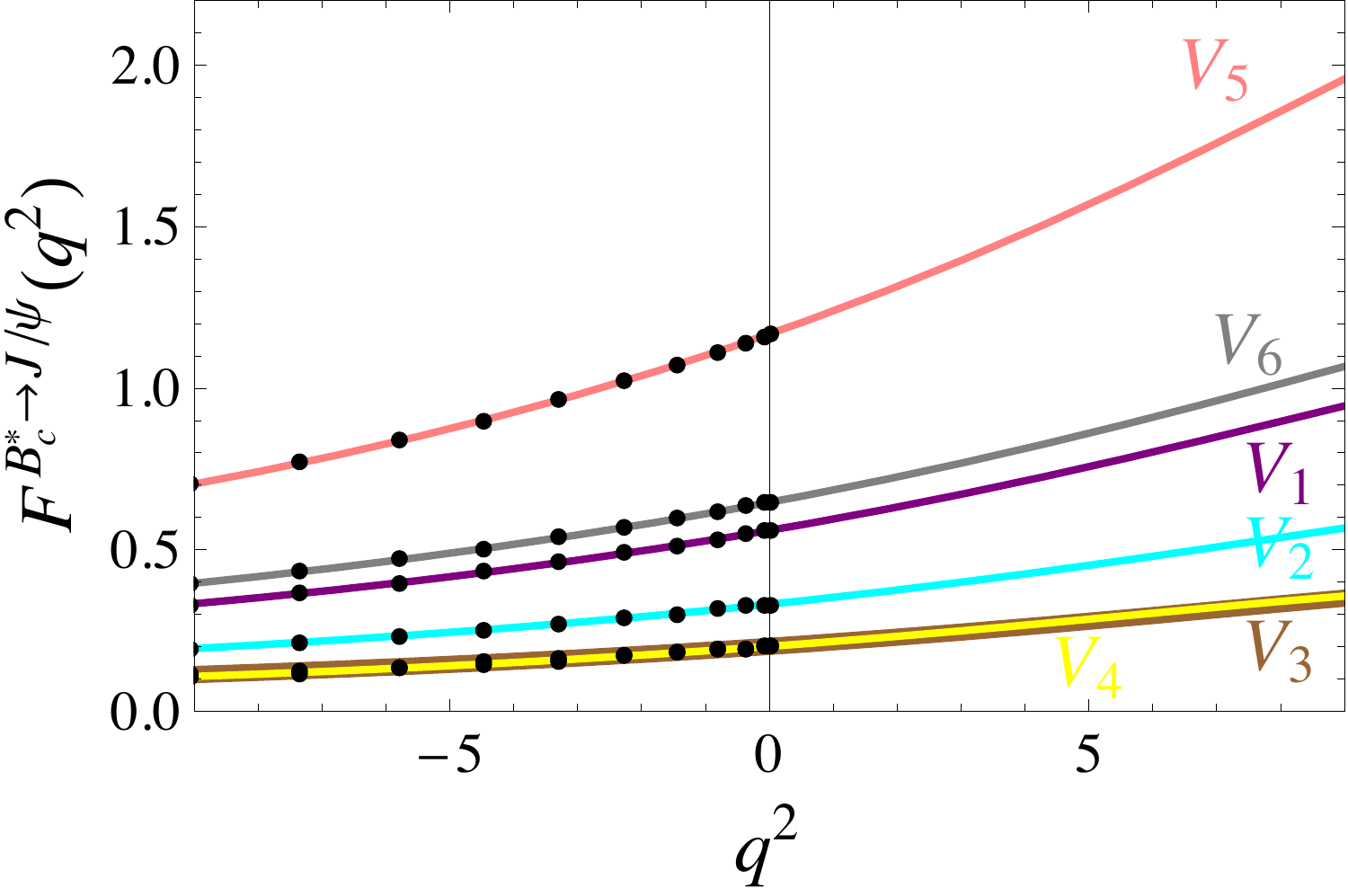}}\,
\subfigure{\includegraphics[scale=0.26]{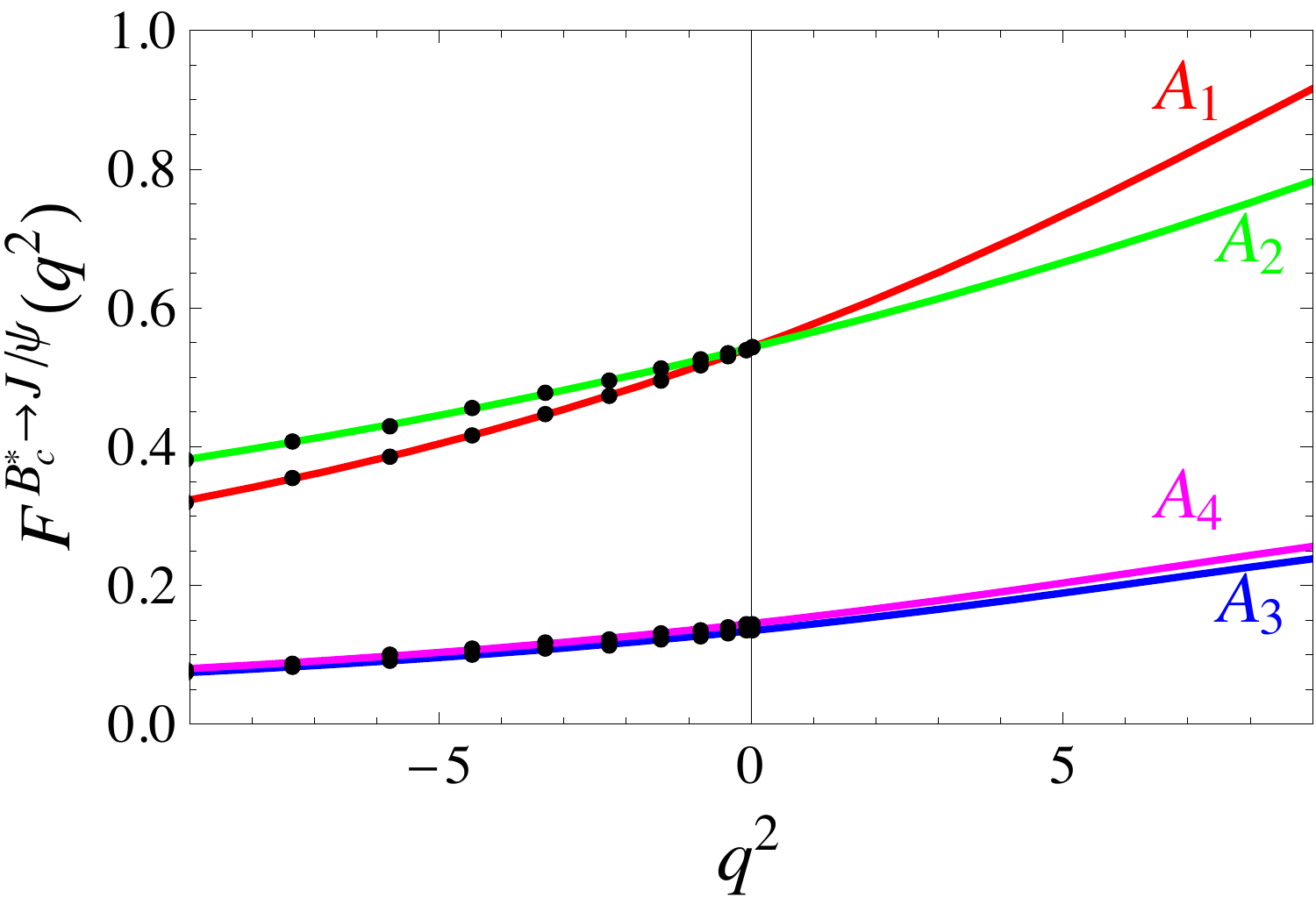}}\,
\subfigure{\includegraphics[scale=0.26]{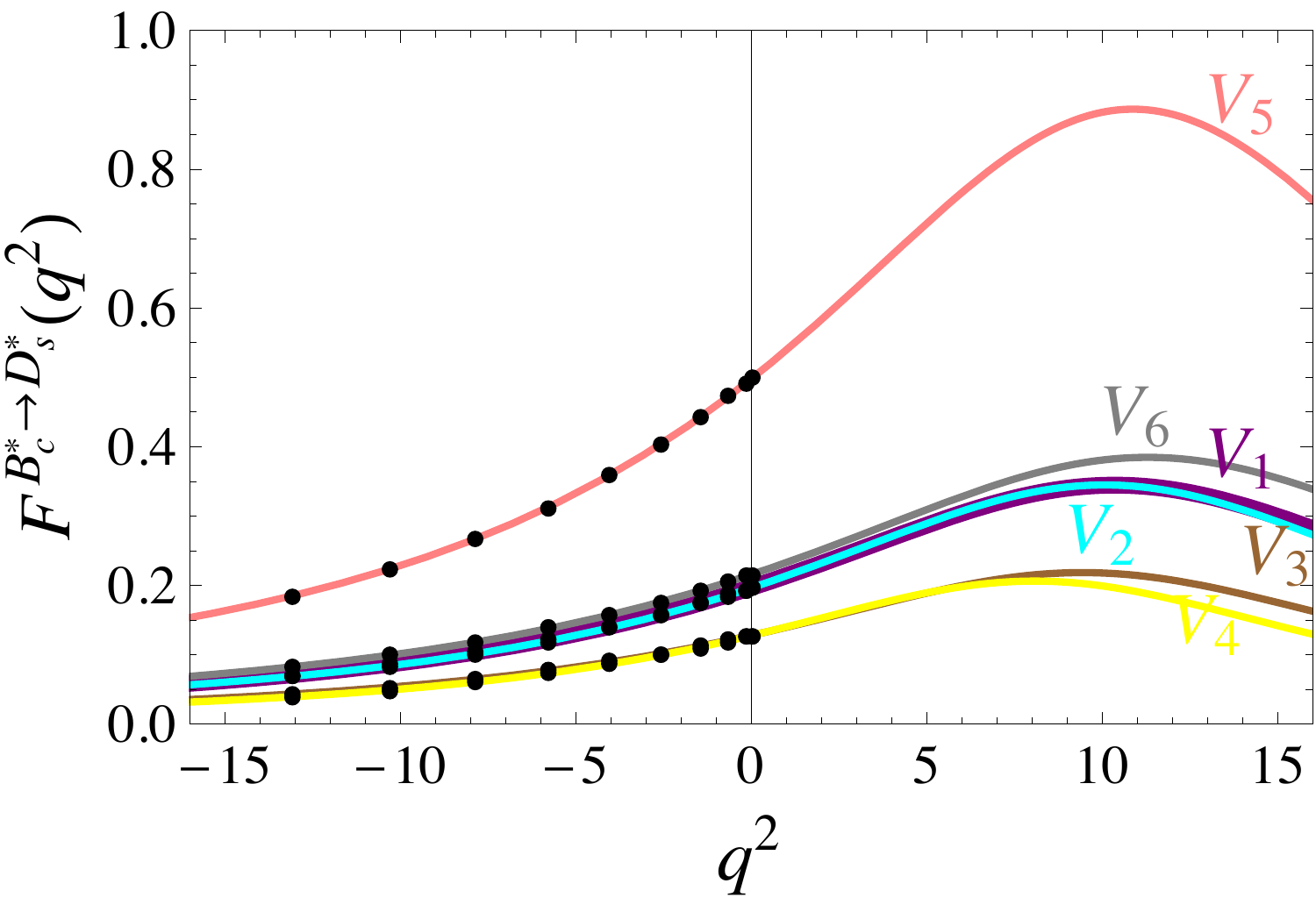}}\,
\subfigure{\includegraphics[scale=0.26]{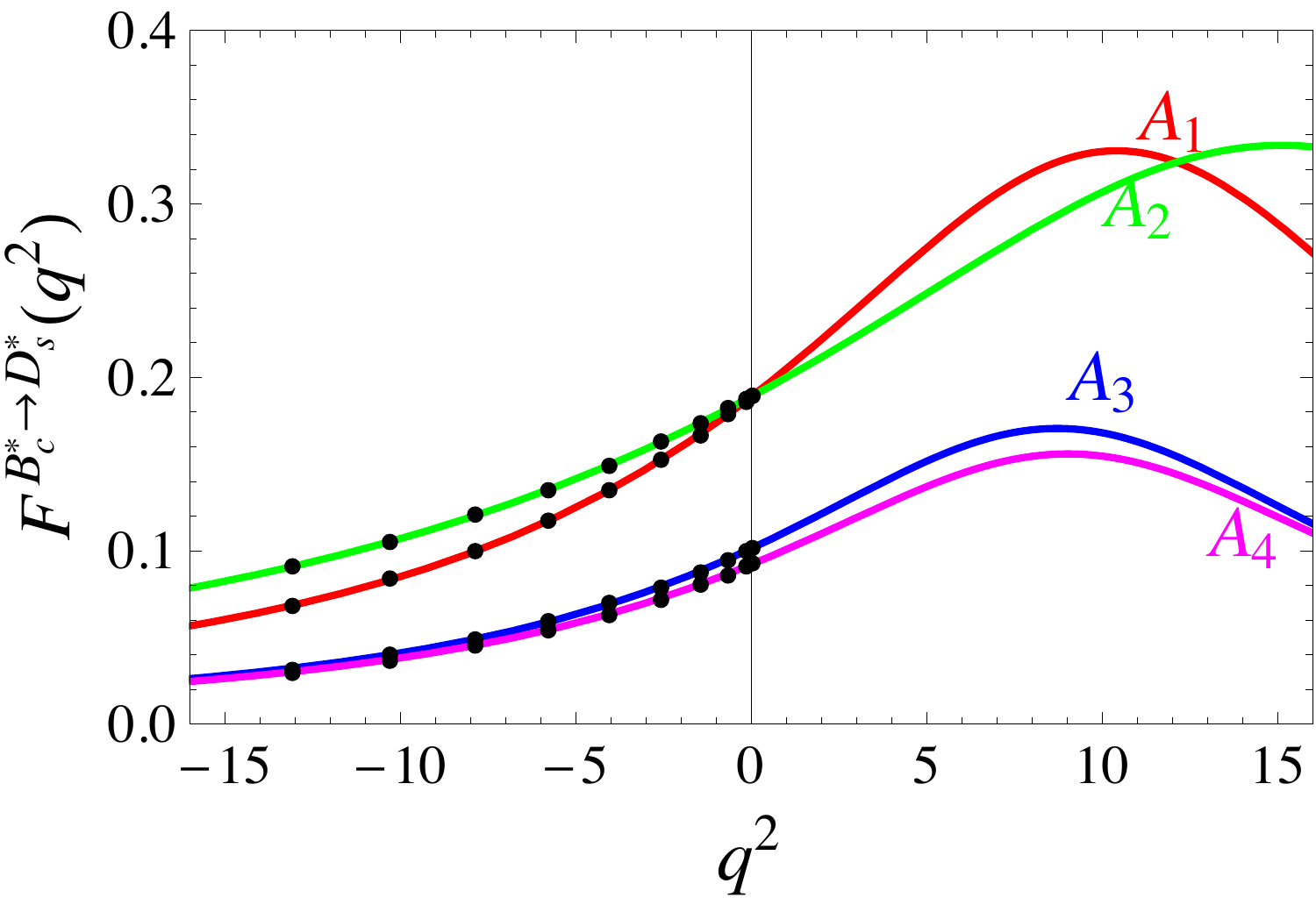}}\\
\subfigure{\includegraphics[scale=0.26]{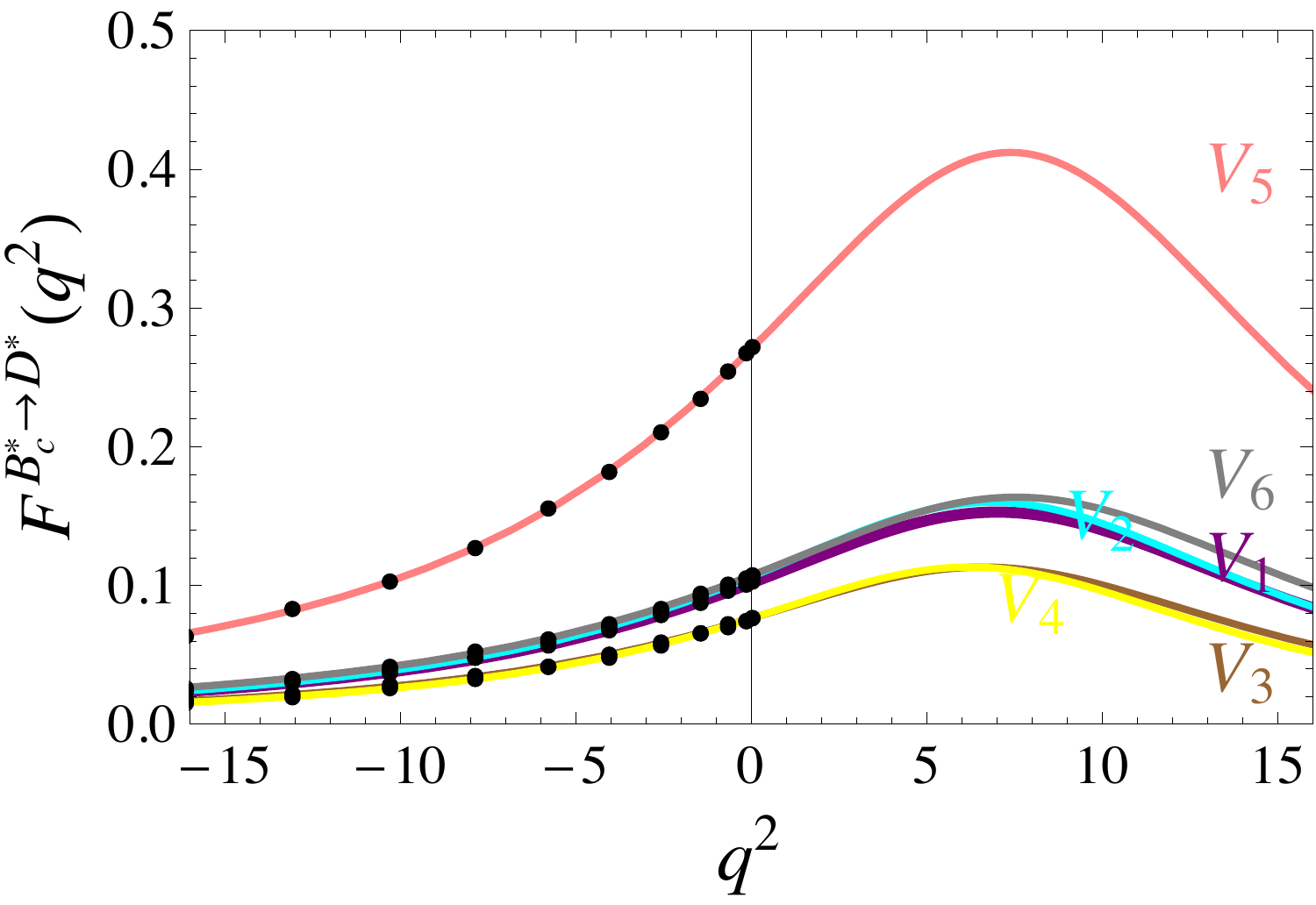}}\,
\subfigure{\includegraphics[scale=0.26]{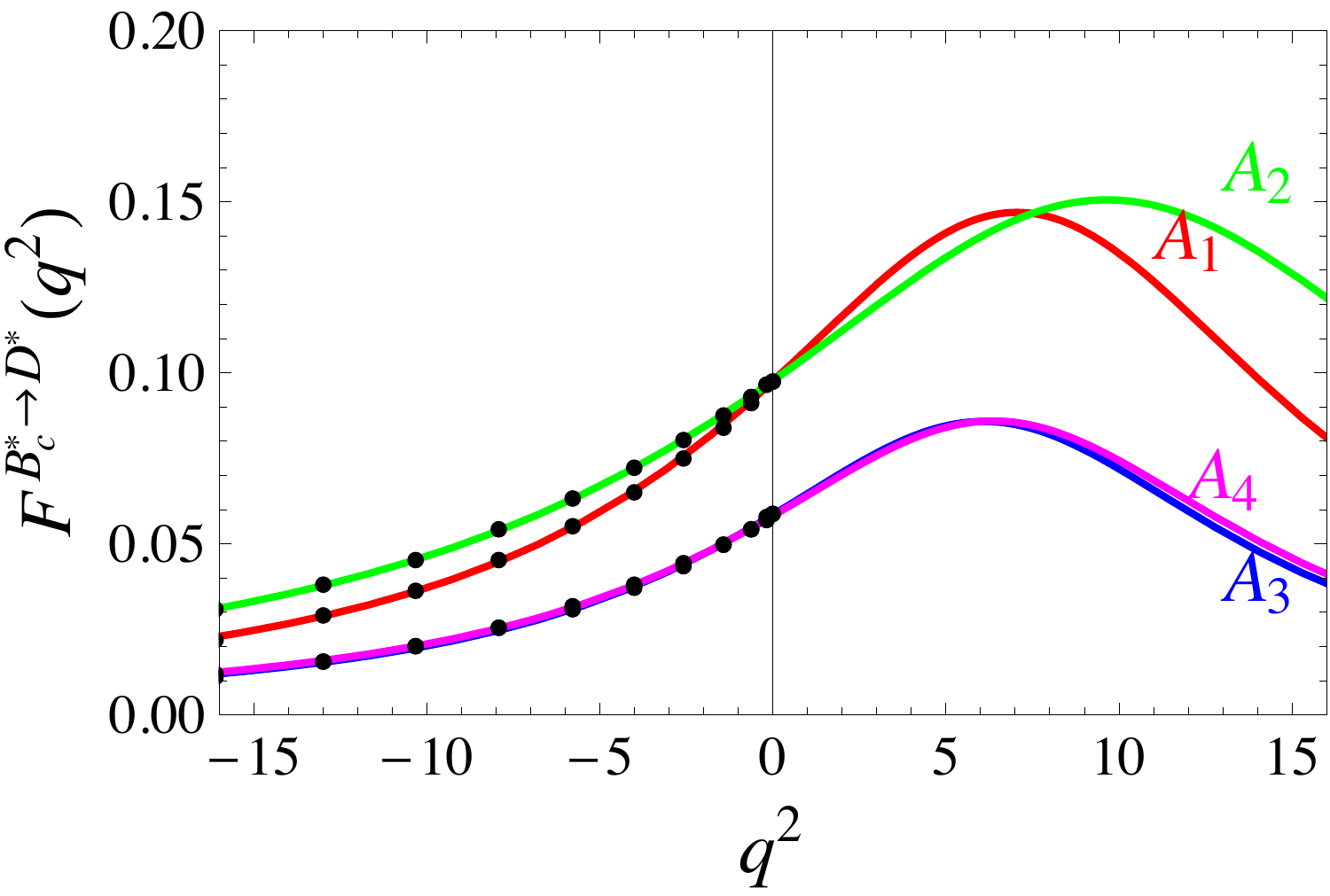}}\,
%\subfigure{\includegraphics[scale=0.26]{BcDV}}\,
%\subfigure{\includegraphics[scale=0.26]{BcDA}}\\
%\subfigure{\includegraphics[scale=0.19]{BcJpsiV}}\,
%\subfigure{\includegraphics[scale=0.28]{BcJpsiA}}\,
\subfigure{\includegraphics[scale=0.24]{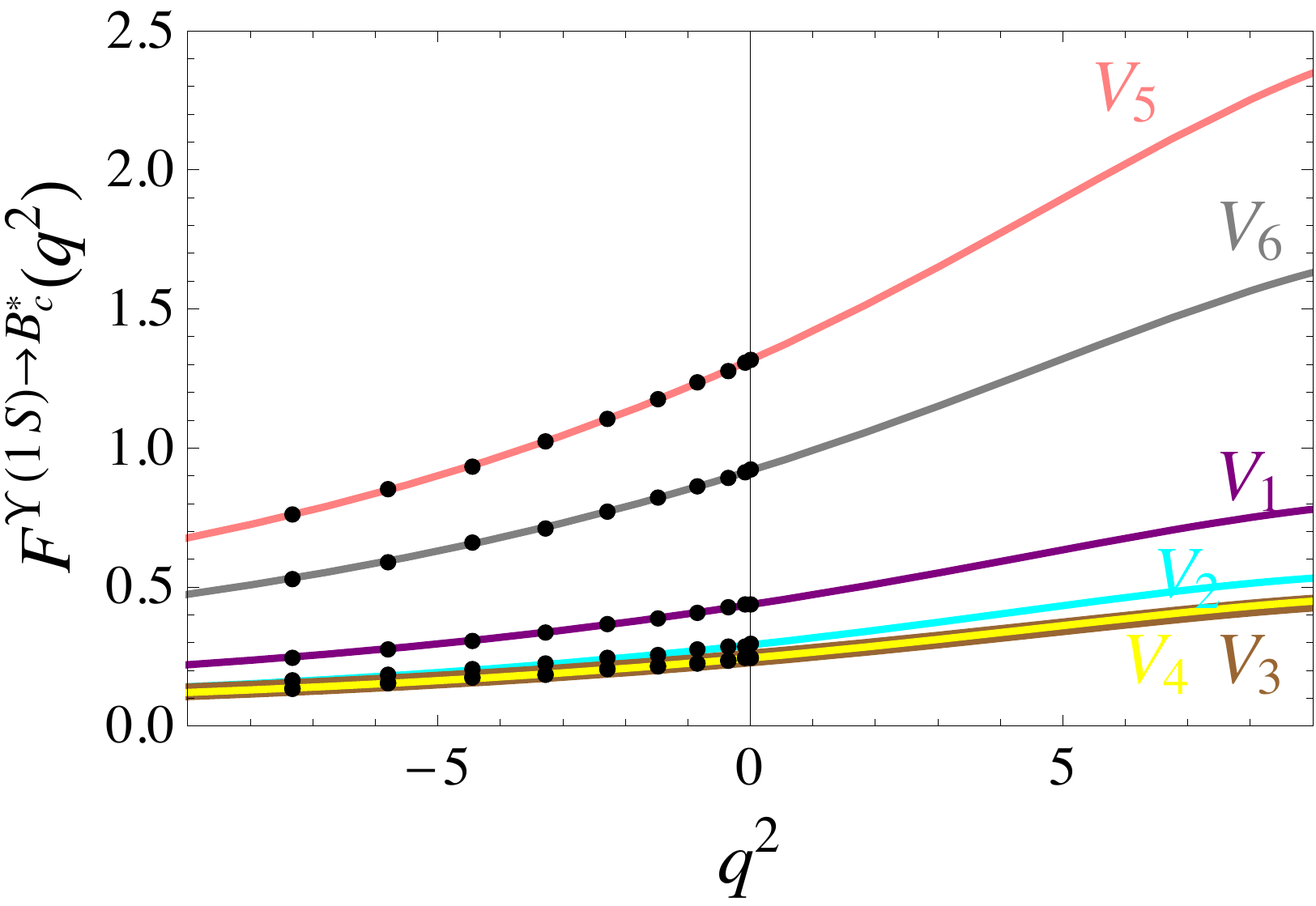}}\,
\subfigure{\includegraphics[scale=0.24]{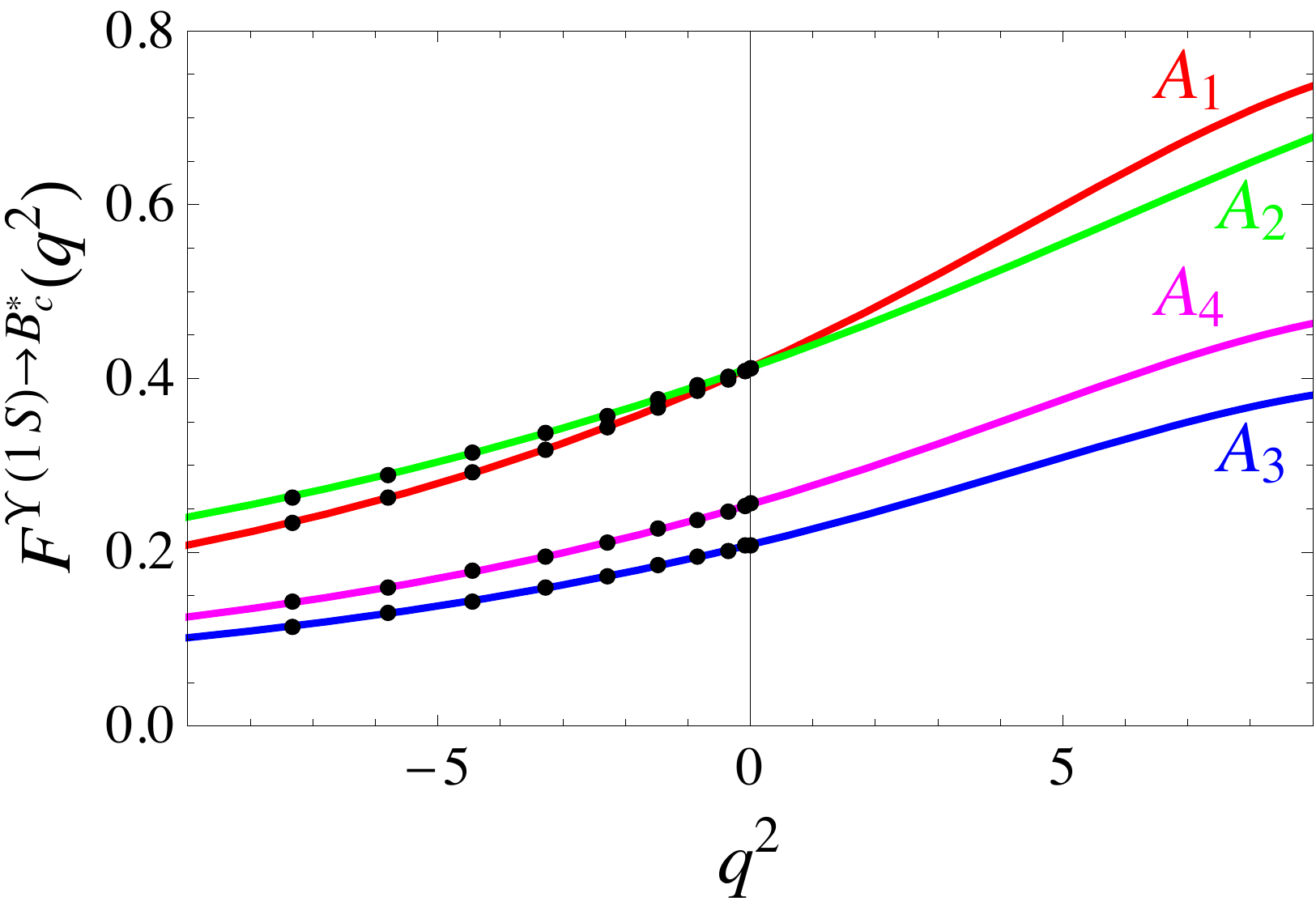}}\\
\subfigure{\includegraphics[scale=0.24]{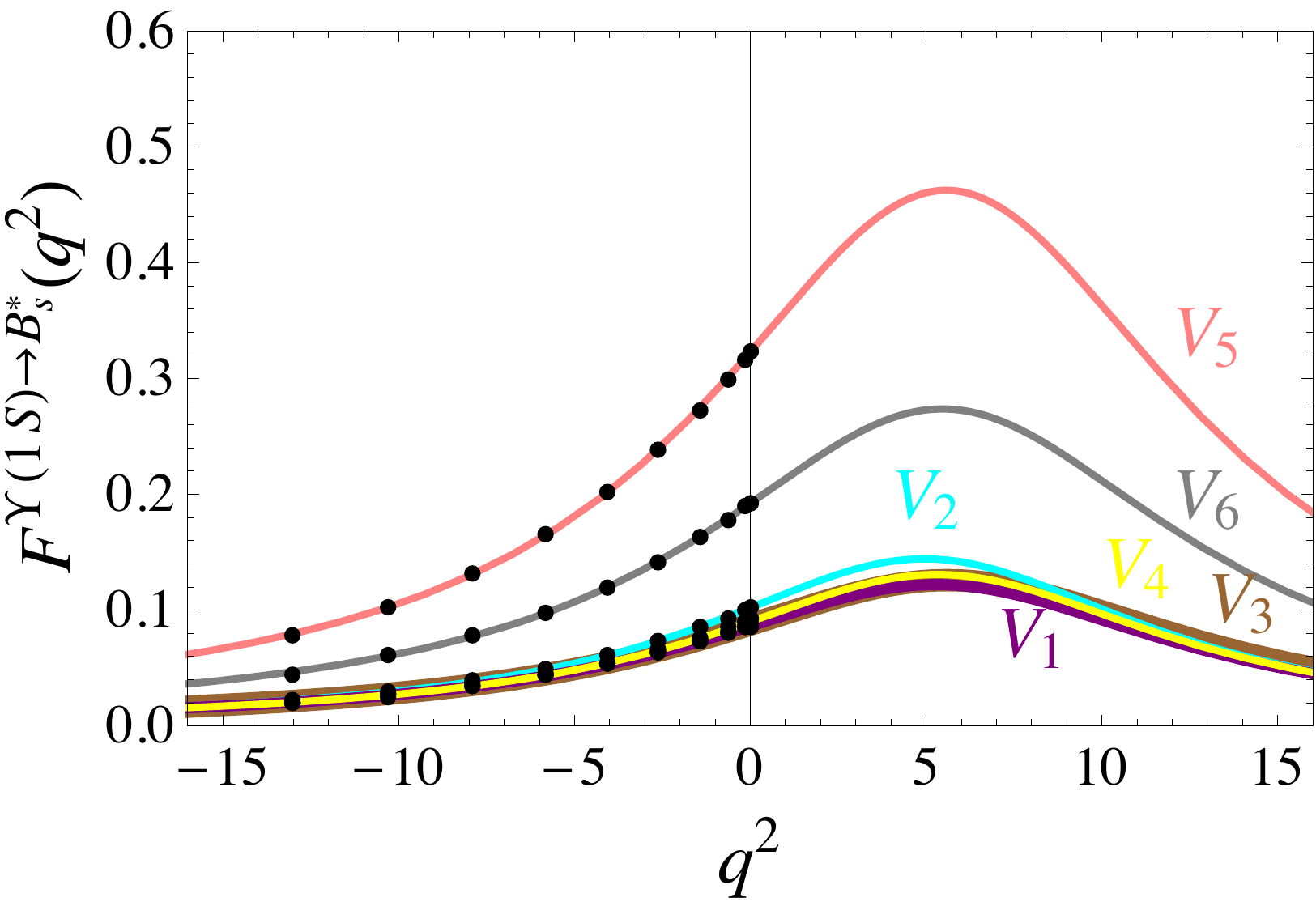}}\,
\subfigure{\includegraphics[scale=0.24]{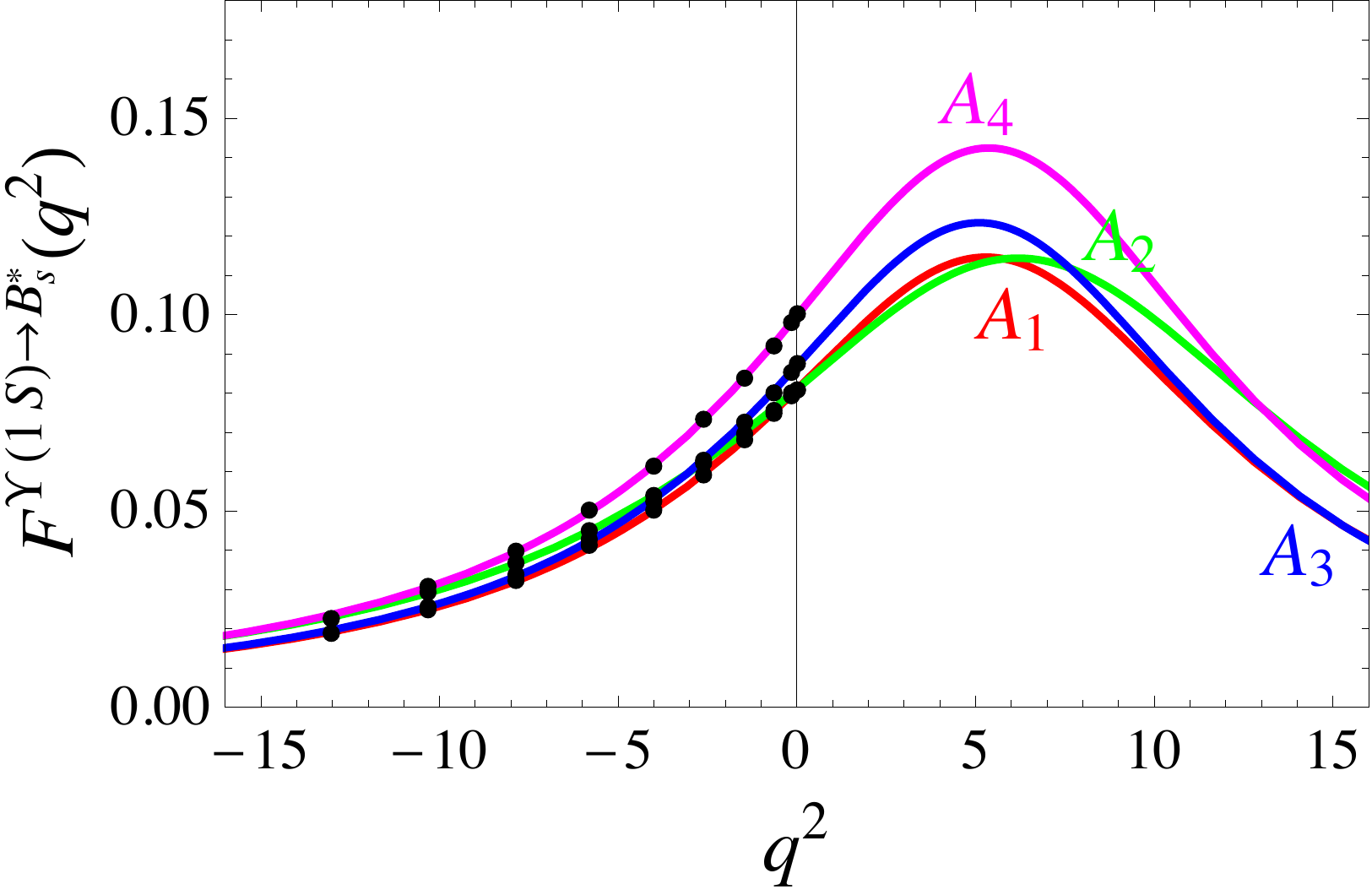}}\,
\subfigure{\includegraphics[scale=0.24]{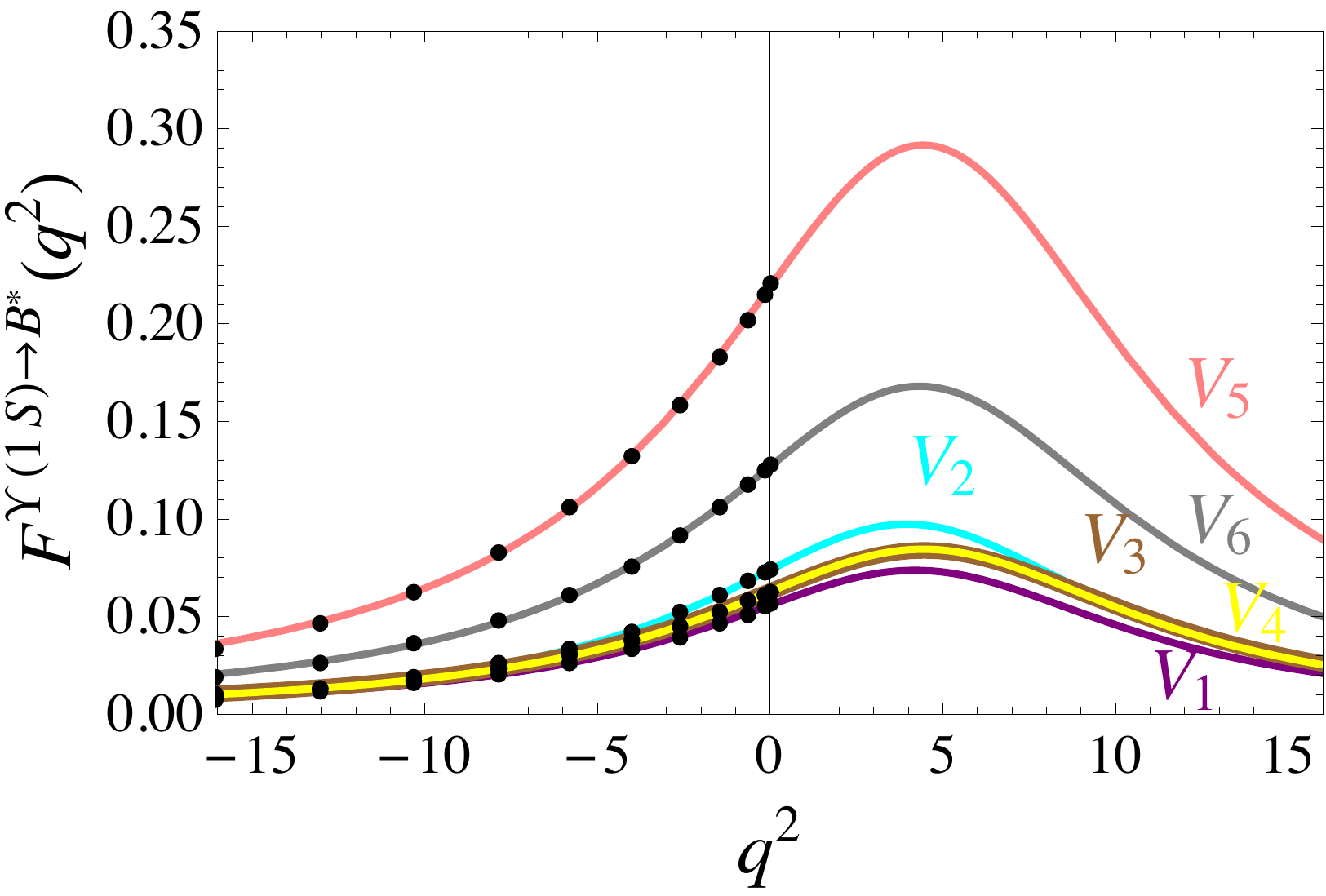}}\,
\subfigure{\includegraphics[scale=0.24]{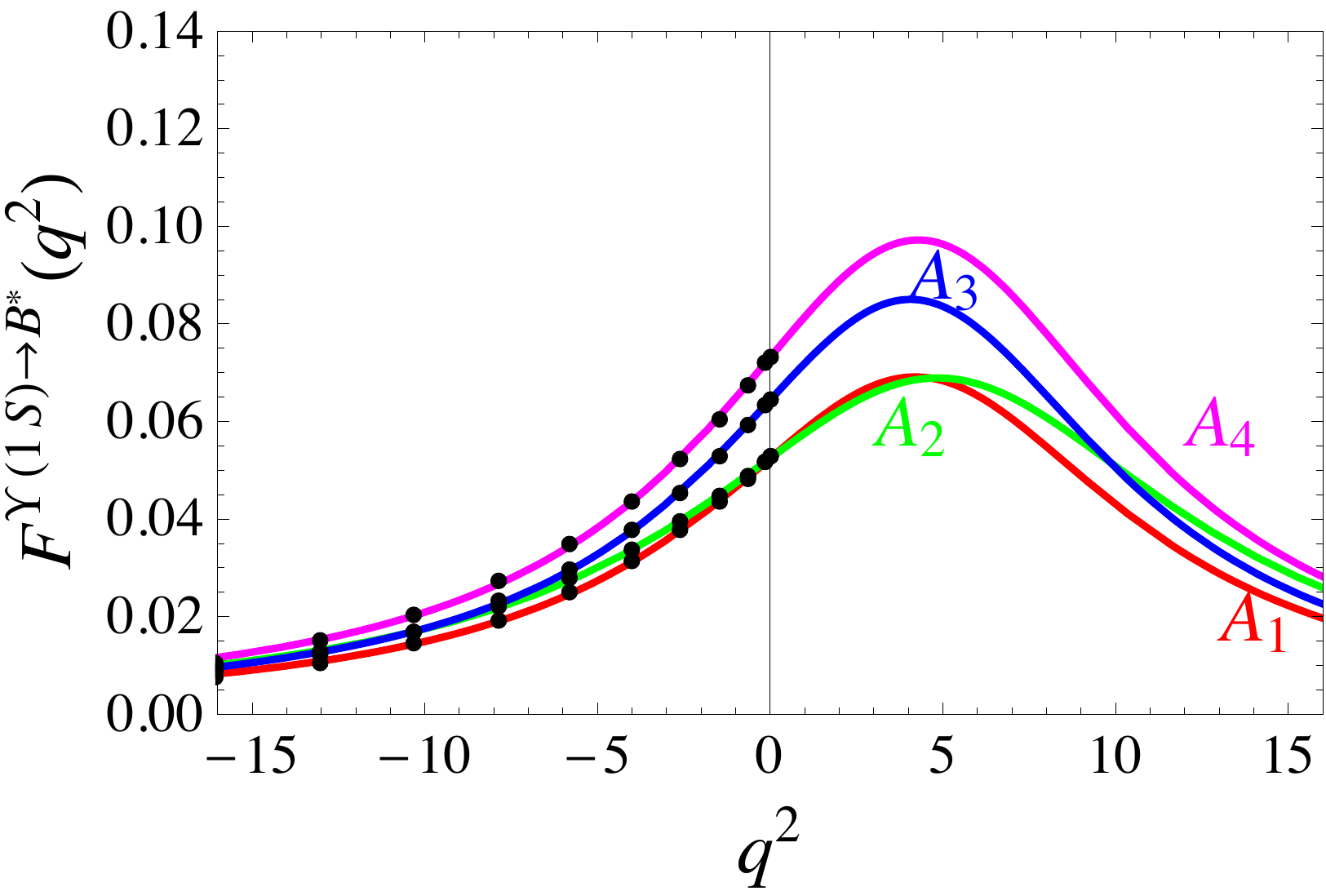}}
\end{center}
\label{fig:sche1b}
\end{figure}

\begin{table}[]
\scriptsize
\begin{center}
\caption{\label{tab:sche1c}  Same as Table~\ref{tab:sche2c} except with the parameterization scheme given by Eq.~\eqref{eq:para1}. The errors are due to parameters $\beta$ and quark masses only (cf. section 4 for a more complete discussion).}
\vspace{0.2cm}
\let\oldarraystretch=\arraystretch
\renewcommand*{\arraystretch}{1.1}
\setlength{\tabcolsep}{5.8pt}
\begin{tabular}{lccc|lccccccc}
\hline\hline
&${\cal F}(0)$  &$a$   &$b$  &&${\cal F}(0)$  &$a$   &$b$
\\\hline
$V_1^{D^*\to\rho}$
&$0.65_{-0.01-0.09}^{+0.01+0.09}$ &$1.27_{-0.01-0.23}^{+0.00+0.31}$ &$0.32_{-0.01-0.07}^{+0.00+0.09}$
&$V_1^{D^*\to K^*}$
&$0.74_{-0.02-0.10}^{+0.01+0.09}$ &$1.28_{-0.01-0.31}^{+0.01+0.36}$
&$0.34_{-0.01-0.15}^{+0.00+0.17}$
\\
$V_2^{D^*\to\rho}$
&$0.51_{-0.02-0.04}^{+0.02+0.03}$ &$1.37_{-0.02-0.29}^{+0.01+0.31}$ &$0.34_{-0.02-0.08}^{+0.01+0.11}$
&$V_2^{D^*\to K^*}$
&$0.43_{-0.03-0.08}^{+0.03+0.03}$ 
&$1.33_{-0.03-0.26}^{+0.03+0.28}$ &$0.33_{-0.02-0.12}^{+0.02+0.15}$
\\
$V_3^{D^*\to\rho}$
&$0.29_{-0.00-0.03}^{+0.00+0.03}$ &$1.54_{-0.03-0.24}^{+0.03+0.36}$ &$0.60_{-0.04-0.13}^{+0.04+0.25}$
&$V_3^{D^*\to K^*}$
&$0.26_{-0.00-0.02}^{+0.00+0.02}$ &$1.54_{-0.03-0.33}^{+0.03+0.37}$ &$0.62_{-0.04-0.12}^{+0.04+0.17}$
\\
$V_4^{D^*\to\rho}$
&$0.29_{-0.00-0.03}^{+0.00+0.03}$ &$1.94_{-0.04-0.46}^{+0.04+0.66}$ &$1.28_{-0.08-0.44}^{+0.08+0.72}$
&$V_4^{D^*\to K^*}$
&$0.26_{-0.00-0.02}^{+0.00+0.02}$ &$1.77_{-0.03-0.31}^{+0.03+0.38}$ &$0.91_{-0.05-0.37}^{+0.03+0.54}$
\\
$V_5^{D^*\to\rho}$
&$1.42_{-0.02-0.17}^{+0.02+0.17}$ &$1.21_{-0.01-0.20}^{+0.01+0.30}$ &$0.26_{-0.01-0.04}^{+0.00+0.05}$
&$V_5^{D^*\to K^*}$
&$1.50_{-0.03-0.16}^{+0.02+0.12}$ &$1.23_{-0.01-0.18}^{+0.01+0.22}$ &$0.28_{-0.01-0.04}^{+0.01+0.05}$
\\
$V_6^{D^*\to\rho}$
&$0.68_{-0.01-0.17}^{+0.01+0.17}$ &$1.13_{-0.01-0.25}^{+0.01+0.35}$ &$0.24_{-0.01-0.09}^{+0.00+0.15}$
&$V_6^{D^*\to K^*}$
&$0.78_{-0.02-0.17}^{+0.02+0.23}$ &$1.19_{-0.02-0.24}^{+0.02+0.33}$ &$0.27_{-0.01-0.10}^{+0.00+0.16}$
\\
$A_1^{D^*\to\rho}$
&$0.59_{-0.01-0.07}^{+0.01+0.07}$ &$1.08_{-0.01-0.19}^{+0.00+0.27}$ &$0.22_{-0.01-0.04}^{+0.01+0.06}$
&$A_1^{D^*\to K^*}$
&$0.69_{-0.01-0.09}^{+0.01+0.08}$ &$1.11_{-0.01-0.17}^{+0.00+0.21}$ &$0.24_{-0.00-0.03}^{+0.01+0.04}$
\\
$A_2^{D^*\to\rho}$
&$0.59_{-0.01-0.07}^{+0.01+0.07}$ &$0.15_{-0.03-0.05}^{+0.03+0.05}$ &$0.04_{-0.01-0.02}^{+0.00+0.02}$
&$A_2^{D^*\to K^*}$
&$0.69_{-0.01-0.09}^{+0.01+0.08}$ &$0.26_{-0.05-0.05}^{+0.04+0.03}$ &$0.02_{-0.01-0.00}^{+0.01+0.01}$
\\
$A_3^{D^*\to\rho}$
&$0.22_{-0.01-0.03}^{+0.01+0.03}$ &$1.52_{-0.02-0.36}^{+0.02+0.43}$ &$0.62_{-0.04-0.25}^{+0.03+0.45}$
&$A_3^{D^*\to K^*}$
&$0.15_{-0.01-0.04}^{+0.01+0.02}$ &$1.48_{-0.05-0.21}^{+0.04+0.18}$ &$0.59_{-0.04-0.06}^{+0.04+0.05}$
\\
$A_4^{D^*\to\rho}$
&$0.24_{-0.01-0.04}^{+0.01+0.06}$ &$1.47_{-0.03-0.34}^{+0.03+0.51}$ &$0.58_{-0.02-0.21}^{+0.04+0.42}$
&$A_4^{D^*\to K^*}$
&$0.22_{-0.01-0.03}^{+0.01+0.05}$ &$1.47_{-0.03-0.30}^{+0.03+0.39}$ &$0.61_{-0.04-0.33}^{+0.04+0.40}$
\\
\hline
%%%%%%%%%%%%%%%%%%%%%%%%%%%%%%%%%%%%%%%%%%
$V_1^{D_s^*\to K^*}$  &$0.58_{-0.02-0.09}^{+0.02+0.10}$ &$1.31_{-0.02-0.17}^{+0.01+0.19}$ &$0.41_{-0.02-0.05}^{+0.02+0.05}$
&$V_1^{D_s^*\to \phi}$ &$0.71_{-0.01-0.10}^{+0.01+0.09}$ &$1.35_{-0.00-0.19}^{+0.01+0.28}$ &$0.45_{-0.01-0.07}^{+0.00+0.03}$
\\
$V_2^{D_s^*\to K^*}$  &$0.42_{-0.04-0.03}^{+0.04+0.02}$ &$1.32_{-0.06-0.03}^{+0.05+0.01}$ &$0.38_{-0.04-0.05}^{+0.04+0.05}$
&$V_2^{D_s^*\to \phi}$ &$0.38_{-0.03-0.11}^{+0.03+0.04}$ &$1.41_{-0.04-0.24}^{+0.03+0.34}$ &$0.43_{-0.03-0.03}^{+0.03+0.04}$
\\
$V_3^{D_s^*\to K^*}$  &$0.30_{-0.01-0.03}^{+0.00+0.03}$ &$1.43_{-0.05-0.31}^{+0.04+0.43}$ &$0.54_{-0.04-0.19}^{+0.04+0.26}$
&$V_3^{D_s^*\to \phi}$ &$0.28_{-0.00-0.02}^{+0.00+0.02}$ &$1.48_{-0.04-0.33}^{+0.03+0.46}$ &$0.57_{-0.04-0.20}^{+0.03+0.26}$
\\
$V_4^{D_s^*\to K^*}$  &$0.30_{-0.01-0.03}^{+0.00+0.03}$ &$1.58_{-0.10-0.37}^{+0.09+0.40}$ &$0.65_{-0.12-0.24}^{+0.14+0.33}$
&$V_4^{D_s^*\to \phi}$ &$0.28_{-0.00-0.02}^{+0.00+0.02}$ &$1.71_{-0.09-0.40}^{+0.08+0.48}$ &$0.77_{-0.13-0.29}^{+0.16+0.31}$
\\
$V_5^{D_s^*\to K^*}$  &$1.36_{-0.04-0.20}^{+0.04+0.19}$ &$1.23_{-0.02-0.24}^{+0.02+0.28}$ &$0.33_{-0.02-0.12}^{+0.02+0.17}$
&$V_5^{D_s^*\to \phi}$ &$1.54_{-0.03-0.16}^{+0.02+0.16}$ &$1.27_{-0.01-0.18}^{+0.01+0.23}$ &$0.35_{-0.01-0.04}^{+0.01+0.02}$
\\
$V_6^{D_s^*\to K^*}$  &$0.68_{-0.03-0.15}^{+0.03+0.21}$ &$1.21_{-0.01-0.26}^{+0.01+0.33}$ &$0.31_{-0.01-0.12}^{+0.02+0.21}$
&$V_6^{D_s^*\to \phi}$ &$0.86_{-0.02-0.19}^{+0.02+0.25}$ &$1.23_{-0.01-0.28}^{+0.01+0.39}$ &$0.34_{-0.01-0.14}^{+0.00+0.23}$
\\
$A_1^{D_s^*\to K^*}$  &$0.52_{-0.02-0.08}^{+0.02+0.07}$ &$1.13_{-0.01-0.14}^{+0.01+0.17}$ &$0.30_{-0.01-0.03}^{+0.01+0.03}$
&$A_1^{D_s^*\to \phi}$ &$0.65_{-0.01-0.09}^{+0.01+0.08}$ &$1.16_{-0.01-0.17}^{+0.00+0.23}$ &$0.32_{-0.01-0.05}^{+0.00+0.04}$
\\
$A_2^{D_s^*\to K^*}$  &$0.52_{-0.02-0.08}^{+0.02+0.08}$ &$0.31_{-0.03-0.08}^{+0.02+0.06}$ &$-0.06_{-0.00-0.02}^{+0.01+0.04}$
&$A_2^{D_s^*\to \phi}$ &$0.65_{-0.01-0.09}^{+0.01+0.08}$ &$0.33_{-0.04-0.05}^{+0.04+0.04}$ &$0.04_{-0.01-0.03}^{+0.00+0.02}$
\\
$A_3^{D_s^*\to K^*}$  &$0.20_{-0.02-0.03}^{+0.02+0.03}$ &$1.35_{-0.07-0.09}^{+0.06+0.09}$ &$0.49_{-0.06-0.07}^{+0.05+0.07}$
&$A_3^{D_s^*\to \phi}$ &$0.16_{-0.01-0.05}^{+0.01+0.02}$ &$1.44_{-0.05-0.33}^{+0.04+0.40}$ &$0.55_{-0.05-0.20}^{+0.05+0.32}$
\\
$A_4^{D_s^*\to K^*}$  &$0.28_{-0.01-0.05}^{+0.01+0.04}$ &$1.37_{-0.05-0.31}^{+0.04+0.42}$ &$0.51_{-0.03-0.18}^{+0.05+0.23}$
&$A_4^{D_s^*\to \phi}$ &$0.27_{-0.01-0.04}^{+0.01+0.05}$ &$1.41_{-0.05-0.22}^{+0.04+0.24}$ &$0.55_{-0.06-0.11}^{+0.06+0.12}$
\\
\hline
%%%%%%%%%%%%%%%%%%%%%%%%%%%%%%%%%%%%%%%%%%
$V_1^{J/\psi\to D^*}$   &$0.55_{-0.02-0.12}^{+0.02+0.12}$ &$1.61_{-0.04-0.13}^{+0.03+0.18}$ &$0.96_{-0.03-0.02}^{+0.05+0.02}$
&$V_1^{J/\psi\to D_s^*}$ &$0.72_{-0.02-0.13}^{+0.01+0.12}$ &$1.44_{-0.05-0.11}^{+0.05+0.13}$ &$0.71_{-0.07-0.03}^{+0.07+0.03}$
\\
$V_2^{J/\psi\to D^*}$   &$0.60_{-0.05-0.10}^{+0.05+0.15}$ &$1.93_{-0.04-0.45}^{+0.03+0.60}$ &$1.28_{-0.02-0.63}^{+0.03+0.82}$
&$V_2^{J/\psi\to D_s^*}$ &$0.68_{-0.06-0.11}^{+0.06+0.14}$ &$1.76_{-0.04-0.50}^{+0.03+0.50}$ &$1.01_{-0.04-0.50}^{+0.03+0.66}$
\\
$V_3^{J/\psi\to D^*}$   &$0.36_{-0.01-0.03}^{+0.01+0.03}$ &$1.68_{-0.04-0.40}^{+0.04+0.51}$ &$0.91_{-0.06-0.36}^{+0.07+0.44}$
&$V_3^{J/\psi\to D_s^*}$ &$0.36_{-0.00-0.02}^{+0.00+0.01}$ &$1.64_{-0.03-0.40}^{+0.03+0.56}$ &$0.83_{-0.04-0.35}^{+0.04+0.46}$
\\
$V_4^{J/\psi\to D^*}$   &$0.36_{-0.01-0.03}^{+0.01+0.03}$ &$1.90_{-0.03-0.43}^{+0.03+0.47}$ &$1.18_{-0.05-0.46}^{+0.05+0.55}$
&$V_4^{J/\psi\to D_s^*}$ &$0.36_{-0.00-0.02}^{+0.00+0.01}$ &$1.91_{-0.02-0.46}^{+0.02+0.56}$ &$1.24_{-0.01-0.48}^{+0.01+0.57}$
\\
$V_5^{J/\psi\to D^*}$   &$1.79_{-0.05-0.34}^{+0.05+0.52}$ &$1.54_{-0.03-0.37}^{+0.03+0.46}$ &$0.82_{-0.03-0.36}^{+0.03+0.44}$
&$V_5^{J/\psi\to D_s^*}$ &$2.16_{-0.04-0.36}^{+0.04+0.51}$ &$1.39_{-0.04-0.32}^{+0.03+0.38}$ &$0.61_{-0.03-0.26}^{+0.03+0.38}$
\\
$V_6^{J/\psi\to D^*}$   &$1.23_{-0.05-0.34}^{+0.04+0.53}$ &$1.61_{-0.08-0.48}^{+0.08+0.36}$ &$0.45_{-0.33-0.40}^{+0.31+0.41}$
&$V_6^{J/\psi\to D_s^*}$ &$1.58_{-0.04-0.40}^{+0.04+0.58}$ &$1.32_{-0.05-0.33}^{+0.04+0.40}$ &$0.59_{-0.05-0.27}^{+0.04+0.39}$
\\
$A_1^{J/\psi\to D^*}$   &$0.47_{-0.01-0.09}^{+0.01+0.09}$ &$1.39_{-0.03-0.12}^{+0.02+0.16}$ &$0.71_{-0.04-0.02}^{+0.03+0.02}$
&$A_1^{J/\psi\to D_s^*}$ &$0.62_{-0.01-0.10}^{+0.01+0.08}$ &$1.25_{-0.03-0.19}^{+0.03+0.12}$ &$0.53_{-0.03-0.02}^{+0.03+0.02}$
\\
$A_2^{J/\psi\to D^*}$   &$0.47_{-0.01-0.09}^{+0.01+0.09}$ &$0.38_{-0.08-0.23}^{+0.07+0.16}$ &$-0.07_{-0.05-0.05}^{+0.05+0.06}$
&$A_2^{J/\psi\to D_s^*}$ &$0.62_{-0.01-0.10}^{+0.01+0.08}$ &$0.90_{-0.08-0.35}^{+0.08+0.34}$
&$-1.75_{-0.48-0.76}^{+0.50+0.72}$
\\
$A_3^{J/\psi\to D^*}$   &$0.38_{-0.01-0.08}^{+0.01+0.11}$ &$1.69_{-0.01-0.43}^{+0.01+0.61}$ &$0.95_{-0.00-0.40}^{+0.01+0.61}$
&$A_3^{J/\psi\to D_s^*}$ &$0.35_{-0.01-0.06}^{+0.01+0.07}$ &$1.66_{-0.02-0.41}^{+0.01+0.55}$ &$0.88_{-0.03-0.50}^{+0.03+0.68}$
\\
$A_4^{J/\psi\to D^*}$   &$0.45_{-0.01-0.11}^{+0.01+0.15}$ &$1.59_{-0.03-0.39}^{+0.03+0.56}$ &$0.85_{-0.04-0.35}^{+0.05+0.56}$
&$A_4^{J/\psi\to D_s^*}$ &$0.45_{-0.01-0.09}^{+0.01+0.13}$ &$1.55_{-0.03-0.40}^{+0.03+0.57}$ &$0.77_{-0.03-0.33}^{+0.05+0.46}$
\\
%%%%%%%%%%%%%%%%%%%%%%%%%%%%%%%%%%%%%%%%%%
\hline
$V_1^{B_c^*\to B^*}$   &$0.52_{-0.02-0.11}^{+0.02+0.11}$   &$1.85_{-0.03-0.18}^{+0.02+0.24}$ &$1.56_{-0.03-0.19}^{+0.02+0.21}$
&$V_1^{B_c^*\to B_s^*}$ &$0.63_{-0.01-0.12}^{+0.01+0.12}$   &$1.61_{-0.01-0.16}^{+0.01+0.20}$ &$1.12_{-0.01-0.13}^{+0.01+0.13}$
\\
$V_2^{B_c^*\to B^*}$   &$1.18_{-0.13-0.28}^{+0.13+0.30}$   &$2.41_{-0.04-0.36}^{+0.03+0.43}$ &$2.51_{-0.04-0.20}^{+0.04+0.31}$
&$V_2^{B_c^*\to B_s^*}$ &$1.06_{-0.11-0.40}^{+0.11+0.20}$   &$2.15_{-0.02-0.35}^{+0.01+0.37}$ &$1.88_{-0.01-0.39}^{+0.02+0.48}$
\\
$V_3^{B_c^*\to B^*}$   &$0.40_{-0.01-0.04}^{+0.01+0.03}$   &$1.88_{-0.07-0.27}^{+0.07+0.26}$ &$1.35_{-0.15-0.26}^{+0.16+0.27}$
&$V_3^{B_c^*\to B_s^*}$ &$0.40_{-0.00-0.03}^{+0.00+0.01}$   &$1.69_{-0.05-0.25}^{+0.05+0.31}$ &$1.02_{-0.03-0.29}^{+0.04+0.11}$
\\
$V_4^{B_c^*\to B^*}$   &$0.40_{-0.01-0.04}^{+0.01+0.03}$   &$2.01_{-0.06-0.40}^{+0.07+0.37}$ &$1.51_{-0.16-0.28}^{+0.17+0.22}$
&$V_4^{B_c^*\to B_s^*}$ &$0.40_{-0.00-0.03}^{+0.00+0.01}$   &$1.79_{-0.06-0.24}^{+0.06+0.24}$ &$1.14_{-0.11-0.22}^{+0.10+0.20}$
\\
$V_5^{B_c^*\to B^*}$   &$3.15_{-0.11-0.82}^{+0.10+1.12}$   &$1.80_{-0.03-0.26}^{+0.02+0.33}$ &$1.43_{-0.02-0.34}^{+0.02+0.41}$
&$V_5^{B_c^*\to B_s^*}$ &$3.52_{-0.06-0.87}^{+0.06+1.12}$   &$1.57_{-0.01-0.22}^{+0.01+0.28}$ &$1.04_{-0.02-0.25}^{+0.01+0.27}$
\\
$V_6^{B_c^*\to B^*}$   &$2.66_{-0.10-0.79}^{+0.10+1.11}$   &$1.79_{-0.03-0.27}^{+0.02+0.35}$ &$1.45_{-0.03-0.36}^{+0.03+0.34}$
&$V_6^{B_c^*\to B_s^*}$ &$3.02_{-0.06-0.85}^{+0.06+1.15}$   &$1.55_{-0.01-0.23}^{+0.01+0.30}$ &$1.05_{-0.01-0.20}^{+0.01+0.29}$
\\
$A_1^{B_c^*\to B^*}$   &$0.43_{-0.02-0.08}^{+0.01+0.07}$   &$1.59_{-0.02-0.16}^{+0.02+0.21}$ &$1.15_{-0.03-0.14}^{+0.02+0.15}$
&$A_1^{B_c^*\to B_s^*}$ &$0.53_{-0.01-0.09}^{+0.01+0.08}$   &$1.61_{-0.01-0.06}^{+0.01+0.10}$ &$1.12_{-0.01-0.13}^{+0.01+0.12}$
\\
$A_2^{B_c^*\to B^*}$   &$0.43_{-0.02-0.08}^{+0.01+0.07}$   &$0.32_{-0.10-0.20}^{+0.11+0.26}$ &$-0.05_{-0.04-0.09}^{+0.04+0.09}$
&$A_2^{B_c^*\to B_s^*}$ &$0.53_{-0.01-0.09}^{+0.01+0.08}$   &$0.19_{-0.08-0.13}^{+0.08+0.13}$ &$-0.74_{-0.15-0.35}^{+0.10+0.39}$
\\
$A_3^{B_c^*\to B^*}$   &$0.81_{-0.03-0.20}^{+0.03+0.25}$   &$1.93_{-0.02-0.35}^{+0.01+0.41}$ &$1.44_{-0.06-0.37}^{+0.05+0.41}$
&$A_3^{B_c^*\to B_s^*}$ &$0.73_{-0.02-0.15}^{+0.02+0.16}$   &$1.99_{-0.03-0.35}^{+0.03+0.50}$ &$1.43_{-0.12-0.36}^{+0.10+0.38}$
\\
$A_4^{B_c^*\to B^*}$   &$0.89_{-0.02-0.24}^{+0.02+0.32}$   &$1.86_{-0.07-0.34}^{+0.06+0.41}$ &$1.35_{-0.12-0.35}^{+0.14+0.40}$
&$A_4^{B_c^*\to B_s^*}$ &$0.85_{-0.01-0.20}^{+0.01+0.25}$   &$1.94_{-0.08-0.37}^{+0.05+0.47}$ &$1.24_{-0.24-0.23}^{+0.20+0.33}$
\\
\hline\hline
\end{tabular}
\end{center}
\end{table}

%%%%%%%%%%%%%%%%%%%%%%%%%%%%%%%%%%%%%%%%%%%%%%%%%%%%%%%%%%%%%%%%%%%%%%%%%%%%%%%%%
\begin{table}[t]
\scriptsize
\begin{center}
\caption{\label{tab:sche1b1} Same as Table~\ref{tab:sche2b1} except with the parameterization scheme given by Eq.~\eqref{eq:para1}. The errors are due to parameters $\beta$ and quark masses only (cf. section 4 for a more complete discussion). }
\vspace{0cm}
\let\oldarraystretch=\arraystretch
\renewcommand*{\arraystretch}{1.3}
\setlength{\tabcolsep}{5.8pt}
\begin{tabular}{lccc|lccc}
\hline\hline
  &${\cal F}(0)$  &$a$   &$b$ & &${\cal F}(0)$  &$a$   &$b$ 
\\\hline
$V_1^{B^*\to\rho}$
&$0.27_{-0.01-0.04}^{+0.01+0.05}$ &$1.58_{-0.02-0.11}^{+0.02+0.15}$ &$0.72_{-0.04-0.02}^{+0.04+0.02}$
&$V_1^{B^*\to K^*}$
&$0.31_{-0.01-0.06}^{+0.01+0.06}$ &$1.59_{-0.01-0.09}^{+0.01+0.10}$ &$0.72_{-0.01-0.01}^{+0.01+0.01}$
\\
$V_2^{B^*\to\rho}$
&$0.25_{-0.01-0.03}^{+0.01+0.03}$ &$1.70_{-0.10-0.23}^{+0.10+0.24}$ &$0.78_{-0.06-0.07}^{+0.06+0.07}$
&$V_2^{B^*\to K^*}$
&$0.26_{-0.01-0.04}^{+0.01+0.03}$ &$1.59_{-0.00-0.08}^{+0.01+0.08}$ &$0.71_{-0.01-0.01}^{+0.01+0.01}$
\\
$V_3^{B^*\to\rho}$
&$0.10_{-0.00-0.01}^{-0.00+0.02}$ &$2.00_{-0.03-0.17}^{+0.03+0.22}$ &$1.60_{-0.07-0.18}^{+0.07+0.25}$
&$V_3^{B^*\to K^*}$
&$0.11_{-0.00-0.02}^{-0.00+0.02}$ &$2.03_{-0.03-0.18}^{+0.02+0.19}$ &$1.58_{-0.06-0.13}^{+0.07+0.12}$
\\
$V_4^{B^*\to\rho}$
&$0.10_{-0.00-0.01}^{+0.00+0.02}$ &$2.23_{-0.05-0.23}^{+0.04+0.14}$ &$2.33_{-0.09-0.19}^{+0.11+0.22}$
&$V_4^{B^*\to K^*}$
&$0.11_{-0.00-0.02}^{+0.00+0.02}$ &$2.17_{-0.03-0.15}^{+0.03+0.19}$ &$1.92_{-0.08-0.08}^{+0.08+0.14}$
\\
$V_5^{B^*\to\rho}$
&$0.57_{-0.01-0.09}^{+0.01+0.10}$ &$1.55_{-0.03-0.11}^{+0.03+0.14}$ &$0.68_{-0.04-0.01}^{+0.04+0.01}$
&$V_5^{B^*\to K^*}$
&$0.65_{-0.02-0.11}^{+0.02+0.11}$ &$1.57_{-0.04-0.10}^{+0.03+0.12}$ &$0.69_{-0.05-0.01}^{+0.04+0.01}$
\\
$V_6^{B^*\to\rho}$&$0.12_{-0.00-0.02}^{+0.00+0.03}$ &$1.50_{-0.02-0.12}^{+0.02+0.16}$ &$0.60_{-0.03-0.02}^{+0.03+0.04}$
&$V_6^{B^*\to K^*}$&$0.15_{-0.01-0.03}^{+0.01+0.04}$ &$1.53_{-0.03-0.11}^{+0.03+0.13}$ &$0.62_{-0.04-0.01}^{+0.04+0.03}$
\\
$A_1^{B^*\to\rho}$&$0.26_{-0.01-0.04}^{+0.01+0.05}$ &$1.57_{-0.02-0.11}^{+0.02+0.15}$ &$0.71_{-0.05-0.01}^{+0.04+0.02}$
&$A_1^{B^*\to K^*}$&$0.31_{-0.01-0.06}^{+0.01+0.06}$ &$1.58_{-0.01-0.10}^{+0.01+0.13}$ &$0.72_{-0.02-0.01}^{+0.01+0.01}$
\\
$A_2^{B^*\to\rho}$
&$0.26_{-0.01-0.04}^{+0.01+0.05}$ &$0.70_{-0.01-0.01}^{+0.00+0.01}$ &$0.08_{-0.01-0.04}^{+0.01+0.03}$
&$A_2^{B^*\to K^*}$
&$0.31_{-0.01-0.06}^{+0.01+0.06}$ &$0.75_{-0.01-0.01}^{+0.01+0.01}$ &$0.08_{-0.01-0.04}^{+0.01+0.04}$
\\
$A_3^{B^*\to\rho}$
&$0.07_{-0.00-0.01}^{+0.00+0.01}$ &$2.13_{-0.00-0.22}^{+0.01+0.27}$ &$1.83_{-0.01-0.26}^{+0.02+0.36}$
&$A_3^{B^*\to K^*}$
&$0.06_{-0.00-0.01}^{+0.00+0.01}$ &$2.13_{-0.02-0.22}^{+0.02+0.25}$ &$1.76_{-0.04-0.27}^{+0.06+0.34}$
\\
$A_4^{B^*\to\rho}$
&$0.06_{-0.00-0.01}^{+0.00+0.01}$ &$2.12_{-0.03-0.20}^{+0.03+0.24}$ &$1.83_{-0.08-0.23}^{+0.09+0.32}$
&$A_4^{B^*\to K^*}$
&$0.07_{-0.00-0.01}^{+0.00+0.01}$ &$2.14_{-0.03-0.17}^{+0.04+0.23}$ &$1.80_{-0.08-0.20}^{+0.08+0.27}$
\\
\hline
%%%%%%%%%%%%%%%%%%%%%%%%%%%%%%%%%%%%%%%%%%%%%%%%%%%%%%%%%%%%%%%%%%%%%%%%%%%%%%%%%%%%%%%%%%%
$V_1^{B^*\to D^*}$ &$0.70_{-0.01-0.11}^{+0.01+0.10}$ &$1.57_{-0.01-0.03}^{+0.01+0.04}$ &$0.58_{-0.01-0.02}^{+0.01+0.04}$
&$V_1^{B_s^*\to K^*}$ &$0.19_{-0.01-0.05}^{+0.01+0.06}$ &$1.95_{-0.05-0.09}^{+0.05+0.10}$ &$1.71_{-0.16-0.04}^{+0.16+0.04}$
\\
$V_2^{B^*\to D^*}$   &$0.35_{-0.01-0.04}^{+0.01+0.02}$ &$1.58_{-0.02-0.10}^{+0.02+0.09}$ &$0.57_{-0.05-0.10}^{+0.04+0.12}$
&$V_2^{B_s^*\to K^*}$ &$0.18_{-0.01-0.04}^{+0.01+0.04}$ &$1.96_{-0.05-0.08}^{+0.04+0.10}$ &$1.72_{-0.16-0.04}^{+0.15+0.05}$
\\
$V_3^{B^*\to D^*}$   &$0.12_{-0.00-0.01}^{+0.00+0.02}$ &$2.09_{-0.02-0.33}^{+0.02+0.29}$ &$1.45_{-0.05-0.44}^{+0.04+0.56}$
&$V_3^{B_s^*\to K^*}$ &$0.09_{-0.01-0.02}^{+0.01+0.02}$ &$2.26_{-0.04-0.13}^{+0.04+0.16}$ &$2.70_{-0.18-0.11}^{+0.20+0.14}$
\\
$V_4^{B^*\to D^*}$   &$0.12_{-0.00-0.01}^{+0.00+0.02}$ &$2.21_{-0.02-0.42}^{+0.02+0.43}$ &$1.65_{-0.03-0.34}^{+0.03+0.33}$
&$V_4^{B_s^*\to K^*}$ &$0.09_{-0.01-0.02}^{+0.01+0.02}$ &$2.45_{-0.03-0.15}^{+0.03+0.17}$ &$3.42_{-0.17-0.11}^{+0.17+0.13}$
\\
$V_5^{B^*\to D^*}$   &$1.19_{-0.01-0.14}^{+0.01+0.09}$ &$1.56_{-0.02-0.04}^{+0.02+0.04}$ &$0.55_{-0.02-0.02}^{+0.03+0.04}$
&$V_5^{B_s^*\to K^*}$ &$0.42_{-0.03-0.10}^{+0.02+0.12}$ &$1.91_{-0.06-0.09}^{+0.05+0.09}$ &$1.61_{-0.13-0.05}^{+0.14+0.04}$
\\
$V_6^{B^*\to D^*}$   &$0.51_{-0.01-0.13}^{+0.01+0.15}$ &$1.56_{-0.03-0.06}^{+0.02+0.05}$ &$0.54_{-0.02-0.02}^{+0.03+0.03}$
&$V_6^{B_s^*\to K^*}$ &$0.10_{-0.01-0.03}^{+0.01+0.03}$ &$1.83_{-0.05-0.09}^{+0.04+0.10}$ &$1.42_{-0.13-0.04}^{+0.13+0.03}$
\\
$A_1^{B^*\to D^*}$   &$0.69_{-0.01-0.11}^{+0.01+0.10}$ &$1.57_{-0.01-0.03}^{+0.01+0.04}$ &$0.57_{-0.01-0.03}^{+0.01+0.04}$
&$A_1^{B_s^*\to K^*}$ &$0.19_{-0.01-0.05}^{+0.01+0.05}$ &$1.91_{-0.02-0.08}^{+0.03+0.10}$ &$1.64_{-0.09-0.04}^{+0.09+0.04}$
\\
$A_2^{B^*\to D^*}$   &$0.69_{-0.01-0.11}^{+0.01+0.10}$ &$0.75_{-0.02-0.01}^{+0.01+0.01}$ &$-0.01_{-0.01-0.04}^{+0.00+0.04}$
&$A_2^{B_s^*\to K^*}$ &$0.19_{-0.01-0.05}^{+0.01+0.05}$ &$1.11_{-0.06-0.03}^{+0.05+0.02}$ &$0.48_{-0.06-0.11}^{+0.07+0.10}$
\\
$A_3^{B^*\to D^*}$   &$0.07_{-0.00-0.01}^{+0.00+0.01}$ &$2.18_{-0.03-0.37}^{+0.02+0.37}$ &$1.57_{-0.05-0.50}^{+0.04+0.57}$
&$A_3^{B_s^*\to K^*}$ &$0.06_{-0.00-0.01}^{+0.00+0.01}$ &$2.34_{-0.10-0.16}^{+0.08+0.11}$ &$2.94_{-0.12-0.19}^{+0.14+0.26}$
\\
$A_4^{B^*\to D^*}$   &$0.08_{-0.00-0.01}^{+0.00+0.01}$ &$2.21_{-0.01-0.33}^{+0.00+0.36}$ &$1.64_{-0.01-0.58}^{+0.01+0.54}$
&$A_4^{B_s^*\to K^*}$ &$0.05_{-0.00-0.01}^{+0.00+0.02}$ &$2.33_{-0.05-0.15}^{+0.04+0.18}$ &$2.90_{-0.19-0.16}^{+0.20+0.21}$
\\\hline
%%%%%%%%%%%%%%%%%%%%%%%%%%%%%%%%%%%%%%%%%%%%%%%%%%%%%%%%%%%%%%%%%%%%%%%%%%%%%%%%%%%%%%%%%%%
$V_1^{B_s^*\to \phi}$  &$0.27_{-0.01-0.06}^{+0.01+0.07}$ &$1.88_{-0.02-0.07}^{+0.02+0.09}$ &$1.43_{-0.08-0.03}^{+0.08+0.02}$
&$V_1^{B_s^*\to D_s^*}$ &$0.69_{-0.01-0.12}^{+0.01+0.11}$ &$1.65_{-0.03-0.01}^{+0.03+0.01}$ &$0.82_{-0.06-0.09}^{+0.06+0.08}$
\\
$V_2^{B_s^*\to \phi}$  &$0.23_{-0.01-0.05}^{+0.01+0.05}$ &$1.89_{-0.03-0.07}^{+0.03+0.09}$ &$1.43_{-0.08-0.05}^{+0.08+0.05}$
&$V_2^{B_s^*\to D_s^*}$ &$0.38_{-0.01-0.03}^{+0.01+0.03}$ &$1.71_{-0.01-0.19}^{+0.01+0.25}$ &$0.85_{-0.01-0.12}^{+0.02+0.16}$
\\
$V_3^{B_s^*\to \phi}$  &$0.11_{-0.00-0.02}^{+0.00+0.02}$ &$2.22_{-0.03-0.12}^{+0.03+0.14}$ &$2.32_{-0.10-0.10}^{+0.10+0.13}$
&$V_3^{B_s^*\to D_s^*}$ &$0.15_{-0.00-0.02}^{+0.00+0.02}$ &$2.11_{-0.03-0.36}^{+0.04+0.35}$ &$1.59_{-0.07-0.54}^{+0.07+0.59}$
\\
$V_4^{B_s^*\to \phi}$  &$0.11_{-0.00-0.02}^{+0.00+0.02}$ &$2.40_{-0.03-0.12}^{+0.03+0.13}$ &$2.91_{-0.12-0.08}^{+0.11+0.06}$
&$V_4^{B_s^*\to D_s^*}$ &$0.15_{-0.00-0.02}^{+0.00+0.02}$ &$2.41_{-0.01-0.43}^{+0.00+0.32}$ &$2.38_{-0.01-0.47}^{+0.01+0.54}$
\\
$V_5^{B_s^*\to \phi}$  &$0.57_{-0.02-0.12}^{+0.02+0.14}$ &$1.85_{-0.04-0.08}^{+0.03+0.08}$ &$1.34_{-0.07-0.04}^{+0.08+0.03}$
&$V_5^{B_s^*\to D_s^*}$ &$1.22_{-0.02-0.16}^{+0.01+0.11}$ &$1.64_{-0.04-0.01}^{+0.03+0.01}$ &$0.78_{-0.05-0.08}^{+0.05+0.09}$
\\
$V_6^{B_s^*\to \phi}$  &$0.15_{-0.01-0.04}^{+0.01+0.05}$ &$1.77_{-0.03-0.08}^{+0.03+0.10}$ &$1.19_{-0.06-0.03}^{+0.06+0.03}$
&$V_6^{B_s^*\to D_s^*}$ &$0.55_{-0.01-0.14}^{+0.01+0.15}$ &$1.58_{-0.04-0.03}^{+0.04+0.02}$ &$0.72_{-0.05-0.05}^{+0.05+0.05}$
\\
$A_1^{B_s^*\to \phi}$  &$0.26_{-0.01-0.06}^{+0.01+0.07}$ &$1.84_{-0.02-0.07}^{+0.03+0.09}$ &$1.36_{-0.07-0.04}^{+0.08+0.03}$
&$A_1^{B_s^*\to D_s^*}$ &$0.68_{-0.01-0.12}^{+0.01+0.11}$ &$1.65_{-0.04-0.01}^{+0.04+0.01}$
&$0.81_{-0.05-0.09}^{+0.05+0.09}$
\\
$A_2^{B_s^*\to \phi}$  &$0.26_{-0.01-0.06}^{+0.01+0.07}$ &$1.05_{-0.03-0.02}^{+0.03+0.03}$ &$0.37_{-0.03-0.09}^{+0.03+0.08}$
&$A_2^{B_s^*\to D_s^*}$ &$0.68_{-0.01-0.12}^{+0.01+0.11}$ &$0.80_{-0.05-0.05}^{+0.04+0.04}$ &$0.13_{-0.02-0.08}^{+0.02+0.06}$
\\
$A_3^{B_s^*\to \phi}$  &$0.07_{-0.00-0.01}^{+0.00+0.01}$ &$2.29_{-0.01-0.17}^{+0.00+0.23}$ &$2.48_{-0.01-0.21}^{+0.00+0.29}$
&$A_3^{B_s^*\to D_s^*}$ &$0.09_{-0.00-0.01}^{+0.00+0.02}$ &$2.21_{-0.00-0.37}^{+0.00+0.39}$ &$1.76_{-0.01-0.50}^{+0.00+0.61}$
\\
$A_4^{B_s^*\to \phi}$  &$0.07_{-0.00-0.02}^{+0.00+0.02}$ &$2.29_{-0.03-0.14}^{+0.03+0.17}$ &$2.52_{-0.11-0.15}^{+0.11+0.18}$
&$A_4^{B_s^*\to D_s^*}$ &$0.09_{-0.00-0.01}^{+0.00+0.01}$ &$2.22_{-0.03-0.38}^{+0.01+0.37}$ &$1.75_{-0.01-0.57}^{+0.01+0.55}$
\\
\hline\hline
\end{tabular}
\end{center}
\end{table}

%%%%%%%%%%%%%%%%%%%%%%%%%%%%%%%%%%%%%%%%%%%%%%%%%%%%%%%%%%%%%%%%%%%%%%%%%%%%%%%%%
\begin{table}[t]
\scriptsize
\begin{center}
\caption{\label{tab:sche1b2} Same as Table~\ref{tab:sche2b2} except with the parameterization scheme given by Eq.~\eqref{eq:para1}. The errors are due to parameters $\beta$ and quark masses only (cf. section 4 for a more complete discussion).}
\vspace{0cm}
\let\oldarraystretch=\arraystretch
\renewcommand*{\arraystretch}{1.3}
\setlength{\tabcolsep}{5.8pt}
\begin{tabular}{lccc|lccc}
\hline\hline
  &${\cal F}(0)$  &$a$   &$b$ & &${\cal F}(0)$  &$a$   &$b$
  \\\hline
$V_1^{B_c^*\to D^*}$   &$0.10_{-0.01-0.04}^{+0.01+0.06}$   &$3.76_{-0.13-0.10}^{+0.12+0.09}$ &$10.56_{-1.12-2.09}^{+1.18+1.97}$
&$V_1^{B_c^*\to D_s^*}$ &$0.20_{-0.02-0.07}^{+0.02+0.10}$   &$2.42_{-0.09-0.06}^{+0.10+0.06}$ &$3.43_{-0.39-0.62}^{+0.44+0.56}$
\\
$V_2^{B_c^*\to D^*}$   &$0.11_{-0.01-0.04}^{+0.01+0.06}$   &$3.85_{-0.10-0.09}^{+0.10+0.08}$ &$11.05_{-0.93-2.20}^{+1.02+2.05}$
&$V_2^{B_c^*\to D_s^*}$ &$0.20_{-0.02-0.06}^{+0.02+0.07}$   &$2.51_{-0.10-0.05}^{+0.10+0.04}$ &$3.15_{-0.29-0.35}^{+0.36+0.59}$
\\
$V_3^{B_c^*\to D^*}$   &$0.08_{-0.01-0.03}^{+0.01+0.04}$   &$3.89_{-0.13-0.07}^{+0.13+0.06}$ &$11.59_{-1.19-1.99}^{+1.25+1.87}$
&$V_3^{B_c^*\to D_s^*}$ &$0.13_{-0.01-0.04}^{+0.01+0.04}$   &$2.57_{-0.09-0.03}^{+0.10+0.03}$ &$4.00_{-0.43-0.59}^{+0.48+0.53}$
\\
$V_4^{B_c^*\to D^*}$   &$0.08_{-0.01-0.03}^{+0.01+0.04}$   &$4.13_{-0.12-0.04}^{+0.12+0.03}$ &$13.07_{-1.24-2.16}^{+1.33+2.01}$
&$V_4^{B_c^*\to D_s^*}$ &$0.13_{-0.01-0.04}^{+0.01+0.04}$   &$2.76_{-0.04-0.05}^{+0.04+0.05}$ &$5.02_{-0.52-0.54}^{+0.43+0.48}$
\\
$V_5^{B_c^*\to D^*}$   &$0.27_{-0.03-0.11}^{+0.03+0.15}$   &$3.67_{-0.13-0.12}^{+0.13+0.10}$ &$9.78_{-1.06-2.08}^{+1.12+1.99}$
&$V_5^{B_c^*\to D_s^*}$ &$0.50_{-0.05-0.16}^{+0.05+0.20}$   &$2.36_{-0.09-0.06}^{+0.10+0.06}$ &$3.19_{-0.37-0.57}^{+0.41+0.57}$
\\
$V_6^{B_c^*\to D^*}$   &$0.11_{-0.01-0.04}^{+0.01+0.06}$   &$3.60_{-0.13-0.13}^{+0.13+0.11}$ &$9.39_{-1.03-2.05}^{+1.09+2.02}$
&$V_6^{B_c^*\to D_s^*}$ &$0.22_{-0.02-0.07}^{+0.02+0.08}$   &$2.28_{-0.10-0.08}^{+0.10+0.07}$ &$2.96_{-0.35-0.50}^{+0.41+0.56}$
\\
$A_1^{B_c^*\to D^*}$   &$0.10_{-0.01-0.04}^{+0.01+0.06}$   &$3.75_{-0.12-0.10}^{+0.13+0.08}$ &$10.49_{-1.11-2.10}^{+1.15+1.98}$
&$A_1^{B_c^*\to D_s^*}$ &$0.19_{-0.02-0.06}^{+0.02+0.08}$   &$2.37_{-0.09-0.06}^{+0.10+0.05}$ &$3.28_{-0.36-0.60}^{+0.43+0.55}$
\\
$A_2^{B_c^*\to D^*}$   &$0.10_{-0.01-0.04}^{+0.01+0.06}$   &$2.87_{-0.15-0.17}^{+0.14+0.19}$ &$5.85_{-0.76-1.79}^{+0.79+1.83}$
&$A_2^{B_c^*\to D_s^*}$ &$0.19_{-0.02-0.06}^{+0.02+0.08}$   &$1.65_{-0.10-0.17}^{+0.11+0.15}$ &$1.58_{-0.24-0.49}^{+0.27+0.48}$
\\
$A_3^{B_c^*\to D^*}$   &$0.06_{-0.01-0.02}^{+0.01+0.03}$   &$4.12_{-0.13-0.02}^{+0.13+0.02}$ &$13.32_{-1.30-1.81}^{+1.37+1.78}$
&$A_3^{B_c^*\to D_s^*}$ &$0.10_{-0.01-0.03}^{+0.01+0.03}$   &$2.68_{-0.09-0.02}^{+0.10+0.02}$ &$4.43_{-0.45-0.59}^{+0.50+0.50}$
\\
$A_4^{B_c^*\to D^*}$   &$0.06_{-0.01-0.02}^{+0.01+0.03}$   &$3.99_{-0.12-0.03}^{+0.13+0.03}$ &$12.35_{-1.22-1.81}^{+1.29+1.78}$
&$A_4^{B_c^*\to D_s^*}$ &$0.09_{-0.01-0.03}^{+0.01+0.03}$   &$2.61_{-0.09-0.01}^{+0.09+0.01}$ &$4.17_{-0.42-0.52}^{+0.48+0.46}$
\\
\hline
%%%%%%%%%%%%%%%%%%%%%%%%%%%%%%%%%%%%%%%%%%%%%%%%%%%%%%%%%%%%%%%%%%%%%%%%%%%%%%%%%%%%%%%%%%%%
$V_1^{B_c^*\to J/\psi}$       &$0.56_{-0.01-0.17}^{+0.01+0.17}$ &$2.38_{-0.05-0.13}^{+0.05+0.16}$&$2.62_{-0.15-0.55}^{+0.15+0.61}$  
&$V_1^{\Upsilon(1S)\to B^*}$ &$0.06_{-0.01-0.02}^{+0.01+0.03}$  &$3.21_{-0.08-0.08}^{+0.08+0.09}$ &$10.68_{-0.83-0.08}^{+0.81+0.06}$
\\
$V_2^{B_c^*\to J/\psi}$       &$0.33_{-0.01-0.04}^{+0.01+0.05}$  &$2.45_{-0.02-0.02}^{+0.02+0.02}$  &$2.76_{-0.07-0.29}^{+0.06+0.25}$
&$V_2^{\Upsilon(1S)\to B^*}$ &$0.08_{-0.01-0.03}^{+0.01+0.03}$  &$3.39_{-0.05-0.11}^{+0.06+0.13}$ &$11.98_{-0.92-0.23}^{+0.96+0.22}$
\\
$V_3^{B_c^*\to J/\psi}$       &$0.20_{-0.00-0.02}^{+0.00+0.02}$  &$2.61_{-0.06-0.05}^{+0.06+0.05}$  &$3.26_{-0.19-0.17}^{+0.20+0.17}$
&$V_3^{\Upsilon(1S)\to B^*}$ &$0.06_{-0.01-0.02}^{+0.01+0.03}$  &$3.13_{-0.08-0.07}^{+0.08+0.07}$ &$9.77_{-0.75-0.13}^{+0.83+0.08}$
\\
$V_4^{B_c^*\to J/\psi}$       &$0.20_{-0.00-0.02}^{+0.00+0.02}$  &$2.84_{-0.07-0.09}^{+0.07+0.11}$ &$4.06_{-0.27-0.16}^{+0.28+0.16}$
&$V_4^{\Upsilon(1S)\to B^*}$ &$0.06_{-0.01-0.02}^{+0.01+0.03}$  &$3.22_{-0.07-0.08}^{+0.07+0.08}$ &$10.11_{-0.75-0.16}^{+0.82+0.11}$
\\
$V_5^{B_c^*\to J/\psi}$       &$1.17_{-0.02-0.29}^{+0.02+0.23}$  &$2.33_{-0.06-0.15}^{+0.05+0.17}$ &$2.46_{-0.13-0.69}^{+0.15+0.60}$
&$V_5^{\Upsilon(1S)\to B^*}$ &$0.22_{-0.02-0.08}^{+0.03+0.11}$  &$3.13_{-0.08-0.07}^{+0.08+0.08}$ &$9.86_{-0.78-0.08}^{+0.76+0.05}$
\\
$V_6^{B_c^*\to J/\psi}$       &$0.65_{-0.01-0.19}^{+0.01+0.20}$  &$2.45_{-0.25-0.31}^{+0.24+0.27}$ &$2.28_{-0.13-0.59}^{+0.13+0.52}$
&$V_6^{\Upsilon(1S)\to B^*}$ &$0.13_{-0.02-0.05}^{+0.01+0.06}$  &$3.15_{-0.08-0.07}^{+0.08+0.08}$ &$10.18_{-0.82-0.03}^{+0.89+0.03}$
\\
$A_1^{B_c^*\to J/\psi}$       &$0.54_{-0.01-0.17}^{+0.01+0.16}$  &$2.38_{-0.05-0.14}^{+0.05+0.16}$ &$2.61_{-0.15-0.70}^{+0.15+0.61}$
&$A_1^{\Upsilon(1S)\to B^*}$ &$0.05_{-0.01-0.02}^{+0.01+0.03}$  &$3.21_{-0.08-0.08}^{+0.08+0.09}$ &$10.64_{-0.83-0.09}^{+0.91+0.06}$
\\
$A_2^{B_c^*\to J/\psi}$       &$0.54_{-0.01-0.17}^{+0.01+0.16}$  &$1.59_{-0.05-0.19}^{+0.05+0.18}$ &$1.09_{-0.07-0.48}^{+0.07+0.43}$
&$A_2^{\Upsilon(1S)\to B^*}$ &$0.05_{-0.01-0.02}^{+0.01+0.03}$  &$2.75_{-0.10-0.03}^{+0.09+0.03}$ &$7.91_{-0.73-0.16}^{+0.80+0.12}$
\\
$A_3^{B_c^*\to J/\psi}$       &$0.13_{-0.00-0.02}^{+0.00+0.03}$  &$2.72_{-0.04-0.64}^{+0.03+0.72}$ &$3.56_{-0.10-1.64}^{+0.08+1.72}$
&$A_3^{\Upsilon(1S)\to B^*}$ &$0.06_{-0.01-0.02}^{+0.01+0.03}$  &$3.33_{-0.08-0.11}^{+0.07+0.12}$ &$11.42_{-0.84-0.16}^{+0.92+0.14}$
\\
$A_4^{B_c^*\to J/\psi}$       &$0.14_{-0.00-0.02}^{+0.00+0.02}$  &$2.72_{-0.06-0.06}^{+0.05+0.09}$ &$3.57_{-0.19-0.09}^{+0.20+0.12}$
&$A_4^{\Upsilon(1S)\to B^*}$ &$0.07_{-0.01-0.03}^{+0.01+0.04}$  &$3.22_{-0.07-0.09}^{+0.07+0.09}$ &$10.45_{-0.77-0.03}^{+0.76+0.03}$
\\
\hline
%%%%%%%%%%%%%%%%%%%%%%%%%%%%%%%%%%%%%%%%%%%%%%%%%%%%%%%%%%%%%%%%%%%%%%%%%%%%%%%%%%%%%%%%%%%%
$V_1^{\Upsilon(1S)\to B_s^*}$
&$0.09_{-0.01-0.03}^{+0.01+0.04}$ &$3.26_{-0.06-0.08}^{+0.06+0.09}$ &$9.04_{-0.51-0.09}^{+0.54+0.07}$
&$V_1^{\Upsilon(1S)\to B_c^*}$
&$0.44_{-0.01-0.13}^{+0.01+0.15}$ &$3.16_{-0.09-0.04}^{+0.09+0.03}$  &$5.36_{-0.35-0.17}^{+0.37+0.14}$
\\
$V_2^{\Upsilon(1S)\to B_s^*}$
&$0.11_{-0.01-0.03}^{+0.01+0.04}$ &$3.46_{-0.02-0.11}^{+0.02+0.13}$ &$10.23_{-0.14-0.22}^{+0.15+0.24}$
&$V_2^{\Upsilon(1S)\to B_c^*}$
&$0.29_{-0.03-0.05}^{+0.03+0.08}$ &$3.38_{-0.04-0.46}^{+0.03+0.47}$  &$6.08_{-0.14-1.85}^{+0.11+1.46}$
\\
$V_3^{\Upsilon(1S)\to B_s^*}$
&$0.09_{-0.01-0.03}^{+0.01+0.04}$ &$3.19_{-0.06-0.07}^{+0.06+0.07}$ &$8.36_{-0.47-0.06}^{+0.50+0.05}$
&$V_3^{\Upsilon(1S)\to B_c^*}$
&$0.24_{-0.01-0.04}^{+0.01+0.03}$ &$3.18_{-0.07-0.14}^{+0.08+0.12}$  &$5.33_{-0.31-0.34}^{+0.33+0.24}$
\\
$V_4^{\Upsilon(1S)\to B_s^*}$
&$0.09_{-0.01-0.03}^{+0.01+0.04}$ &$3.43_{-0.21-0.22}^{+0.20+0.20}$ &$9.47_{-1.27-0.70}^{+1.10+0.71}$
&$V_4^{\Upsilon(1S)\to B_c^*}$
&$0.24_{-0.01-0.04}^{+0.01+0.03}$ &$3.29_{-0.06-0.12}^{+0.06+0.06}$  &$5.63_{-0.28-0.97}^{+0.33+0.49}$
\\
$V_5^{\Upsilon(1S)\to B_s^*}$
&$0.32_{-0.02-0.11}^{+0.02+0.14}$ &$3.18_{-0.06-0.07}^{+0.05+0.07}$ &$8.36_{-0.48-0.06}^{+0.50+0.06}$
&$V_5^{\Upsilon(1S)\to B_c^*}$
&$1.31_{-0.04-0.33}^{+0.04+0.33}$ &$3.08_{-0.07-0.05}^{+0.08+0.04}$  &$5.00_{-0.33-0.13}^{+0.35+0.11}$
\\
$V_6^{\Upsilon(1S)\to B_s^*}$
&$0.19_{-0.01-0.06}^{+0.01+0.09}$
&$3.19_{-0.06-0.07}^{+0.06+0.08}$ &$8.57_{-0.49-0.19}^{+0.52+0.14}$
&$V_6^{\Upsilon(1S)\to B_c^*}$
&$0.92_{-0.03-0.25}^{+0.03+0.35}$ &$3.06_{-0.09-0.34}^{+0.09+0.34}$  &$4.97_{-0.27-1.14}^{+0.36+1.27}$
\\
$A_1^{\Upsilon(1S)\to B_s^*}$
&$0.08_{-0.01-0.03}^{+0.01+0.04}$ &$3.25_{-0.05-0.08}^{+0.06+0.09}$ &$9.01_{-0.50-0.09}^{+0.54+0.06}$
&$A_1^{\Upsilon(1S)\to B_c^*}$
&$0.41_{-0.01-0.12}^{+0.01+0.14}$ &$3.16_{-0.09-0.04}^{+0.09+0.04}$  &$5.35_{-0.34-0.16}^{+0.37+0.14}$
\\
$A_2^{\Upsilon(1S)\to B_s^*}$
&$0.08_{-0.01-0.03}^{+0.01+0.04}$ &$2.77_{-0.07-0.03}^{+0.07+0.03}$ &$6.56_{-0.43-0.09}^{+0.46+0.08}$
&$A_2^{\Upsilon(1S)\to B_c^*}$
&$0.41_{-0.01-0.12}^{+0.01+0.14}$
&$3.46_{-0.11-0.05}^{+0.11+0.05}$  &$3.21_{-0.32-0.07}^{+0.34+0.07}$
\\
$A_3^{\Upsilon(1S)\to B_s^*}$
&$0.09_{-0.01-0.03}^{+0.01+0.03}$ 
&$3.39_{-0.05-0.09}^{+0.06+0.09}$ &$9.75_{-0.52-0.08}^{+0.54+0.08}$
&$A_3^{\Upsilon(1S)\to B_c^*}$
&$0.21_{-0.01-0.04}^{+0.01+0.05}$ &$3.35_{-0.02-0.49}^{+0.01+0.53}$  &$5.95_{-0.06-1.69}^{+0.11+1.32}$
\\
$A_4^{\Upsilon(1S)\to B_s^*}$
&$0.10_{-0.01-0.03}^{+0.01+0.04}$ &$3.28_{-0.05-0.08}^{+0.06+0.09}$ &$8.97_{-0.50-0.04}^{+0.51+0.03}$
&$A_4^{\Upsilon(1S)\to B_c^*}$
&$0.25_{-0.01-0.04}^{+0.01+0.03}$ &$3.30_{-0.08-0.29}^{+0.07+0.20}$  &$5.76_{-0.32-1.08}^{+0.34+0.99}$
\\
\hline\hline
\end{tabular}
\end{center}
\end{table}

\end{appendix}

\newpage

\section*{Acknowledgements}
We would like to thank Ho-Meoyng Choi at KNU, Jun-Feng Sun and Nan Li at HNNU, Xin-Qiang Li at CCNU and Yue-Long Shen at OU  for helpful discussions. This work is supported by the National Natural Science Foundation of China (Grant Nos. 11875122 and 11475055) and the Program for Innovative Research Team in University of Henan Province (Grant No.19IRTSTHN018).

\end{document}